\documentclass[ece,dissertation]{puthesis}
\usepackage{amsmath}
\usepackage{multicol}
\usepackage{subfigure}
\usepackage{float}
\usepackage{caption}
\usepackage{amsmath}
\usepackage{amssymb}
\usepackage{dsfont}
\usepackage{graphicx}
\graphicspath{{./figs/}}
\usepackage{epstopdf}
\DeclareGraphicsExtensions{.pdf,.jpeg,.png,.eps}
\usepackage[boxruled]{algorithm2e}
\usepackage{algorithm2e,setspace}
\title{Real-Time Stochastic Predictive Control for Hybrid Vehicle Energy Management}

\author{Kyle R. Williams}{Williams, Kyle R.}

\pudegree{Doctor of Philosophy}{Ph.D.}{May}{2018}

\majorprof{Monika Ivantysynova, School of Mechanical Engineering}

\campus{West Lafayette}

\begin{document}
	
\newcommand{\btau}{\boldsymbol{\tau}}
\newcommand{\Tr}{^\mathsf{T}}
\newcommand{\x}{\mathbf{x}}
\newcommand{\bw}{\mathbf{w}}
\newcommand{\bu}{\mathbf{u}}
\renewcommand{\u}{\mathbf{u}}
\newcommand{\blambda}{\boldsymbol{\lambda}}
\newcommand{\rfrac}[2]{{}^{#1}\!/_{#2}}	
\renewcommand{\(}{\left(}
\renewcommand{\)}{\right)}

\volume

%
%
%
%
%

\begin{statement}
	\entry{Dr.~Monika M. Ivantysynova, Chair}{School of Mechanical Engineering}
	\entry{Dr.~Gregory M. Shaver}{School of Mechanical Engineering}
	\entry{Dr.~Kartik B. Ariyur}{School of Mechanical Engineering}
	\entry{Dr.~Andrea Vacca}{School of Mechanical Engineering}
	\approvedby{Dr.~Jay P. Gore}{Head of the School Graduate Program}
\end{statement}

\begin{dedication}
  For Sara and India.
\end{dedication}

\begin{acknowledgments}
  First and foremost I would like to thank my advisor, Prof. Monika Ivantysynova.  The concepts developed in this work are the result of her encouragement and direction provided during my time at the Maha Fluid Power Research Center.  Monika's outstanding guidance has made a profound impact on my career, I feel very fortunate to have been part of her research group.  I would also like to thank my advisory committee members Profs. Shaver, Vacca and Ariyur for their invaluable input, particularly during my preliminary defense.   
  
  I am truly indebted to Ryan, Anthony, Leo and Mateus for their tremendous support during the experimental phase of this work.  Ryan setup the entire test rig, and together with Leo and Mateus, worked through every issue the rig encountered.  I cannot emphasize enough how much I appreciated Anthony's unparalleled commitment to the lab.  From start to finish Anthony was always there moving this test rig forward.  Without the help of these individuals the experiments would not have happened.  I also wanted to thank my fellow Maha researchers, past and present, for providing many insightful discussions and questions along the way.  I am grateful for the help I have received from the staff within the Maha Lab and the ME grad office.  Specifically, Susan Gauger, Julayne Moser, Connie McMindes, and Cathy Elwell have helped me so very much over the past five years.   
  
  Finally, I need to thank my wife, Sara, and my daughter, India.  Working full time for a large corporation while independently pursuing a PhD over the past five years was difficult.  I never would have finished without Sara's strength and constant support.  Thank you, India, for your patience.  I started this four months before you were born and I am so happy to be done before you get a moment older.  Moving forward, I will never again have to miss a zoo trip with you to go work on my PhD.  
\end{acknowledgments}


\tableofcontents

\listoftables

\listoffigures





\begin{abstract}
  This work presents three computational methods for real time energy management in a hybrid hydraulic vehicle (HHV) when driver behavior and vehicle route are not known in advance.  These methods, implemented in a receding horizon control (aka model predictive control) framework, are rather general and can be applied to systems with nonlinear dynamics subject to a Markov disturbance.  State and input constraints are considered in each method.  A mechanism based on the steady state distribution of the underlying Markov chain is developed for planning beyond a finite horizon in the HHV energy management problem.  Road elevation information is forecasted along the horizon and then merged with the statistical model of driver behavior to increase accuracy of the horizon optimization.  The characteristics of each strategy are compared and the benefit of learning driver behavior is analyzed through simulation on three drive cycles, including one real world drive cycle.  A simulation is designed to explicitly demonstrate the benefit of adapting the Markov chain to real time driver behavior.  Experimental results demonstrate the real time potential of the primary algorithm when implemented on a processor with limited computational resources.
\end{abstract}
%
%
%

\chapter{INTRODUCTION AND STATE OF THE ART}

\section{Introduction}
\begin{figure}[h!]
	\centering	
	\begin{minipage}{0.5\textwidth}
		\includegraphics [trim = 5mm 0 0 0, clip,width=1\textwidth]{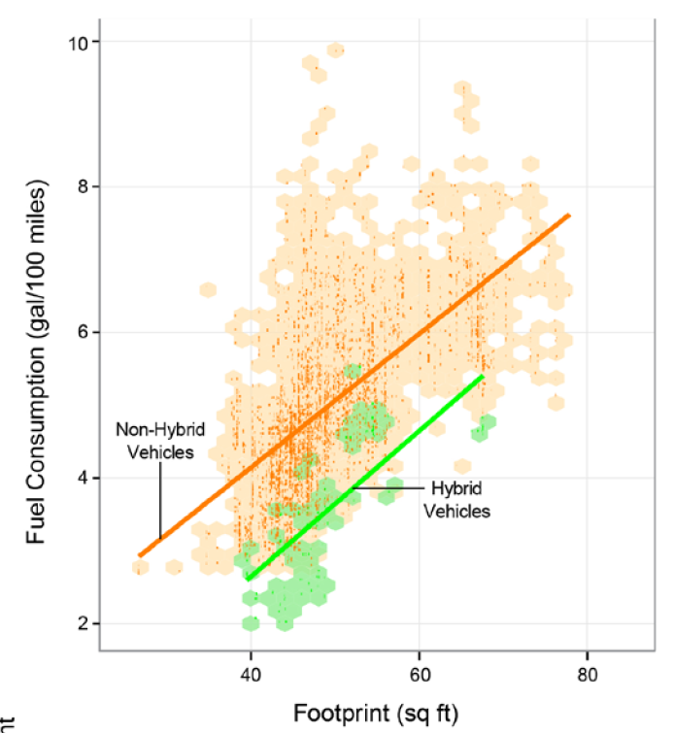}
	\end{minipage}
	\caption{Source: U.S. Department Of Energy, 2014.} 
	\label{fig:EPA}
\end{figure}

The hybrid vehicle offers a solution for personal, public and commercial transportation vehicles which can significantly reduce fuel consumption and engine emissions output in comparison to conventional vehicle solutions.  Figure \ref{fig:EPA} shows fuel consumption vs. vehicle size in square feet for conventional and hybrid vehicles.  
Typically, for the same size vehicle the hybrid solution offers significantly reduced fuel consumption.  By incorporating a reversible energy storage device on-board the hybrid vehicle, kinetic energy conventionally dissipated as heat during braking can be recovered during a process known as regenerative braking.  As a secondary benefit, the hybrid vehicle offers greater flexibility in engine management than a conventional vehicle.  The uncertain nature of driver behavior and driving environment presents one of the biggest challenges in hybrid vehicle control.  In both hybrid electric vehicles (HEVs) and hydraulic hybrid vehicles (HHVs), the control system must ensure proper charge of the reversible energy storage to ensure future driver demands can be satisfied while also observing system constraints and maximizing overall system efficiency.  As such, the challenge of optimally managing the engine and reversible energy sources has been an area of active research over the past two decades.  This challenge has focused on the development of control strategies which minimize an objective function based on fuel consumption and/or engine emissions while maintaining vehicle drivability and satisfying system constraints.  The development of these strategies has included, but is not limited to, modeling driver behavior, modeling changes in the driving environment, creating an objective function to reflect the optimization goal, incorporating real time telematics information, and developing control methods which incorporate all mentioned models and information to optimize the given objective.

\begin{figure}[h!]
	\centering
	\captionsetup{justification=centering}	
	\begin{minipage}{0.7\textwidth}
		\includegraphics [width=1\textwidth]{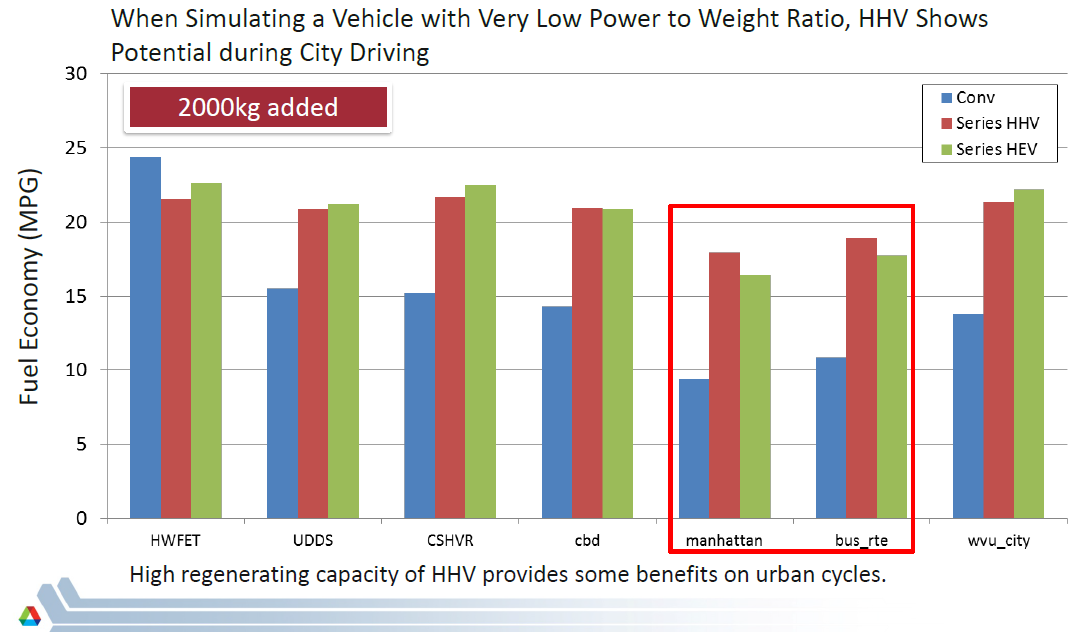}
	\end{minipage}
	\caption{Series HHV vs HEV.  Source: U.S. Department Of Energy, 2012.}
	\label{fig:HHVvsHEV}
\end{figure}

\begin{figure}[h!]
	\centering	
	\begin{minipage}{0.7\textwidth}
		\includegraphics [width=1\textwidth]{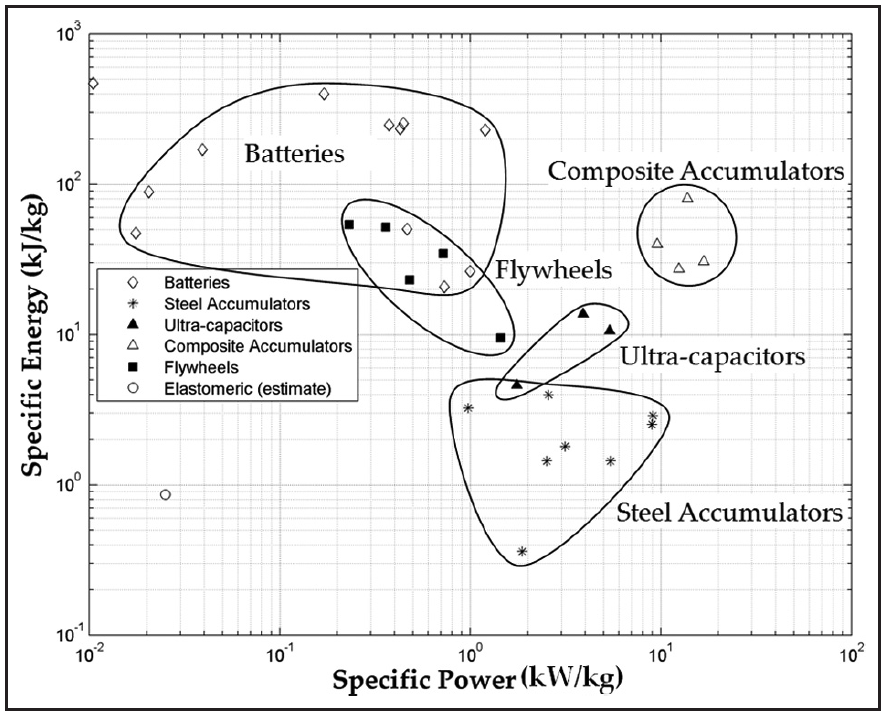}
	\end{minipage}
	\caption{Specific energy versus specific power of various energy storage devices \cite{midgley2012comparison}.} 
	\label{fig:Ragone}
\end{figure}

Hydraulic hybrid vehicles can be competitive with and even outperform HEVs in terms of fuel savings at a reduced cost \cite{SDP4}.  Figure \ref{fig:HHVvsHEV} compares fuel economy of a series HHV compared to a series HEV in city driving when the power to weight ratio of the vehicle is low.  
The series HHV has an advantage of the series HEV in urban routes when rapid energy transfer to and from the energy storage device is required.  The benefit of the HHV can be explained with a plot of energy density vs. power density as shown in Fig. \ref{fig:Ragone}.  Although batteries typically have greater energy density than hydraulic accumulators, the greater power density of a hydraulic accumulator means the HHV can potentially store and reuse energy much more quickly.

\section{State of the Art}
The state of the art in hybrid vehicle energy management is reviewed.  No significant differentiation between control strategies for HEV vs. HHV is made, since any given strategy can typically be applied to either HEV or HHV with straight-forward adjustment.

\subsection{Heuristic Policies and Instantaneous Optimization}
Energy management for hybrid vehicles (both HEVs and HHVs)  is an old problem.  Early solutions involved finite horizon dynamic programming (DP) simulations for predefined drive cycles.  By creating a time-varying value function $V(\x,t)$, dynamic programming can determine a globally optimal open loop control trajectory for a given drive cycle.  A major drawback is the resulting open loop control trajectories are only valid for the specific drive cycle under investigation.  To generate an implementable controller, heuristic feedback policies were extracted from the DP results in an attempt to replicate the properties of optimal open loop control trajectories \cite{jalil1997rule,RizzoniReview}.  A downside of heuristic strategies is they must optimistically hope the cycle being driven resembles the training cycle (that is, the cycle(s) on which the heuristic rules were formed).  Instantaneous optimization strategies were developed to alleviate the need for human-formed rules.  These methods perform real time optimization, producing control inputs which instantaneously minimize fuel consumption or emissions in response to the present operating condition of the vehicle \cite{paganelli2001control,KumarInstOpt2008,WilliamsInstOpt2008}.  An interesting connection between a type of instantaneous optimization called equivalent consumption minimization strategy (ECMS) \cite{paganelli2002equivalent,musardo2005ecms} and Pontraygin's Minimum Principle (PMP) is presented in \cite{serrao2009ecms}.  A method for real time energy management based explicitly on PMP is developed in \cite{kim2011optimal}.  Here, the authors fix a co-state value associated with the real time solution of PMP which influences fuel consumption results.  The challenge with this approach is pairing the best co-state value for the cycle being driven in order to minimize fuel.  

\subsection{Stochastic Methods}
A completely different solution category for energy management is developed when a statistical model of driver behavior is incorporated into the solution strategy.  A statistical model known as a Markov chain has proven an effective approach for capturing driver behavior \cite{PentlandLiu,FilevKolmanovskyMCModels}.  Stochastic dynamic programming (SDP) methods \cite{PutermanMDP} work directly with the Markov chain to formulate globally optimal time-varying control policies $\u=\boldsymbol{\mu}(\x,t)$ which consider driver statistics, minimizing the expected or average running cost of the objective function over a time horizon.  In \cite{SDP1,SDP2,SDP4}, energy management strategies based on SDP in an infinite horizon setting are developed.  Infinite horizon SDP mathematically formulates a time-invariant value function $V(\x)$ based on statistics of several drive cycles, from which globally optimal state-feedback control policy $\u=\boldsymbol{\mu}(\x)$ can be constructed.  A major advantage of the infinite horizon SDP approach is the state-feedback control policy can be implemented in a lookup table manner for real time vehicle control.  In the relatively recent work of \cite{opila2013real}, experiments are carried out on a modified Volvo S-80 HEV using a state-feedback control policy based on SDP.  An interesting comparison between finite horizon DP and infinite horizon SDP as applied to a hydraulic hybrid vehicle is discussed in \cite{MahaKumar2010investigation}.

Like its deterministic counterpart, SDP scales poorly to problems involving large state spaces and becomes computationally intractable for very large problems.  Neuro-Dynamic Programming (NDP) \cite{Ber96,SuttonBarto,PowellADP} alleviates the scaling issue through the use of neural networks.  In NDP, the value function is represented as a parameterized neural network, $\hat V(\x,\boldsymbol{\theta})$, and then tuned by adjusting parameters $\boldsymbol{\theta}$ in order to satisfy the associated Bellman equations.  As a result, the value of many states can be adjusted at once by adjusting a single parameter.  Using neural networks in this way allows NDP to efficiently handle significantly larger state spaces than SDP since not every state must be visited during construction of the value function.  Neuro-Dynamic Programming is employed in \cite{Raji} to minimize an impressively complex objective function comprising fuel consumption and engine emissions in a HHV.  

A shortcoming of computationally intensive stochastic methods such as SDP and NDP is that the resulting control policies are based on models of driver behavior which are typically not adapted in real time.  The findings in \cite{asher2017prediction} suggest that stochastically robust methods such as SDP may not provide optimal fuel economy in hybrid vehicles when cycle mispredictions exist.  Such mispredictions can be caused, for example, when the Markov chain model used in the SDP formulation is not representative of the actual drive cycle, emphasizing the need for adaptation of the statistical model if stochastic methods are to be employed. 

\subsection{Model Predictive Control Methods}
Model predictive control (MPC) \cite{MPCbook} is fundamentally characterized by the \textit{fast} computation of a finite horizon optimization at every time step.  The underlying solver can be based on DP, PMP, SDP, quadratic programming (QP), or other general nonlinear programming type methods \cite{Betts}.  At each timestep, MPC generates an open loop control trajectory $\{\u_0,\u_1,\dots, \u_{N-1}\}$.  The first control input $\u_0$ is applied to the system and then the finite horizon optimization re-starts with up-to-date system information.  One of the biggest advantages to the MPC method is that real time information can be incorporated to make immediate changes to the problem formulation, resulting in an control trajectory that is more closely tuned to present driving conditions.  In \cite{ExpDecDriverDemand}, model predictive control is used for energy management of an HEV with driver torque demand modeled as an exponentially decreasing process along the horizon according to $\tau_{n+1} = \alpha\tau_n$ with $0<\alpha<1$.  In \cite{CCEFPHybridMPC}, MPC is used for energy management of a HHV with driver demand assumed constant along the horizon.  The finite horizon optimization is solved using Newton's method with logarithmic barrier functions \cite{BoydCVX}.

Model predictive control can incorporate forecasted information provided by on-board telematics such as a global positioning system.  The authors of \cite{katsargyri2009optimally} use path forecasting in the form of previewed vehicle speed and road grade in a hybrid electric vehicle.  In a similar approach, road grade is previewed along a horizon assuming constant vehicle speed in a conventional vehicle in \cite{Hellstrom}.  Since the state and action spaces are low in \cite{katsargyri2009optimally} and \cite{Hellstrom}, dynamic programming is used to perform the finite horizon optimization.

\subsection{Predictive Methods Under Uncertainty}
\subsubsection{Stochastic Model Predictive Control}
Stochastic model predictive control (SMPC) methods \cite{Cannon2007,Mesbah2016} combine the statistical decision making associated with SDP and NDP with the real time computation of MPC.  A unique challenge to SMPC is the development of computationally efficient solvers which can handle the computational burden associated with stochastic optimization.  A stochastic QP solver for Markov Jump Linear Systems with transition probability estimation is presented in \cite{StochDriverLearning}.  Here, driver behavior is represented as a Markov chain and Monte Carlo sampling is used to generate several driver demand paths with relatively high likelihood.  To reduce computational burden sample paths with low likelihood are not considered in the problem formulation.  A key feature of the method is that the Markov transition probabilities are adapted in real time to the actual drive cycle.  The developed strategy performs nearly as well as a benchmark strategy which has full access to the drive cycle and significantly outperforms a strategy incorporating no learning mechanism, indicating that significant benefit can be achieved when the Markov chain is adapted in real time.  A method for predicting road grade is incorporated in the framework of SMPC in \cite{SDPRoadGrade}.  In addition to driver behavior, road grade is modeled as a Markov chain and the subsequent stochastic optimization is performed with finite horizon SDP with reported execution times between 10 and 100 seconds.

\subsubsection{Neural Network Predictors}
Neural networks (NN) are used to predict driver acceleration demand and vehicle velocity along a finite horizon in \cite{sun2015velocity}.  An MPC formulation based on \cite{falcone2008hierarchical} is used to carry out the finite horizon optimization.  A major finding in \cite{sun2015velocity} is that an MPC strategy based on NN-based velocity predictions outperforms the same strategy incorporating Markov chain-based velocity predictors.  The NN and Markov chain were both trained on a large data set and evaluated on a separate data set.  It is worth noting the authors of \cite{sun2015velocity} explicitly state no learning mechanisms were used to estimate the parameters of the Markov chain in real time, possibly eliminating one of the most flexible and useful attributes of the Markov chain driver modeling approach for hybrid vehicle energy management.

\section{Research Goals and Contributions}
The primary goal of this research is to develop a control algorithm for hybrid vehicle energy management, with the ultimate goal of maximizing fuel economy.  Since driver actions are largely uncertain, the algorithm should be able to consider consequences of possible future driver actions during planning of the state and control trajectories.  The algorithm needs to be flexible enough to adapt in real time to driver behavior, and additionally, incorporate real time telematics information in order to reduce uncertainty during planning.  A secondary goal of this research is to determine the degree to which learning driver behavior and incorporating real time telematics information can improve fuel economy.  A third and final goal of this research is to experimentally demonstrate the algorithm is capable of controlling a hybrid powertrain using a resource limited processor.

\subsection{Contributions}
The primary contributions of this work are:
\begin{itemize}
	\item Three novel computational methods for real time energy management in a HHV when driver behavior and vehicle route are not known in advance are developed in Chapter \ref{section:PredEnergyManage}.  These methods, implemented in a receding horizon control (aka model predictive control) framework, are rather general and can be applied to systems with nonlinear dynamics subject to a Markov disturbance.  State and control constraints are considered in each method.
	\item A novel mechanism for planning beyond a finite horizon in the HHV energy management problem is investigated.  This mechanism is based on the steady state distribution of the underlying Markov chain model describing driver behavior.  The method is initially discussed in Section \ref{section:LongTermBehavior} and incorporated into HHV energy management in Section \ref{section:StochControlFormulations}.
	\item Road elevation information is forecasted along the horizon and for the first time is merged with the statistical model of driver behavior to increase accuracy of the horizon optimization.  The method of incorporating road grade information is developed in Section \ref{section:RoadGradeForecasting}.  
	\item The impact of incorrect statistical information, and the required time to adapt to correct statistical information, is for the first time investigated in Section \ref{section:CrossTraining}.
	\item Real time potential of the novel computational methods is assessed for the first time through an experimental setup discussed in Chapter \ref{section:Experiment}.
\end{itemize} 

\section{Organization of Chapters}
The next chapter summarizes several of the underlying concepts and methods of optimal control and reinforcement learning which have been widely used in vehicle control applications.  Several of these concepts lead to the development of the algorithms in Chapter 5.  Chapter 3 presents an overview of hybrid vehicles and hybrid vehicle dynamics.  

In chapter 4, a statistical model of driver behavior based on a Markov chain is presented.  The Markov chain is adapted in real time to the drive cycle according to a simple filtering process described in \cite{StochDriverLearning}.  The Markov multi-step transition probability matrix is analyzed as a mechanism to model driver actions along a horizon.  Driver behavior from three drive cycles, including one cycle obtained from real-world driving measurements, is analyzed.  The steady state distribution of the Markov chain model is presented as a way to plan beyond a finite horizon.  

Chapter 5 presents three novel methods for real time energy management of an HHV when driver behavior and vehicle route are not known in advance.  A simplified discrete-time model of the system dynamics is explained.  Two benchmark methods are also created, one is a theoretically best-achievable controller and the second is a simplified strategy based on instantaneous optimization.  

In Chapter 6, simulations are carried out.  The characteristics of each strategy are compared and the benefit of learning driver behavior is analyzed.  A simulation is designed to explicitly demonstrate the benefit of adapting the Markov chain to real time driver behavior.  The statistical driver model is initialized on incorrect cycle statistics, then allowed to adapt to the driven cycle.  Learning typically converges in 2-3 runs of the given cycle, corresponding to 20 to 60 minutes.  

An experiment is performed on a series HHV test rig setup in Chapter 7.  The purpose of the experiment is to (1) demonstrate that the computationally intensive algorithms developed in Chapter 5 can run in real time on a processor with limited computational resources and (2) demonstrate the algorithm can successfully control a series hybrid using a simplified control-oriented model of the real physics.  


\section{Notation}
\begin{tabular}{ll}
	\textbf{symbol} & \textbf{meaning}\\
	\hline 
		$\x$ & vector  \\ 
		$\x(t)$ & vector at time $t$ \\ 
		$\x_n$ & vector at timestep $n$ \\
		$\x_n\Tr$ & vector at timestep $n$ transposed \\
		$x_{i,n}$ & the $i^{th}$ element of a vector $\x$ at timestep n\\
		$\vec{\x}$ & $=\{\x_0,\x_1,\dots,\x_{N-1}\} = \(\x_n\)_{n=0}^{N-1}$ a sequence of vectors \\
		$\vec{\x}^{[k]}$ & the $k^{th}$ iteration of vector sequence $\vec{\x}$\\
		$J(\x,\u)$ & a function evaluated at $\x,\u$\\
		$J^{(x)}(\x_n,\u_n)$ & partial of $J$ wrt argument $\x$, evaluated at $\x_n,\u_n$\\
	\hline 
\end{tabular} 
\chapter{BACKGROUND}
This chapter summarizes several of the underlying concepts and methods of optimal control and reinforcement learning which have been widely used in vehicle control applications.  


\section{Deterministic Optimal Control}
Consider the discrete time dynamic system described by
\begin{align}\label{eq:2SysDT}
\x_{n+1} = F_n(\x_n,\bu_n),~~n=0,1,\dots
\end{align}
where $\x_n\in {\mathbb{R}^{dimX}}$ and $\bu_n \in {\mathbb{R}^{dimU}}$ are the system state and control input vectors, respectively, and $\x_0$ is given.  The dynamics described by Equation (\ref{eq:2SysDT}) can represent a large class of systems, including the discrete time evolution of an inherently continuous time process\footnote{The notation $\dot \x$ represents $\frac{d\x(t)}{dt}$} $\dot \x(t) = f(\x(t),\bu(t),t)$ according to
\begin{align*}
F_n(\x_n,\bu_n) = \x_n + \int\limits_{n\Delta t}^{(n+1)\Delta t}f(\x(\tau),\bu(\tau)) d\tau
\end{align*}
where $t \geq 0$ and $\x(0)$ is given.  The horizon cost 
\begin{align}\label{eq:J0}
J_0 &= h(\mathbf x_N)+\sum_{n=0}^{N-1} g_n(\x_n,\bu_n)
\end{align}
is the sum of a terminal cost $h(\x_N)$ and a time-varying running cost $g_n(\x_n,\bu_n)$ which is affected by the state and control input at each stage in the horizon.  The goal of optimal control is to design an appropriate control sequence $\vec{\bu} = \(\bu_n\)_{n=0}^{N-1}$ which minimizes the receding horizon cost $J_0$ when the system starts from initial state $\x_0$ at time $n=0$ and is subjected to the control sequence $\vec\bu$ along the horizon.  Additionally, control and state constraints must be satisfied at all points along the horizon.  The minimization problem is formally stated as 
\begin{subequations}\label{eq:FiniteHorizonProblemDet}
	\begin{align}
	\min_{\bu_0,\bu_1,\dots,\bu_{N-1}} &\left\{h(\mathbf x_N)+\sum_{n=0}^{N-1} g_n(\x_n,\bu_n)\right\}\\
	\text{subject to}~~& \x_{n+1} = F_n(\x_n,\bu_n) \\
	&\x_n \in \mathbf X \\
	&	\bu_n \in \mathbf U		\\
	&n=0,1,\dots,N-1
	\end{align}
\end{subequations}
where $\mathbf X$ and $\mathbf U$ are the constrained state and control sets, respectively.

\subsection{Nonlinear Programming}
Perhaps the most straightforward and popular approach for solving Equation (\ref{eq:FiniteHorizonProblemDet}) is by transforming the problem into a nonlinear program \cite{Betts}.  Nonlinear programming refers to the general process of solving an optimization problem subject to equality and inequality constraints in the decision variables.  The most common nonlinear programming method used by far in optimal control is \textit{quadratic programming} (QP).  A typical QP problem is formulated as
\begin{align*}
\min_{\mathbf z} ~~&\frac{1}{2} \mathbf z\Tr Q \mathbf  z + q\Tr \mathbf z\\
\text{subject to}~~& A \mathbf z \leq \mathbf b \\
&  D\mathbf z = \mathbf c
\end{align*}
The finite horizon optimal control problem Equation (\ref{eq:FiniteHorizonProblemDet}) can be transformed into a QP problem by approximating the horizon cost with a quadratic function and linearizing the system dynamics about some nominal trajectory $(\hat \x_n,\hat \bu_n)_{n=0}^{N-1}$.  Neglecting for simplicity the terminal cost $h(\x_N)$, the horizon cost Equation (\ref{eq:J0}) can be approximated with the quadratic function
\begin{subequations}
\begin{align}
	J_0 &\approx \sum_{n=0}^{N-1} \frac{1}{2}\mathbf z_n \Tr Q_n \mathbf{z_n} + q_n\Tr \mathbf{z_n}\\
	\text{where} \nonumber\\
	Q_n &=
	\begin{bmatrix}
	g^{(xx)}_n & g^{(xu)}_n \\
	g^{(ux)}_n & g^{(uu)}_n
	\end{bmatrix}_{(\hat \x_n, \hat \bu_n)} ~~~
	q_n = 
	\begin{bmatrix}
		g^{(x)}_n \\
	    g^{(u)}_n
	\end{bmatrix}_{(\hat \x_n, \hat \bu_n)} \\
	\mathbf{z_n} &=
	\begin{bmatrix}
		\delta \x_n \\
		\delta \bu_n
	\end{bmatrix}
\end{align}
\end{subequations}
The system dynamics can be linearized according to 
\begin{subequations}
\begin{align}
	\delta \x_{n+1} &= A_n \delta \x_n + B_n \delta \bu_n\\
	A_n &= F_n^{(x)}(\hat \x_n, \hat \bu_n) \\
	B_n &= F_n^{(u)}(\hat \x_n, \hat \bu_n)
\end{align}
\end{subequations}
where $\(\delta \x_n,\delta \bu_n\)_{n=0}^{N-1}$ is a small perturbation from the nominal trajectory.  The equivalent QP problem can then be described by
\begin{subequations}\label{eq:2QPproblem}
	\begin{align}
		\min_\mathbf{z}~ &\frac{1}{2}\mathbf z \Tr Q \mathbf{z} + q\Tr \mathbf{z} \\		
		&Q =  
		\begin{bmatrix}
		Q_0 & & & \\
		 & Q_1 & &  \\
		& & \ddots & \\
		& & & Q_{N-1}
		\end{bmatrix} \label{eq:2QPSparseMatrix}\\
		&q \Tr= 
		\begin{bmatrix}
		q_0 & q_1 & \dots &	q_{N-1}
		\end{bmatrix}\\
		&\mathbf z\Tr = 
		\begin{bmatrix}
		\delta \x_0 &\delta \bu_0 &	\delta \x_1 &	\delta \bu_1 \dots \delta \x_{N-1} &	\delta \bu_{N-1}
		\end{bmatrix}
		\end{align}
		\begin{align}	
		\text{subject to} \nonumber&\\
		&\delta \x_{n+1} - A_n \delta \x_{n} - B_n \delta \bu_{n} = \mathbf 0\\
		&\mathbf a - \hat \x_n \leq \delta \x_{n} \leq \mathbf{b}- \hat \x_n\\
		&\mathbf c - \hat \bu_n\leq \delta \bu_{n} \leq \mathbf{d}- \hat \bu_n
	\end{align}	
\end{subequations} 
Once the QP problem Equation (\ref{eq:2QPproblem}) has been solved, the nominal trajectory is updated according to $\(\hat \x_n,\hat \bu_n\)_{n=0}^{N-1} \leftarrow \(\hat \x_n+\delta \x_n,\hat \bu_n+\delta \bu_n\)_{n=0}^{N-1}$ and the process is restarted.  The broad use of the quadratic programming approach for solving the finite horizon optimal control problem can perhaps be attributed to the availability of powerful tools which can efficiently solve Equation (\ref{eq:2QPproblem}) by exploiting the underlying sparsity of the equivalent problem due to matrix Equation (\ref{eq:2QPSparseMatrix}) \cite{Betts}.

\subsection{The Minimum Principle}
Unlike the nonlinear programming approach, the minimum principle solves Equation (\ref{eq:FiniteHorizonProblemDet}) using a variational approach.  For fixed $\x_0$, let the finite horizon cost of control sequence $\vec \bu$ be given by 
\begin{align}
J_0(\vec\bu,\x_0) &= h(\x_N)+\sum_{n=0}^{N-1} g_n(\x_n,\bu_n) 
\end{align}
Define the \textit{Hamiltonian} 
\begin{align} \label{eq:2Hamiltonian}
	H(\x,\bu,\lambda) =  g(\x,\bu) + \lambda\Tr F(\x,\bu)
\end{align}
where $\lambda$ serves as a dynamic Lagrange multiplier ensuring the system dynamics constraint $\x_{n+1} = F_n(\x_n,\bu_n)$ is satisfied.  The horizon cost becomes
\vspace{-1mm}
\begin{align*}
J_0(\vec\bu,\x_0) &=h(\x_N)+\sum_{n=0}^{N-1} \Big[H(\x_n,\bu_n,\lambda_{n+1}) - \lambda_{n+1}\Tr \x_{n+1}\Big] \\
&=h(\x_N)-\lambda_N\Tr \x_N+\lambda_0\Tr \x_0 +\sum_{n=0}^{N-1}\Big[ H(\x_n,\bu_n,\lambda_{n+1}) - \lambda_{n}\Tr \x_{n}\Big]
\end{align*}
The variation in $J_0(\vec\bu,\x_0)$ due to small variations $\delta \bu_n$ about the nominal control sequence $\vec\bu$ is
\begin{align}\label{eq:2CovdeltaU}
\delta J_0(\vec\bu,\x_0) &=	\left[\frac{\partial h(\x_N)}{\partial \x_N} 
- \lambda_N\right]\Tr\delta \x_N
+ \lambda_0\Tr \delta \x_0 \nonumber \\
&~~~~~+\sum_{n=0}^{N-1}\Bigg\{\Bigg[\frac{\partial H(\x_n,\bu_n,\lambda_{n+1})}{\partial \x_n} - \lambda_n\Tr\Bigg]\delta \x_n 
+ \frac{\partial H(\x_n,\bu_n,\lambda_{n+1})}{\partial \bu_n} \delta \bu_n \Bigg\}
\end{align}
To enforce the condition $\frac{\partial J_n}{\partial \x_n} = 0$ along an optimal trajectory, the Lagrange multipliers are chosen to satisfy
\begin{align}\label{eq:2CovLambda}
\lambda_n = \dfrac{\partial H(\x_n,\bu_n,\lambda_{n+1})}{\partial \x_n} ~~~~
\lambda_N = \dfrac{\partial h(\x_N)}{\partial \x_N}
\end{align} 
Noticing that $\delta \x_0 = 0 $ (since the initial state is fixed) and substituting Equation (\ref{eq:2CovLambda}) into Equation (\ref{eq:2CovdeltaU})
\begin{align*}
\delta J_0(\vec\bu,\x_0) &=	\sum_{n=0}^{N-1}\frac{\partial H(\x_n,\bu_n,\lambda_{n+1})}{\partial \bu_n} \delta \bu_n
\end{align*}
Assuming any control constraints $\mathbf U_n$ are convex, the following necessary condition for local optimality of $\vec\bu^*$ is established \cite{BerVolI}
\begin{align}\label{eq:1PMPLocalOptCond}
\frac{\partial H(\x_n,\bu_n^*,\lambda_{n+1})}{\partial \bu_n} \left(\bu_n-\bu_n^*\right) \geq 0~,~~n=0,...,N-1,\forall \bu_n \in \mathbf U_n
\end{align}

\subsubsection{Global Optimality}
If the system dynamics $F(\x,\bu)$ are linear in $\bu$ and the running cost $g_n$ is convex in $\bu$, the Hamiltonian $H(\x,\bu,\lambda) =  g(\x,\bu) + \lambda\Tr F(\x,\bu)$ is convex in $\bu_n$.   In this case, local necessary condition Equation (\ref{eq:1PMPLocalOptCond}) is equivalent to the stronger necessary and sufficient condition
\begin{align}\label{eq:1PMPGlobalOptCond}
\bu_n^* = \arg\min_{\bu_n} \left\{  H(\x_n,\bu_n,\lambda_{n+1}) \right\}
\end{align}
Convexity of $H$ ensures any local optimum is a global optimum, but can only be established with the restrictions on $F(\x,\bu)$ and $g(\x,\bu)$ mentioned above.  

\subsubsection{Constraints}
Incorporating state and control constraints in the framework of the minimum principle is more challenging than in the nonlinear programming approach.  See \cite{stengel2012optimal} for a treatment of applying inequality constraints within the framework of the minimum principle.

\subsection{Dynamic Programming}
The minimum principle finds a locally optimal control sequence which minimizes the finite horizon cost Equation (\ref{eq:J0}), and under certain restrictions, this control sequence is globally optimal. Dynamic programming (DP), alternatively, always finds a globally optimal state-feedback control \textit{policy}, a mapping from the states and time to control inputs $\pi:{\mathbf X}\times{T}\rightarrow {\mathbf U}$, regardless of restrictions on $F$ and $g$.  Dynamic programming exploits Bellman's \textit{principle of optimality}, which states that if a given state-action sequence is optimal, and we remove the first state and action, the remaining sequence is also optimal (with the second state of the original sequence now acting as the initial state).  Under this principle, the problem of minimizing Equation (\ref{eq:J0}) is broken down into many smaller problems in a stage-wise manner.  Dynamic programming constructs a \textit{state value function} $V_n(\x)$, a record of the optimal cost-to-go from any state $\x$ at any time $n$ to the end of the horizon
\begin{align}
V_n(\x) &= \min_{\bu_n,...,\bu_{N-1}} J_n \big| \x_n=\x \nonumber \\
&=  \min_{\bu_n,...,\bu_{N-1}} \Bigg[ h(\x_N)+\sum_{k=n}^{N-1} g_k(\x_k,\bu_k) \Big| \x_n=\x \Bigg]
\end{align}
The state value function can be described recursively as
\begin{align}
V_n(\x) &= \min_{\bu_n \in \mathbf U}\Bigg\{g_n(\x,\bu_n) + \underbrace{\min_{\bu_{n+1},...,\bu_{N-1}} \left[ h(\x_N)+\sum_{k=n+1}^{N-1} g_k \Big| \x_{n+1}=F_n(\x,\bu_n) \right]}_{V_{n+1}(F_n(\x,\bu_n))} \Bigg\} \nonumber\\
 &= \min_{\bu_n \in \mathbf U} \Big[ g_n(\x,\bu_n)+V_{n+1}(F_n(\x,\bu_n)) \Big] \label{eq:2DeterministicBellmanEq}
\end{align}
with boundary condition 
\begin{align}
V_N(\x) &= h(\x) \label{eq:2DeterministicBellmanEqTerminalCond}
\end{align}
The optimal state-feedback control policy can be inferred directly from the state value function through
\begin{align}
\pi_n^*(\x) & = \arg\min_{\bu_n\in {U}}\Big[g_n(\x,\bu_n)+V_{n+1}(F_n(\x,\bu_n))\Big]  \label{eq:2DeterministicBellmanOptPolicy}    
\end{align}
Equations (\ref{eq:2DeterministicBellmanEq}) and (\ref{eq:2DeterministicBellmanOptPolicy}) are referred to as the Bellman equations.  These equations can be solved recursively by working backwards along the horizon from boundary condition Equation (\ref{eq:2DeterministicBellmanEqTerminalCond}).  At each horizon stage $V_n(\x)$ is computed for every state $\x$ using Equation (\ref{eq:2DeterministicBellmanEq}) starting from boundary condition $V_N(\x) = h(\x)$.  For computational feasibility, the state space is usually discretized  and the state dynamics are projected onto the discretization according to $\x_{n+1} = \mathrm{proj} \left[F_n(\x_n,\bu_n)\right]$.  

\subsubsection{Constraints}
State and input constraints are easily handled with dynamic programming.  Any $\x \notin \mathbf X$ is assigned an arbitrarily large value $V_n(\x)$, preventing the state feedback control policy $\pi$ from ever designing a control input that will lead to any $\x \notin \mathbf X$.  Control constraints are handled by restricting the optimization Equation (\ref{eq:2DeterministicBellmanEq}) to only search through feasible controls $\bu_n \in \mathbf U$, and restricting the  control policy Equation (\ref{eq:2DeterministicBellmanOptPolicy}) to choose from feasible $\bu_n$.

\subsubsection{Computational complexity}
It is worth mentioning that the computational effort associated with dynamic programming grows considerably with the size of the state space.  Quantizing each dimension of the state space $\mathbf X \subset \mathbb R^{dimX}$ into $\mathrm{quant_x}$ levels produces a state space of size $|\mathbf X| = \mathrm{quant_x}^{dimX}$.  For an $N$ length horizon problem, this amounts to performing $O(N\cdot |\mathbf X|) = O(N\cdot \mathrm{quant_x}^{dimX})$ optimization problems of the form Equation (\ref{eq:2DeterministicBellmanEq}).  Although dynamic programming is still much more efficient than exploring every possible state path which amounts to $O(|\mathbf X|^N)$ evaluations, performing dynamic programming quickly in even a moderately sized state space presents a considerable challenge.  Because of this famous \textit{curse of dimensionality}, dynamic programming is, for the most part, real time prohibitive.  

\subsection{DDP / iLQR}\label{section:DDP/iLQR}
Differential dynamic programming (DDP) \cite{Jacobson1970DDP,tassa2008receding} and the closely related iterative linear quadratic regulator (iLQR) \cite{li2004iterative,tassa2012synthesis} are dynamic programming methods in which a quadratic approximation to the value function Equation (\ref{eq:2DeterministicBellmanEq}) is created at each point along the horizon.  In creating this quadratic approximation, DDP uses a second order expansion of the system dynamics while iLQR uses a first order expansion.  The associated benefit of DDP over iLQR is improved convergence at the expense of additional computation, however, depending on the application, it can be beneficial to choose faster computation over improved convergence (e.g. in a model predictive control setting in which convergence will never actually happen and computation is a premium).   Like the minimum principle, these methods generate a locally optimal control sequence rather than a globally optimal control policy as in dynamic programming.  Defining the \textit{state-control value function} $Q_n(\x,\bu)$ as 
\begin{align}\label{eq:2DDPQ}
Q_n(\x,\bu) &= g_n(\x,\bu) + V_{n+1}\big(F_n(\x,\bu)\big)
\end{align}
the state value function can be expressed as
\begin{align}
V_N(\x) &= h(\x)\\
V_n(\x) &= Q_n(\x,\bu^*)
\end{align}
where $\bu^*=\arg\min_\bu Q_n(\x,\bu)$ is the value that minimizes Equation (\ref{eq:2DDPQ}).  Given a nominal trajectory, $\(\hat \x_n,\hat \bu_n\)_{n=0}^{N-1}$, a local quadratic model of $Q_n$ can be constructed as
\begin{align}\label{eq:2DDPQapprox}
&Q_n(\hat \x_n+\delta \x_n,\hat \bu_n+\delta \bu_n)  \nonumber \\
&~~~~~~~~~~\approx Q_n^{(0)} + Q_n^{(\x)}\delta \x_n + Q_n^{(u)}\delta \bu_n + \frac{1}{2}\left[\delta \x_n\Tr~ \delta \bu_n\Tr\right]
\begin{bmatrix}
Q_n^{(xx)} & Q_n^{(xu)}\\
Q_n^{(ux)} & Q_n^{(uu)}
\end{bmatrix}
\begin{bmatrix}
\delta \x_n \\
\delta \bu_n
\end{bmatrix}
\end{align}
For given $\hat \x_n, \hat \bu_n,\delta \x_n$, the value of $\delta \bu_n$ which minimizes this local model of $Q_n$ is given by
\begin{align}\label{eq:2DDPuopt}
\delta \bu_n^* = \arg\min_{\delta \bu_n} Q_n = -\left(Q_n^{(uu)}\right)^{-1}\left(Q_n^{(u)}+Q_n^{(ux)}\delta \x_n\right)
\end{align}
The various partial derivatives $Q_n^{(\cdot)} = \nabla_{(\cdot)}Q(\hat \x_n,\hat \bu_n)$ are determined considering Equation (\ref{eq:2DDPQ})
\vspace{-1mm}
\begin{equation}\label{eq:2DDPQpartials}
\begin{tabular}{cc}	
	$\begin{aligned}
	Q_n^{(0)} &= g_n + V_{n+1} \\
	Q_n^{(x)} &= g_n^{(x)} + V_{n+1}^{(x)}  F_n^{(x)} \\
	Q_n^{(u)} &= g_n^{(u)} + V_{n+1}^{(x)}  F_n^{(u)}
	\end{aligned}$ 
	& ~~~
	$\begin{aligned}
	Q_n^{(xx)} &= g_n^{(xx)} +  F_n^{(x)\mathsf{T}} V_{n+1}^{(xx)}  F_n^{(x)} + V_{n+1}^{(x)} \cdot F_n^{(xx)}\\
	Q_n^{(ux)} &= g_n^{(ux)} + F_n^{(u)\mathsf{T}} V_{n+1}^{(xx)}  F_n^{(x)} + V_{n+1}^{(x)} \cdot F_n^{(ux)}\\
	Q_n^{(uu)} &= g_n^{(uu)} + F_n^{(u)\mathsf{T}} V_{n+1}^{(xx)}  F_n^{(u)} + V_{n+1}^{(x)} \cdot F_n^{(uu)}
	\end{aligned}$
\end{tabular}
\end{equation}

where the $ij^{th}$component of each matrix in the last three equations is defined as
\begin{subequations}\label{eq:2SecondOrderDynamicsExpansion}
	\begin{align}
	\left(V_{n+1}^{(x)} \cdot F_n^{(xx)}\right)_{ij} &= V_{n+1}^{(x)}\cdot \frac{\partial^2 F_n}{\partial x_i\partial x_j} \\
	\left(V_{n+1}^{(x)} \cdot F_n^{(ux)}\right)_{ij} &= V_{n+1}^{(x)}\cdot \frac{\partial^2 F_n}{\partial u_i\partial x_j} \\
	\left(V_{n+1}^{(x)} \cdot F_n^{(uu)}\right)_{ij} &= V_{n+1}^{(x)}\cdot \frac{\partial^2 F_n}{\partial u_i\partial u_j}
	\end{align}
\end{subequations}
The second order terms of Equation (\ref{eq:2SecondOrderDynamicsExpansion}) are ignored in iLQR, while in DDP they are included.  Substituting $\delta \u_n^*$ from Equation (\ref{eq:2DDPuopt}) into the local model Equation (\ref{eq:2DDPQapprox}) and simplifying gives a local model for $V_n(\x_n)$ about $\x_n=\hat \x_n + \delta \x_n$ 
\begin{align}\label{eq:2DDPVnModel}
V_n(\hat \x_n + \delta \x_n) &\approx 
Q_n^{(0)}-\frac{1}{2} Q_n^{(u)\mathsf{T}}(Q_n^{(uu)})^{-1}Q_n^{(u)}
+ \left[Q_n^{(x)}-Q_n^{(u)}(Q_n^{(uu)})^{-1}Q_n^{(ux)}\right]\delta \x_n \nonumber\\
&~~~~~~+ \frac{1}{2}\delta \x_n\Tr\left[Q_n^{(xx)}-Q_n^{(xu)}(Q_n^{(uu)})^{-1}Q_n^{(ux)}\right]\delta \x_n
\end{align}
Equating terms in the Taylor series expansion for $V_n(\x_n)$ gives an update for the partial derivatives of the value function
\begin{subequations}\label{eq:2DDPVpartials}
	\begin{align}
	V_n^{(0)}(\hat \x_n) &= Q_n^{(0)}-\frac{1}{2} Q_n^{(u)\mathsf{T}}(Q_n^{(uu)})^{-1}Q_n^{(u)}\\
	V_n^{(x)}(\hat \x_n) &=Q_n^{(x)}-Q_n^{(u)}(Q_n^{(uu)})^{-1}Q_n^{(ux)} \\
	V_n^{(xx)}(\hat \x_n) &=Q_n^{(xx)}-Q_n^{(xu)}(Q_n^{(uu)})^{-1}Q_n^{(ux)}
	\end{align}
\end{subequations}
A new trajectory $\(\x_n,\u_n\)_{n=0}^{N-1}$ is simulated using the current measurement of the system state according to 
\vspace{-1mm}
\begin{subequations}\label{eq:2DDPForwardPass}
	\begin{align}
	\x_{0} &= \x_0^{\text{meas}} \\
	\u_{n}^* &= \hat \u_n \underbrace{-\(Q_n^{(uu)}\)^{-1}Q_n^{(u)}-\(Q_n^{(uu)}\)^{-1}Q_n^{(ux)}\(\x_n - \hat \x_n\)}_{\delta \u_n^*} \\
	\x_{n+1} &=F_n(\x_n,\u_n^*)
	\end{align}
\end{subequations}
Starting from initial condition $V_N(\hat \x_N) = h(\hat \x_N)$  Equations (\ref{eq:2DDPQpartials}) and (\ref{eq:2DDPVpartials}) are solved backwards in time from $n=N$ to $n=0$ constituting the \textit{backwards pass}.  Starting from initial condition $\x_{0} = \x_0^{\text{meas}}$, a new system trajectory is then simulated according to Equation (\ref{eq:2DDPForwardPass}) which constitutes the \textit{forward pass}.  This simulated trajectory is then used as the new nominal trajectory $\(\hat \x_n,\hat \u_n\)_{n=0}^{N-1} := \(\x_n,\u_n\)_{n=0}^{N-1}$, and the process is restarted.  

By creating a local model of the value function through differentials, DDP and iLQR solve two major issues associated with dynamic programming.  For one, DDP / iLQR work directly with a continuous state space, so there is no need to artificially discretize the state.  Secondly, DDP and iLQR converge much faster than DP as they do not require a visit to each state in the state space during the backward sweep.  A stochastic variant of differential dynamic programming suitable for real time computation is proposed in section \ref{section:ASDDP}.

\subsubsection{Constraints}
Choosing an optimal control input which minimizes  Equation (\ref{eq:2DDPQapprox}) at each stage $n$ in the horizon amounts to a stage-wise quadratic programming problem.  Formulating this stage-wise QP problem in the context of DDP remains an active area of research.  Box-bounded control input constraints are addressed in \cite{tassa2014control}, general state and control inequality constraints are considered in the recent work of \cite{xie2017differential}.  In this work, state and input constraints are addressed in the stochastic setting in Section \ref{section:ASDDP}.


\section{Systems with Stochastic Dynamics}
This section discusses the basis principles of stochastic systems as relevant to optimal control and reinforcement learning problems.  The stochastic systems considered here can be described by the difference equation
\begin{subequations}\label{eq:2StochSysDT}
	\begin{align}
	\x_{n+1} &= F_n(\x_n,\bu_n,\bw_n),~n=0,1,\dots \\
	\x_0 &:~given\\
	\bw_0 &:~given
	\end{align}
\end{subequations}
where $\bw_n \in \mathbf W$ is a stochastic disturbance input to the system.

\subsection{Stochastic Optimization}
Stochastic optimization refers to a collection of methods for minimizing an objective function when a stochastic effect is present \cite{nemirovski2009robust}.  Consider the objective function $J(\boldsymbol{\theta},\vec\bw)$ which depends on the decision parameter $\boldsymbol{\theta}$ and the sequence of stochastic disturbances $\vec{\bw} = \{\bw_0,\bw_1,\dots\}$.  The parameter $\boldsymbol{\theta}$ is quite general and can represent the terms of a control sequence or the parameters of a parameterized control policy.   The goal of stochastic optimization is then to minimize the expected value 
\begin{align}\label{eq:2StochOptMinEJ}
	\min_{\boldsymbol{\theta}} \mathbb{E}[J(\boldsymbol{\theta},\vec\bw)] = \sum\limits_{\vec{{\mathsf{w}}}} J(\boldsymbol \theta, \vec{{\mathsf{w}}}) \mathrm{Pr}\left[\vec \bw = \vec{{\mathsf{w}}} \right]
\end{align}

\subsubsection{Sample Average Approximation}
Typically, the stochastic optimization Equation (\ref{eq:2StochOptMinEJ}) cannot be solved directly due to the combinatorial difficulty of computing $\mathbb{E}[J(\boldsymbol{\theta},\vec\bw)]$.  An alternative is to first compute the sample average
\begin{align}
	\hat J(\boldsymbol{\theta}) &=\sum\limits_{k=1}^{K} J(\boldsymbol{\theta},\vec\bw^{[k]})
\end{align}
and then minimize $	\hat J(\boldsymbol{\theta})$ through nonlinear programming methods.  The approximation $\hat J(\boldsymbol{\theta})$ improves as the number of samples $K$ increases in accordance to the Central Limit Theorem which states the difference between the sample average and true average convergence to a zero mean Normal distribution whose variance depends directly on the number of samples $K$
\begin{align}
\hat J(\boldsymbol{\theta}) - \mathbb{E}[J(\boldsymbol{\theta},\vec\bw)] \xrightarrow{d} \mathcal{N}\left(0,\frac{\sigma^2}{K}\right)
\end{align}
where $\sigma^2$ is the variance of $J(\boldsymbol{\theta},\vec\bw)$.

\subsubsection{Stochastic Approximation} \label{section:SA} 
Stochastic approximation is an iterative method which uses noisy measurements to find the root of a function, $H(\boldsymbol{\theta}^*)=\mathbf 0$, when $H(\boldsymbol{\theta})$ cannot be computed directly\footnote{ It may be the case that $H(\boldsymbol{\theta})$ is inaccessible, or it may be too expensive to compute directly.} but noisy sample observations $\mathbf y^{[k]} = H(\boldsymbol{\theta^{[k]}})+\boldsymbol\eta^{[k]}$
are available.  It is assumed that $\boldsymbol \eta^{[k]}$ is a zero-mean noise process so that $\mathbf y^{[k]}$ is an unbiased estimate of $H(\boldsymbol\theta^{[k]})$ in the sense that $\mathbb E[\mathbf y^{[k]}] = H(\boldsymbol\theta^{[k]})$. 
The stochastic approximation iteration is
\begin{align} \label{eq:2SA}
\boldsymbol\theta^{[k+1]} = \boldsymbol\theta^{[k]} + \alpha^{[k]} \mathbf y^{[k]}
\end{align}
with the two following conditions imposed on the learning rate $\alpha^{[k]}$
\begin{align} \label{eq:2SALearningRates}
\sum_{k=0}^\infty \alpha^{[k]} &= \infty ~~~~~~~~
\sum_{k=0}^\infty \left(\alpha^{[k]}\right)^2 < \infty
\end{align}
Roughly speaking, the first condition ensures the sequence is non-terminating so that asymptotic convergence properties hold, while the second condition ensures the noise in the samples does not dominate algorithm progress.  An intuitive justification of Equation (\ref{eq:2SALearningRates}) in the context of mean estimation is provided in \cite{MLMohri}.  The aggregate behavior of Equation (\ref{eq:2SA}) with learning rates Equation (\ref{eq:2SALearningRates}) can be assessed through the \textit{ODE method} \cite{Ber96,Borkar}, which states Equation (\ref{eq:2SA}) asymptotically tracks the  ordinary differential equation
\begin{align} \label{eq:2OdeMethod}
\dot {\boldsymbol\theta}(t) &= H\left(\boldsymbol\theta(t)\right)
\end{align}

\subsubsection{Gradient Descent Form of Stochastic Approximation}\label{section:2SAGradForm}
When $H(\boldsymbol\theta) = -\nabla_{\boldsymbol\theta} \mathbb{E}[J(\boldsymbol\theta, \vec \bw)]$, stochastic approximation finds a local solution to Equation (\ref{eq:2StochOptMinEJ}) using noisy gradient observations $\mathbf y^{[k]}=-\nabla_{\boldsymbol{\theta}} J(\boldsymbol{\theta},\vec\bw^{[k]})$ forming a process known as \textit{stochastic gradient descent} \cite{Borkar}.  Convergence of Equation (\ref{eq:2OdeMethod}) can be shown by constructing the Lyapunov function $V(\boldsymbol\theta) = \mathbb E[J(\boldsymbol\theta,\vec\bw)]$ and showing $\frac{dV}{dt}<0$ through
\begin{align*}
\frac{dV}{dt} &= \nabla_{\boldsymbol\theta} \mathbb E[J(\boldsymbol\theta(t),\vec \bw)] \cdot \dot{\boldsymbol\theta}(t) =-\left(\nabla_{\boldsymbol\theta} \mathbb E[J(\boldsymbol\theta(t),\vec 
\bw)]\right)^2 <0
\end{align*}
Convergence to $\boldsymbol\theta^*$ implies $\mathbf 0=H(\boldsymbol\theta^*)=-\nabla_{\boldsymbol\theta} \mathbb{E}[J(\boldsymbol\theta^*, \vec \bw)]$, satisfying the necessary conditions for a local minimum assuming $\boldsymbol\theta$ is unconstrained  (convergence proofs of stochastic gradient descent can be found in \cite{OnlineLearning,Ber96,Borkar}).  The algorithm proposed in Section \ref{section:SGDM} is based on stochastic gradient descent.

\subsubsection{Fixed Point Form of Stochastic Approximation}\label{section:2SAFixedPtForm}
A central concept in online learning is the fixed point form of stochastic approximation \cite{Borkar} in which
\begin{align}\label{eq:2SAFixedPtForm}
H(\boldsymbol\theta) = F(\boldsymbol\theta) - \boldsymbol\theta 
\end{align}
and $F$ is contractive so that $||F(\boldsymbol{\theta}_a)-F(\boldsymbol{\theta}_b)||_2 \leq \lambda ||\boldsymbol{\theta}_a-\boldsymbol{\theta}_b||_2$ for $0 \leq \lambda < 1$.  
Convergence of Equation (\ref{eq:2OdeMethod}) to equilibrium $\boldsymbol\theta^*$ is shown by constructing a Lyapunov function $V(\boldsymbol\theta) = \frac{1}{2}||\boldsymbol\theta-\boldsymbol\theta^*||_2^2$ and employing the ODE method Equation (\ref{eq:2OdeMethod}) 
\vspace{-1mm}
\begin{align*}
	\frac{dV}{dt} &=(\boldsymbol\theta(t) - \boldsymbol\theta^*) \cdot \dot{\boldsymbol\theta}(t) \\
	&= (\boldsymbol\theta(t) - \boldsymbol\theta^*) \cdot \left( F(\boldsymbol\theta(t)) - F(\boldsymbol\theta^*) \right) + 
			(\boldsymbol\theta(t) - \boldsymbol\theta^*) \cdot \left( F(\boldsymbol\theta^*) - \boldsymbol\theta(t) \right) \\
	&\leq ||\boldsymbol\theta(t) - \boldsymbol\theta^*||_2~|| F(\boldsymbol\theta(t)) - F(\boldsymbol\theta^*)||_2 - ||\boldsymbol\theta(t)-\boldsymbol\theta^*||_2^2 \\
	&\leq -(1-\lambda)~||\boldsymbol\theta(t) - \boldsymbol\theta^*||_2^2 
\end{align*}
Convergence to $\boldsymbol\theta^*$ implies $\mathbf 0=H(\boldsymbol\theta^*)= F(\boldsymbol\theta^*) - \boldsymbol\theta^*$.  Convergence can also be proved for general norms $||\cdot||_p, ~~p \geq 1$ \cite{Borkar}.

\subsection{Markov Decision Processes}
When the disturbance term $\bw_n$ in Equation (\ref{eq:2StochSysDT}) obeys the \textit{Markov property}, which roughly states that future behavior of a system is influenced only by the present state, ignoring the sequence of events that lead to the present state, the system dynamics take on a particularly simplified form known as a Markov decision process (MDP) in which the state transitions are given in terms of a controlled distribution  
\begin{align}
\x_{n+1} \sim p(\x'|\x,\bu)
\end{align}
When the state space $\mathbf X$ is continuous, the distribution is a density function defined by
\begin{align}
\int_{\x'} p(\mathbf s|\x,\u) d\mathbf s =  \mathrm{Pr}[\x_{n+1} \in \x' | \x_n = \x, \u_n=\u ]
\end{align}
If a discrete space is assumed, the distribution simplifies to a mass function
\begin{align}
p(\x'|\x,\u)= \mathrm{Pr}[\x_{n+1}=\x'|\x_n=\x,\u_n=\u]
\end{align}
In this context, the state vector now includes all deterministic and modeled stochastic states.  The model $p(\x'|\x,\u)$ can be determined empirically through direct interaction with the environment or through first principles modeling assuming some form of the stochastic effect.

\section{Stochastic Dynamic Programming}
As in the deterministic setting, stochastic dynamic programming (SDP) uses a model of the environment to construct a state-value function, a record of the optimal cost-to-go from each state in the state space.  The goal of SDP is to minimize
\begin{align}\label{eq:FiniteHorizonProblemStochMDP}
\min_{\bu_0,\bu_1,\dots,\bu_{N-1}} &\mathbb{E}\left[h(\mathbf x_N)+\sum_{n=0}^{N-1} g_n(\x_n,\bu_n)\Big | \x_0 = \x\right]
\end{align}
where system dynamics are governed according to the Markov decision process $\x_{n+1} \sim p(\x'|\x,\bu)$.

\subsection{Finite Horizon SDP}
The finite horizon cost of following policy $\bu_n=\pi_n(\x_n)$ is given by
\begin{align}
	J^\pi = h(\x_N) + \sum_{n=0}^{N-1} g_n(\x_n,\pi_n(\x_n))
\end{align}
The expected cost-to-go of following policy $\u_n=\pi_n(\x_n)$ starting from state $\x$ at time $n$ until time $N$ is represented by the \textit{policy value}, $V_n^\pi(\x)$
\begin{align}
V_N^\pi(\x) &=  \mathbb{E}\left[h(\x_N) \Big| \x_N=\x \right] = h(\x)\\
V_n^\pi(\x) &= \mathbb{E}\left[ \sum_{k=n}^{N-1} g_n(\x_k,\pi(\x_k,k)) + h(\x_N) \Big| \x_n=\x \right] \nonumber \\
&=  g(\x,\pi_n(\x))+\left.\mathbb{E}\Big[V_{n+1}^\pi(\x_{n+1})\right| \x_n=\x\Big] 
\end{align}
The \textit{state value function} $V_n(\x)$ results from following the optimal time-varying state-feedback policy $\pi_n^*(\x)$, which must satisfy the \textit{Bellman equation}
\begin{align}
	V_n(\x) 	&=  \min_{\u\in {\mathbf U}} \left\{g_n(\x,\u)+\left.\mathbb{E}\Big[V_{n+1}(\x_{n+1})\right|\x_n=\x\Big]\right\}  \label{eq:2BellmanEq}
\end{align}
The optimal policy can be inferred directly from the state value function through
\begin{align}
    \pi_n^*(\x) & = \arg\min_{\u\in {\mathbf U}}\Big\{g_n(\x,\u)+\left.\mathbb{E}\Big[V_{n+1}(\x_{n+1})\right|\x_n=\x\Big]\Big\}  \label{eq:2BellmanOptPolicy}    
\end{align}
 These two equations provide a means to recursively compute the state value function exactly by working backward through time starting from $N$, using a model of the environment $p(\x'|\x,\u)$
\begin{align}
	V_N(\x) &= h(\x)  \\
	V_n(\x) 	&= \min_{\u\in {\mathbf U}} \Big\{g_n(\x,\u)+ \sum_{\x'\in \mathbf X}p(\x'|\x,\u) V_{n+1}(\x')\Big\}
\end{align}
\begin{align}
    \pi_n^*(\x) & = \arg\min_{\u\in {\mathbf U}}\Big\{g_n(\x,\u)+\sum_{\x'\in X}p(\x'|\x,\u) V_{n+1}(\x')\Big\}
\end{align}
For this reason, finite horizon dynamic programming in both the deterministic and stochastic case is often referred to as backward dynamic programming.
\subsection{Infinite Horizon SDP}
Finite horizon dynamic programming constructs a state value function which explicitly depends on time, even if the instantaneous cost and system dynamics are independent of time.  As a result, the optimal state feedback policy, which is inferred directly from the state value function, also depends explicitly on time.  The benefit of working in an infinite horizon is that the state value function and therefore the state feedback policy is invariant with time, as long as the instantaneous cost $g(\x,\u)$ and process dynamics $p(\x'|\x,\u)$ are independent of time \cite{BerVolI}.  The discounted infinite horizon cost of following policy $\u_n=\pi(\x_n)$ is given by
\begin{align}
J^\pi = \sum_{k=0}^{\infty} \gamma^k g(\x_k,\pi(\x_k))
\end{align}
where the discount factor $0 < \gamma < 1$ serves to reduce the impact of costs incurred far into the future on the immediate cost prediction.  The expected cost-to-go following policy $\u_n=\pi(\x_n)$ from state $\x$ starting at arbitrary time $n$ is given by the policy value
\begin{align}
	V^\pi(\x)  &=  \mathbb{E}\left[ \sum_{k=n}^{\infty}  \gamma^k g(\x_k,\pi(\x_k)) \Big| \x_n=\x \right] \nonumber\\
	&=  g(\x,\pi(\x))+\gamma\left.\mathbb{E}\Big[V^\pi(\x_{n+1})\right| \x_n=\x\Big] \label{eq:2PolicyValueDpInfHor}
\end{align}
The state value function must satisfy the infinite horizon Bellman equation 
\begin{align} \label{eq:2BellmanEqInfHor}
		V(\x) 		&=  \min_{\u\in {\mathbf U}} \left\{g(\x,\u)+\left.\gamma \mathbb{E}\Big[V(\x_{n+1})\right|\x_n=\x\Big]\right\} 
\end{align}
and the optimal policy can be inferred directly from the state value function through
\begin{align} \label{eq:2DPOptPolicyInfHor}
\pi^*(\x) & = \arg\min_{\u\in {\mathbf U}}\Big\{g(\x,\u)+\left.\gamma \mathbb{E}\Big[V(\x_{n+1})\right|\x_n=\x\Big]\Big\}
\end{align}
Constructing the state value function is less straight-forward in the infinite horizon case, as working backwards through time is not possible since a terminal time does not exist.  Rather, an approximation to state value function, $\hat V(\x)$, can be solved for iteratively in a process called \textit{value iteration}, treating the resulting Bellman equation
\begin{align}\label{eq:2BellmanEqInfHorDiscX}
\hat V(\x) &=  \min_{\u\in {\mathbf U}} \Big\{g(\x,\u)+\gamma \sum_{\x'\in \mathbf X}p(\x'|\x,\u) \hat V(\x')\Big\}
\end{align}
as a consistency condition.  Under mild conditions the operation
\vspace{-1mm}
\begin{align} \label{eq:2ValueIterationT}
	T[\hat V(\x)] = \min_{\u\in {\mathbf U}} \Big\{g(\x,\u)+\gamma \sum_{\x'\in \mathbf X}p(\x'|\x,\u) \hat V(\x')\Big\}
\end{align}
is a contraction mapping, so the fixed point iteration 
\begin{align} \label{eq:2ValueIterationFixedPtIt}
	\hat V^{[k+1]} = T[\hat V^{[k]}]
\end{align}
converges to $V$ as long as each $\x\in \mathbf X$ is repeatedly visited.  The approximation error after $K$ iterations is bounded by $||\hat V^{[k]}-V|| \leq \lambda^K || \hat V^{[0]} - V || $ for the norm $||V||=\max_{\x} V(\x)$.  Equations (\ref{eq:2ValueIterationT}) and (\ref{eq:2ValueIterationFixedPtIt}) together form value iteration.  The state value function can also be found in a process known as \textit{policy iteration}, in which the policy value is solved for exactly at each iteration by solving the system of linear equations
\begin{align}\label{eq:2PolicyIterationV}
V^{\pi}(\x) &=  g(\x,\pi(\x))+\gamma \sum_{\x'\in \mathbf X}p(\x'|\x,\pi(\x)) V^\pi(\x')~~~~\forall \x\in \mathbf X 
\end{align}
and making the policy update 
\begin{align}\label{eq:2PolicyIterationPi}
\pi(\x) \leftarrow \arg\min_{\u\in {\mathbf U}}\Big\{g(\x,\u)+\gamma\sum_{\x'\in \mathbf X}p(\x'|\x,\u) V^\pi(\x')\Big\}~~~~\forall \x\in \mathbf X
\end{align}
Equations (\ref{eq:2PolicyIterationV}) and (\ref{eq:2PolicyIterationPi}) form policy iteration.  Convergence is guaranteed since $||V^\pi||$ must decrease on every iteration \cite{PutermanMDP}.  When the control space $\mathbf U$ is finite, convergence occurs in a finite number of iterations since there are only finitely many policies in a discrete action and state space.  A third process known as \textit{modified policy iteration} combines value iteration with policy iteration.  Rather than being solved for exactly, the policy value  is updated for several iterations through the update
\begin{align}\label{eq:2ModPolicyIterationV}
\hat V^{\pi}(\x) \leftarrow  g(\x,\pi(\x))+\gamma \sum_{\x'\in \mathbf X}p(\x'|\x,\pi(\x)) \hat V^\pi(\x')~~~~\forall \x\in \mathbf X 
\end{align}
After several iterations of Equation (\ref{eq:2ModPolicyIterationV}), the policy is updated according to Equation (\ref{eq:2PolicyIterationPi}).  Value iteration is easier to implement than policy iteration, as solving Equation (\ref{eq:2PolicyIterationV}) exactly is computationally expensive especially when the state space $\mathbf X$ is large.  However, policy iteration typically converges faster than value iteration since $V^\pi(\x)$ is exact at each policy update.  Modified policy iteration lies somewhere in-between, removing the need to exactly compute the system of Equations (\ref{eq:2PolicyIterationV}) but providing a better estimate of $V^\pi(\x)$ at each iteration.


\section{Reinforcement Learning: Model-Free Value Function Methods}
In the infinite horizon setting, dynamic programming provides a framework for iteratively computing the value of a policy which requires a model of the environment, $p(\x'|\x,\bu)$.  \textit{Reinforcement learning methods}, unlike dynamic programming, do not require a model of the environment to compute the value of a policy.  
\subsection{Monte Carlo Estimation} \label{section:MC}
Monte Carlo estimation provides a method of policy evaluation based on samples of the discounted infinite horizon cost.  Following policy $\pi$ the cost occurred at each time step is recorded.  In the infinite horizon setting the final estimate can be based on the truncated series
\begin{align}
\tilde J^\pi(\x_0) &=\sum_{n=0}^{N-1} \gamma^{n} g(\x_n,\pi(\x_n)) 
\end{align}
The full series can be decomposed into the the truncated portion and a bounded term
\begin{align*}
	J^\pi(\x_0) &= \sum_{n=0}^{\infty} \gamma^{n} g(\x_n,\pi(\x_n)) = \sum_{n=0}^{N-1} \gamma^{n} g(\x_n,\pi(\x_n)) + 
	\underbrace{\gamma^N \sum_{n=0}^{\infty} \gamma^{n} g(\x_{n+N},\pi(\x_{n+N}))}_{\leq \dfrac{\gamma^N}{1-\gamma} g_{max}}	
\end{align*}
Since all quantities above are positive, the series truncation error is bounded by
\[
\left| J^\pi(\x_0) - \tilde J^\pi(\x_0) \right| \leq \frac{\gamma^N}{1-\gamma} g_{max}
\]
The Monte Carlo update to the state policy value is given by
\begin{align}\label{eq:2MCupdate}
 \hat V^\pi(\x) \leftarrow \hat V^\pi(\x) + \alpha(\x) \Big[\tilde J^\pi(\x)  -  \hat V^\pi(\x) \Big]
\end{align}
The learning rate $\alpha$ can be a function of the number of visits $k$ to state $\x$, \\${\alpha(\x) = \frac{1}{k(\x)}}$, so that Equation (\ref{eq:2MCupdate}) constructs the empirical average or it can be a fixed value $0< \alpha < 1$ so that Equation (\ref{eq:2MCupdate}) creates a noisy average with exponentially fading memory.

\subsection{Temporal-Difference Learning}\label{section:TD}
Like SDP, temporal difference (TD) algorithms provide a means for policy evaluation.  Unlike SDP, TD methods do not rely on a model of the environment $p(\x'|\x,\pi(\x))$ to compute the policy value.  Assuming a finite state space, the TD policy value update is
\begin{align}\label{eq:2TDupdate}
	\hat V^\pi(\x) & \leftarrow \hat V^\pi(\x) + \alpha(\x)\Big[\underbrace{g(\x,\pi(\x)) + \gamma \hat V^\pi(\x') - \hat V^\pi(\x)}_{\text{temporal difference}} \Big]
\end{align}
Here, $\x$ is the state value at time $n$ and $\x'$ is the observed state value at time $n+1$.  The learning rate $\alpha(\x)$ is a function of the number of visits $k$ to state $\x$ and satisfies Equation (\ref{eq:2SALearningRates}).  The \textit{temporal difference} is the difference between the one-sample estimate of the cost-to-go from state $\x'$ at time $n+1$ and the current policy value estimate of state $\x$ at time $n$.  In view of stochastic approximation of Section \ref{section:SA}, this is a stochastic fixed point iteration with $y = g(\x,\pi(\x)) + \gamma \hat V^\pi(\x') - \hat V^\pi(\x)$ providing an unbiased estimate of $g(\x,\pi(\x)) + \gamma \mathbb{E}[\hat V^\pi(\x_{n+1})|\x_n=\x] - \hat V^\pi(\x)$.  Convergence of the TD update Equation (\ref{eq:2TDupdate}) implies $\hat V^\pi(\x) = g(\x,\pi(\x)) + \gamma \mathbb{E}[\hat V^\pi(\x_{n+1})|\x_n=\x]$ which satisfies the dynamic programming based policy evaluation Equation (\ref{eq:2PolicyValueDpInfHor}), indicating $\hat V^\pi(\x)$ converges to $ V^\pi(\x)$ for all $\x$ as long as each state is repeatedly visited ($\alpha(\x)$ must tend to zero for each $\x$).  

\subsection{Q-learning}
If a record of the state-control value function $Q(\x,\u)$ was available such that 
$V(\x)=\min_\u Q(\x,\u)$, finding the optimal control policy would no longer require a model of the environment as Equation (\ref{eq:2DPOptPolicyInfHor}) reduces to
\begin{align}
\pi^*(\x) & = \arg\min_{\u}\Big\{g(\x,\u)+\left.\gamma \mathbb{E}\Big[V(\x_{n+1})\right|\x_n=\x\Big]\Big\}\nonumber \\
&=  \arg\min_{\u}Q(\x,\u)  
\end{align}
where the state-control value function satisfies
\begin{align}
Q(\x,\u) &=   g(\x,\u)+\gamma \mathbb{E}[V(\x_{n+1})\Big|\x_n=\x]\nonumber\\
&=  g(\x,\u)+\gamma \mathbb{E}\Big[\min_{\mathbf v} Q(\x_{n+1},\mathbf v)\Big|\x_n=\x\Big]
\end{align}
More generally, the state-control policy function associated with following policy $\pi(\x)$, $Q^\pi(\x,\u)$, satisfies
\begin{align}
Q^\pi(\x,\u) &=  g(\x,\u)+\gamma \mathbb{E}\Big[Q^\pi(\x_{n+1},\pi(\x_{n+1}))\Big|\x_n=\x\Big]
\end{align}
as $V^\pi(\x)=Q^\pi(\x,\pi(\x))$. The breakthrough known as Q-learning constructs such a state-control value function through trial and error interaction with the environment.  The Q-learning update 
\begin{align}
	\hat Q(\x,\u) & \leftarrow \hat Q(\x,\u) + \alpha(\x,\u)\left[g(\x,\u) + \gamma \min_{\mathbf v} \hat Q(\x',\mathbf v) - \hat Q(\x,\u) \right]
\end{align}
is based on stochastic fixed point iteration of Section \ref{section:SA} and can be viewed as a generalization of temporal difference learning. Here $\x$ and $\u$ are the state and control at time $n$, and $\x'$ is the state observed at time $n+1$.  The learning rate $\alpha(\x,\u)$ is a function of the number of visits $k$ to state-control pair $(\x,\u)$ and satisfies Equation (\ref{eq:2SALearningRates}).  According to stochastic approximation theory, $\hat Q(\x,\u)$  converges to $Q(\x,\u)$ for all $(\x,\u)$ provided that all controls continue to be tried from all states, and each state is repeatedly visited ($\alpha(\x,\u)$ must tend to zero for each $(\x,\u)$).

\section{Value Function Approximation}
Working with a state space $\mathbf X$ that is finite (i.e. discrete and bounded) admits \textit{tabular solutions}, in which the state value function can be described by a simple lookup table representation.   Value function approximation (VFA) (also known as approximate dynamic programming, adaptive dynamic programming, and neuro dynamic programming) provides a means to approximately construct the value function when $\mathbf X$ becomes large or even infinite (i.e. if $\mathbf X$ is continuous), in which case filling out entries of a tabular representation of the value function becomes computationally intractable.  Value function approximation is concerned with the weighted least squares problem 
\begin{subequations}
	\begin{align}\label{eq:2StateVFALeastSquaresProblem}
	\min_{\boldsymbol\theta} \mathbb{E}\left[\frac{1}{2} d(\x,\boldsymbol\theta)^2 \right] =	\min_\theta \sum_{\x}\rho(\x)\frac{1}{2}\Big[\underbrace{\tilde V^\pi(\x) - \hat V(\x,\boldsymbol\theta)}_{\triangleq d(\x,\boldsymbol\theta)}\Big]^2
	\end{align}
	when a state value function is learned and
	\begin{align}\label{eq:2StateControlVFALeastSquaresProblem}
	\min_{\boldsymbol\theta} \mathbb{E}\left[\frac{1}{2} d(\x,\u,\boldsymbol\theta)^2 \right] =	\min_{\boldsymbol\theta} \sum_{\x,\u}\rho(\x,\u)\frac{1}{2}\Big[\underbrace{\tilde Q^\pi(\x,\u) - \hat Q(\x,\u,\boldsymbol\theta)}_{\triangleq d(\x,\u,\boldsymbol\theta)}\Big]^2
	\end{align}
\end{subequations}
when a state-control value function is learned.	 Here $\rho(\cdot)$ is some distribution among the states or state-control pairs, $\tilde V^\pi(\x)$ and $\tilde Q^\pi(\x,\u)$ are sample estimates of policy values $V^\pi(\x)$ and $Q^\pi(\x,\u)$ based on available information (and are not retained in memory), and $\hat V(\x,\boldsymbol\theta)$, $\hat Q(\x,\u,\boldsymbol\theta)$ are parameterized approximations of $\tilde V^\pi(\x)$ and $\tilde Q^\pi(\x,\u)$ (which are retained in memory).  These approximations can be constructed as radial basis functions, neural networks, etc.   A popular choice is the linear approximation\footnote{It is worth noting that lookup table methods are a special case of linear approximation with as many elements of $\boldsymbol\theta$ as states (state VFA) or state-control pairs (state-control VFA), with the basis function serving as an indicator function.} with (possibly nonlinear) basis function $\boldsymbol\phi$, 
\begin{align}
\hat V(\x,\boldsymbol\theta) = \boldsymbol \phi(\x)^\mathsf{T}\boldsymbol\theta ~~~~~~~~~
\hat Q(\x,\u,\boldsymbol\theta) = \boldsymbol\phi(\x,\u)^\mathsf{T}\boldsymbol\theta
\end{align}
There are typically many more states or state-control pairs than elements of $\boldsymbol\theta$, so changing one element of $\boldsymbol\theta$ changes the estimated value of many states or state-control pairs.  The least squares problem Equation (\ref{eq:2StateVFALeastSquaresProblem}) can be solved through stochastic gradient descent \cite{SuttonBarto} according to the update
\begin{align}\label{eq:2VFASGDUpdateV}
\boldsymbol\theta  &\leftarrow \boldsymbol\theta - \alpha(\x)\nabla_{\boldsymbol\theta} \frac{1}{2}d(\x,\boldsymbol\theta)^2 \nonumber\\
&= \boldsymbol\theta + \alpha(\x) \Big[\tilde V^\pi(\x) - \hat V(\x,\boldsymbol\theta)\Big]	\nabla_{\boldsymbol\theta} \hat V(\x,\boldsymbol\theta) 
\end{align}
where learning rate $\alpha(\x)$ is a function of the number of visits to $\x$ and satisfies Equation (\ref{eq:2SALearningRates}).  Similarly,  Equation (\ref{eq:2StateControlVFALeastSquaresProblem}) can be solved through the update \cite{SuttonBarto}
\begin{align}\label{eq:2VFASGDUpdateQ}
\boldsymbol\theta  &\leftarrow \boldsymbol\theta + \alpha(\x,\u) \Big[\tilde Q^\pi(\x,\u) - \hat Q(\x,\u,\boldsymbol\theta)\Big]	\nabla_{\boldsymbol\theta} \hat Q(\x,\u,\boldsymbol\theta) 
\end{align}
The least squares problem can be solved through a variety of other methods including batch least squares such as averaged steepest descent and Gauss-Newton iteration or incremental least squares such as Kalman or Extended Kalman filtering \cite{Ber96}.  If a state policy value function is learned, the optimal policies can be formed with a model of the environment through
\begin{align}
\pi(\x) &= \arg\min_\u \left\{ g(\x,\u) +\gamma \mathbb{E}\left[\hat V(\x_{n+1},\boldsymbol\theta)|\x_n =\x \right]\right\}	
\end{align}
whereas learning a state-control policy value allows construction of the optimal policy in a model-free setting
\begin{align}
\pi(\x) &= \arg\min_\u \hat Q(\x,\u,\boldsymbol\theta)
\end{align}
The sample estimates are often formed through dynamic programming or temporal difference updates according to 
\begin{subequations}\label{eq:2VFATDs}
	\begin{align}
	&\text{DP estimate of $V^\pi$:~~~~}\tilde V^\pi(\x) =  g(\x,\pi(\x)) +\left.\gamma \mathbb{E}\Big[\hat V(\x_{n+1},\boldsymbol\theta)\right|\x_n=\x\Big] \nonumber \\
	&~~~~~~~~~~~~~~~~~~~~~~~~~~~~~~~~~~~~~= g(\x,\pi(\x)) + \gamma\int_{\x'}p(\x'|\x,\pi(\x))  \hat V(\x',\boldsymbol\theta) ~d\x'\\
	&\text{TD estimate of $V^\pi$:~~~~~} \tilde V^\pi(\x) = g(\x,\pi(\x)) +\gamma \hat V(\x',\boldsymbol\theta) \\
	&\text{TD estimate of $Q^\pi$:~~~~~} \tilde Q^\pi(\x,\u) =  g(\x,\u) +\gamma \hat Q(\x',\pi(\x'),\boldsymbol\theta)
	\end{align}
\end{subequations}
At each time step $(\x,\u)$ is drawn from distribution $\rho$ or generated through direct interaction with the environment, and $\x'$ is drawn from model $p(\x'|\x,\pi(\x))$ or through direct interaction with the environment.   
%

In general, solving the least squares problem does not guarantee the approximation converges to the policy value.  For one, there is no guarantee the chosen approximation architecture is capable of accurately representing the policy value.  Secondly, the sample estimates given in Equation (\ref{eq:2VFATDs}) are \textit{biased} estimates of policy values $V^\pi$ or $Q^\pi$ since each sample incorporates the policy value approximation $\hat V$ or $\hat Q$.   

\chapter{HYBRID VEHICLE MODEL}  

\section{Hybrid Vehicle Background} 
\begin{figure}[h!]
	\centering
	\begin{minipage}{.5\textwidth}
		\includegraphics [width=1\textwidth]{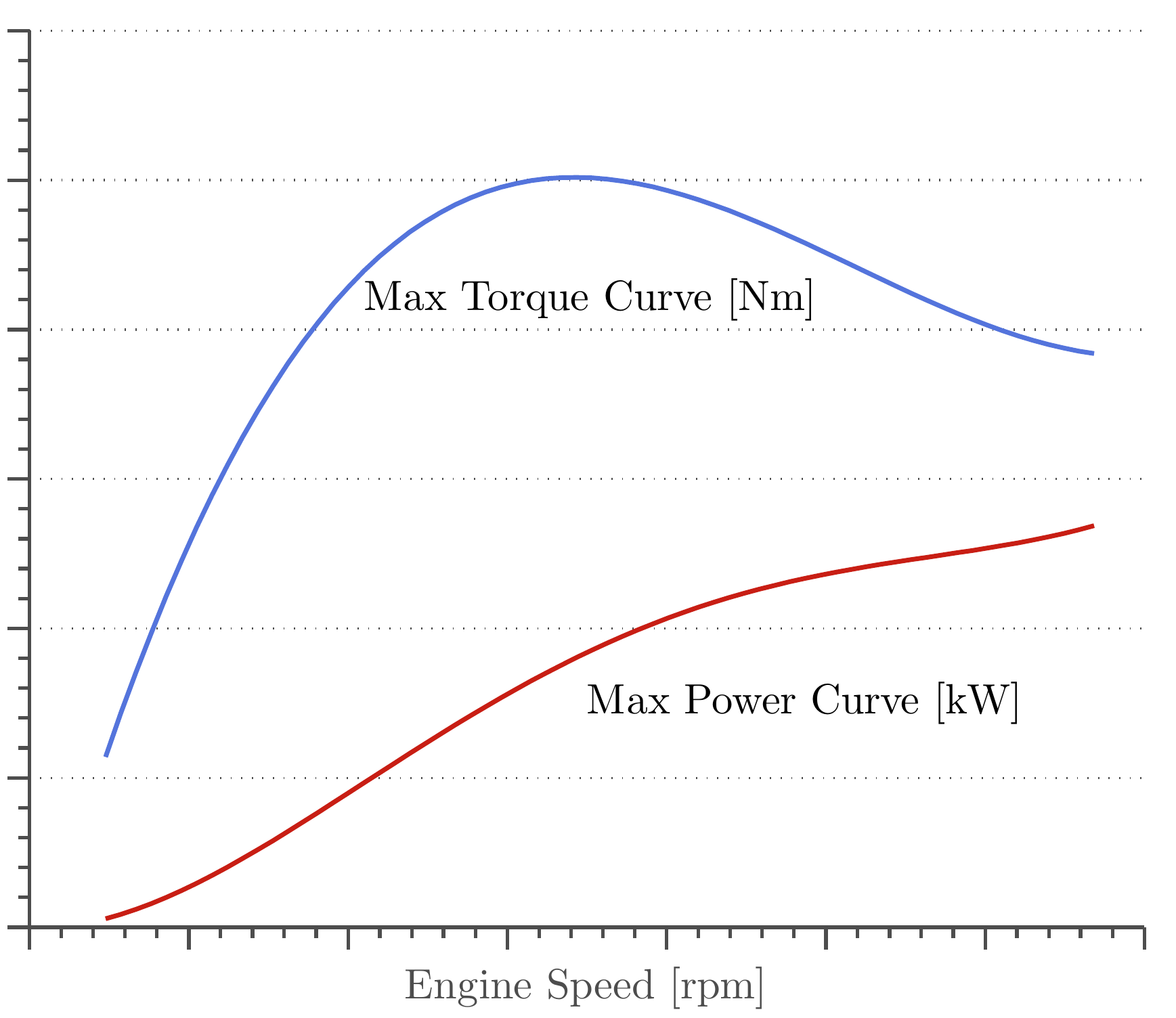}
	\end{minipage}~
	\begin{minipage}{.5\textwidth}
		\includegraphics [width=1\textwidth]{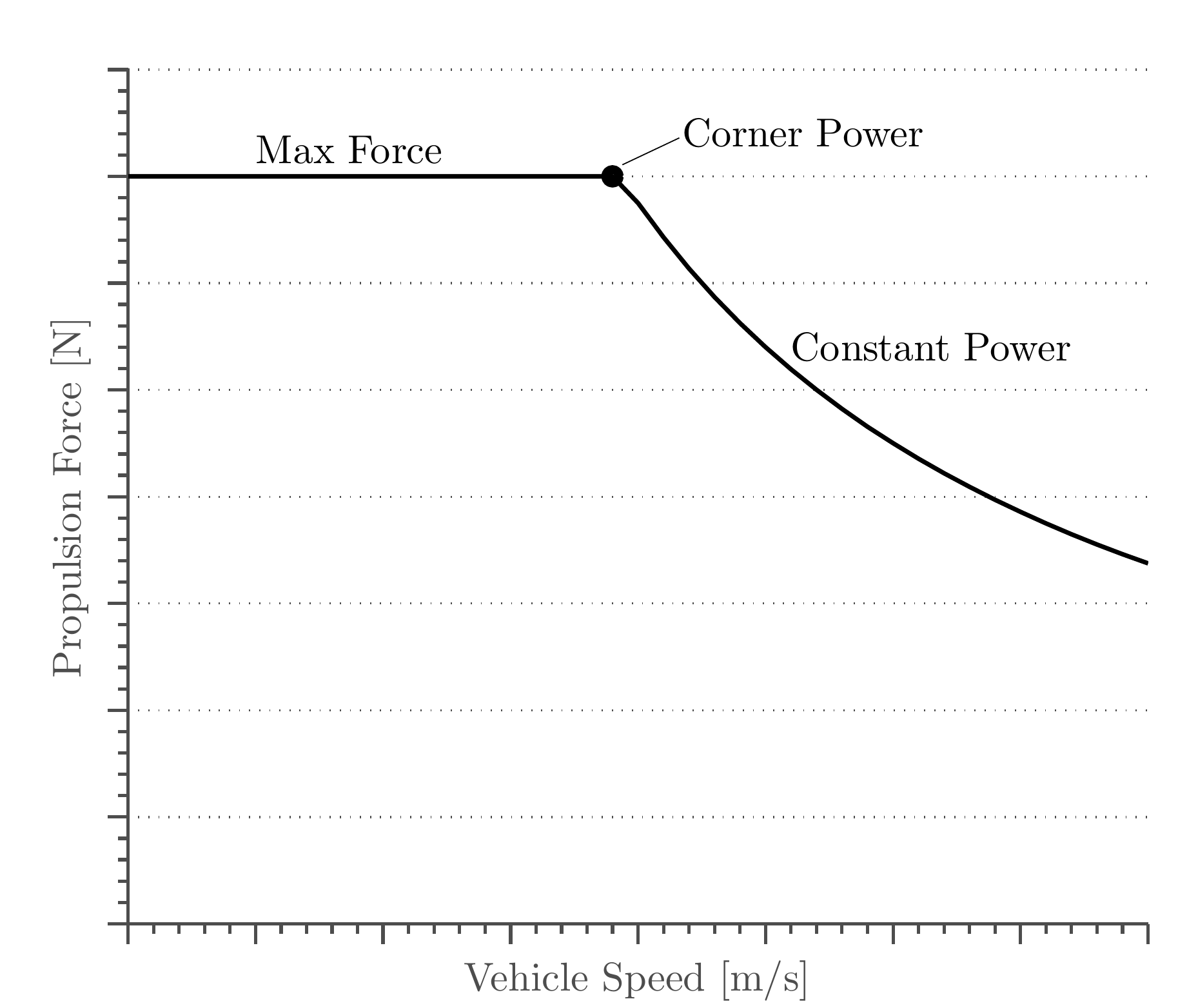}
	\end{minipage}
	\caption{Engine capabilities and vehicle propulsion requirements.}
	\label{fig:ApplicationCurves}
\end{figure}

The requirements of a given vehicle application are specified by speed and propulsion limits, and the vehicle transmission matches these requirements to engine capabilities.  Figure \ref{fig:ApplicationCurves} shows typical engine torque and power curves on the left, and a typical vehicle propulsion requirement curve on the right.  The engine is typically sized to deliver a specified minimum torque and power over a range of speeds.  In any given application, there is typically some maximum propulsion force required as indicated in the right figure of Fig. \ref{fig:ApplicationCurves}.  The corner power location is set by the maximum power available from the engine.  The maximum available propulsion force decreases along a curve of constant power past the corner power location.
Hydraulic hybrid vehicles (HHV) consist of a primary power path originating from an internal combustion engine and a secondary power path originating from a hydraulic accumulator.  The arrangement of the primary and secondary power paths can be divided into three architectures: parallel in which the secondary power path is in parallel with the primary, series in which the secondary power path is in series with the primary, and series-parallel which combines features of the series and parallel arrangements.  One of the defining features of all HHV architectures is regenerative braking, a process by which vehicle kinetic energy is transfered to the hydraulic accumulator to be released during a subsequent propulsion event. 
\subsection{Accumulator Energy Storage}
\begin{figure}[h!]
	\centering
	\begin{minipage}{.3\textwidth}
		\includegraphics [trim = 150mm 130mm 110mm 0mm, clip,width=1\textwidth]{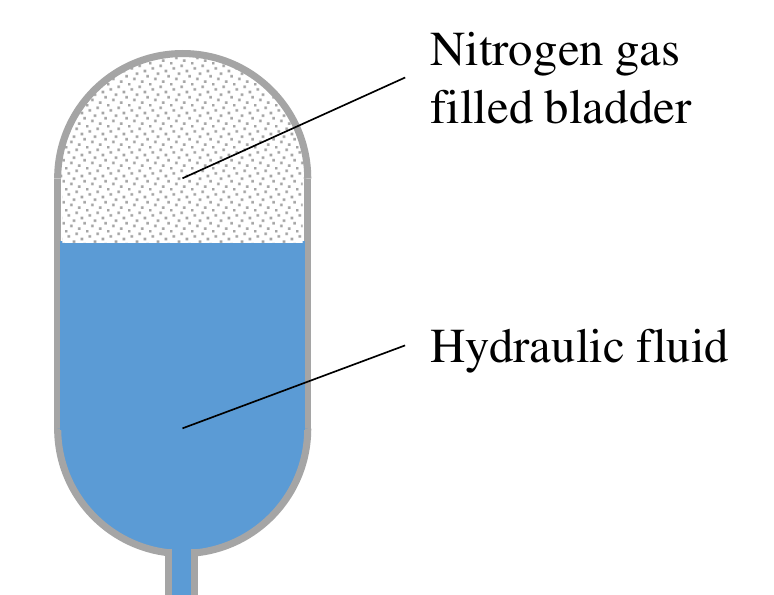}
	\end{minipage}~
	\begin{minipage}{.6\textwidth}
		\includegraphics [width=1\textwidth]{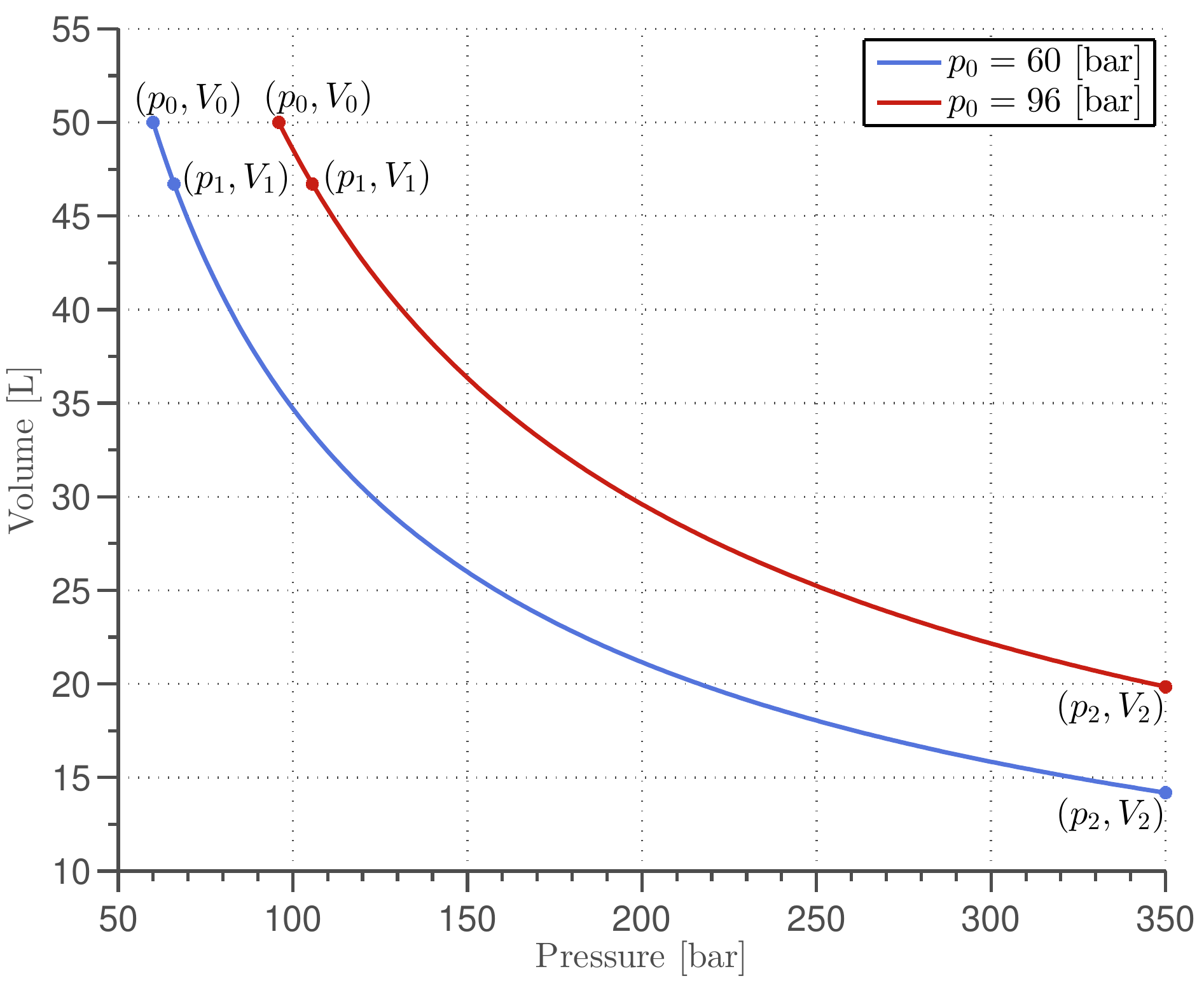}
	\end{minipage}
	\caption{Bladder type hydraulic accumulator.  Left: Schematic, Right: $p-V$ curves for two precharge pressures, for a fixed $V_0=50\times 10^{-3}~ m^3$.}
	\label{fig:Accum}
\end{figure}

\begin{figure}[h!]
	\centering
	\begin{minipage}{.8\textwidth}
		\includegraphics [width=1\textwidth]{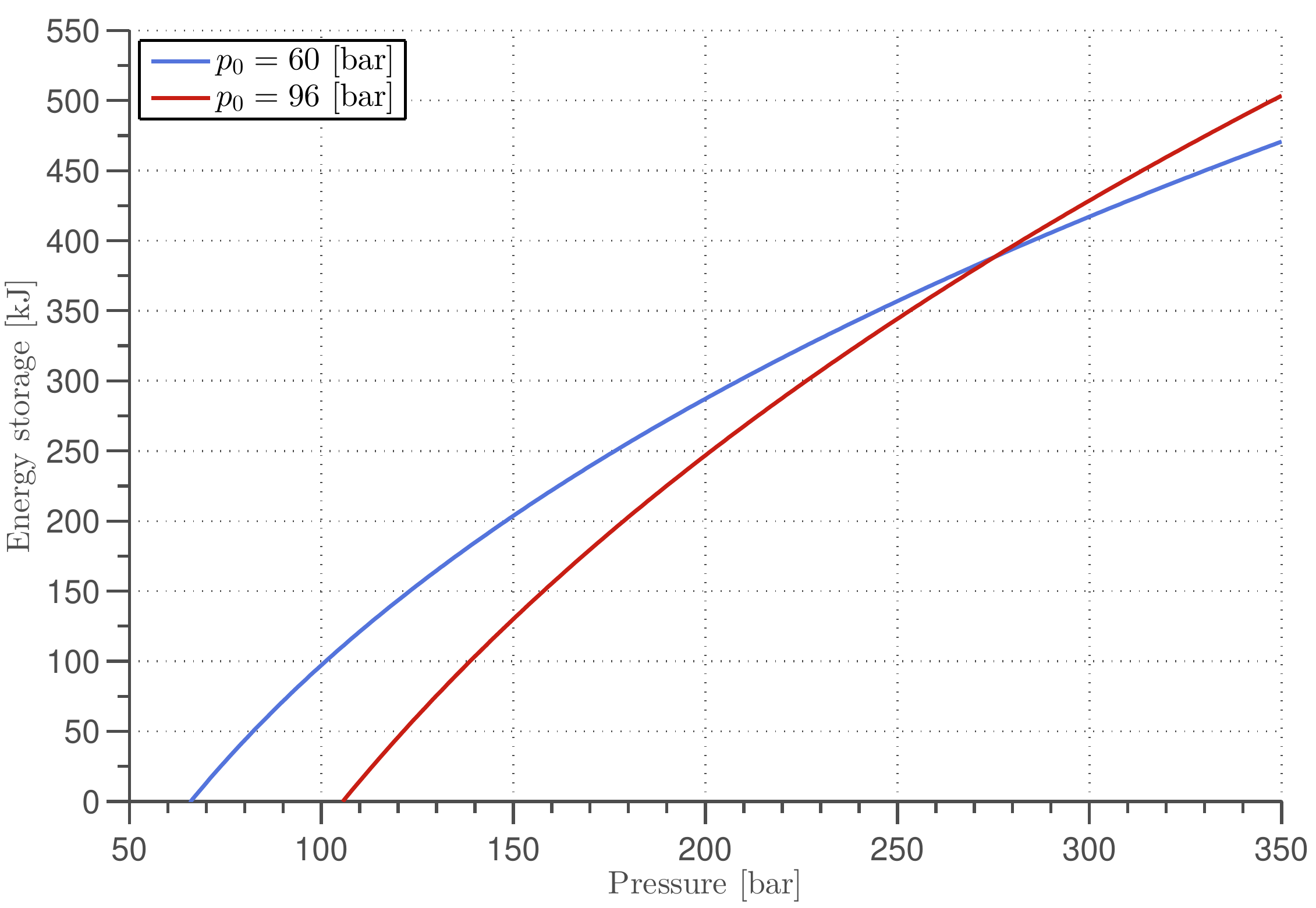}
	\end{minipage}
	\caption{Hydraulic accumulator energy storage curves for $V_0=50\times 10^{-3} ~m^3$, $p_1=1.1\times p_0$.}
	\label{fig:AccumPE}
\end{figure}

Energy storage in a hydraulic hybrid is accomplished through a hydraulic accumulator, typically of the bladder-type as shown in Fig. \ref{fig:Accum}.
The top portion contains a bladder filled with Nitrogen gas.  Hydraulic fluid can enter and exit the accumulator through a port on the lower side of the accumulator.  The {precharge pressure}, $p_0$, is the gas pressure when no hydraulic fluid is present in the accumulator.  The {minimum operating pressure}, $p_1$, is typically set to $110 \%$ of $p_0$ and, as its name suggests, is the lowest allowable operating pressure ensuring safe accumulator operation.  The maximum allowable operating pressure is shown in the $p-V$ diagram of Fig. \ref{fig:Accum} as $p_2$. Assuming the Nitrogen mass transfer between the tank and accumulator occurs under isentropic conditions (i.e. without heat transfer, corresponding to mass transfer with a perfectly insulated accumulator), the thermodynamic relationships within the Nitrogen gas bladder are 
\begin{align*}
pV&=mRT~(\text{Ideal gas law}) ,~~
\frac{p}{p_0} = \left(\frac{T}{T_0}\right)^{\gamma/(\gamma-1)} 
\end{align*}
where $\gamma=1.4$ is the specific heat ratio of Nitrogen.  Since the Nitrogen mass is constant throughout accumulator operation, these two equations can be combined to yield the pressure-volume relationship for the Nitrogen gas, 
\begin{align}\label{eq:3pVpolytropic}
	pV^\gamma = p_0V_0^\gamma = c
\end{align}
The energy stored in the accumulator between points 1 and 2 on the $p-V$ diagram of Fig. \ref{fig:Accum} is given by,
\begin{align}\label{eq:3AccumEnergyStorage}
E_{12}=
\int_{1}^{2}p dV=
-\frac{c^{1/\gamma}}{\gamma}\int_{1}^{2} p^{-1/\gamma} dp=
\frac{p_0^{1/\gamma}V_0}{(1-\gamma)}\left(p_2^{(1-1/\gamma)}-p_1^{(1-1/\gamma)}\right)
\end{align}
where we have used the fact $0 = d\left(pV^\gamma \right) = V^\gamma dp + \gamma p V^{\gamma-1} dV$ as evident from Equation (\ref{eq:3pVpolytropic}).  Accumulator energy storage curves are shown in Fig. \ref{fig:AccumPE}.  Here, it is assumed that energy storage is in reference to point 1 on the $p-V$ curve (i.e. energy storage is zero at $(p_1,V_1)$), since this is the lowest pressure allowable during accumulator operation.  
An interesting observation is the curves for $p_0=60$ bar and $p_0=96$ bar terminate at nearly the same energy storage level, yet the curve associated with $p_0=60$ bar accomplishes a given energy level at a lower associated pressure for a larger range of operation.  The accumulator can be designed by considering the energy storage required for a given application.  For a given vehicle speed $v_{veh}$, the kinetic energy $E_{K}(v_{veh}) = \frac{1}{2}m_{veh}v_{veh}^2$ represents the maximum available energy that can be transfered to the accumulator.  Precharge pressure $p_0$ and accumulator size $V_0$ can therefore be chosen by setting Equation (\ref{eq:3AccumEnergyStorage}) equal to the desired value of $E_K$, using the constraint $p_1=1.1\times p_0$.

\subsection{Architectures}
This subsection reviews basic operation and design considerations for the parallel, series, and series-parallel hydraulic hybrids.  A comprehensive comparison between various series-parallel configurations is discussed in \cite{MahaCarl2006comparison}.  An interesting use of a high-speed flywheel as the secondary energy source in a series-parallel configuration is found in \cite{MahaKumar2007study}.  A novel concept known as the blended hybrid, whereby the hydraulic accumulator is passively disconnected when system differential pressure rises below accumulator pressure, is discussed in \cite{MahaSprengel2013investigation,MahaBleazard2015optimal,MahaSprengel2015implementation}.  The intention of the blended hybrid is to allow the transmission to operate at lower pressures than accumulator pre-charge, lowering losses in the hydraulic circuit.  The blended hybrid architecture is adapted to the series and series-parallel configurations in \cite{MahaSprengel2014recent}.

\subsubsection{Parallel HHV}
\begin{figure}[h!]
	\centering
	\begin{minipage}{.65\textwidth}
		\includegraphics [trim = 0mm 50mm 160mm 0mm, clip,width=1\textwidth]{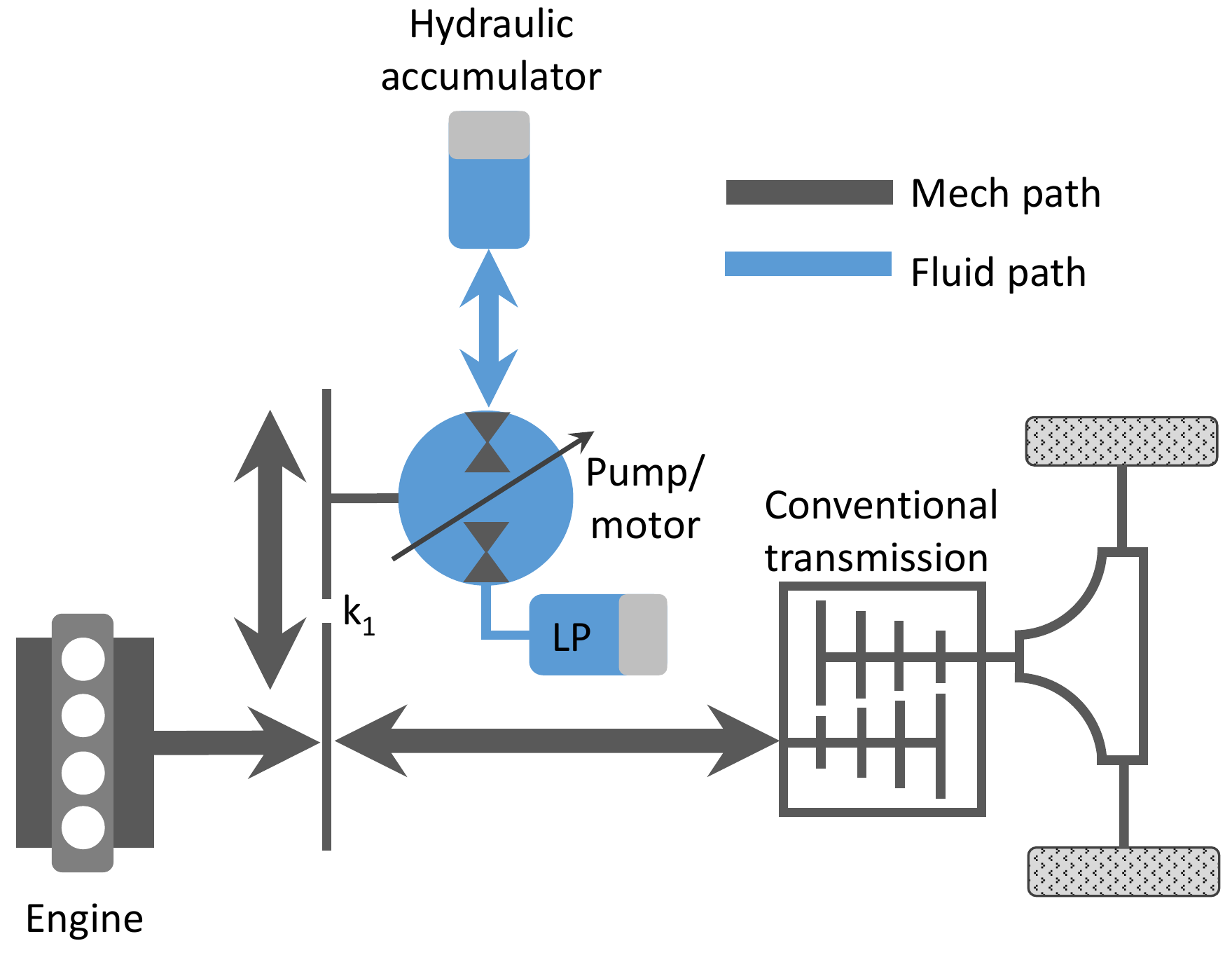}
	\end{minipage}
	\caption{Parallel hybrid hydraulic vehicle.}
	\label{fig:parallelHHV}
\end{figure}

The schematic of the parallel HHV is shown in Fig. \ref{fig:parallelHHV}.  Power from the engine can be supplemented with hydraulic power from a pump/motor unit, which can transfer power to/from the hydraulic accumulator.  
Essentially, the parallel HHV is a conventional power train augmented with a secondary power path.  As such, the engine can be downsized in the sense that maximum power can be achieved by supplementing available engine power with hydraulic power from the pump/motor unit.  
\subsubsection{Series HHV}
\begin{figure}[h!]
	\centering
	\begin{minipage}{.80\textwidth}
		\includegraphics [trim = 0mm 60mm 120mm 0mm, clip,width=1\textwidth]{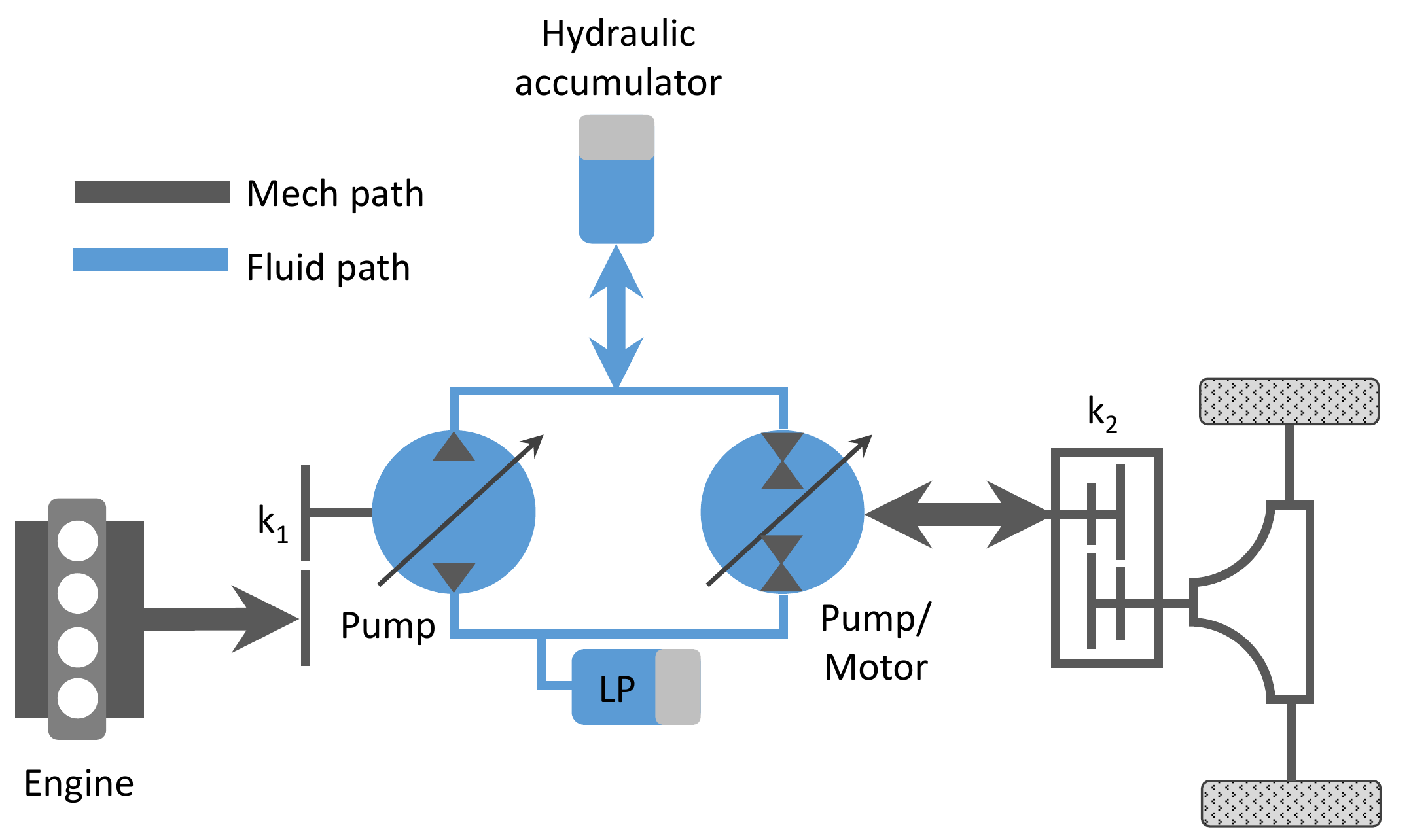}
	\end{minipage}
	\caption{Series hybrid hydraulic vehicle.}
	\label{fig:seriesHHV}
\end{figure}

The schematic of the series HHV with a two-stage output gearbox is shown in Fig. \ref{fig:seriesHHV}.  
Power from the engine is transmitted to a hydraulic pump which converts the mechanical power into pressurized fluid flow.  A hydraulic pump/motor unit converts the pressurized fluid flow into mechanical power as the source of vehicle propulsion.  During regenerative braking, the pump/motor units operates as a pump charging the hydraulic accumulator by transferring fluid from low pressure to the accumulator.   At low speeds the effective ratio between the motor and wheels, including the drive axle, is $k_2=k_{2,lo}$, while at higher speeds this ratio changes to $k_2=k_{2,hi}$.  Positive net flow between the pump and motor is transferred into the hydraulic accumulator, while negative net flow indicates fluid is being transferred from the hydraulic accumulator.  The pump/motor unit is designed such that, at maximum displacement $V_{m}^{max}$, maximum propulsive force can be achieved in low gear at some nominal system differential pressure $p^{\circ}$
\begin{align}
	F_{p}^{max} = \frac{p^{\circ} V_m^{max}}{2\pi}\frac{k_{2,lo}}{r_{tire}}
\end{align}
where system differential pressure is the difference between the hydraulic accumulator and low pressure, $p = p_{ha}-p_{lp}$.  The pump unit can be designed to deliver required flow rate at high speed through the following flow balance
\begin{align}
	n_{eng}^{max}k_1V_p^{max} = \frac{v_{veh}^{max}}{r_{tire}}k_{2,hi}V_m^{max}
\end{align}
Alternatively, the pump unit can be designed such that the engine can be loaded to maximum torque at some nominal system differential pressure $p^*$ through the following torque balance
\begin{align}
	p^* V_p^{max}\frac{k_1}{2\pi} = T_{eng}^{max}
\end{align}
\subsubsection{Series-Parallel HHV}
\begin{figure}[h!]
	\centering
	\begin{minipage}{.95\textwidth}
		\includegraphics [trim = 0mm 60mm 90mm 0mm, clip,width=1\textwidth]{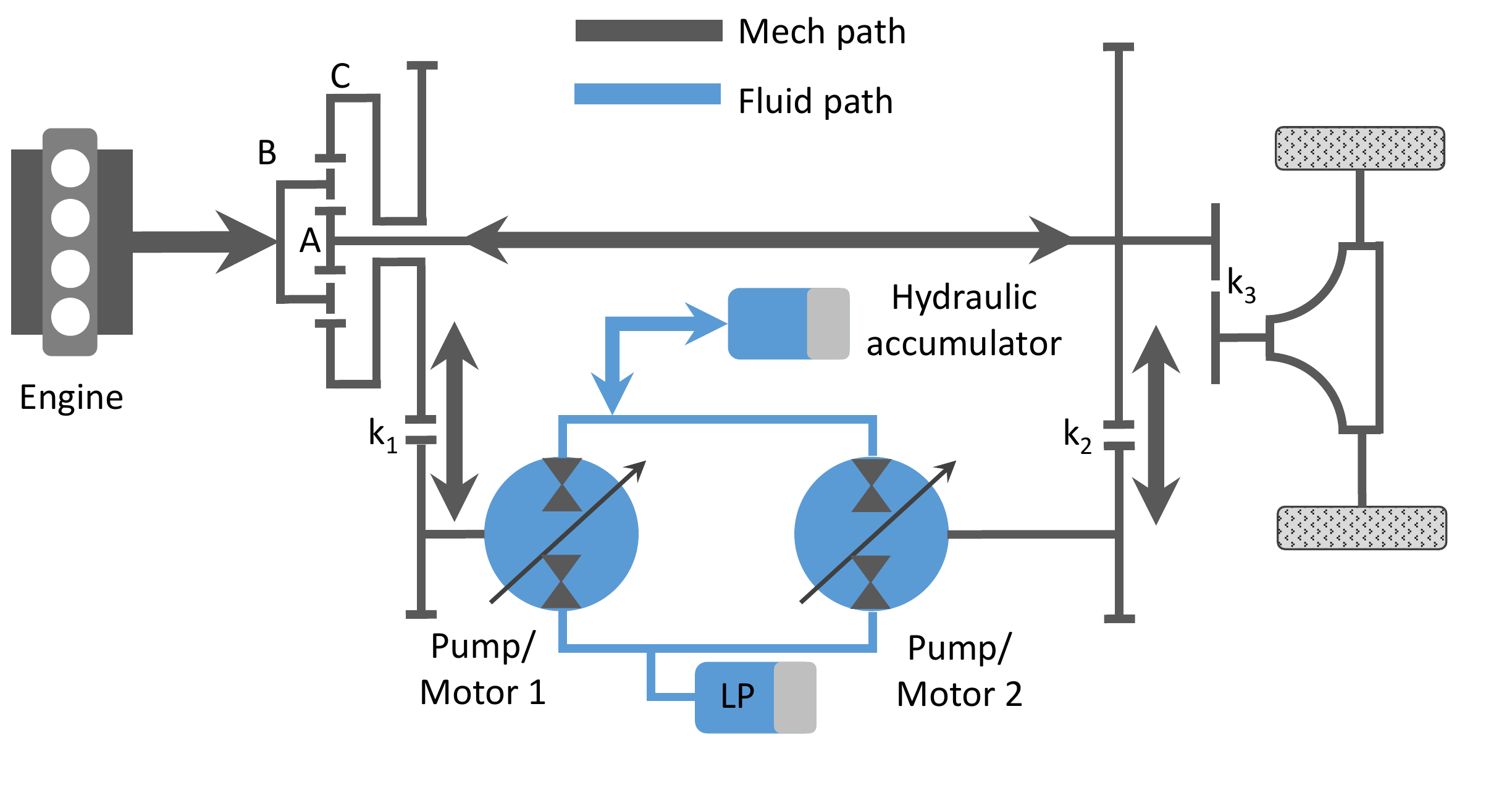}
	\end{minipage}
	\caption{Series-parallel hybrid hydraulic vehicle.}
	\label{fig:series-parallelHHV}
\end{figure}

\begin{figure}[h!]
	\centering
	\begin{minipage}{.8\textwidth}
		\includegraphics [width=1\textwidth]{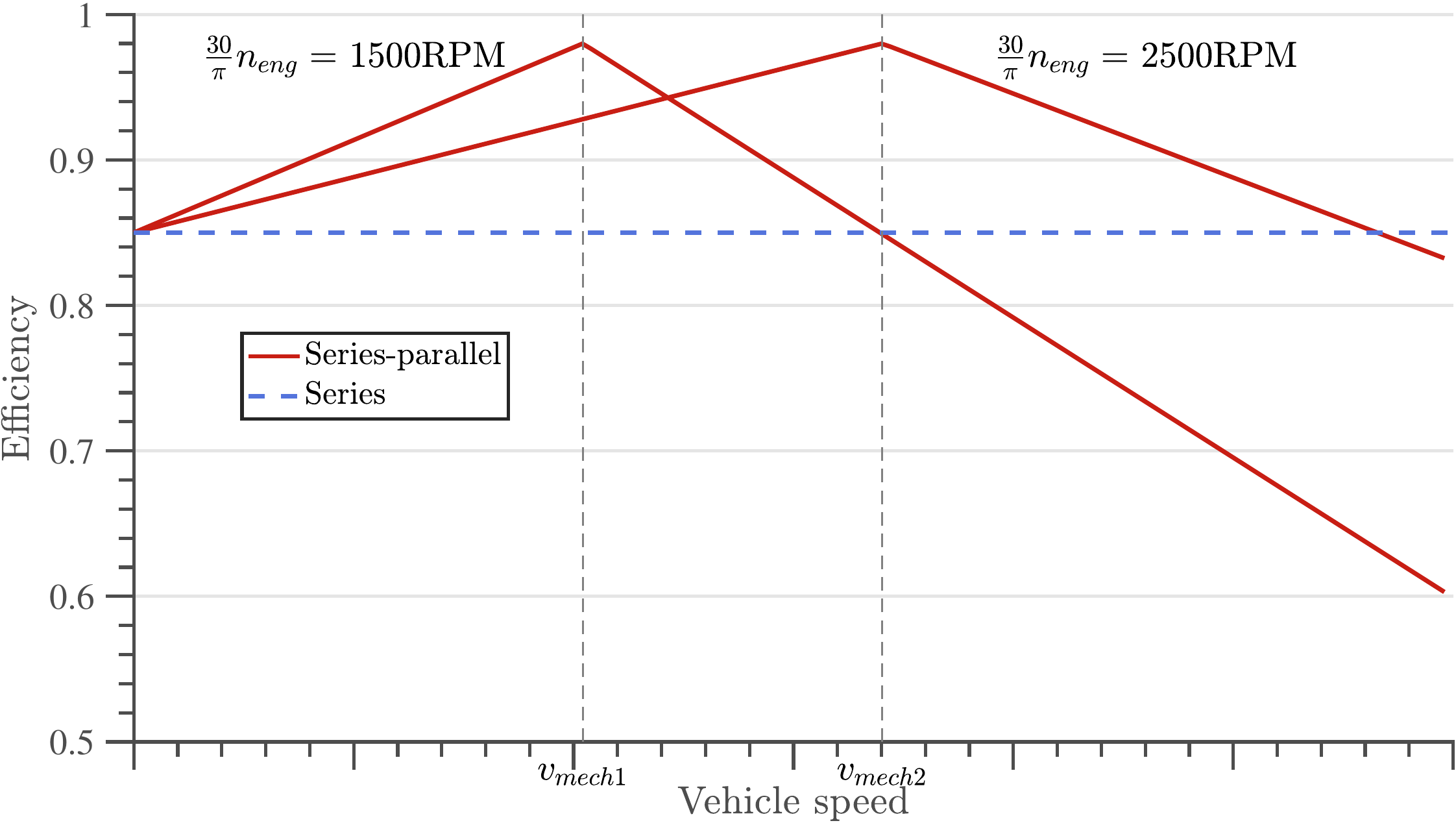}
	\end{minipage}
	\caption{Series-parallel vs. series HHV efficiency.}
	\label{fig:series-parallelEfficiency}
\end{figure}

The schematic of the series-parallel HHV is shown in Fig. \ref{fig:series-parallelHHV}. 
A defining feature of the series-parallel HHV is the planetary gear connected to the engine.  The planetary gear allows for power splitting between two separate paths.  The engine connects to the planetary gear via carrier gear B, while the hydraulic pump/motor unit 1 connects via ring gear C, and the output shaft and hydraulic pump/motor unit 2 are connected via sun gear A.  The planetary gear behavior is defined through the following speed, torque and power relationships between members $A,B,C$
\vspace{-1mm}
\begin{subequations}\label{eq:3PlanetaryGearEqs}
	\begin{align}
		& n_A-(1-k_0)n_B-k_0n_C = 0 \\
		& T_A = \frac{1}{k_0-1} T_B\\
		& T_C = \frac{-k_0}{k_0-1}T_B \\
		\text{mech power path:}~~~& P_A = \frac{1}{k_0-1}\frac{n_A}{n_B}P_B\\
		\text{hyd power path:}~~~& P_C = \left(1+\frac{1}{k_0-1}\frac{n_A}{n_B}\right) P_B
	\end{align}	
\end{subequations}
where planetary gear ratio $k_0$ is determined by the geometry of the planetary gear.  The last two equations indicate that the power split between the mechanical path and the hydraulic path is determined by the ratio of vehicle speed to engine speed as indicated by the term $\frac{n_A}{n_B}$.  From the last equation in Equation (\ref{eq:3PlanetaryGearEqs}), the power through the hydraulic path becomes zero when the ratio of sun gear speed to carrier gear speed becomes $	\frac{n_A}{n_B} = 1-k_0$.  This condition produces the most efficient point of power transfer within the series-parallel HHV known as the full-mechanical speed point given by
\begin{align}\label{eq:3series-parallelFullMech}
	v_{mech} = {(1-k_0)n_{eng}}\frac{r_{tire}}{k_3}
\end{align}
Efficiency vs. vehicle speed of the series-parallel HHV compared to the series HHV, assuming a fixed hydraulic path efficiency of 85\%, is shown in Fig. \ref{fig:series-parallelEfficiency}.
Efficiency of the series-parallel HHV declines past speed $v_{mech}$.  As such, the planetary gear ratio $k_0$ and drive gear ratio $k_3$ can be designed according to Equation (\ref{eq:3series-parallelFullMech}) so $v_{mech}$ occurs at some desired engine speed.  Gear ratio $k_2$ can be designed such that the speed of unit II is limited to some maximum value considering the maximum vehicle speed
\begin{align}
	n_\text{II}^{max} = \frac{v_{veh}^{max}k_3k_2}{r_{tire}}
\end{align}
Gear ratio $k_1$ can be designed considering maximum engine speed at zero vehicle speed, at which point unit I reaches maximum speed
\begin{align}
	n_\text{I}^{max} = \left.\frac{n_A-(1-k_0)n_{eng}^{max}}{k_0}k_1\right|_{n_A=0}
\end{align}
Hydraulic unit II can be designed so that maximum propulsion force is achieved at some nominal system differential pressure, $p^\circ$
\begin{align}
	F_{p}^{max} = p^{\circ}\frac{ V_\text{II}^{max}}{2\pi}\frac{k_{2}k_3}{r_{tire}}
\end{align}
Finally, hydraulic unit I can be designed so the engine can be loaded to its maximum torque capability at some nominal system differential pressure $p^*$
\begin{align}
	T_{eng}^{max} = p^*\frac{V_\text{I}^{max}}{2\pi}\frac{1-k_0}{k_0}k_1
\end{align}
\section{Series HHV Dynamics}\label{section:SeriesHHVDynamics}
\begin{figure}[h!]
	\centering
	\begin{minipage}{0.9\textwidth}
		\includegraphics [width=1\textwidth]{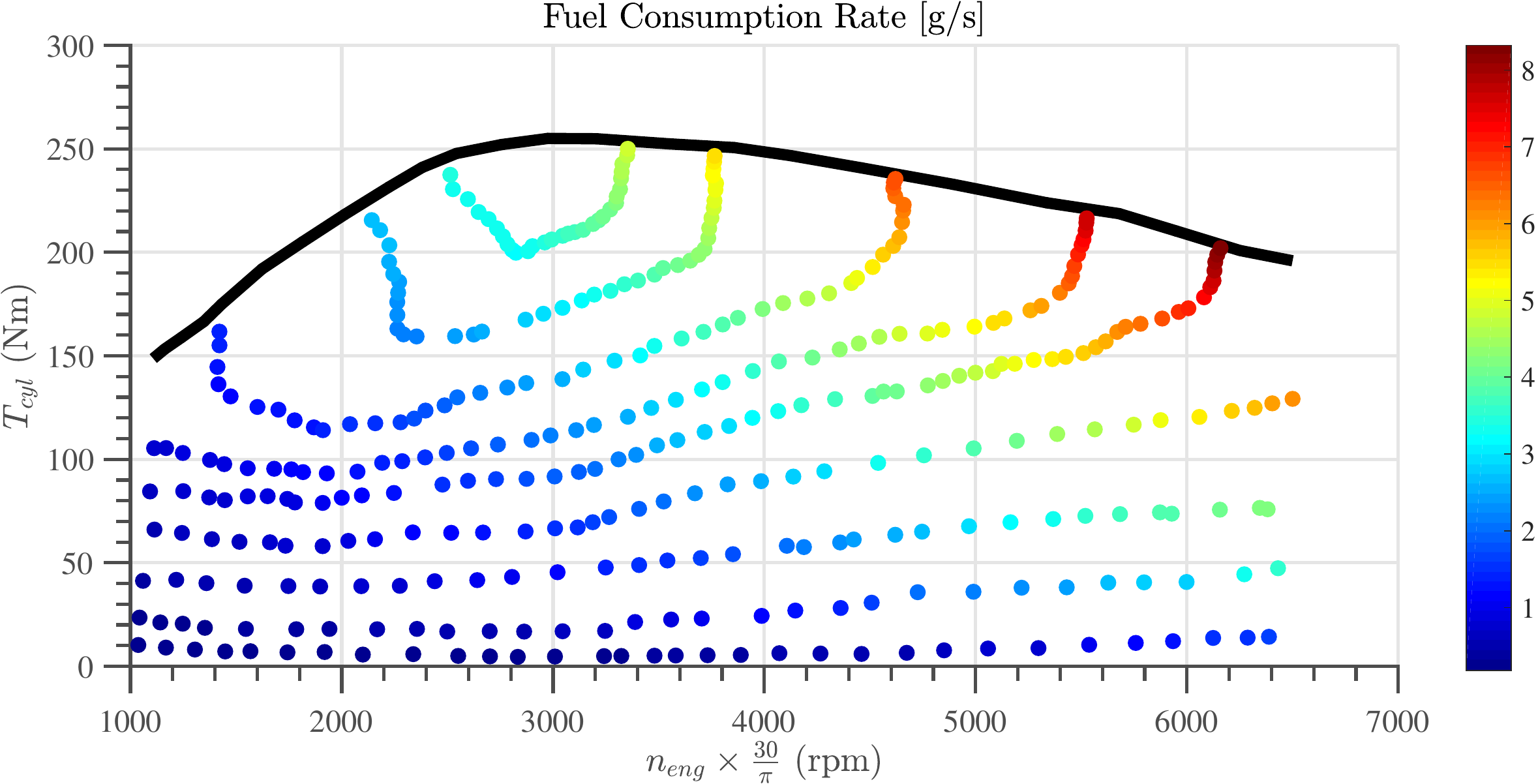}
	\end{minipage}
	\caption{Engine fuel consumption rate, $b_f(n_{eng},T_{cyl})$, and maximum torque curve, $T_{cyl}^{max}(n_{eng})$.}
	\label{fig:EngMap}
\end{figure}

\begin{figure}[h!]
	\centering
	\begin{minipage}{.4\textwidth}
		\includegraphics [trim = 0mm 0mm 0mm 10mm, clip,width=1\textwidth]{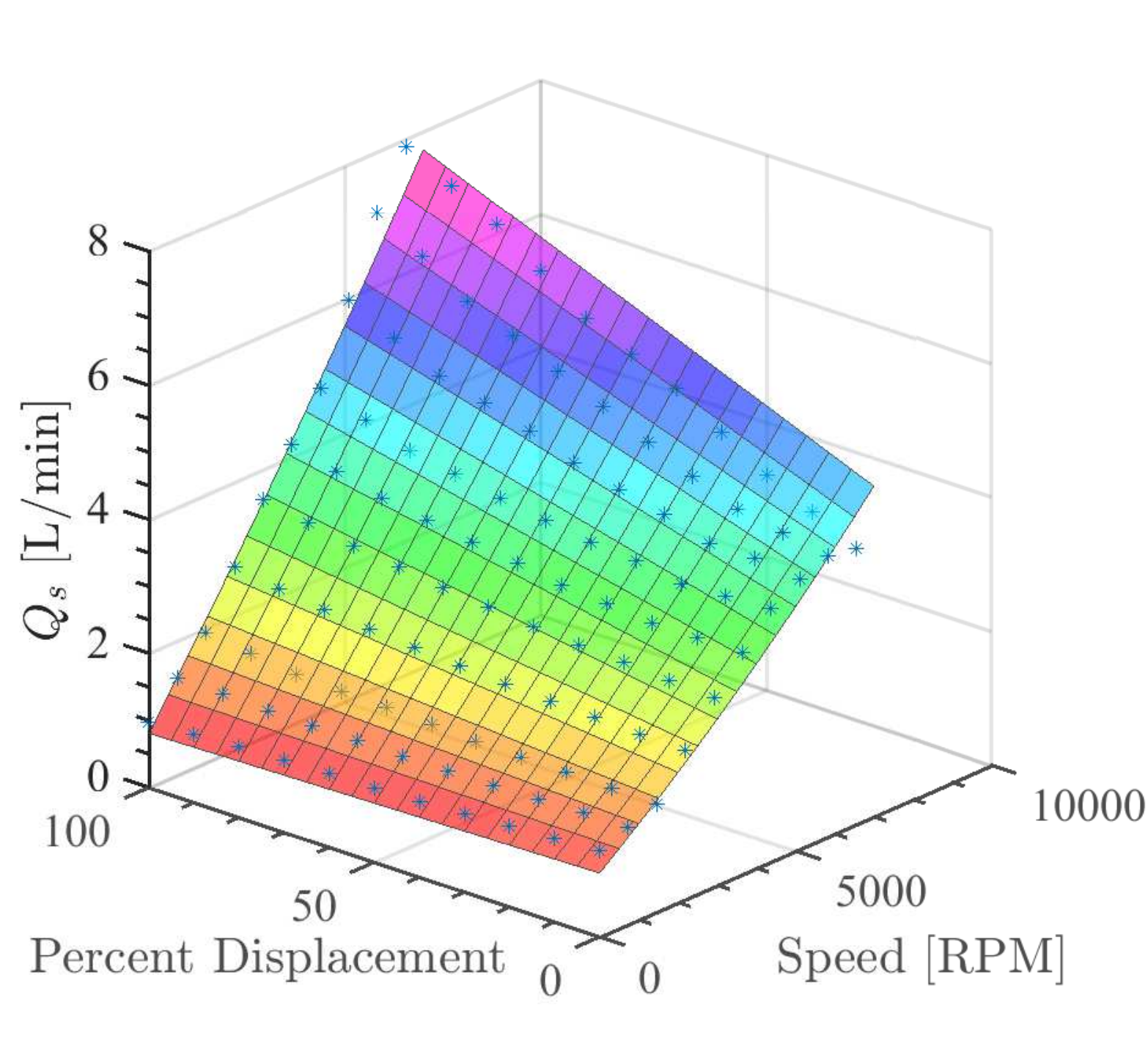}
	\end{minipage}~
	\begin{minipage}{.4\textwidth}
		\includegraphics [trim = 0mm 0mm 0mm 10mm, clip,width=1\textwidth]{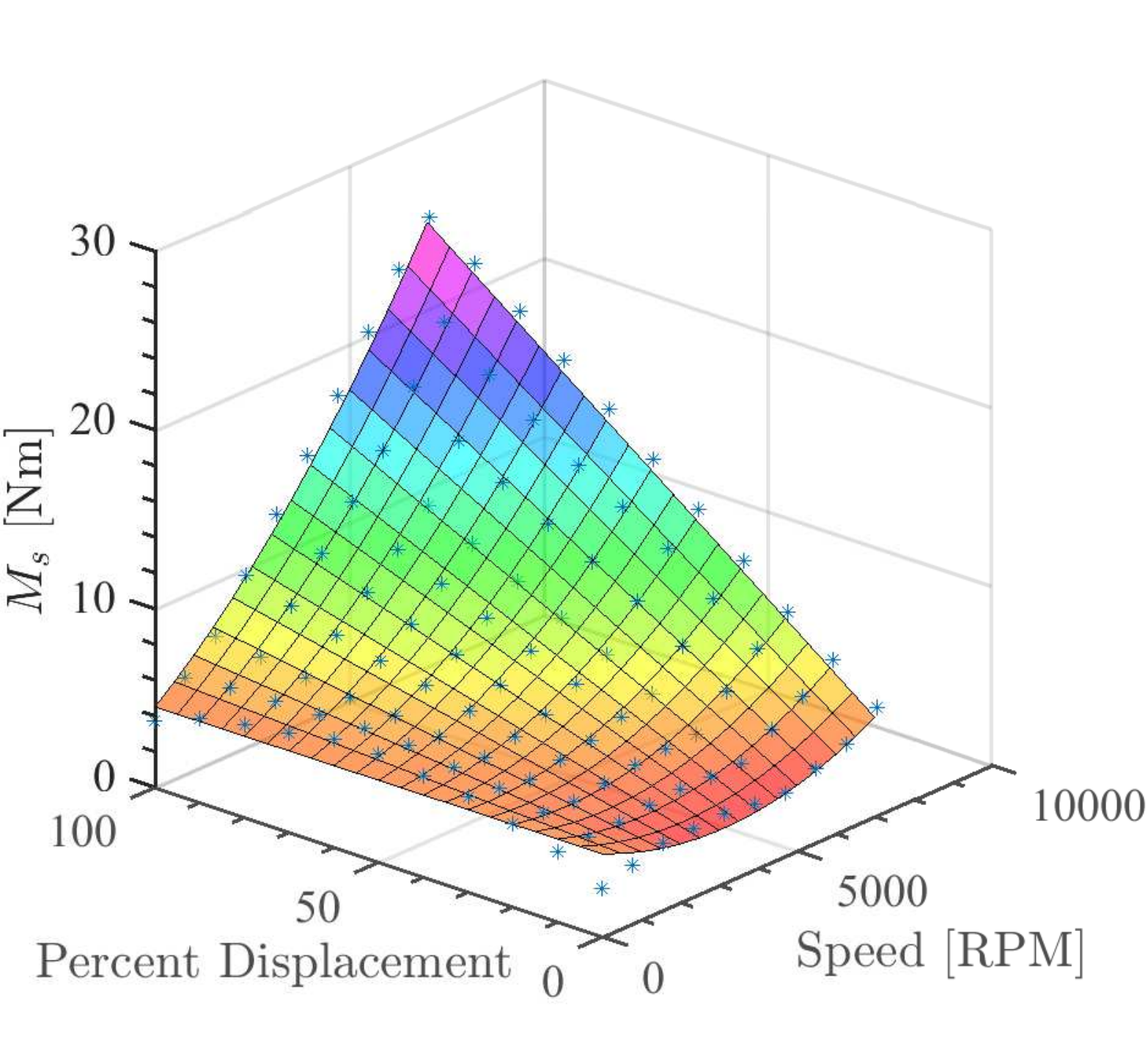}
	\end{minipage}
	\caption{$Q_{s,p}$ and $M_{s,m}$, $p=250$ bar for 60 cc/rev max displacement volume hydraulic unit. Data points in blue markers, second order polynomial fits $\hat Q_{s,p}$ and $\hat M_{s,m}$ shown as shaded surface.}
	\label{fig:QsMs}
\end{figure}

Due to its simple design and superior engine management capabilities, this work focuses on designing an optimal control strategy for the series hybrid shown in Fig. \ref{fig:seriesHHV}.  The vehicle velocity dynamic is given by
\begin{align}\label{eq:VehicleDynamics}
\dot v_{veh}(t) &= \frac{1}{m_{veh}}\Big[ F_p(t) - \tfrac{1}{2}C_d\rho_{air}v_{veh}(t)^2  - m_{veh}g\Big(C_r cos(\phi(t))+sin(\phi(t))\Big) \Big]
\end{align}
where $m_{veh}$ is vehicle mass, $\rho_{air}$ is air density and $g$ is the gravitational constant.  The terms $C_d$ and $C_r$ are drag and rolling resistance coefficients associated with the vehicle, where $\phi$ represents the road grade. The propulsive force $F_p$ is dependent on the differential system pressure\footnote{More generally, for safety, traditional friction brakes can be added so that the propulsion force becomes $F_p = \left(\frac{V_m}{2\pi} p - M_{s,m}\right)\frac{k_2}{r_{tire}} - {F_{brake}}$.  In this work, the friction brake force term $F_{brake}$ is neglected as its role in the drive cycles investigated was negligible.}, $p$, motor displacement volume $V_m$, motor torque losses $M_{s,m}$, and is limited by the maximum displacement volume of the motor, $V_m^{max}$
\begin{align}
F_p &= \left(\frac{V_m}{2\pi} p - M_{s,m}\right)\frac{k_2}{r_{tire}} \\
& \leq \left(\frac{V_m^{max}}{2\pi} p - M_{s,m}\right)\frac{k_2}{r_{tire}} \\
&= F_p^{max}(p)
\end{align}
The displacement volume of the hydraulic motor, $V_m$, is determined based on the applied force commanded by the driver, $F_p^{cmd}$
\begin{align}\label{eq:MotorDispCalc}
V_m &= \frac{2\pi}{p}\left(\frac{F_p^{cmd}r_{tire}}{k_2}+\hat M_{s,m}\right)
\end{align}
The term $\hat M_{s,m}$ is a polynomial approximation of the hydraulic motor torque loss term $M_{s,m}$.  In general, hydraulic system losses tend to increase as the system differential pressure increases.  As such, $p$ must be managed carefully as to satisfy driver propulsion demands ensuring $F_p^{cmd} \leq F_p^{max}(p)$ while simultaneously minimizing the losses experienced by the hydraulic system.  The dynamics of engine speed $n_{eng}$ and intake manifold pressure $p_{im}$ are given by \cite{KolmanovskyEngTrq,Heywood}
\begin{align}
\dot n_{eng}(t) &= \frac{1}{I_{eng}}\left[T_{cyl}(t) - \frac{k_1}{2\pi} V_p(t) p(t) - k_1 M_{s,p}(t) \right] \\
\dot p_{im}(t) &=\frac{R T_{im}}{V_{im}}\left[W_{thr}(t) -\frac{\eta_vV_d}{4\pi R T_{im}}n_{eng}(t)p_{im}(t)\right]
\end{align}
Here, $W_{thr}$ is throttle mass flow rate, $R$ is the ideal gas constant for air, $T_{im}$ is the intake manifold temperature, $\eta_v$ is volumetric efficiency of the engine, $V_d$ and $V_{im}$ are the volumes of the engine displacement and intake manifold.  The torque produced by the engine cylinders, $T_{cyl}$, is determined from the engine thermal efficiency, $\eta_t$, the lower heating value of the fuel $Q_{lhv}$, the air-fuel ratio in the cylinders, $AFR$, and the inducted air mass in the cylinders $m_{cyl}$ \cite{Heywood}
\begin{align}
m_{cyl} &= \frac{\eta_v V_d}{R T_{im}}p_{im} \\
T_{cyl} &= \frac{\eta_tQ_{lhv}}{4\pi AFR}m_{cyl} = \frac{\eta_t\eta_vQ_{lhv}V_d}{4\pi R T_{im} AFR}p_{im} \label{eq:Tcyl}
\end{align}
The maximum capability of the engine in this work is 125 kW, as the engine speed is limited to 5000 RPM.  The maximum torque curve as a function of engine speed and fuel consumption rate, $b_f$, as a function of engine speed and torque are described by Fig. \ref{fig:EngMap}
The dynamic of the hydraulic differential system pressure $p$ is 
\begin{align}
\dot p(t) &=\frac{1}{C_h(p)} \Big[\frac{k_1}{2\pi}V_p(t)n_{eng}(t)-\frac{k_2}{2\pi r_{tire}}V_m(t)v_{veh}(t)-Q_{s,p}(t)-Q_{s,m}(t)\Big] 
\end{align}
where $Q_{s,p}, Q_{s,m}$ are the flow losses of the pump and motor, $k_1, k_2$ are gear ratios, and $V_p$ is the displacement volume of the hydraulic pump.  It is assumed here that low pressure is nearly constant.  The capacitance of the hydraulic system \cite{Rahmfeld} is
\begin{align}
C_h(p) = \frac{V_{ha}p_{ha}^{1/\gamma_{gas}}}{\gamma_{gas}(p+p_{lp})^{1+1/\gamma_{gas}}} + \frac{V_L}{K_L}
\end{align}
where $V_{ha}, p_{ha}$ are the pre-charge volume and pressure of the hydraulic accumulator, $\gamma_{gas}$ is the specific heat ratio of the pressurized gas within the accumulator, $p_{lp}$ is the pressure of the low-pressure system and $V_L, K_L$ are the volume and bulk modulus of the hydraulic lines.  Example hydraulic losses and their second order polynomial approximations are shown in Fig. \ref{fig:QsMs} 

\chapter{STATISTICAL MODEL OF DRIVER BEHAVIOR} 

\section{Driver Behavior as a Markov Process} \label{section:DriverDynamics}
Driver behavior is characterized in terms of an acceleration demand, $w$, which can be inferred from the driver's propulsive force command $F_p^{cmd}$ through\footnote{It is assumed $F_p^{cmd}$ can be inferred, for example, from driver foot pedal position.  During simulation and experiments in this work $F_p^{cmd}$ is the output of a PI feedback process used to track a vehicle speed reference.}
\begin{align}\label{eq:wMeasured}
w &= \frac{1}{m_{veh}}\Big[F_{p}^{cmd} -  \tfrac{1}{2}C_d\rho_{air}v_{veh}^2 - m_{veh}g\left(C_r cos(\phi)+sin(\phi)\right)\Big] 
\end{align}  
where $\phi$ is the road grade, assumed available from measurement or estimation.  If the driver acceleration demand $w$ can be forecast along a horizon to some statistical accuracy, then a control strategy which incorporates an underlying statistical model can be designed.  It is well known that driver behavior can be modeled effectively as a Markov process \cite{PentlandLiu,SDP1,StochDriverLearning}, a type of stochastic process which adheres to the \textit{Markov property}.  The Markov property roughly states that future behavior of the process is influenced only by the present state, unaffected by the sequence of events that lead to the present state.  More specifically, the stochastic process $\{w_0,w_1,w_2,\dots\}$ is Markovian if
\begin{align}\label{eq:MarkovCondition}
	\mathrm{Pr}[w_{n+1} = w^j| \mathcal F_n] = \mathrm{Pr}[w_{n+1}=w^j | w_n=w^i]
\end{align}
where each $w_n\in W$ is a random variable and $w^i$ and $w^j$ are realizations of the random variables $w_n$ and $w_{n+1}$, respectively.  Equation (\ref{eq:MarkovCondition}) states that the probability of the next transition given all prior information up to time $n$ is the same as the probability of the next transition given the information only of the previous state\footnote{A discrete time deterministic dynamic system described by dynamics $F(x_n,u_n)$ and some initial condition can be viewed as obeying condition Equation (\ref{eq:MarkovCondition}), where $\mathrm{Pr}[x_{n+1}=F(x,u)|x_n=x]=1$,  $\mathrm{Pr}[x_{n+1}\neq F(x,u)|x_n=x]=0$}.  If, in addition to satisfying the Markov property, the process is also time invariant then 
\begin{align}\label{eq:MarkovConditionTimeInv}
	\mathrm{Pr}[w_{n+1}=w^j | w_n=w^i] = \mathrm{Pr}[w_{m+1}=w^j | w_m=w^i]
\end{align}
for any $n,m \geq 0$.  The benefit of working with assumptions Equation (\ref{eq:MarkovCondition}) and Equation (\ref{eq:MarkovConditionTimeInv}) is that all subsequent computations in the energy management strategies developed in the next chapter are greatly simplified.  

In this work the driver acceleration demand $w$ is modeled as a discrete state discrete time Markov process.  Each transition is described by the probability distribution matrix $(P_{ij})$ whose elements are defined as
\begin{align}	\label{eq:MarkovPij}
P_{ij} &\triangleq \mathrm{Pr}[w_{n+1} = w^j | w_n = w^i]
\end{align}
The multi-step probability $P_{ij}^{(n)}$ describes the probability of a demand at time $n$ given the value of the demand at time $0$ 
\begin{align} \label{eq:MarkovMultiStepPij}
P_{ij}^{(n)} \triangleq \mathrm{Pr}[w_{n} = w^j | w_0 = w^i]
\end{align}
and, as the notation suggests, is computed by raising matrix  $(P_{ij})$ to the exponent $n$ and selecting the $ij^{th}$ element \cite{Lawler2006StochProcesses}.  The multi-step distribution will be used extensively in the development of a stochastic strategy described in Section \ref{section:ASDDP}.

Driver acceleration demand $w$ is quantized evenly into 19 levels, $w^i$, $i=1,2,\dots,19$, between $-3$ to $3~ m/s^2$.  Any acceleration demand lower than $-3~ m/s^2$ is associated with $w^1$ while any acceleration demand greater than $3~ m/s^2$ is associated with $w^{19}$.

\begin{figure}[h!]
	\centering
	\begin{minipage}{0.8\textwidth}
		\includegraphics [trim = 0mm 0mm 0mm 0mm, clip,width=1\textwidth]{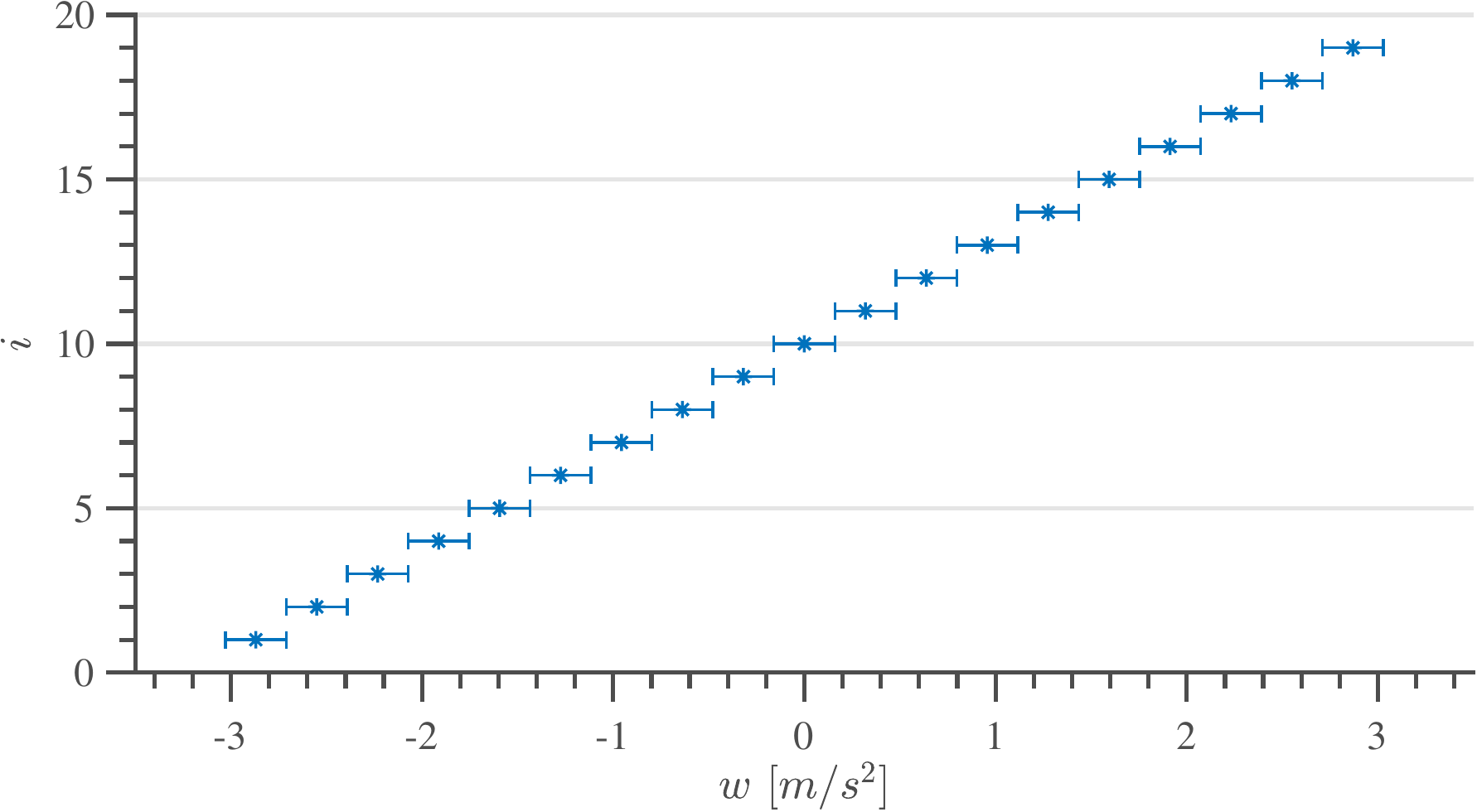}
	\end{minipage} 	
	\caption{Quantization of driver acceleration demand.}
	\label{fig:wDiscretization}
\end{figure}


\section{Learning Driver Behavior} \label{section:LearningDriverBeahvior}

\begin{figure}[h!]
	\centering
	\begin{minipage}{\textwidth}
		\includegraphics [trim = 0mm 0mm 0mm 0mm, clip,width=1\textwidth]{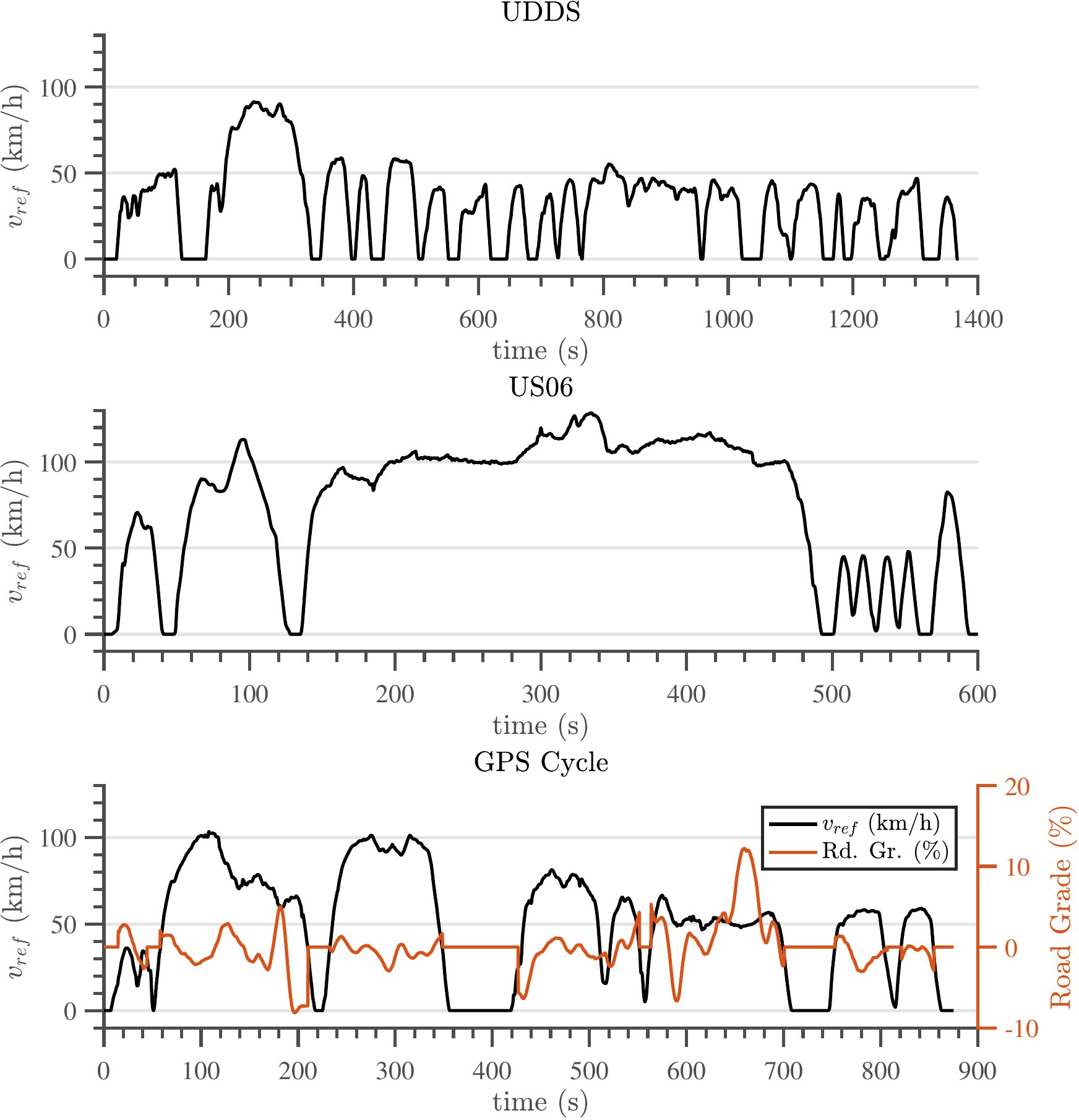}
	\end{minipage} 	
	\caption{Drive cycles investigated.}
	\label{fig:DriveCycles}
\end{figure}

In this work three primary drive cycles are considered, shown in Fig. \ref{fig:DriveCycles}.  
The first drive cycle is the EPA's Urban Dynamometer Driving Schedule (UDDS), a representative urban drive cycle with frequent stops having an average speed of 31.5 km/h and a total run time of approximately 23 minutes.   The second drive cycle is the EPA's aggressive urban drive cycle (US06).  Having an average speed of 78 km/h with a short runtime of 10 minutes, the US06 cycle was developed by the EPA in response to criticism of the UDDS cycle's inability to represent aggressive, high speed and/or high acceleration driving with rapid speed fluctuations.  The third drive cycle, referred to as the GPS cycle, is moderate traffic city driving data from West Lafayette, IN and includes altitude data collected by an on-board GPS device.  The GPS cycle has a total runtime of approximately 15 minutes.

A sequence of driver acceleration demands $\{w_0,w_1,w_2,\dots\}$ is created from Equation (\ref{eq:wMeasured}) according to $w_n = w(n\Delta t)$, with sampling rate $\Delta t=1$ second.  Estimates of the Markov single-step transition probabilities at time step $n$, $n=1,2,3,\dots$, denoted $\hat P_{ij}^{[n]}$, are determined through a first order filtering process according to \cite{StochDriverLearning}
\begin{align}\label{eq:PijEstimator}
\hat{P}_{ij}^{[n]} &= \left\{
\begin{array}{ll}
\alpha \mathds{1}_{ij}^{[n]} + (1-\alpha)\hat{P}_{ij}^{[n-1]} & \text{if~}  w_{n}=w^i\\
\hat{P}_{ij}^{[n-1]}  & \text{if~} w_n \neq w^i
\end{array}
\right.
\end{align}
where $\hat{P}_{ij}^{[0]}$ is an arbitrary initialization and the indicator function $\mathds{1}_{ij}^{[n]}$ is defined by
\begin{align}\label{eq:PijEstimatorIndicator}
\mathds{1}_{ij}^{[n]}&=\left\{
\begin{array}{ll}
1& \text{if~} w_{n+1}=j,w_n=w^i\\
0&\text{if~} w_{n+1}\neq j , w_n=w^i
\end{array}
\right.
\end{align}
The updates described by Equation (\ref{eq:PijEstimator}) and Equation (\ref{eq:PijEstimatorIndicator}) are performed for all $w^i,w^j\in {W}$ at each time step $n$.  The parameter $\alpha \in [0,1]$ is the learning rate that determines the exponential rate at which the dependence on past information is decreased.  This estimation process produces an unbiased estimate as now shown.  Let $(n_k,~k=1,2,3,\dots)$ be an indexed sequence of time steps in which the chain is in state $w^i \in W$, and assume that each state $w^i$ is visited an infinite number of times (i.e., $k\rightarrow \infty$)
\begin{align*}
\mathbb E[\hat P_{ij}^{[n_k]}] &= \alpha \mathbb{E}[\mathds{1}_{ij}^{[n_k]}] + (1-\alpha) \mathbb E[\hat P_{ij}^{[n_{k-1}]}] \\
&= \alpha \mathbb E[\mathds{1}_{ij}^{[n_k]}] + (1-\alpha) \Big[\alpha \mathbb E[\mathds{1}_{ij}^{[n_{k-1}]}] \\
&~~~~~~~+ (1-\alpha) \mathbb E[\hat P_{ij}^{[n_{k-2}]}]\Big] \\
&= \alpha \mathbb E[\mathds{1}_{ij}] \underbrace{\sum_{m=0}^{k-2} (1-\alpha)^m}_{\rightarrow \frac{1}{\alpha}} + \underbrace{(1-\alpha)^{k-1}}_{\rightarrow 0} \mathbb E[\hat P_{ij}^{[n_1]}]\\
& \rightarrow \mathbb E[\mathds{1}_{ij}] = \mathrm{Pr}[w_{n+1}=w^j|w_n=w^i] = P_{ij}\text{ as $k \rightarrow \infty$}
\end{align*}
In the third equality above, it is noted that $\mathbb E[\mathds{1}_{ij}^{[n_k]}]=E[\mathds{1}_{ij}]$ for every $n_k$ since each $\mathds{1}_{ij}^{[n_k]}$ is a iid copy of the random variable $\mathds{1}_{ij}$ for each fixed $w^i \in W$.  Since  $\mathbb E[\hat P_{ij}] \rightarrow P_{ij}$
, the estimator is \textit{unbiased}.  By a slight abuse of notation, $P_{ij}$ and estimate $\hat P_{ij}$ are used interchangeably throughout the remainder of this work.  

\begin{figure}[h!]
	\centering
	\begin{minipage}{.48\textwidth}
		\includegraphics [width=.99\textwidth]{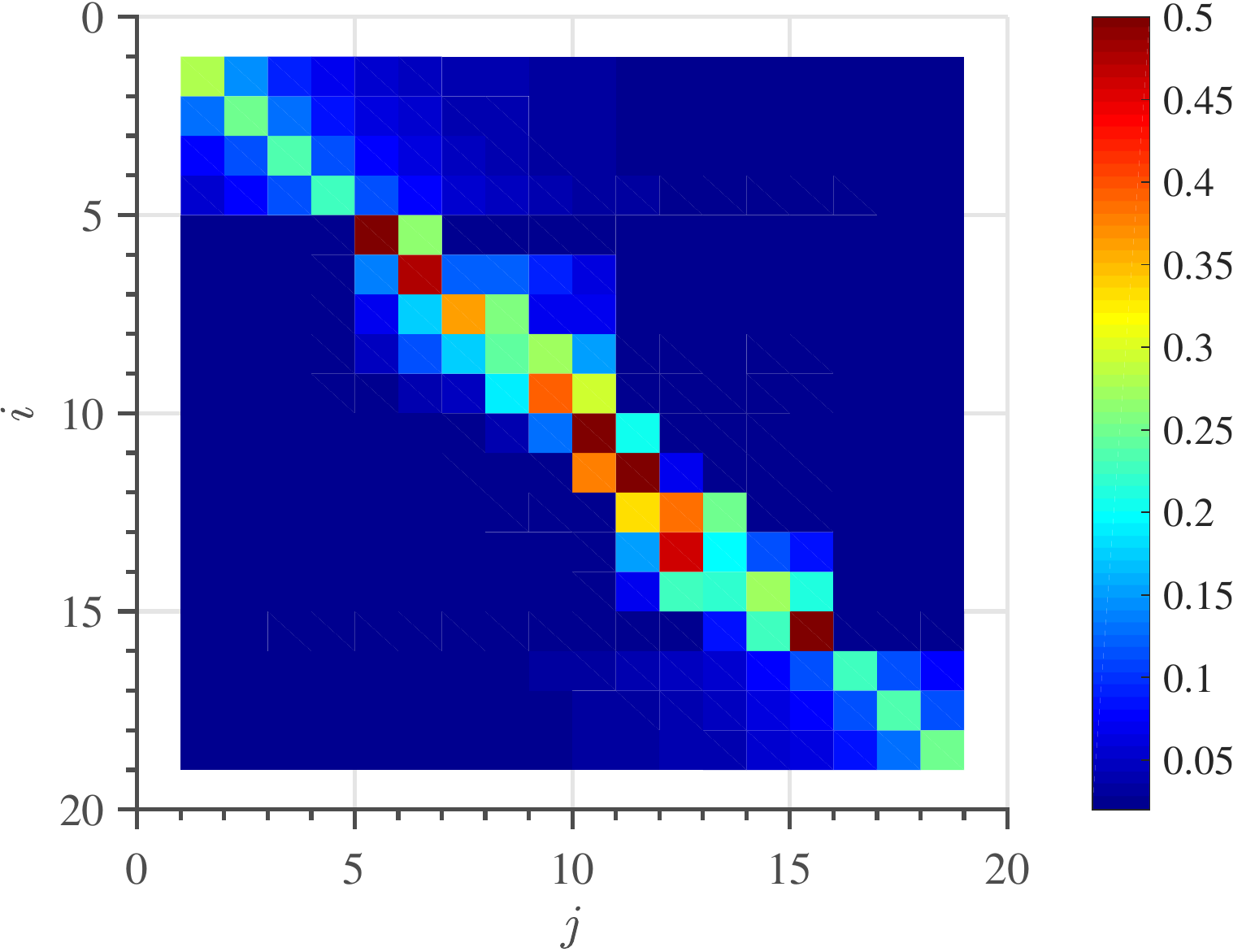}
	\end{minipage}~
	\begin{minipage}{.48\textwidth}
		\includegraphics [width=.99\textwidth]{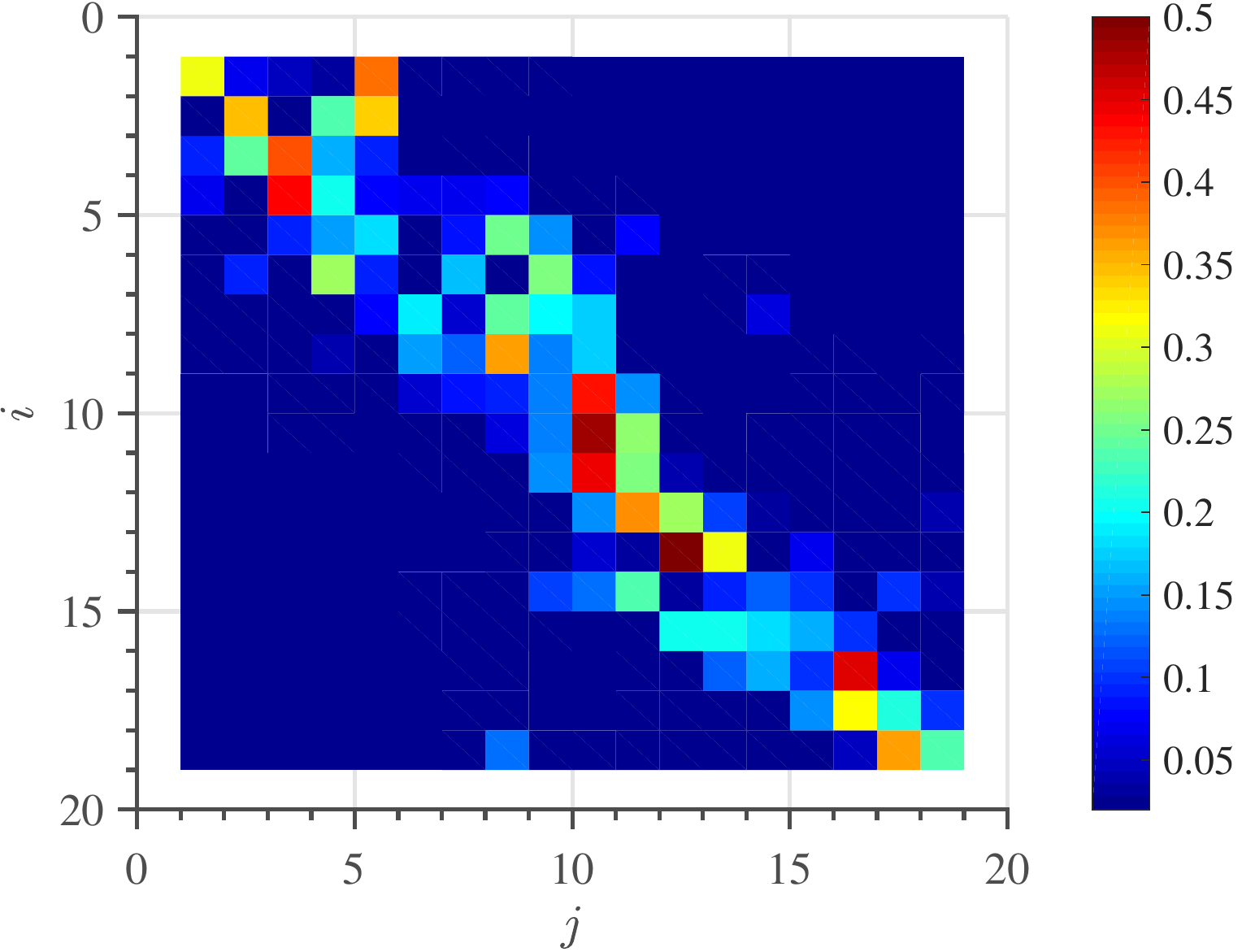}
	\end{minipage} 
	\begin{minipage}{0.49\textwidth}
		\includegraphics [width=0.99\textwidth]{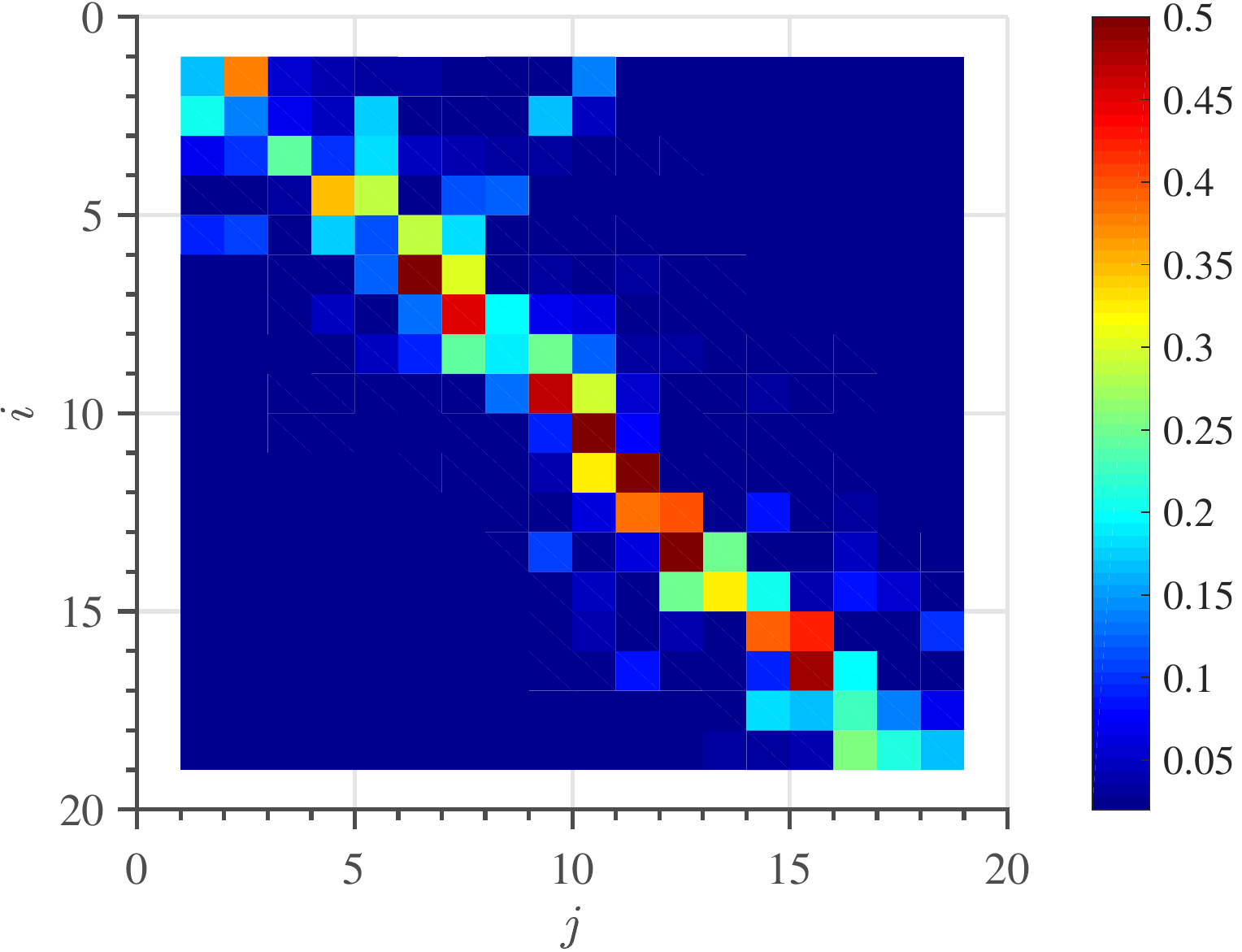}
	\end{minipage} 	
	\caption{$(P_{ij})$ for UDDS drive cycle (upper left), US06 drive cycle (upper right), and GPS drive cycle (lower).}
	\label{fig:DriveCyclePdT}
\end{figure}

The transition probability matrix $(P_{ij})$ is learned according to Equation (\ref{eq:PijEstimator}) and Equation (\ref{eq:PijEstimatorIndicator}) for each drive cycle described in Fig. \ref{fig:DriveCycles}.  The learning rate is chosen as $\alpha = 0.025$ so that only 20\% of the initial estimate $\hat P_{ij}^{[0]}$ is retained in memory after 60 transitions from $i$ to $j$ (the influence of $\hat P_{ij}^{[0]}$ on $\hat P_{ij}^{[n]}$ is $\hat P_{ij}^{[0]}(1-\alpha)^n$).  The matrices $(P_{ij})$ shown in Fig. \ref{fig:DriveCyclePdT} are color coded so that dark red indicates a transition probability that is greater than 0.5, while dark blue indicates a value near 0.  All three matrices show a somewhat similar pattern along the diagonal, in that the driver tends to demand an acceleration level at the next time step that is near his or her current demand.  However, the degree to which the driver chooses a slightly higher or lower demand at the next time step varies greatly with the drive cycle.  In the UDDS cycle the driver has a strong preference to operate along the diagonal, while in the US06 cycle the driver is much more likely to choose an off-diagonal transition.  During the GPS cycle, driver behavior appears to be somewhat of a mixture of behavior from UDDS and US06 cycles.

\begin{figure}[h!]
	\centering
	\begin{minipage}{0.5\textwidth}
		\includegraphics [width=1\textwidth]{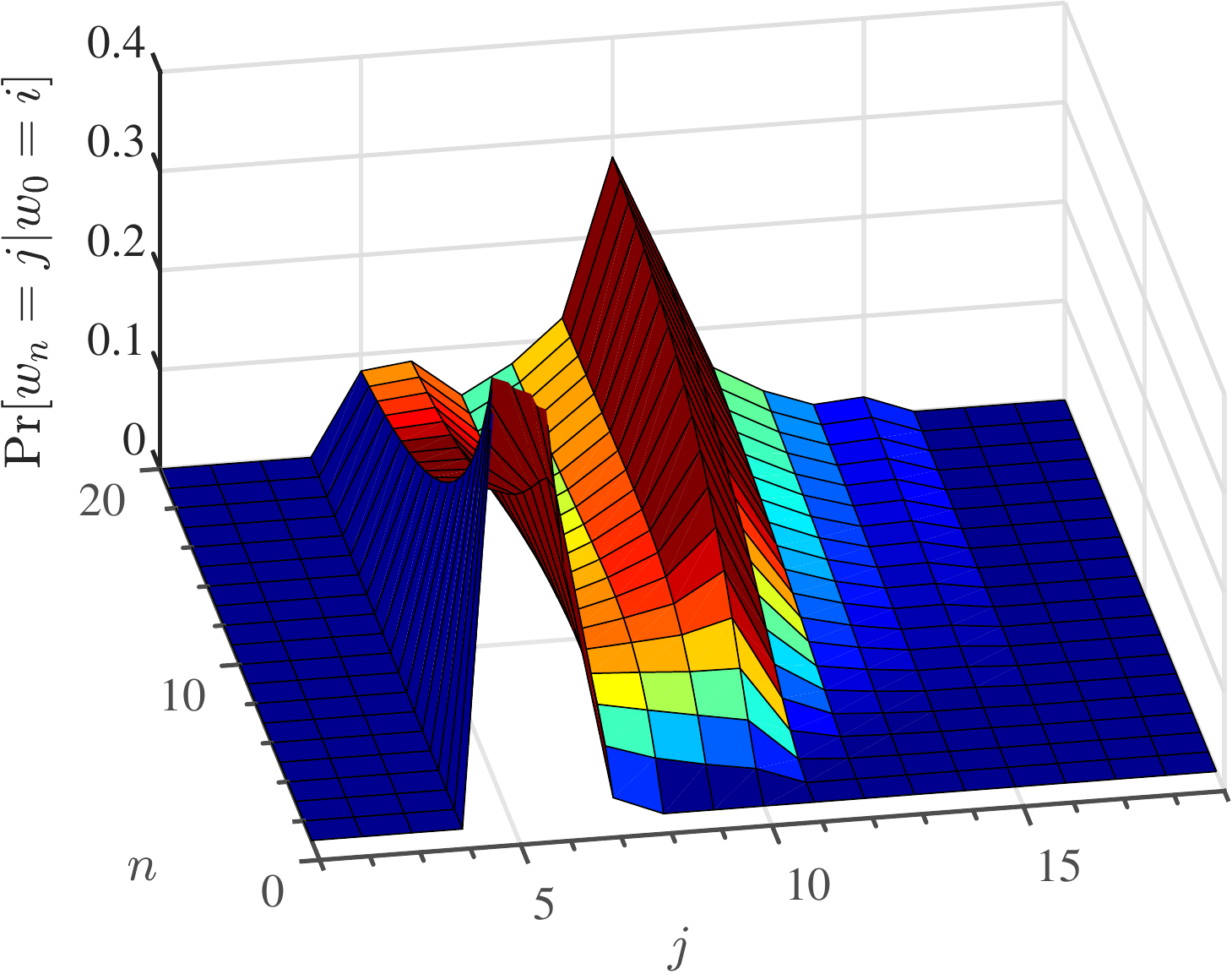}
	\end{minipage}~
	\begin{minipage}{0.5\textwidth}
		\includegraphics [width=1\textwidth]{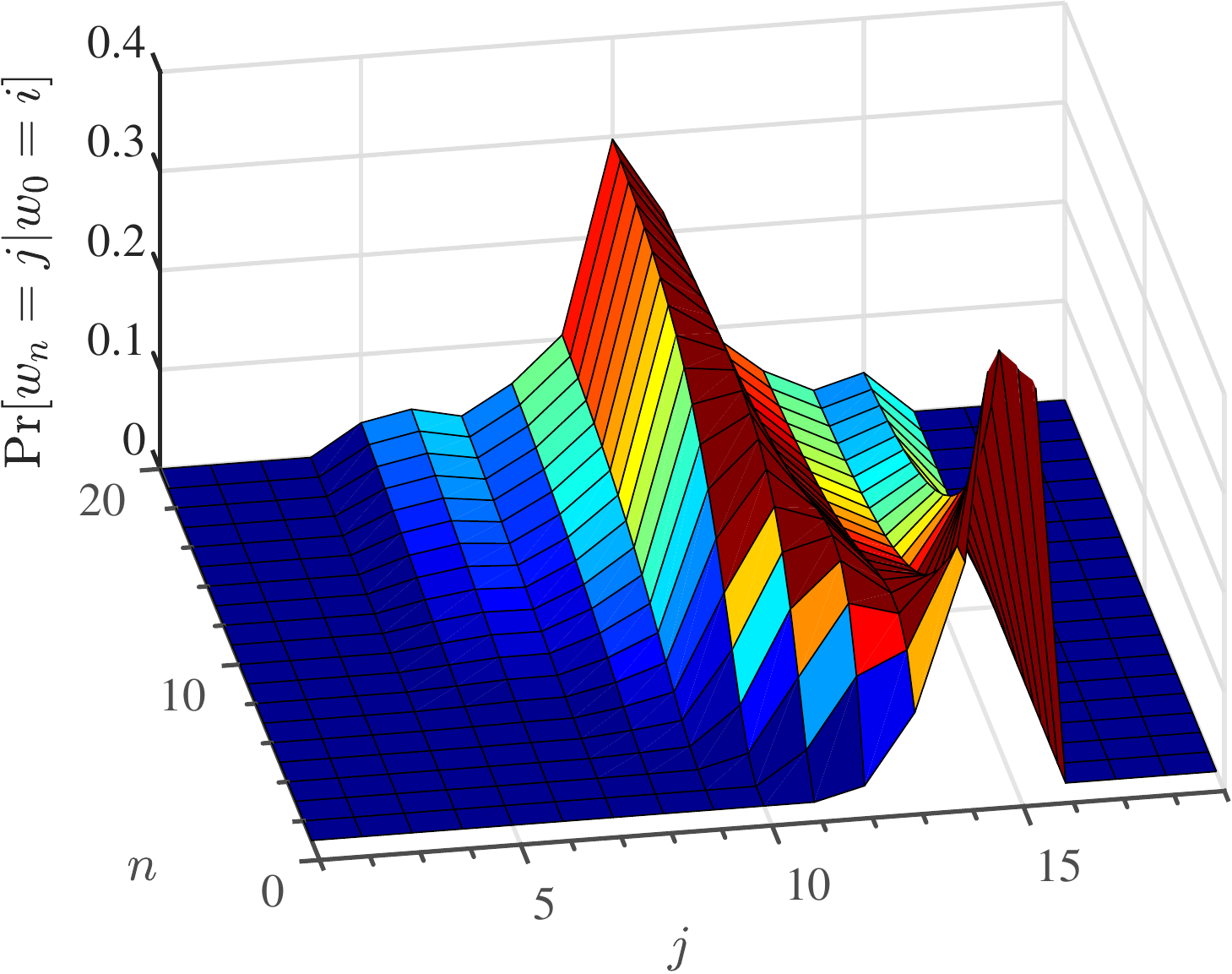}
	\end{minipage} 	
	\begin{minipage}{0.5\textwidth}
		\includegraphics [width=1\textwidth]{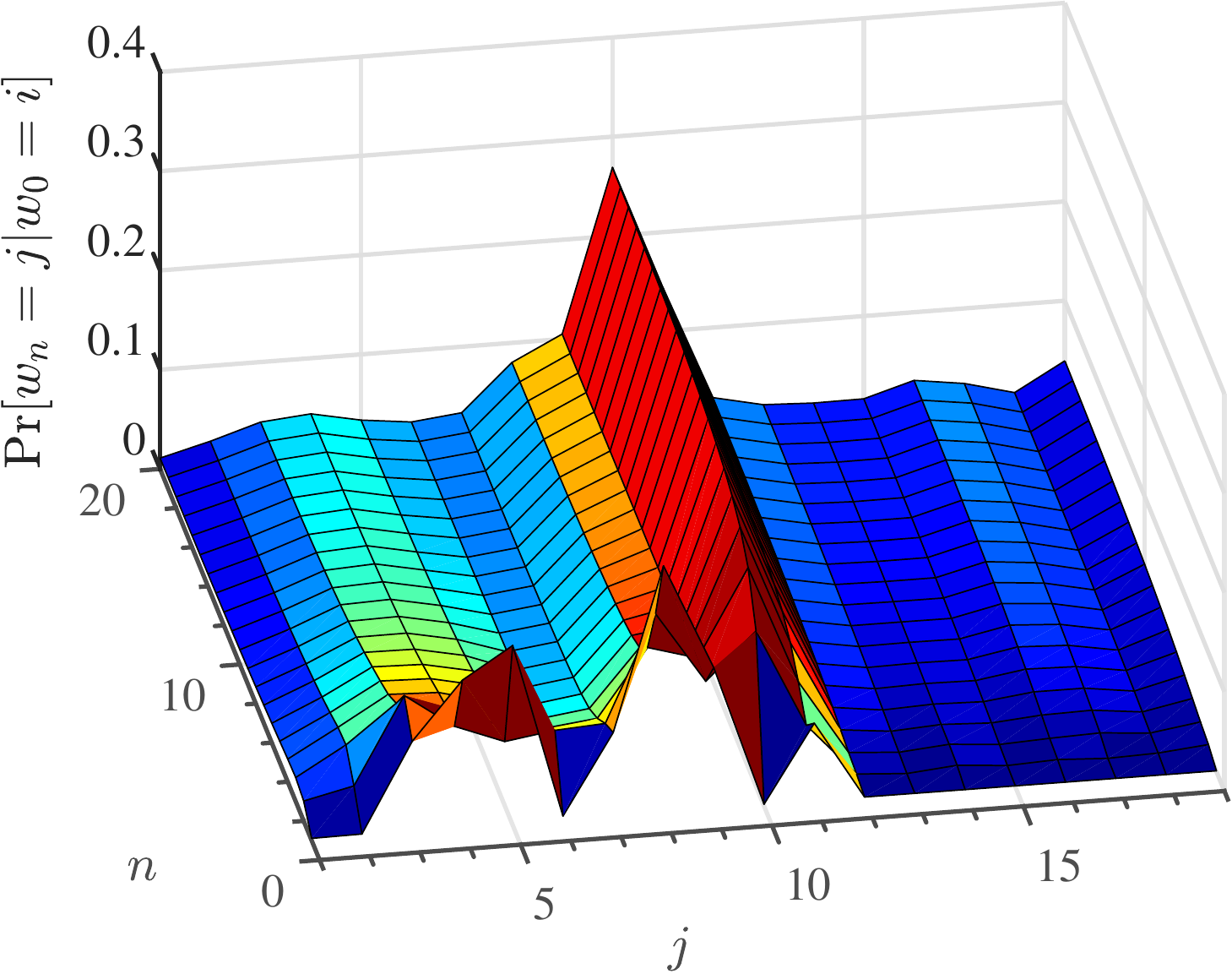}
	\end{minipage}~
	\begin{minipage}{0.5\textwidth}
		\includegraphics [width=1\textwidth]{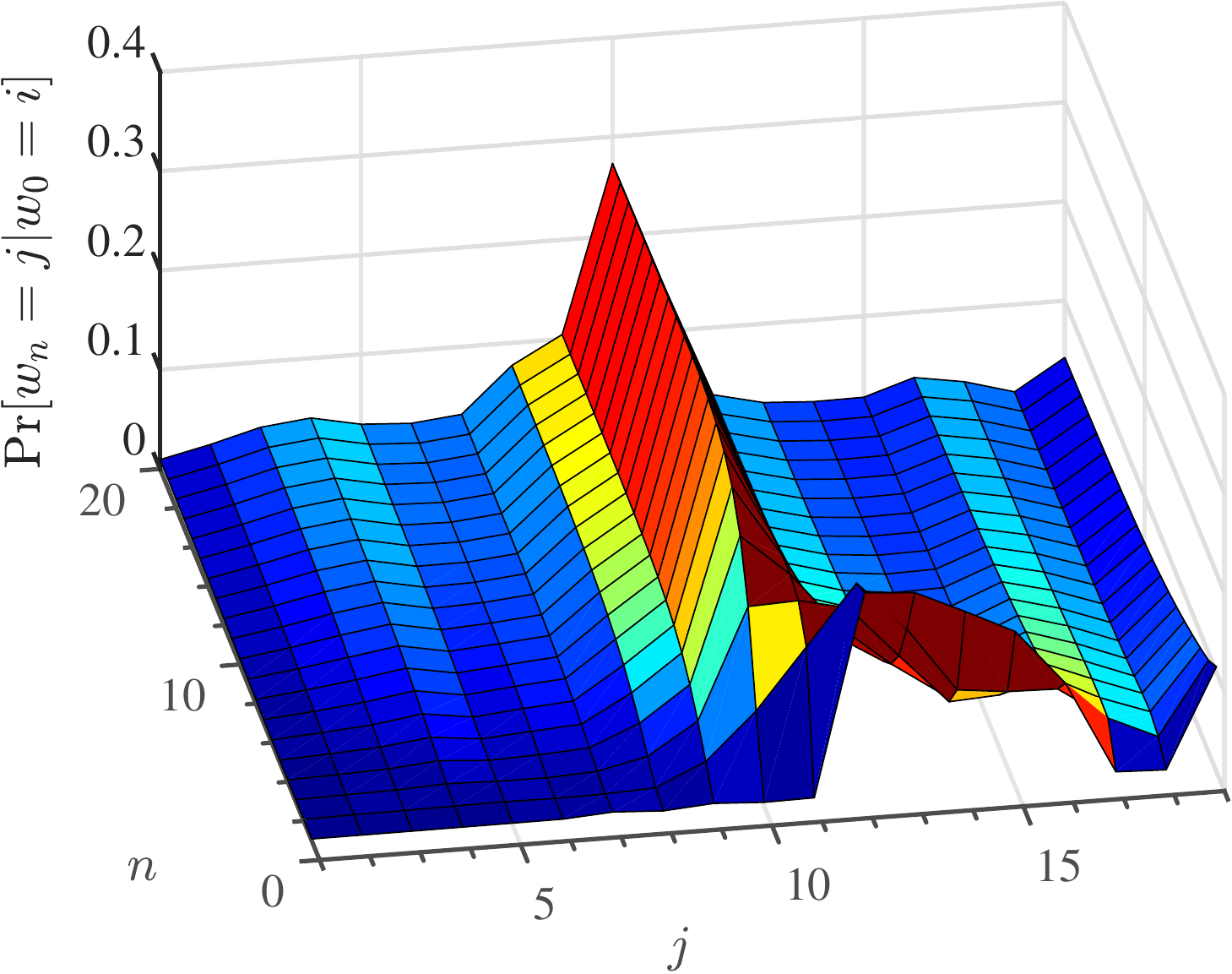}
	\end{minipage} 	
	\begin{minipage}{0.5\textwidth}
		\includegraphics [width=1\textwidth]{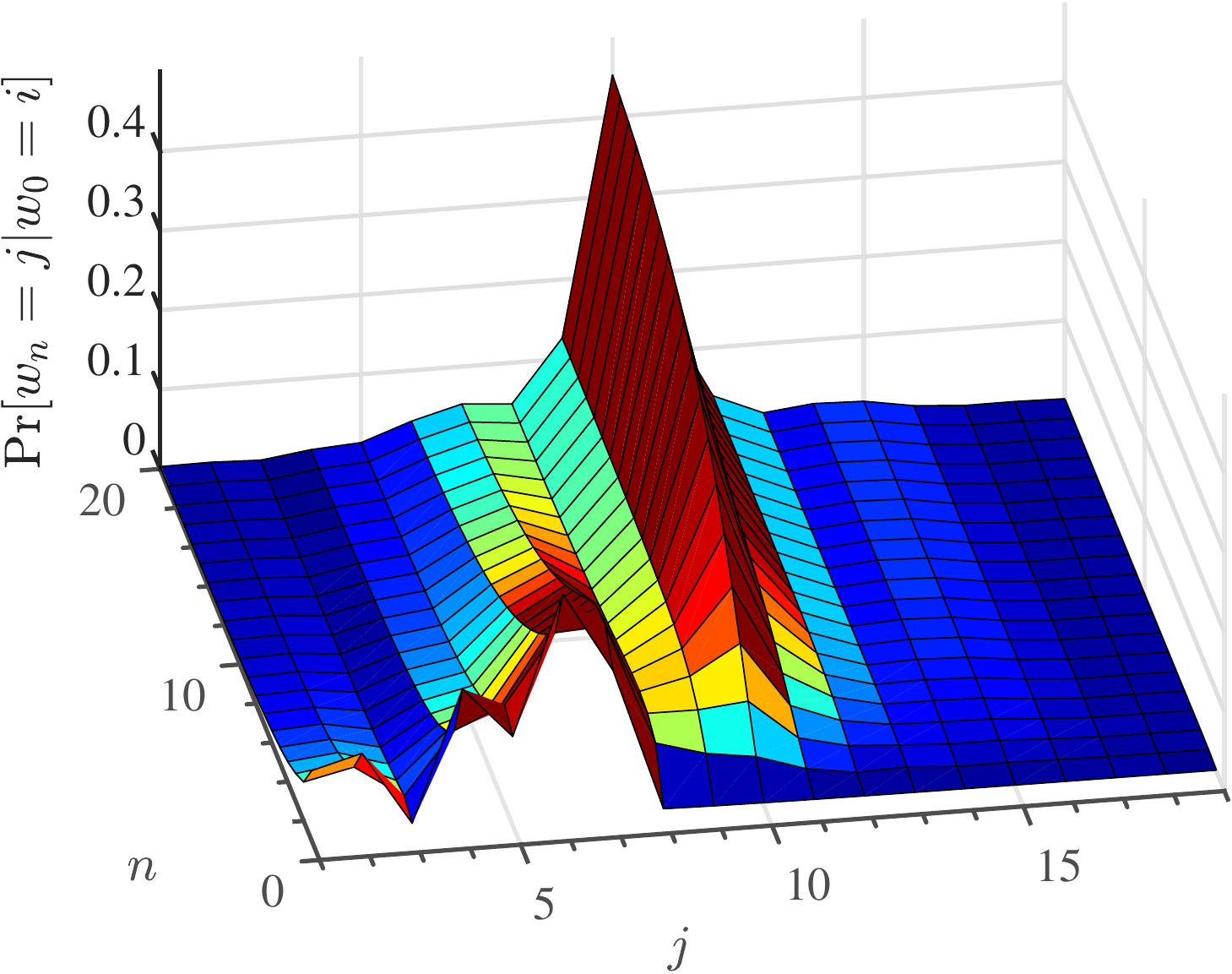}
	\end{minipage}~
	\begin{minipage}{0.5\textwidth}
		\includegraphics [width=1\textwidth]{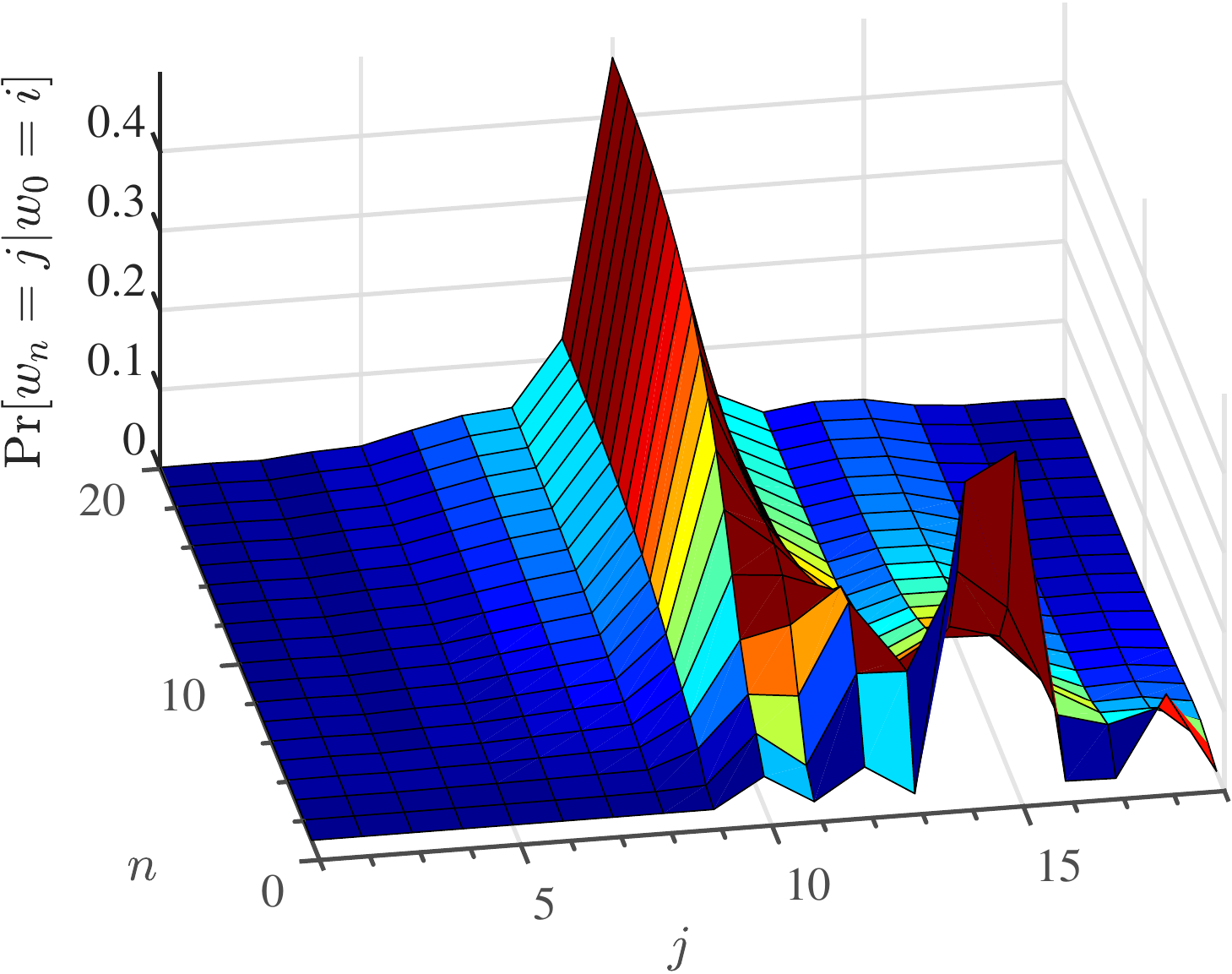}
	\end{minipage}	
	\caption{Propagation of $\mathrm{Pr}[w_n=w^j|w_0=w^i]$ for $i = 5$ (left column) and  $i = 15$ (right column).  Driver statistics from UDDS cycle (top row), US06 cycle (middle row) and GPS cycle (bottom row).}
	\label{fig:PnPropagate}
\end{figure}

The transition probabilities shown in Fig. \ref{fig:DriveCyclePdT} give insight into the singe-step behavior of the driver.  However, for the purposes of planning along a horizon it is desirable to understand driver behavior several seconds into the horizon.  The multi-step distribution Equation (\ref{eq:MarkovMultiStepPij}), $P_{ij}^{(n)} = \mathrm{Pr}[w_n=w^j|w_0=w^i]$, provides this information.  The propagation of the multi-step distribution for each cycle is shown in Fig \ref{fig:PnPropagate}.  Two initial demands are shown, the left column corresponding to the driver initially demanding a moderately negative acceleration and the right column corresponding to the driver initially demanding a moderately positive acceleration.  The effect of small differences in the $(P_{ij})$ matrices shown in Fig. \ref{fig:DriveCyclePdT} are immediately apparent.  For one, the single-step distribution (corresponding to $n=1$) is very different for each cycle.  Secondly, the paths along which the various transition probabilities grow and decay differs from one drive cycle to another. 

\begin{figure}[h!]
	\centering
	\begin{minipage}{1\textwidth}
		\includegraphics [width=1\textwidth]{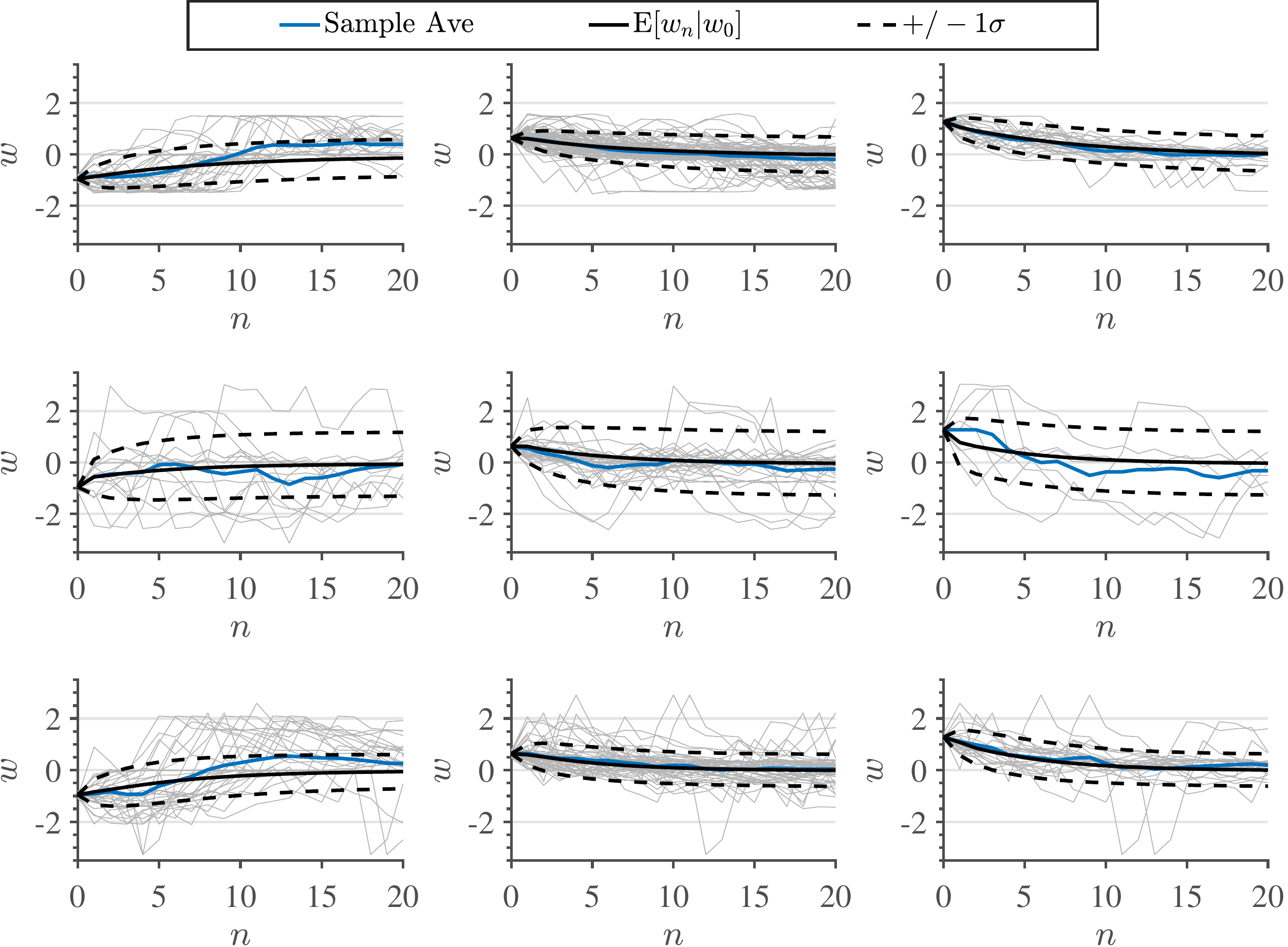}
	\end{minipage} 	
	\caption{Propagation of $\mathbb{E}[w_n|w_0=w^i]$.  Sample paths shown in light grey.  Top row: UDDS cycle, middle row: US06 cycle, bottom row: GPS cycle.  Left column: $w_0=-1~m/s^2$, middle column: $w_0=0.6~m/s^2$, right column: $w_0=1.3~m/s^2$.}
	\label{fig:EwnPropagate}
\end{figure}

From the multi-step distribution, the expected value and variance of the driver acceleration demand sequence, for $n=0,1,\dots,N$, can be computed according to
\begin{align}
\mathbb{E}[w_n|w_0=w^i] &= \sum_{j \in W} P_{ij}^{(n)} w^j \label{eq:Ewn} \\
\text{Var}[w_n|w_0=w^i] &= \sum_{j \in W} P_{ij}^{(n)} (w^j)^2 - \left(\sum_{j \in W} P_{ij}^{(n)} w^j\right)^2 \label{eq:Varwn}
\end{align}
The expected path of driver acceleration demand given by Equation (\ref{eq:Ewn}) is compared to the sample average for each drive cycle in Fig. \ref{fig:EwnPropagate} for three initial demands.  Also shown are the standard deviation of driver acceleration demand for each cycle, calculated as $\sigma = \sqrt{\text{Var}[w_n|w_0=w^i]}$ from Equation (\ref{eq:Varwn}).
These quantities provide indication as to the degree to which driver behavior can be anticipated along the horizon, and will be used further in Section \ref{section:StochControlFormulations}.

\newpage
\section{Long Term Driver Statistics} \label{section:LongTermBehavior}
\begin{figure}[h!]
	\centering
	\begin{minipage}{1\textwidth}
		\includegraphics [width=1\textwidth]{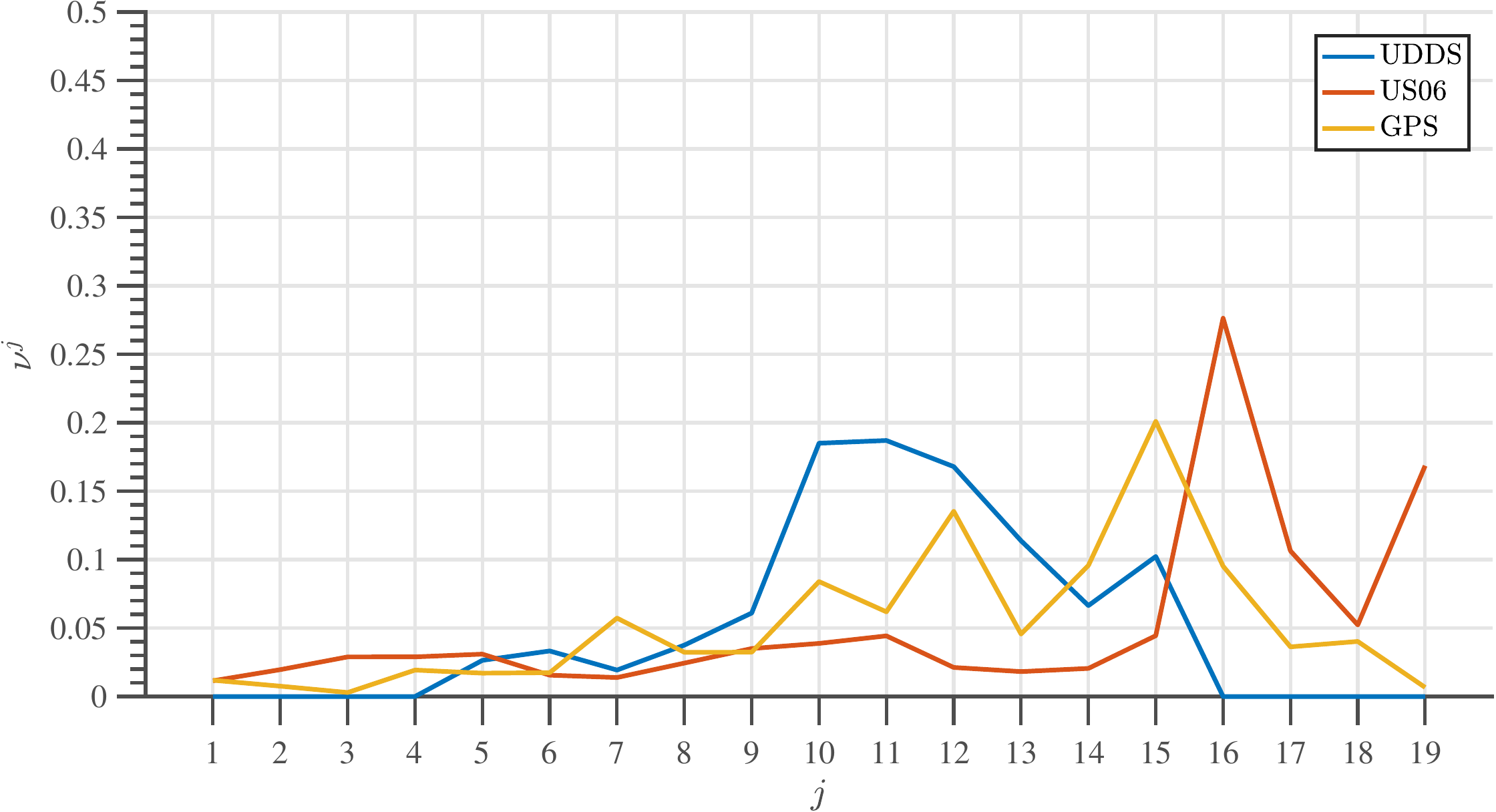}
	\end{minipage}
\caption{Long term driver behavior $\nu^i$.  Statistics at low vehicle speeds $< 10 m/s$.}
\label{fig:fig_wInftyDistLo}
\end{figure}

\begin{figure}
	\centering
	\begin{minipage}{1\textwidth}
		\includegraphics [width=1\textwidth]{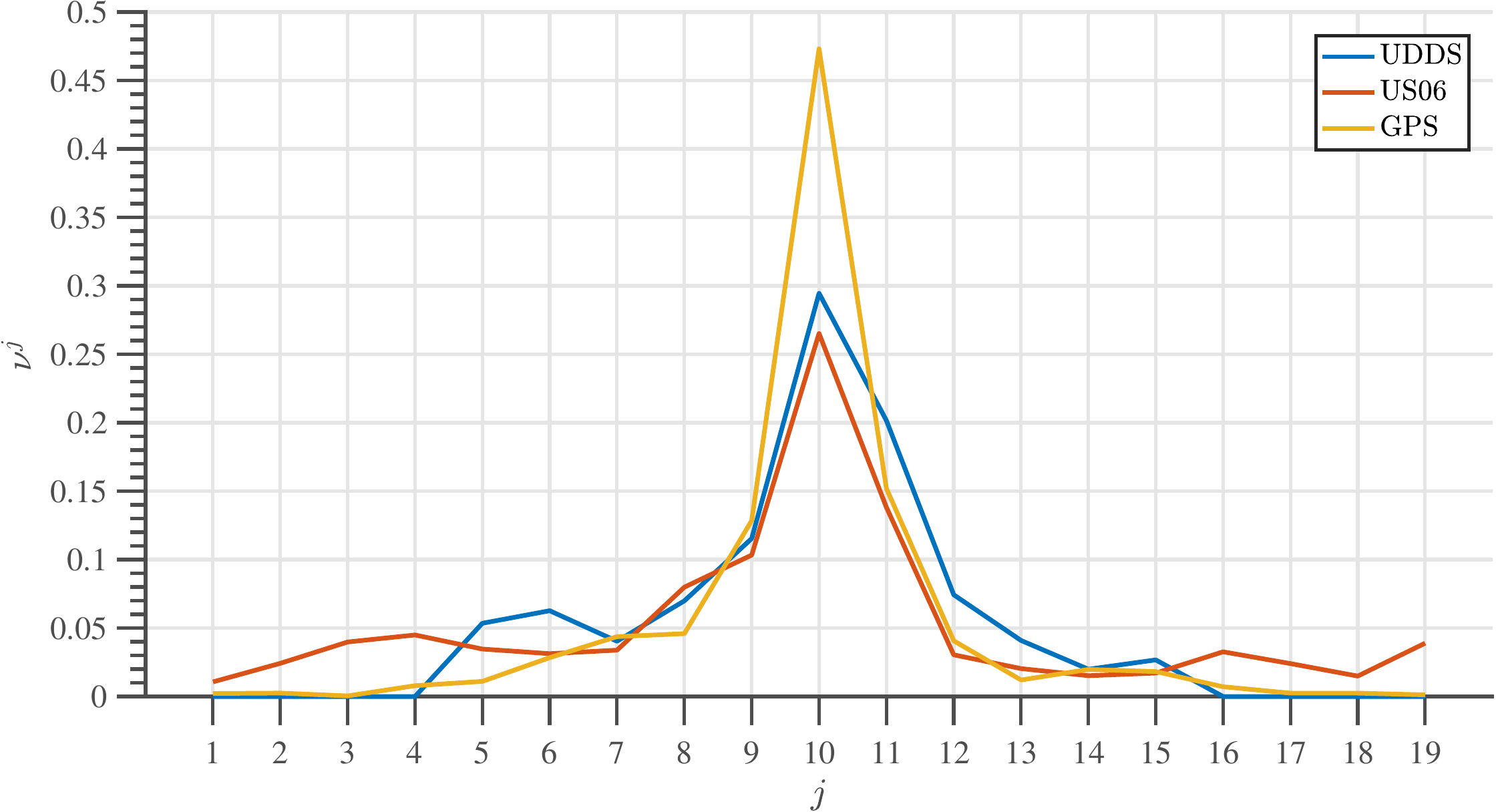}
	\end{minipage} 	
	\caption{Long term driver behavior $\nu^i$.  Aggregate statistics, independent of speed.}
	\label{fig:fig_wInftyDist}
\end{figure}

It was shown in Section \ref{section:LearningDriverBeahvior} that the multi-step distribution $P_{ij}^{(n)}$ may be used to generate a reasonable estimation of expected driver behavior along a horizon, given an initial condition corresponding to the driver's immediate demand.  The multi-step distribution can also provide valuable information about the driver's longer term statistical behavior.  Let the distribution 
\begin{align}
\nu^{ij} = \lim_{n\rightarrow \infty}\frac{1}{n}\sum_{k=1}^{\infty}\mathds{1}_{\{w_k=w^j|w_0=w^i\}}
\end{align}
denote the long run fraction of time the chain visits state $w^j$ when starting in state $w^i$.  It can be shown, see for instance, \cite{Lawler2006StochProcesses}, that this limit exists for all finite state Markov Chains.  Assuming the chain is irreducible\footnote{Roughly speaking, a Markov Chain is said to be \textit{irreducible} if any state of the chain can be reached, eventually, from any initial state.  The chain describing driver behavior is clearly irreducible.}, then $\nu^{ij} = \nu^j$ for each $i$ so that convergence is independent of the initial state.  In this case, $\nu^j$ may be interpreted as the fraction of time the driver demands acceleration $w^j$.  Assuming furthermore that the chain is also aperiodic\footnote{Roughly speaking, state $i$ is said to be periodic if $i$ can only be revisited cyclically with period $d>1, d\in \mathbb{N}$, so that $P_{ii}^{(n)}>0$ whenever $n$ is a multiple of $d>1$ and $P_{ii}^{(n)}=0$ otherwise.  Clearly, if a periodic state exists in the chain, convergence of $P^{(n)}$ is not possible since $\lim_{k\rightarrow \infty} P^{(kd)} \neq \lim_{k\rightarrow \infty} P^{(kd+1)}$.  The chain describing driver behavior is not periodic since any state can be revisited immediately at the next timestep, so that each state has period $d=1$. }, $\nu^j$ can be computed directly from the multi-step distribution through
\begin{align}\label{eq:nuj}
\nu^j = \lim_{n\rightarrow \infty} P_{ij}^{(n)}
\end{align}
During numerical experiments it was found that the driver tends to exhibit behavior during low speed driving which differs from behavior during higher speed driving.  As a result, two separate models for $(P_{ij})$ are learned: an aggregate model which is independent of speed and another model specifically for low speed driving below 10 m/s (approximately 23 mph).  The distributions of $\nu^j$ are shown for each of the three drive cycles in Figs. \ref{fig:fig_wInftyDistLo} and \ref{fig:fig_wInftyDist}.  Interestingly, the long term driver behavior distribution shows significant cycle to cycle differences during low speed driving.  The aggressive behavior of the driver during the US06 cycle is immediately apparent as more than 54\% of low speed driving occurs at high acceleration ($i\geq 16$).   In contrast, 43\% of low speed driving occurs near coasting ($9 \leq i \leq 11$) during the UDDS cycle.  

\chapter{PREDICTIVE ENERGY MANAGEMENT}\label{section:PredEnergyManage}

Having established models for vehicle and driver dynamics, a model-based predictive energy management strategy can be designed.  The goal is to minimize fuel consumption while meeting driver propulsion demands by solving the following finite horizon stochastic optimization problem 
\begin{subequations}\label{eq:FHSOCP}
	\begin{align}
	\min_{\bu_0,\bu_1,\dots,\bu_{N-1}} &\mathbb{E}\left[\sum_{n=0}^{N-1} g_n(\x_n,\x_{n+1},\bu_n,w_n)\Big|\x_0,w_0\right]\\
	\text{subject to}~~& \x_{n+1} = F_n(\x_n,\bu_n,w_n) \\
	&\x_n \in \mathbf X \\
	&	\bu_n \in \mathbf U	
	\end{align}
\end{subequations}
Complimentary methods for approximately solving Equation (\ref{eq:FHSOCP}) are developed.  The method developed in Section \ref{section:SGDM} performs stochastic optimization based on Monte Carlo sampling, while the methods developed in Sections \ref{section:ASDDP} and \ref{section:APDDP} rely on a dynamic programming approach using the multi step distributions discussed in Section \ref{section:LearningDriverBeahvior}.

\section{Embedded System Model}\label{section:SystemModel}
A simplified model of the system dynamics described in Section \ref{section:SeriesHHVDynamics} is now developed.  This simplified model will serve as the model accessible by various control algorithms developed in subsequent sections.  The continuous time embedded system model is defined as
\vspace{-1mm}
\begin{subequations}\label{eq:PlantAlgorithmModel}
	\begin{align}
	\dot \ell &= v_{veh} \\
	\dot v_{veh} &= w \\
	\dot n_{eng} &= \frac{1}{I_{eng}}\left[T_{cyl} - \frac{k_1}{2\pi} V_p p - k_1 \hat M_{s,p} \right] \\
	\dot p &=\frac{1}{C_h(p)} \Big[\frac{k_1}{2\pi}V_pn_{eng}-\frac{k_2}{2\pi r_{tire}}V_mv_{veh}-\hat Q_{s,p}-\hat Q_{s,m}\Big] 
	\end{align}
\end{subequations}
Compared to the model described in Section \ref{section:SeriesHHVDynamics}, the engine intake manifold dynamics have been neglected and all hydraulic losses are replaced by second order polynomial approximations $\hat Q_{s,p}, \hat Q_{s,m}, \hat M_{s,p}, \hat M_{s,m}$.  Additionally, the vehicle acceleration dynamic is represented directly by the driver acceleration demand $w$.  The motor displacement volume is once again calculated according to Equation (\ref{eq:MotorDispCalc})
\begin{align*}
V_m &= \frac{2\pi}{p}\left(\frac{F_p^{cmd}r_{tire}}{k_2}+\hat M_{s,m}\right)
\end{align*}
where $F_p^{cmd}$ is determined from the driver's acceleration demand $w$ by rearranging Equation (\ref{eq:wMeasured})
\begin{align} \label{eq:FpCmd}
F_p^{cmd}  &= m_{veh}w + \tfrac{1}{2}C_d\rho_{air}v_{veh}^2 + m_{veh}g\left[C_r cos(\phi)+sin(\phi)\right]
\end{align}
The system state and control vectors are defined as 
\begin{align}
	\x&=
	\begin{bmatrix}
	\ell\\ v_{veh}\\ n_{eng}\\ p
	\end{bmatrix} ,~~~~
	\bu=
	\begin{bmatrix}
	m_1^{-1}T_{cyl} \\ m_2^{-1} V_p
	\end{bmatrix}
\end{align}
respectively.  The control inputs are non-dimensionalized versions of cylinder torque and pump displacement volume, with 
\vspace{-1mm}
\begin{subequations}\label{eq:InputScaling}
	\begin{align}
	T_{cyl}&=m_1u_1 \\
	V_p&=m_2u_2
	\end{align}
\end{subequations}
The dynamics of  Equation (\ref{eq:PlantAlgorithmModel}), represented compactly as $\dot \x=f(\x,\bu,w,t)$, are numerically integrated using time step $\Delta t$ by carrying out a Taylor Series Expansion to second order according to
\begin{align}
\x(t+\Delta t) &= \x(t) + \Delta t \dot \x(t) + \frac{\Delta t^2}{2}\ddot \x(t)+o(\Delta t^2) \label{eq:TaylorExpand}
\end{align}
The coefficients $\dot \x(t)$ and $\ddot \x(t)$ are determined as follows\footnote{For simplicity, it is assumed $\frac{\partial f}{\partial t}=0$} with $w$ and $\bu$ assumed as piecewise constant in the interval $[t,t+\Delta t]$ 
\begin{align*}
\dot \x(t) &= \frac{d\x(t)}{dt} = f(\x,\bu,w,t)\\
\ddot \x(t) &= \frac{d\dot \x(t)}{dt} = \left.\frac{\partial f}{\partial \x}\right|_t f(\x,\bu,w,t) 
\end{align*}
The expansion Equation (\ref{eq:TaylorExpand}) is defined in discrete time with timestep $\Delta t$ as\footnote{The quantity $f(\x(t),\bu(t),w(t),t)$ is represented by shorthand as $ f(t)$}
\begin{align}
\x_{n+1} &= F_n\left(\x_n,\bu_n,w_n\right) \triangleq \x_n + \Delta t f(t)+ \frac{\Delta t^2}{2}\left.\frac{\partial f}{\partial \x}\right|_t f(t) \label{eq:2ndOrderPropagation}
\end{align}
In this work the embedded system model timestep is chosen as $\Delta t = 1$ second.  The horizon length is chosen as $N=12$ so that the prediction horizon is 12 seconds.  It was found through numerical experiments that increasing the horizon beyond 12 timesteps had little to no effect other than increasing computation time.

\section{Road Grade Forecasting}\label{section:RoadGradeForecasting}
Successful predictive energy management is ultimately limited by the ability to forecast the driver's propulsion force command described in Section \ref{section:SystemModel},
\begin{align*} \tag{\ref{eq:FpCmd}}
F_p^{cmd}  &= m_{veh}w + \tfrac{1}{2}C_d\rho_{air}v_{veh}^2 + m_{veh}g\left[C_r cos(\phi)+sin(\phi)\right]
\end{align*}
The largest source of uncertainty is the driver's acceleration demand $w$, which is modeled as a Markov process and identified in Section \ref{section:LearningDriverBeahvior}.  The vehicle speed $v_{veh}$ can then be anticipated as a result of the forecasted acceleration demand through numerical simulation of the model described by Equation (\ref{eq:PlantAlgorithmModel}).  What remains to be addressed in Equation (\ref{eq:FpCmd}) is the road grade $\phi$.  

One approach is to model road grade as an independent Markov process as in \cite{SDPRoadGrade}.  The authors of \cite{SDPRoadGrade} employ stochastic dynamic programming in a finite horizon setting to solve the resulting stochastic optimization problem with reported execution times of 10 to 100 seconds.  However, the uncertainty in forecasting $F_p^{cmd}$ along a horizon can be reduced significantly if forecasted road grade incorporated some geometric information as provided by telematics instrumentation, such as a GPS.  An assessment on the effect of terrain preview as applied to hybrid electric vehicle control is presented in \cite{GPSsurvey}.  Katsargyri \cite{GeorgiaGPS} uses path forecasting in the form of previewed vehicle speed and road grade in a hybrid electric vehicle.  In a similar approach, road grade is previewed along a horizon assuming constant vehicle speed in a conventional vehicle in \cite{Hellstrom}.  Since the state and action spaces are low in \cite{GeorgiaGPS} and \cite{Hellstrom}, deterministic dynamic programming is used in a finite horizon setting to generate the optimal control trajectory in a model predictive control setup.  

\begin{figure}[h!]
	\centering
	\begin{minipage}{1\textwidth}
		\includegraphics [trim = 0mm 0mm 0mm 0mm, clip,width=1\textwidth]{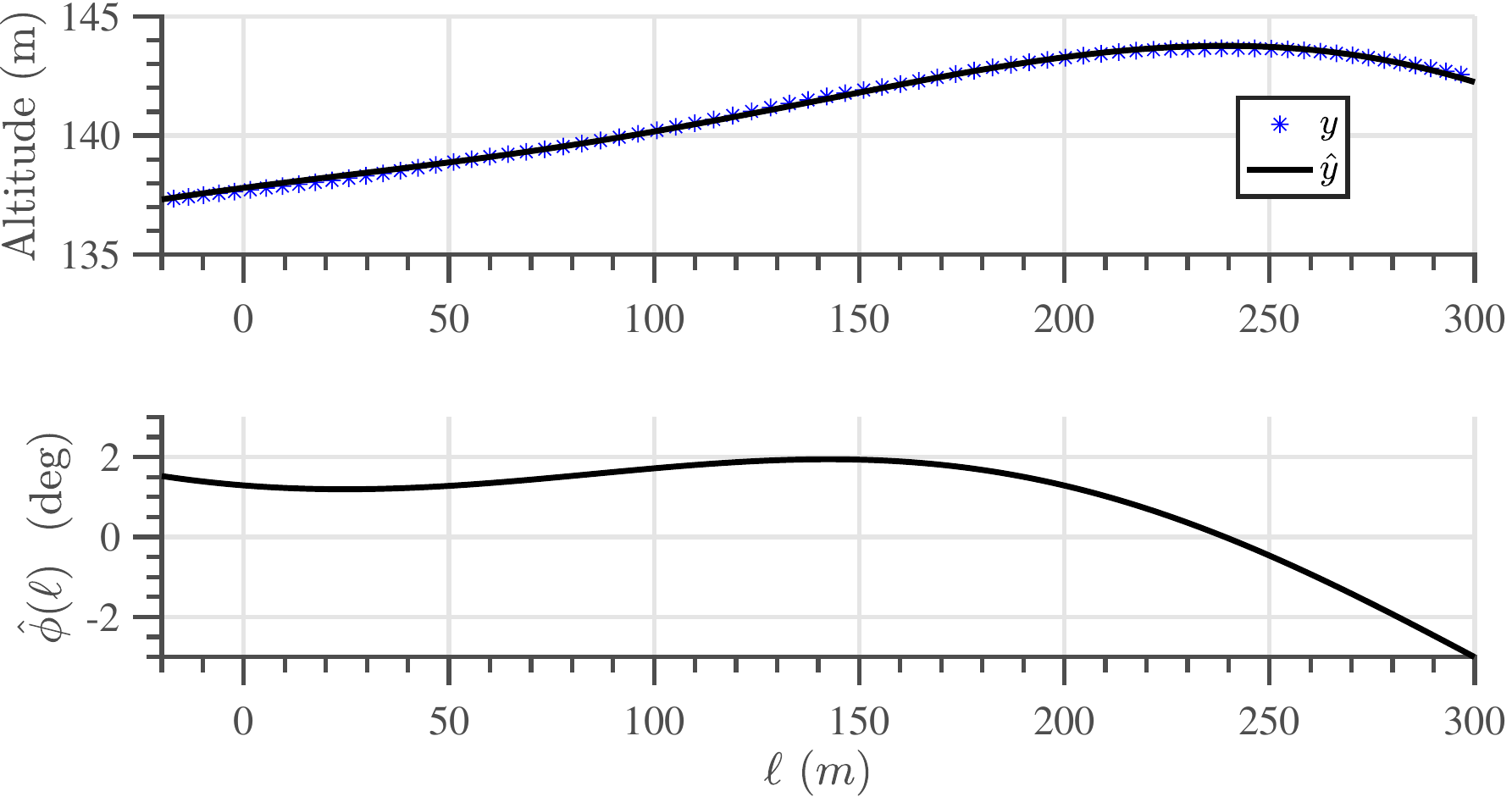}
	\end{minipage}
	\caption{GPS data taken from route in West Lafayette, IN.  Positions $r_i$ are set every 20m, with knots $c_i$ placed every 40m, $r_1$ and $c_1$ are placed at -20m while $r_{n_\ell}$ and $c_{n_k}$ are placed at 300m, as referenced to the vehicle's current position.  $\zeta = 7.5e-5$.}
	\label{fig:GPSFit}
\end{figure}

The approach taken here incorporates spatially distributed GPS information to develop road grade as a function of vehicle position along the prediction horizon.  Unlike previous approaches, future vehicle speed is not assumed known.  The segment of road directly ahead of the vehicle is discretized into a grid of $n_\ell$ equally spaced positions, $r_i, i=1,2,\dots, n_\ell$, so that a sequence of coordinates $(r_i,y_i)_{i=1}^{n_\ell}$ is obtained, where $y$ is the road altitude.  A fit $\hat y$ is applied to these coordinates in the form of a multiquadric radial basis function (RBF) with knots $c_i, i=1,2,\dots, n_k$, where $n_k<n_\ell$
\begin{align}\label{eq:yhat_roadmodel}
\hat y(\ell) &= a_0 + \sum\limits_{i=1}^{n_k}a_i  \sqrt{1+\zeta (\ell-c_i)^2} 
\end{align}
The radial basis function is ideal for this application as its nonlinear basis allows for a high accuracy approximation of road altitude, while the optimal coefficients of its linear weighting structure can be determined efficiently using a least squares projection.  The multiquadric form of RBF is specifically chosen as it is differentiable everywhere \cite{Orr96introductionto,Fornberg06thegibbs}, which will prove valuable when computing road grade.
\begin{figure}[h!]
	\centering
	\begin{minipage}{0.7\textwidth}
		\includegraphics [trim = 20mm 40mm 140mm 20mm, clip,width=.99\textwidth, page=14]{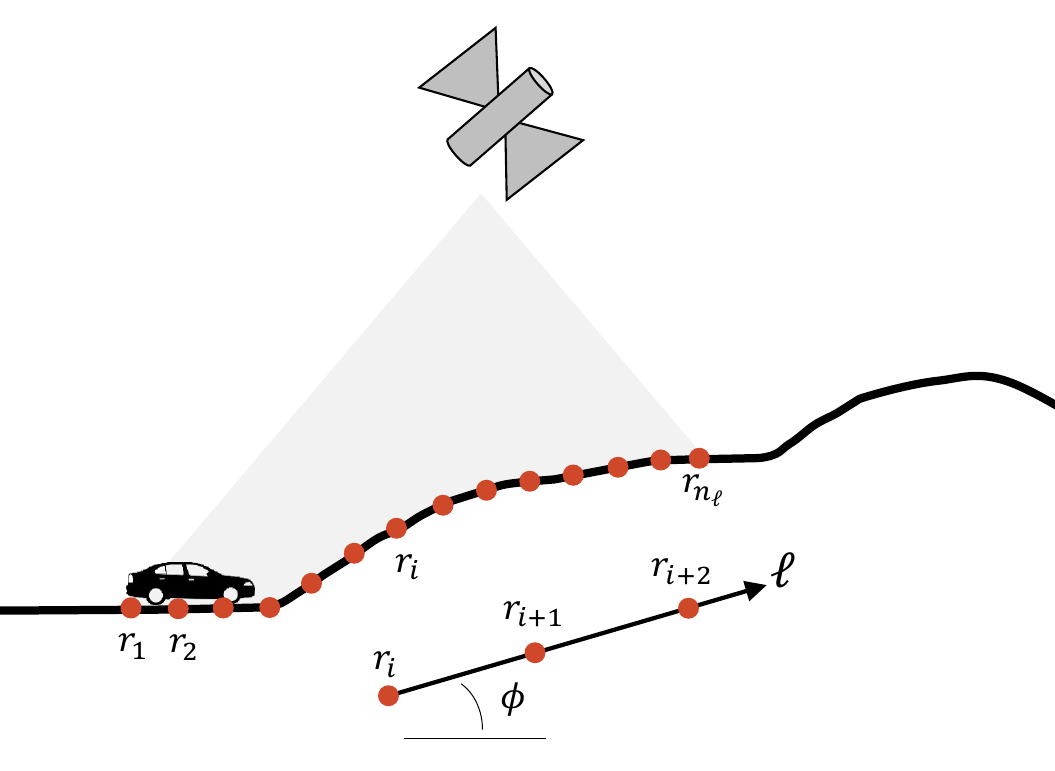}
	\end{minipage}
	\caption{Forecasting road grade along horizon with deterministic, spatially distributed GPS information.}
	\label{fig:GPS}
\end{figure}
Here, $c_i$ are chosen equally spaced along the grid $r_i$ so that $c_1$ and $c_{n_k}$ correspond with $r_1$ and $r_{n_\ell}$, respectively, and $\zeta$ is a fixed parameter which determines the influence each knot has on the RBF output.  The fitting coefficients $a_i$ are calculated in real time using a least squares projection so that the sum of square error $\sum_{i=1}^{n_\ell}\(y_i-\hat y(r_i)\)^2$ is minimized.  Taking the analytical derivative of $\hat y$ from Equation (\ref{eq:yhat_roadmodel}) with respect to position $\ell$ gives rate of change in altitude with respect to position
\begin{align}
\frac{d\hat y}{d\ell} &= \sum\limits_{i=1}^{n_k}a_i \frac{\zeta (\ell-c_i)}{\sqrt{1+\zeta (\ell-c_i)^2}}
\end{align}
from which the road grade model can be computed by taking the inverse sine,
\begin{align}
\hat \phi(\ell) &= \sin^{-1}\left( \frac{d\hat y}{d\ell}\right)
\end{align}
Forecasting road grade along the prediction horizon as a function of time is discussed in Sections \ref{section:SGDM} and \ref{section:ASDDP}.
An example of the road grade estimation applied to real GPS data along a segment of road is shown in Fig. \ref{fig:GPSFit}.

\section{Stochastic Control Formulations} \label{section:StochControlFormulations}

The running cost function used in Equation (\ref{eq:FHSOCP}) is constructed as 
\begin{align}\label{eq:gn}
g_n(\x_n,&\x_{n+1},\bu,w_n) = \nonumber \\
&~~~~K_1\left(x_{3,n+1}-x_{3,n}\right)^2  + K_2~ \hat b_f(x_{3},u_{1}) 
+K_3\left(x_4-p^*\right)^2\times \mathds{1}_{x_4<p^*}
\end{align}
Indicator functions are defined as $\mathds{1}_{a>b} = 1$ if $a>b$, $\mathds{1}_{a>b} = 0$ otherwise.  The first component of $\tilde L$ prevents the engine speed from changing excessively between time steps to prevent undesirable engine operation. The second component is the fuel consumption rate model, $\hat b_f$, a polynomial  approximation to the actual fuel consumption rate shown in Fig. \ref{fig:EngMap}.  
The final term ensures driver demands are satisfied by penalizing system pressures which are lower than a minimum allowable pressure, $p^*$, which is calculated according to 
\begin{align} \label{eq:pdem}
p^* &= \max\{p_{req},p_{set}\}
\end{align}
The value $p_{req}$ is the pressure required to satisfy driver propulsion force command along the horizon
\begin{align}\label{eq:preq}
p_{req} &= \frac{2\pi}{V_m^{max}}\left(\frac{r_{tire}F_p^{cmd}}{k_2}+\hat M_{s,m}\right) 
\end{align}
 Equation (\ref{eq:preq}) is obtained by rearranging the calculation for motor displacement volume Equation (\ref{eq:MotorDispCalc}) and substituting max volume for $V_m$.    Driver propulsion force command $F_p^{cmd}$ is calculated considering the stochastic driver acceleration demand $w$ and resistive forces according to Equation (\ref{eq:FpCmd}), 
\begin{align} \tag{\ref{eq:FpCmd}}
F_p^{cmd}  &= m_{veh}w + \tfrac{1}{2}C_d\rho_{air}v_{veh}^2 + m_{veh}g\left[C_r cos(\phi)+sin(\phi)\right]
\end{align}
Since $F_p^{cmd}$ is linear in $w$, it is evident that the statistical model which describes $w$ will directly influence the forecast of driver propulsion force demand and ultimately $p_{req}$ along the horizon.

\begin{figure}[h!]
	\centering
	\begin{minipage}{1\textwidth}
		\includegraphics [width=1.0\textwidth]{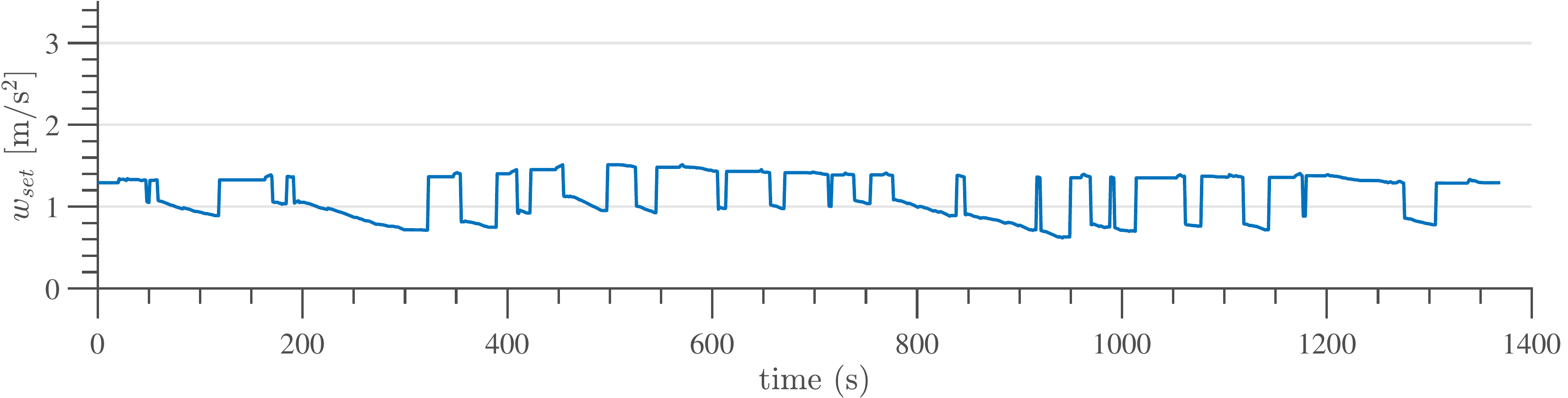}
	\end{minipage}\\
\begin{minipage}{1\textwidth}
	\includegraphics [width=1.0\textwidth]{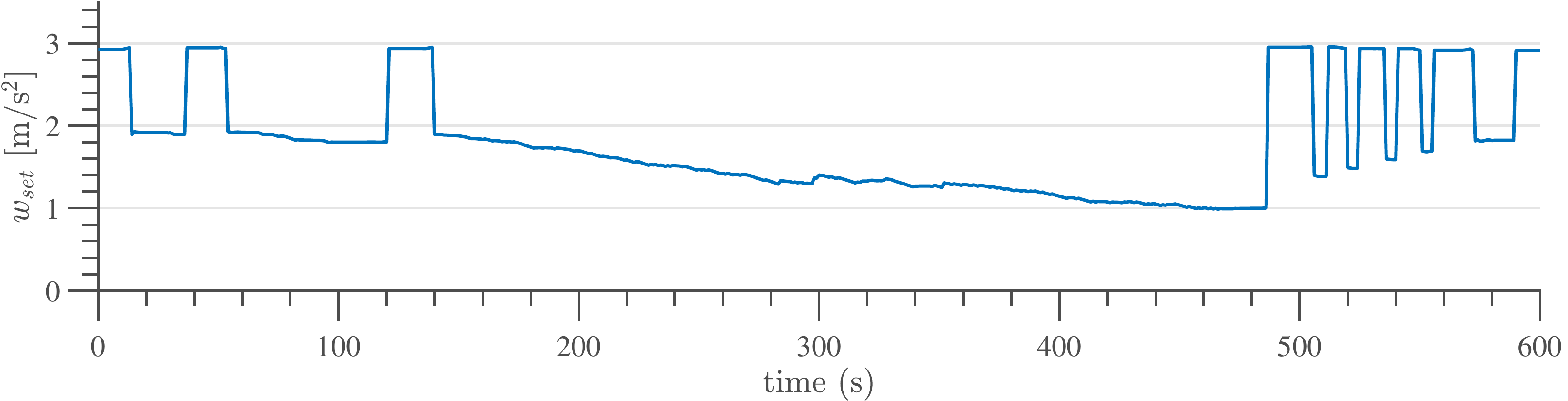}
\end{minipage}\\
	\begin{minipage}{1\textwidth}
		\includegraphics [width=1.0\textwidth]{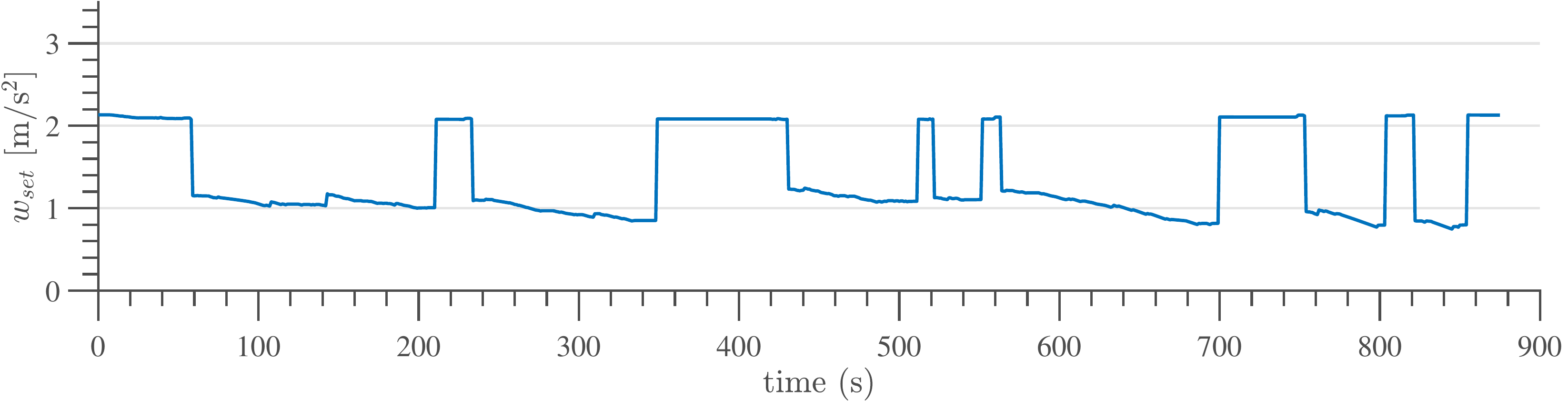}
	\end{minipage}
	\caption{$w_{set}$ for UDDS (top), US06 (middle) and GPS (bottom) cycles.}
	\label{fig:wSet}
\end{figure}

Satisfying a stochastic driver demand as forecast along a finite horizon can lead to short-sighted planning due to variance in the driver's acceleration demand sequence $\{w_n\}_{n=0}^{N-1}$ and sensitivity of this sequence to the initial demand $w_0$.  By leveraging the long term driver statistics explored in Section \ref{section:LongTermBehavior}, the value $p_{set}$ in Equation (\ref{eq:pdem}) provides a pressure target which is independent of initial demand $w_0$ and does not vary along the horizon thereby allowing for planning beyond the horizon.  Recall that $\nu^j$ represents the fraction of time the driver demands acceleration $w^j$, and is calculated from Equation (\ref{eq:nuj}).  The average and standard deviation of non-negative accelerations demands can be determined through
\begin{align}
w_{ave}^+ &= \frac{\sum_j\nu^jw^j}{\sum_j\nu^j} ~,~~j\in j^+\\
w_{std}^+ &= \sqrt{\frac{\sum_j\nu^j(w^j)^2}{\sum_j\nu^j} - {w_{ave}^+}^2} ~,~~j\in j^+
\end{align}
where $j^+ = \{ j | w^j \geq 0 \}$ is the index set of all non-negative acceleration demands.  An acceleration setpoint is now established taking the weighted sum 
\begin{align}\label{eq:wset}
w_{set} = \alpha w_{ave}^++\beta w_{std}^+
\end{align}  
In this work, the weights are set as $\alpha=1$ and $\beta=1.25$.  The value of $w_{set}$ along each drive cycle is shown in Fig. \ref{fig:wSet}.
The value of $w_{set}$ is observed to jump whenever vehicle speed increases (decreases) above (below) 10 m/s, since two separate Markov chains are retained in memory (one is active at speeds below 10 m/s and a second is active for speeds above 10 m/s) as discussed in Section \ref{section:LongTermBehavior}.

The intent of this setpoint is to represent a statistically significant driver acceleration demand, so that as a minimum requirement, a differential system pressure should be maintained so that $w_{set}$ can be satisfied instantly, without needing to increase differential system pressure.  To this end, the minimum pressure setpoint used in Equation (\ref{eq:pdem}) is designed as
\begin{subequations}
	\begin{align} 
	F_{p}^{set}  &= m_{veh}w_{set} \\
	p_{set} &= \frac{2\pi}{V_m^{max}}\left(\frac{r_{tire}F_p^{set}}{k_2}+\hat M_{s,m}\right) \label{eq:pset}
	\end{align}
\end{subequations}
A simple metric for quantifying how well driver demand is met along a drive cycle is discussed in Section \ref{section:PerformanceMetrics}.

\subsection {Stochastic Gradient Descent with Momentum (SGDM)} \label{section:SGDM}
This section develops a method to approximately solve Equation (\ref{eq:FHSOCP}) based on Monte Carlo sampling.  The problem is re-formulated as 
\begin{subequations}\label{eq:FHSOCP_SGDM}
	\begin{align}
	\min_{\bu_0,\bu_1,\dots,\bu_{N-1}} &\mathbb{E}\left[\sum_{n=0}^{N-1} g_n(\x_n,\x_{n+1},\bu_n,w_n)\Big|\x_0,w_0\right]\\
	\text{subject to}~~& \x_{n+1} = F_n(\x_n,\bu_n,w_n)
	\end{align}
\end{subequations}
State-control constraints are handled with SGDM through penalty functions.  The running cost Equation (\ref{eq:gn}) is augmented with penalty function $B(\x,\bu)$ 
\begin{align}
g_n(\x_n,&\x_{n+1},\bu,w_n) = \nonumber \\
&~~~~K_1\left(x_{3,n+1}-x_{3,n}\right)^2  + K_2~ \hat b_f(x_{3},u_{1}) +K_3\left(x_4-p^*\right)^2\times \mathds{1}_{x_4<p^*} +B(\x,\bu) 
\end{align} 
where 
\begin{align}\label{eq:SGDMPenaltyTerm}
B(\x,\bu) &= b_0\left(x_{3}-x_3^{max}\right)^2\times \mathds{1}_{x_{3}>x_3^{max}} +b_0\left(x_{3}-x_3^{min}\right)^2\times \mathds{1}_{x_{3}<x_3^{min}} \nonumber \\
&~~~~~~~+b_1\left(\bu-\bu^{max}\right)^2\times \mathds{1}_{\bu>\bu^{max}} +b_1\left(\bu-\bu^{min}\right)^2\times \mathds{1}_{\bu<\bu^{min}} \nonumber \\
&~~~~~~~+b_2\left(u_1-T_{cyl}^{max}(x_3)\right)^2\times \mathds{1}_{u_1>T_{cyl}^{max}(x_3)} 
\end{align}
The first component in Equation (\ref{eq:SGDMPenaltyTerm}) penalizes engine speeds which are outside allowable limits, and likewise, the second component penalizes control inputs which outside physical limits.  The final component provides the algorithm with information regarding the maximum torque capabilities of the engine as shown in Fig. \ref{fig:EngMap}.  The intent is to discourage the algorithm from choosing engine torque commands which are beyond the engine's ability, dependent on engine speed.  

For convenience we define the horizon cost 
\begin{align}\label{eq:RandomJ}
	J(\vec\bu,\vec w) &= \sum_{n=0}^{N-1} g_n(\x_n,\x_{n+1},\bu_n,w_n)
\end{align}
which is a function of the control input sequence $\vec{\u} = \{\u_0,\u_1,\dots,\u_{N-1}\}$ and the random disturbance input sequence $\vec w = \{w_0,w_1,\dots w_{N-1}\}$.  In all that follows, it is assumed that $\x_0$ and $w_0$ are given so that all expectation computations are conditioned on given values of $\x_0,w_0$.  The goal now is to minimize
\begin{align}\label{eq:MinExpectationRandomJ}
\min_{\bu_0,\bu_1,\dots,\bu_{N-1}} &\mathbb{E}\left[J(\vec\bu,\vec w)\right]
\end{align}
For a given control sequence, the expected value in Equation (\ref{eq:MinExpectationRandomJ})  is 
\begin{align}\label{eq:ExpectedValueJ}
	\mathbb{E}\left[J(\vec\bu,\vec w)\right] &= \sum\limits_{\vec {\mathsf{w}}} J(\vec \bu, \vec{\mathsf{w}}) \mathrm{Pr}\left[\vec w = \vec {\mathsf{w}} \right]
\end{align}
Conceptually, if $\nabla_{\vec\bu} \mathbb{E}[J(\vec\bu,\vec w)]$ could be computed directly, a descent with stepsize $\gamma^{[k]}$ of the form 
\begin{align}\label{eq:ExactDescent}
	\vec\bu^{[k+1]} &=\vec\bu^{[k]} - \gamma^{[k]} S^{[k]} \nabla_{\vec\bu} \mathbb{E}[J(\vec\bu^{[k]},\vec w)]
\end{align}
could be employed, where the order of descent is dependent on
the matrix $S^{[k]}$ \cite{Luen08}.  Unfortunately, explicitly computing $\nabla_{\vec\bu} \mathbb{E}[J(\vec\bu,\vec w)]$ is generally intractable due to a large number\footnote{The number of potential outcomes is $|W|^{N-1}$, where $|W|$ is the number of discrete states in the Markov Chain.} of potential outcomes of the sequence $\vec w$, so implementing Equation (\ref{eq:ExactDescent}) directly is generally not possible.  

One approach is to minimizing Equation (\ref{eq:MinExpectationRandomJ}) is by approximating Equation (\ref{eq:ExpectedValueJ}) with the sample average approximation
\begin{align}\label{eq:SampleAveJ}
\hat J(\vec\bu) = \frac{1}{K}\sum\limits_{k=1}^{K}J(\vec\bu,\vec w^{[k]})
\end{align}
where each $J(\vec\bu,\vec w^{[k]})$ is a Monte Carlo sample of the random variable $J(\vec\bu,\vec w)$.  In general, the approximation $\hat J(\vec\bu)$ improves as the number of Monte Carlo samples $K$ increases in accordance with a law of large numbers argument.  In \cite{StochDriverLearning}, Quadratic Programming is employed to minimize Equation (\ref{eq:SampleAveJ}) as applied to the hybrid electric vehicle (HEV) energy management problem.  The computational challenge with this approach is $K$ trajectories of any relevant system information must be stored in memory, and the subsequent optimization must be performed considering the entire sample set in the spirit of batch optimization \cite{OnlineLearning,MPCbook}.  To reduce the computational burden, \cite{StochDriverLearning} removes Monte Carlo samples with comparatively low probability of occurrence from the batch optimization.  

The stochastic gradient descent (SGD) update
\begin{align}\label{eq:StochOptProblemSGD}
	\vec{ \bu}^{[k+1]} &=\vec \bu^{[k]} - \gamma^{[k]} \nabla_\bu J(\vec\bu^{[k]},\vec w^{[k]})
\end{align}
is a stochastic form of the idealized descent of Equation (\ref{eq:ExactDescent}), and is exactly the gradient form of stochastic approximation from Section \ref{section:SA}.  Stochastic gradient descent finds a locally optimal solution $\vec\bu^*$ which asymptotically (locally) minimizes the original problem Equation (\ref{eq:MinExpectationRandomJ}) \cite{OnlineLearning}.  With SGD, only one Monte Carlo sample of the gradient $\nabla_\bu J(\vec\u^{[k]},\vec w^{[k]})$ is required at each iteration offering significantly reduced computational overhead, allowing SGD to process more samples than batch processing in a fixed amount of time.  In this way, SGD is competitive with and can even outperform second-order batch optimization methods   \cite{OnlineLearning2},\cite{OnlineLearning3}.  The benefit of the sequential optimization approach can understood considering stochastic optimization based on Monte Carlo sampling is as much an estimation problem as it is an optimization problem \cite{OnlineLearning4}.  The total solution error is a combination of optimization error, which measures an algorithm's ability to determine the optimal solution for the given sampling set, and estimation error, which measures the effect of minimizing an empirical average Equation (\ref{eq:SampleAveJ}) rather than expected cost Equation (\ref{eq:ExpectedValueJ}).  If $\vec \bu^*$ is the locally optimal solution determined by a given algorithm, then the total solution error is 

\begin{align*}
&\underbrace{\hat J(\vec\bu^*) - \min_{\vec \bu}\mathbb{E}[J(\vec\bu,\vec w)]}_{\varepsilon_{tot}} = \underbrace{\hat J(\vec\bu^*) - \min_{\vec \bu}\hat J(\vec\bu)}_{\varepsilon_{opt}} + \underbrace{\min_{\vec \bu}\hat J(\vec\bu) - \min_{\vec \bu}\mathbb{E}[J(\vec\bu,\vec w)]}_{\varepsilon_{est}}
\end{align*}
The estimation error generally decreases inversely with $K$, therefore the total solution error depends on the number of samples that can be processed in the alloted time.   The step size sequence $\{\gamma^{[k]}\}_{k\geq 1}$, $\gamma^{[k]}\in \mathbb{R}$ must satisfy the rules given in Section \ref{section:SA} for stochastic approximation:
\begin{align}\tag{\ref{eq:2SALearningRates}}
\sum_{k=1}^\infty\gamma^{[k]} = \infty, ~~~~ 
\sum_{k=1}^\infty\left(\gamma^{[k]}\right)^2 < \infty 
\end{align}
The step size schedule chosen here is
\begin{align}
\gamma^{[k]} = \frac{\gamma_0}{1+(k-1)\epsilon},~k=1,2,...
\end{align}  
where $\epsilon>0$ is called the decay rate.  In this work, we use a slightly modified version of SGD known as stochastic gradient descent with momentum (SGDM) based on Nesterov's Accelerated Gradient (NAG) \cite{Momentum2}
\begin{subequations} \label{eq:SGDM}
	\begin{align}
	\vec{\mathbf v}^{[k+1]} &= \mu \vec{\mathbf v}^{[k]} - \gamma^{[k]}\nabla_{\vec{\bu}}J(\vec\bu^{[k]}+\mu \vec{\mathbf v}^{[k]},\vec w^{[k]}) \label{eq:SGDMa}\\
\vec{\mathbf u}^{[k+1]} &=\vec{\mathbf u}^{[k]} + \vec{\mathbf v}^{[k+1]}
	\end{align}
\end{subequations}
The quantity $\mathbf v \in \mathbb{R}^{dimU}$ is referred to as the velocity term and decays at a rate according to $\mu\in[0,1)$, known as the momentum parameter.  The effect of momentum is to continue pushing the parameter update in directions of previous updates, averaging out oscillations in areas of a rapidly changing gradient.  Simultaneously, if several past updates are approximately aligned, the velocity term will act to propel the parameter update faster than if momentum was absent.  The net result is that SGDM tends to move more rapidly towards a local minimum than classical SGD \cite{OnlineLearning3,Ber96,Momentum,Momentum2}.  An attractive feature of NAG is the gradient computation performed in Equation (\ref{eq:SGDMa}) considers a projected estimate of the control sequence, $\vec{\bu}^{[k]}+\mu \vec{\mathbf v}^{[k]}$, based on the most recent velocity sequence $ \vec{\mathbf v}^{[k]} = (\mathbf{v}_n^{[k]})_{n=0}^{N-1}$.  This projected estimate is in some respect not unlike predictor-corrector methods used to improve stability in numerical solution of ordinary differential equations.  The result is improved stability compared to classical momentum, in which the gradient is computed considering only the current value of the control parameter array, particularly when $\mu \approx 1$ \cite{Momentum2}.

\subsubsection{Computing the Gradient}
This sections proposes a method to iteratively compute the gradient $\nabla_{\vec{\bu}} J$ used in the control sequence update Equation (\ref{eq:SGDM}) based on a piecewise linear approximation to the system dynamics along the horizon.  The gradient \begin{align}\label{eq:GradUJ}
	\nabla_{\vec{\bu}} J &= \left[\nabla_{\bu_0}J ~~ \nabla_{\bu_1}J ~~ \dots ~~ \nabla_{\bu_{N-1}}J\right] \in \mathbb{R}^{dimU\times N}
\end{align}
has individual components given by
\begin{align}\label{eq:GradUnJ}
\nabla_{\bu_n}J &=\sum\limits_{k=0}^{N-1}  \left[\frac{\partial g_k}{\partial \x_k} \frac{d \x_k}{d \bu_n} +  \frac{\partial g_k}{\partial \x_{k+1}} \frac{d \x_{k+1}}{d \bu_n} \right]\Tr + \frac{\partial g_n}{\partial \bu_n}\Tr
\end{align}
where $\frac{\partial g_k}{\partial \x_k} \in \mathbb{R}^{1\times dimX}$, $ \frac{\partial \x_k}{\partial \bu_n}\in \mathbb{R}^{dimX\times dimU}$, $\frac{\partial g_n}{\partial \bu_n}\in\mathbb{R}^{1\times dimU}$.   In evaluating Equation (\ref{eq:GradUnJ}), it will be helpful to define the following matrix
\begin{align}
	\mathcal{C}_n\triangleq\begin{bmatrix}
	\dfrac{d \x_n}{d \bu_0} & 	\dfrac{d \x_n}{d \bu_1} &\dots & 	\dfrac{d \x_n}{d \bu_{N-1}}
	\end{bmatrix}
	\in \mathbb{R}^{dimX \times N dimU}
\end{align}
An efficient recursion for $\mathcal{C}_n$ which can be updated iteratively along the horizon is now developed.  Carrying out the first few $\mathcal{C}_n$ gives
\vspace{-1mm}
\begin{align*}
	\begin{array}{llllll}
	\mathcal{C}_1:~~\dfrac{d \x_1}{d \bu_0} = \dfrac{\partial F_0}{\partial \bu_0} ~~& \mathbf{0}~~& \mathbf{0}~~&\mathbf{0} &\cdots&\mathbf{0}\vspace{0.5em}\\
	\mathcal{C}_2:~~\dfrac{d \x_2}{d \bu_0} = \dfrac{\partial F_1}{\partial \x_1}\dfrac{d \x_1}{d \bu_0} ~~& 	\dfrac{d \x_2}{d \bu_1} =  \dfrac{\partial F_1}{\partial \bu_1} ~~& \mathbf{0}~~&\mathbf{0} &\cdots&\mathbf{0} \vspace{0.5em}\\
	\mathcal{C}_3:~~\dfrac{d \x_3}{d \bu_0} = \dfrac{\partial F_2}{\partial \x_2}\dfrac{d \x_2}{d \bu_0} ~~& 	\dfrac{d \x_3}{d \bu_1} =  \dfrac{\partial F_2}{\partial \x_2} \dfrac{d\x_2}{d \bu_1}~~& \dfrac{d \x_3}{d \bu_2} =  \dfrac{\partial F_2}{\partial \bu_2}~~&\mathbf{0} &\cdots&\mathbf{0} \vspace{0.5em}
	\end{array}
\end{align*}
By inspection, a recursion for $\mathcal{C}_n$ is given by 
\begin{align}
	\mathcal{C}_{n+1} &=\dfrac{\partial F_n}{\partial \x_n} \mathcal{C}_n + 
	\begin{bmatrix}
	\underbrace{\mathbf{0} ~~ \cdots ~~ \mathbf{0}}_{\text{$n$ blocks}} & \dfrac{\partial F_n}{\partial \bu_n} & \underbrace{\mathbf{0} ~~ \cdots ~~ \mathbf{0}}_{\text{$N-1-n$ blocks}}
	\end{bmatrix}~,~~~ n=0,\dots,N-1\\	
	\mathcal{C}_0 &= 
	\begin{bmatrix}
	\mathbf{0} & \mathbf{0} & \mathbf{0} &\cdots & \mathbf{0}
	\end{bmatrix}\nonumber \\
		\mathbf{0} &\in \mathbb{R}^{dimX \times dimU} \nonumber 
\end{align}
In this way, $\mathcal{C}_n$ is updated incrementally at each time step $n$ along the horizon.  The individual partial derivatives are calculated considering the system dynamics Equation (\ref{eq:2ndOrderPropagation})
 \begin{subequations}\label{eq:CnPartials}
	\begin{align}
	\dfrac{\partial F}{\partial \x} &= I + h \frac{\partial f}{\partial \x} + \frac{h^2}{2} \left(\frac{\partial f}{\partial \x}\right)^2 \\
	\dfrac{\partial F}{\partial \bu} &= I + h \frac{\partial f}{\partial \bu} + \frac{h^2}{2} \frac{\partial f}{\partial \x} \frac{\partial f}{\partial \bu}
	\end{align}
\end{subequations}
In deriving Equation (\ref{eq:CnPartials}) all second order partial derivatives of the form $\frac{\partial^2 f}{\partial \x ^2}$ and $\frac{\partial^2 f}{\partial \x \partial \bu}$ have been ignored. The gradient Equation (\ref{eq:GradUJ}) can now be evaluated with $\mathcal{C}_n$ through
\begin{align}
	\nabla_{\vec{\bu}}J &= \mathrm{reshape}\left\{\sum\limits_{n=0}^{N-1} \left[\frac{\partial g_n}{\partial \x_n} \mathcal{C}_n + \frac{\partial g_n}{\partial \x_{n+1}} \mathcal{C}_{n+1}\right]\right\} + \frac{\partial J}{\partial \vec{\bu}}
\end{align}
where the function \textit{reshape} is used to convert the $1\times NdimU$ row vector into a $dimU \times N$ matrix and
\vspace{-1mm}
\begin{align}
	\frac{\partial J}{\partial \vec{\bu}} &= 
	\begin{bmatrix}
	\dfrac{\partial g_0}{\partial \bu_0}\Tr & \dfrac{\partial g_1}{\partial \bu_1}\Tr & \cdots & \dfrac{\partial g_{N-1}}{\partial \bu_{N-1}}\Tr
	\end{bmatrix} \in \mathbb{R}^{dimU \times N}
\end{align}
Finally, $\nabla_{\vec{\bu}}J$ is updated iteratively at each time step along the horizon through
\begin{align}\label{eq:GradUJIterativeCalc}
	\nabla_{\vec{\bu}}J &\leftarrow \nabla_{\vec{\bu}}J + \mathrm{reshape}\left\{\frac{\partial g_n}{\partial \x_n} \mathcal{C}_n + \frac{\partial g_n}{\partial \x_{n+1}} \mathcal{C}_{n+1}\right\} + \dfrac{\partial g_n}{\partial \bu_n}\Tr\times \mathds{1}_n\\
	n&=0,\dots,N-1 \nonumber
\end{align}
where $\mathds{1}_n$ is a $N$-element row vector such that the $k^{th}$ element is given by
\begin{align}
	\mathds{1}_n(k) &=
	\begin{cases}
	1~\text{if $k = n+1$} \\
	0~\text{if $k \neq n+1$}
	\end{cases}
\end{align} 
The update Equation (\ref{eq:GradUJIterativeCalc}) is initialized with $	\nabla_{\vec{\bu}}J = 	\begin{bmatrix}
\mathbf{0} & \cdots & \mathbf{0}
\end{bmatrix} \in \mathbb{R}^{dimU\times N}$.

\subsubsection{Monte Carlo Sampling and Variance Reduction}
Each Monte Carlo sample $J(\vec\bu,\vec w^{[k]})$ is created by randomly generating the sequence $\{ w_n\}_{n=0}^{N-1}$ drawn from the single-step distribution $P_{ij}$ according to
\begin{align*}
w_{n+1} \sim P_{ij}~,~\text{where $w^i \triangleq  w_n$}
\end{align*}
The process of drawing $w_{n+1}$ from $P_{ij}$ is as follows.  A sequence of random numbers $\{\omega_0,\omega_1,\dots,\omega_{N-2}\}$ is generated, where each $\omega_n \in [0,1]$ is an independent uniform random number.  The initial value of $w_0$ is given and, at each stage $n=0,\dots,N-2$, $w^i$ is reset according to $w^i \triangleq  w_n$.  The value assigned to $w_{n+1}$ is then determined from $\omega_n$ according to 

\begin{align*}
\begin{array}{rl}
0 < \omega_n \leq P_{i1} : & w_{n+1}=w^1 \\
P_{i1} < \omega_n \leq P_{i1}+P_{i2} : &  w_{n+1}=w^2 \\
P_{i1}+P_{i2} < \omega_n \leq P_{i1}+P_{i2}+P_{i3} : &  w_{n+1}=w^3 \\
\vdots &  
\end{array}
\end{align*}
The general rule for assigning the specific value $w^j$ to $w_{n+1}$ is 
\begin{align}\label{eq:MonteCarloAssignmentRule}
\begin{array}{rl}
P_{i1} + \cdots + P_{ij-1} < \omega_n \leq P_{i1}+\cdots + P_{ij} : &  w_{n+1}=w^{j}
\end{array}
\end{align}
The assignment rule Equation (\ref{eq:MonteCarloAssignmentRule}) is performed for $n=0,\dots,N-2$.
Variance reduction is accomplished with a technique known as  PEGASUS \cite{ng2000pegasus}, in which the Monte Carlo sampling of Equation (\ref{eq:MonteCarloAssignmentRule}) is performed using the same sets of random numbers.  A set of $K$ random number sequences is generated before the algorithm is started
\begin{align*}
\vec{\omega}^{[1]} &=\{\omega_0,\dots,\omega_{N-2}\}^{[1]}\\
&~\vdots \\
\vec{\omega}^{[k]} &=\{\omega_0,\dots,\omega_{N-2}\}^{[k]}
\end{align*}
At iteration $k$ of SGDM, the $k^{th}$ sequence of random numbers $\vec{\omega}^{[k]}$ is used in the Monte Carlo sampling Equation (\ref{eq:MonteCarloAssignmentRule}).  After $K$ iterations, a new point $(\x_0,w_0)$ is measured and brought in as the new initial condition and the process is restarted using the same $K$ sets of random number sequences.  The benefit is that for a fixed $\(\x_0,w_0\)$ initial condition the optimization process reduces to a completely deterministic optimization,  resulting in significantly reduced variance in the control sequence between executions of SGDM.  

\subsubsection{Scaling and Final Algorithm}
Performance of SGDM is improved significantly by properly scaling the control inputs.  The scaling factors $m_1, m_2$ from Equation (\ref{eq:InputScaling}) are determined empirically so that $\nabla_{\vec{\bu}}J$ has components of approximately equal magnitude along each dimension, which is a common approach in numerical solution of optimal control problems \cite{Betts}.  The final algorithm is shown in Algorithm \ref{Alg:SGDM}.  Maximum algorithm iterations is set to $K=200$.  For the first 50 iterations the stepsize is held constant at $\gamma=0.2$,  afterwards a decay of $\epsilon=0.1$ is used.  The momentum parameter is set as $\mu=0.95$.  These parameters were finely tuned to deliver optimum performance from SGDM.

\begin{algorithm}[h!]	
	\setstretch{1.2}
	\caption{SGDM} \label{Alg:SGDM}
	\KwIn{$\x_0,w_0,\vec\bu,\vec{\mathbf v}$}
	\KwData{$N,\epsilon,\mu,\gamma_0,K,\{\vec\omega^{(1)},\dots,\vec\omega^{[k]}\}$}	
	\For{$k=1:K$}{		
		Given $w_0$, generate sample $\{ w_1,\dots, w_{N-1}\}(\vec\omega^{[k]})$\\
		$\nabla_{\vec{\bu}}J = 	\mathbf{0} \in \mathbb{R}^{dimU\times N}$\\
		$\mathcal{C}_0 = \mathbf{0}\in\mathbb{R}^{dimX\times NdimU}$\\
		$\vec{\bu} := \vec{\bu} + \mu \vec{\mathbf{v}}$ \\
		\For{$n=0:N-1$}{
			$\x_{n+1} = F_n\left(\x_n,\bu_n, w_n\right)$  \vspace{.2em} \\
			Compute $\frac{\partial F_n}{\partial \x_n}$,$\frac{\partial F_n}{\partial \bu_n}$,$\frac{\partial g_n}{\partial \x_n}$,$\frac{\partial g_n}{\partial \x_{n+1}}$ \\		
			$\mathcal{C}_{n+1} =\frac{\partial F_n}{\partial \x_n} \mathcal{C}_n + 
			\begin{bmatrix}
			\underbrace{\mathbf{0} ~~ \cdots ~~ \mathbf{0}}_{\text{$n$ blocks}} & \frac{\partial F_n}{\partial \bu_n} & \underbrace{\mathbf{0} ~~ \cdots ~~ \mathbf{0}}_{\text{$N-1-n$ blocks}}
			\end{bmatrix}$\\
			$\nabla_{\vec{\bu}}J \leftarrow \nabla_{\vec{\bu}}J + \mathrm{reshape}\left\{\frac{\partial g_n}{\partial \x_n} \mathcal{C}_n + \frac{\partial g_n}{\partial \x_{n+1}} \mathcal{C}_{n+1}\right\} + \dfrac{\partial g_n}{\partial \bu_n}\Tr\times \mathds{1}_n$\vspace{0.2em}\\
				}			
			$\gamma = \dfrac{\gamma_0}{1+(k-1)\epsilon}$\vspace{.2em}\\
			$\vec{\mathbf v} \leftarrow \mu \vec{\mathbf v} - \gamma\nabla_{\vec{\bu}} J$\vspace{.2em}\\
			$\vec{\bu} \leftarrow \vec{\bu} + \vec{\mathbf v}$\\
			}		
			\KwOut{$\vec{\bu},\vec{\mathbf v}$}
			\end{algorithm}

\newpage

\subsection{Approximate Stochastic Differential Dynamic Programming (ASDDP)}\label{section:ASDDP}
This sections develops \textit{approximate stochastic differential dynamic programming} (ASDDP), a stochastic variant of the classic differential dynamic programming algorithm described in Section \ref{section:DDP/iLQR}, to approximately solve Equation (\ref{eq:FHSOCP}).  The problem is re-formulated as
\begin{subequations}\label{eq:FHSOCP_ASDDP}
	\begin{align}
	\min_{\bu_0,\bu_1,\dots,\bu_{N-1}} &\mathbb{E}\left[\sum_{n=0}^{N-1} g_n(\x_n,\x_{n+1},\bu_n,w_n)\Big|\x_0,w_0\right]\\
	\text{subject to}~~& \x_{n+1} = F_n(\x_n,\bu_n,w_n) \\
	&\bar \x_{n+1} = \sum_{j}P_{ij}^{(n)}F_n(\bar \x_n,\bu_n,w^j) \label{eq:ASDDP_AvePath}\\
	&D_x \bar \x_{n+1} \leq \mathbf c_x \label{eq:ASDDP_XConstraint}\\
	&D_u \bu_n   \leq \mathbf c_u \label{eq:ASDDP_UConstraint}
	\end{align}
\end{subequations}
Equation (\ref{eq:ASDDP_AvePath}) is the expected state trajectory along the horizon.  Equations (\ref{eq:ASDDP_XConstraint}) and (\ref{eq:ASDDP_UConstraint}) are linear constraints on the expected state and control input trajectories.  The state value function is defined as (the derivation can be found in Appendix \ref{section:VFASDDP})
\begin{align}
V_n(\x_n) &\triangleq   \min_{\bu_n,...,\bu_{N-1}} \mathbb{E}\Big[ h(\x_N)+\sum_{k=n}^{N-1} g_k(\x_k,\bu_k,w_k) \Big| \x_n, w_0=w^i\Big] \nonumber \\
&=  \min_{\bu_n}\mathbb{E}\Big[ g_n(\x_n,\bu_n ,w_n) + V_{n+1}\big( F_n(\x_n,\bu_n,w_n)\big)\big|\x_n,w_0=w^i\Big]\\
&= \min_{\bu_n} \sum_j P_{ij}^{(n)} \Big[g_n(\x_n,\bu_n,w^j) + V_{n+1}\big(F_n(\x_n,\bu_n,w^j)\big)\Big] \label{eq:Vn}
\end{align}
With this state value function, the expectation is conditioned on fixed disturbance information available at the start of the horizon, $w_0=w^i$.  As a result, the transition probabilities change along the horizon according to the multi-step transition probability $P_{ij}^{(n)}$.  The value function $V_n$ can also be given in terms of the  state-control value function $Q_n$ according to $V_n(\x) = Q_n(\x,\bu^*) $ where $\bu^*=\arg\min_\bu Q_n(\x,\bu)$ and $Q_n$ is defined in a manner consistent with Equation (\ref{eq:Vn})  
\begin{align}
Q_n(\x_n,\bu_n)  &= \mathbb{E}\Big[ g_n(\x_n,\bu_n ,w_n) + V_{n+1}\big( F_n(\x_n,\bu_n,w_n)\big)\big|\x_n,w_0=w^i\Big] \nonumber \\
&=\sum_{j}P_{ij}^{(n)} \Big[ g_n(\x_n,\bu_n , w^j) + V_{n+1}\big( F_n(\x_n,\bu_n,w^j)\big) \Big] \label{eq:Qn_ASDDP}
\end{align}
Given a nominal trajectory $\(\hat \x_n,\hat \bu_n\)_{n=0}^{N-1}$ a local model of $Q_n$ to second order is constructed as
\begin{align}\label{eq:LocalQnModel}
&Q_n(\hat \x_n+\delta \x_n,\hat \bu_n+\delta \bu_n) \approx \nonumber \\
&~~~~~~~~~~~Q_n^{(0)} + Q_n^{(x)}\delta \x_n + Q_n^{(u)}\delta \bu_n +\frac{1}{2}\left[\delta \x_n\Tr~ \delta \bu_n\Tr\right]
\begin{bmatrix}
Q_n^{(xx)} & Q_n^{(xu)}\\
Q_n^{(ux)} & Q_n^{(uu)}
\end{bmatrix}
\begin{bmatrix}
\delta \x_n \\
\delta \bu_n
\end{bmatrix} 
\end{align}
Here, $\delta \x_n$ and $\delta \bu_n$ are small perturbations in the state and control vectors at time $n$ and  $Q_n^{(0)} \triangleq Q_n(\hat \x_n,\hat \bu_n)$.  The partial derivatives $Q_n^{(x)},Q_n^{(u)},Q_n^{(xx)},Q_n^{(uu)},Q_n^{(ux)}$ centered about $(\hat \x_n,\hat \bu_n)$ are determined considering Equation (\ref{eq:Qn_ASDDP})
\begin{subequations} \label{eq:QnPartials}
	\begin{align} 
	Q_n^{(a)} &= \sum_{j}P_{ij}^{(n)} \left[g_n^{(a)}(\hat q_n) + V_{n+1}^{(x)}(\x') F_n^{(a)}(\hat q_n)\right] \\
	Q_n^{(ab)} &= \sum_{j}P_{ij}^{(n)} \Big[g_n^{(ab)}(\hat q_n) + F_n^{(a)\mathsf{T}}(\hat q_n) V_{n+1}^{(xx)}(\x') F_n^{(b)}(\hat q_n) \Big]
	\end{align}
\end{subequations} 
where $\hat q_n \triangleq (\hat \x_n, \hat \bu_n, w^j)$ and $\x' \triangleq F_n(\hat \x_n,\hat \bu_n,w^j)$.  To reduce computational burden, the second order derivatives $F_n^{(xx)},F_n^{(ux)},F_n^{(uu)}$ have been neglected in the last equation of (\ref{eq:QnPartials}).  For given $\hat \x_n, \hat \bu_n,\delta \x_n$, the unconstrained value of $\delta \bu_n$ which minimizes the local model Equation (\ref{eq:LocalQnModel}) is 
\begin{align}\label{eq:unLocalMinUnconstrained}
\delta \bu_n^* = \arg\min_{\delta \bu_n} Q_n = -\left(Q_n^{(uu)}\right)^{-1}\left(Q_n^{(u)}+Q_n^{(ux)}\delta \x_n\right)
\end{align}
Substituting $\delta \bu_n^*$ into the local model Equation (\ref{eq:LocalQnModel}) and simplifying gives a local second order model for $V_n(\x)$ about the nominal trajectory $(\hat \x_n)_{n=0}^{N-1}$ for arbitrary $\x$ where $\delta \x_n = \x - \hat\x_n$
\begin{align}\label{eq:LocalVnModel}
V_n&(\x) \approx 
Q_n^{(0)}-\frac{1}{2} Q_n^{(u)\mathsf{T}}(Q_n^{(uu)})^{-1}Q_n^{(u)}+ \left[Q_n^{(x)}-Q_n^{(u)}(Q_n^{(uu)})^{-1}Q_n^{(ux)}\right](\x-\hat \x_n)  \nonumber\\
&~~~~~~~~~~+ \frac{1}{2}(\x_n-\hat \x_n)\Tr\left[Q_n^{(xx)}-Q_n^{(xu)}(Q_n^{(uu)})^{-1}Q_n^{(ux)}\right](\x-\hat \x_n) 
\end{align}
For fixed $\hat \x_n$, the partial derivatives of Equation (\ref{eq:LocalVnModel}) are evaluated at arbitrary $\x$ according to
\begin{subequations}\label{eq:VnPartials}
	\begin{align}
	V_N^{(x)}(\x) &= h^{(x)}(\x) \\
	V_N^{(xx)}(\x) &= h^{(xx)}(\x) \\
	V_n^{(x)}(\x) &=[Q_n^{(x)}-Q_n^{(u)}(Q_n^{(uu)})^{-1}Q_n^{(ux)}] +  [Q_n^{(xx)}-Q_n^{(xu)}(Q_n^{(uu)})^{-1}Q_n^{(ux)}] (\x-\hat \x_n) \\
	V_n^{(xx)}(\x) &=Q_n^{(xx)}-Q_n^{(xu)}(Q_n^{(uu)})^{-1}Q_n^{(ux)}
	\end{align}
\end{subequations}
Starting from initial condition $V_N(\hat \x_N) = h(\hat \x_N)$, Equation (\ref{eq:QnPartials}) and Equation (\ref{eq:VnPartials}) are evaluated backwards in time along the horizon about the nominal trajectory $(\hat \x_n,\hat \bu_n)_{n=0}^{N-1} $ which constitutes the \textit{backward pass}.  

The next step is to update the nominal trajectory $(\hat \x_n,\hat \bu_n)_{n=0}^{N-1}$ by simulating the system forward in time along the horizon, which constitutes the \textit{forward pass}.  Unlike the classic deterministic case of DDP, the forward pass is uncertain in the stochastic setting as state trajectory $(\x_n)_{n=0}^{N-1}$ depends on the realization of the stochastic disturbance trajectory $(w_n)_{n=0}^{N-1}$ .   The expected nominal state trajectory is generated for a given control sequence considering disturbance information available at the beginning of the horizon according to
\vspace{-1mm}
\begin{align}
\bar \x_{n+1} &= \mathbb{E}[F_n(\bar \x_n,\bu_n,w_n)|\bar \x_n,w_0 = w^i]\nonumber \\
&= \sum_{j}P_{ij}^{(n)}F_n(\bar \x_n,\bu_n,w^j)
\end{align}
Starting from initial condition $\x_0 = \x_0^\text{meas}$, a new system trajectory is simulated forward in time along the horizon $n=0,\dots,N-1$ according to Equation (\ref{eq:FowardPass}) which represents the \textit{forward pass}

\begin{subequations}\label{eq:FowardPass}
	\begin{align}
	\bar \x_0 &= \x_0^{\text{meas}}~,~w_0=w_0^{\text{meas}}\\
	\bu_{n}^* &= \hat \bu_n \underbrace{-\left(Q_n^{(uu)}\right)^{-1}\left[Q_n^{(u)}-Q_n^{(ux)}\left(\bar \x_n - \hat \x_n\right)\right]}_{\delta \bu_n^*} \label{eq:unUpdate}\\
	\bar \x_{n+1} &= \sum_{j}P_{ij}^{(n)}F_n(\bar \x_n,\bu_n^*,w^j)
	\end{align}
\end{subequations}
The new nominal trajectory is updated according to $\left\{\hat \x_n,\hat \bu_n\right\}_{n=0}^{N-1} := \left\{\bar \x_n,\bu_n^*\right\}_{n=0}^{N-1}$ and the process is restarted.

\subsubsection{State - Control Constraints}
Minimizing the local model of $Q_n$ given by Equation (\ref{eq:LocalQnModel}) is an unconstrained quadratic optimization problem, whose solution is given by Equation (\ref{eq:unLocalMinUnconstrained}).  However, with some modification the problem of minimizing Equation (\ref{eq:LocalQnModel}) subject to state and control input constraints in a stochastic environment can be addressed.  A first order expansion about $(\bar \x_n,\hat \bu_n)$ is taken to produce an approximation to the system dynamics that is linear in the control input 

\begin{align}
\bar \x_{n+1} &= \sum_{j}P_{ij}^{(n)}F_n(\bar \x_n,\bu_n,w^j) \nonumber \\
&\approx \sum_{j}P_{ij}^{(n)} \left[ F_n(\bar \x_n,\hat \bu_n,w^j) + F_n^{(u)}(\bar \x_n,\hat \bu_n,w^j) \delta \bu_n \right]
\end{align}
The state and control vectors are constrained according to
\begin{subequations}
	\begin{align}
	D_x \bar \x_{n+1} &\leq \mathbf c_x \\
	D_u \left[\hat \bu_n + \delta \bu_n\right]  &\leq \mathbf c_u
	\end{align}
\end{subequations}
Combining these equations leads to the following constrained quadratic programming problem, which is solved with an active set strategy \cite{Nocedal2006}

\begin{subequations} \label{eq:QuadProgASDDP}
	\begin{align}
	&\min_{\delta \bu_n} ~~~~~~~~~~~\frac{1}{2} \delta \bu_n \Tr Q_n^{(uu)}\delta \bu_n + \left(Q_n^{(u)} + \delta \x_n \Tr Q_n^{(xu)}\right) \delta \bu_n \\
	&\text{subject to} ~~~~D\delta \bu_n \leq \mathbf c \\
	&~~~~~~~~~~~~~~~~~D = \begin{bmatrix}
	D_x \sum_{j}P_{ij}^{(n)}  F_n^{(u)}(\bar \x_n,\hat \bu_n,w^j) \\
	D_u
	\end{bmatrix} \\
	&~~~~~~~~~~~~~~~~~~\mathbf c = \begin{bmatrix}
	\mathbf c_x - \sum_{j}P_{ij}^{(n)}  F_n(\bar \x_n,\hat \bu_n,w^j) \\
	\mathbf c_u - D_u \hat \bu_n 
	\end{bmatrix}
	\end{align}
\end{subequations}
Solving the quadratic programming problem described by Equation (\ref{eq:QuadProgASDDP}) constrains the expected state trajectory along the horizon considering control input constraints.  

\subsubsection{Modification for Global Convergence}
A standard modification is made to ensure the Hessian matrix $Q_n^{(uu)}$ is positive definite at all stages along the horizon.  In this way, convergence occurs even far from the solution when $Q_n^{(uu)}$ may not be positive definite.  A simple method is used based on Hessian modification in standard Newton iteration \cite{Shoemaker1991,Nocedal2006},
\begin{subequations} \label{eq:HessianMod}
	\begin{align}
	Q_n^{(uu)} &:=  Q_n^{(uu)} + \tau I
	\end{align}
	where
	\begin{align}
	\tau = \left\{
	\begin{array}{rr}
	\delta - \lambda_{min}\left(Q_n^{(uu)}\right) ,& ~~~ \delta >  \lambda_{min}\left(Q_n^{(uu)}\right) \\ 
	0, &~~~ \delta \leq  \lambda_{min}\left(Q_n^{(uu)}\right) 
	\end{array} \right.
	\end{align}
\end{subequations}
The modification performed by Equation (\ref{eq:HessianMod}) ensures the smallest eigenvalue of $Q_n^{(uu)}$ is no less than $\delta > 0$, which in this work is set to $\delta = 0.003$.  It is worthing noting that the same control input scalings $m_1$ and $m_2$ used in Section \ref{section:SGDM} are used for the ASDDP algorithm.  The benefit of using input scalings here is that the eigenvalues of $Q_n^{(uu)}$ have approximately the same magnitude.  The ASDDP algorithm is summarized in Algorithm \ref{Alg:ASDDP}.

\begin{algorithm}[h!]	
	\setstretch{1.2}
	\caption{ASDDP} \label{Alg:ASDDP}
	\KwIn{$\x_0,w_0, \(\hat \x_n,\hat \bu_n\)_{n=0}^{N}$}		
	$\hat\x_0:=\x_0, w^i:=w_0$\\
	-----\textit{Backward Pass}-----\\
	$\{Q^{(x)}_n,Q^{(u)}_n,Q^{(xx)}_n,Q^{(uu)}_n,Q^{(ux)}_n\} = 0$ \\
	\For{$n=N-1:0$}{	
		\ForEach{$w^j\in W$}				
		{$\x_{n+1} = F_n(\hat \x_n,\hat \bu_n,w^j)$ \\
			\If{n=N-1}{
				$V_{n+1}^{(x)} = h^{(x)}(\x_{n+1})$, ~
				$V_{n+1}^{(xx)} = h^{(xx)}(\x_{n+1})$ \\}
			\Else{
				$V_{n+1}^{(x)} = A + B\big[\x_{n+1}-\hat \x_{n+1}\big]$,~
				$V_{n+1}^{(xx)} = B$
			}	
			$Q^{(x)}_n  = Q^{(x)}_n  + P_{ij}^{(n)}\big[g_n^{(x)}  + V_{n+1}^{(x)}F_n^{(x)}\big]$\\
			$Q^{(u)}_n  = Q^{(u)}_n  + P_{ij}^{(n)}\big[g_n^{(u)}  + V_{n+1}^{(x)}F_n^{(u)}\big]$\\
			$Q^{(xx)}_n  = Q^{(xx)}_n  + P_{ij}^{(n)}\big[g_n^{(xx)}  + F_n^{(x)T} V_{n+1}^{(xx)}F_n^{(x)}\big]$\\
			$Q^{(uu)}_n  = Q^{(uu)}_n  + P_{ij}^{(n)}\big[g_n^{(uu)}  + F_n^{(u)T} V_{n+1}^{(xx)}F_n^{(u)}\big]$\\
			$Q^{(ux)}_n  = Q^{(ux)}_n  + P_{ij}^{(n)}\big[g_n^{(ux)}  + F_n^{(u)T} V_{n+1}^{(xx)}F_n^{(x)}\big]$\\
			$Q^{(xu)}_n = Q^{(ux)T}_n$
		}	
		Modify $Q^{(uu)}_n$ according to Equation (\ref{eq:HessianMod})\\
		$A=Q_n^{(x)}-Q_n^{(u)}\big[Q_n^{(uu)}\big]^{-1}Q_n^{(ux)}$,   $B=Q_n^{(xx)}-Q_n^{(xu)}\big[Q_n^{(uu)}\big]^{-1}Q_n^{(ux)}$ 
	}		
	-----\textit{Forward Pass}----- \\
	$\bar \x_0 := \x_0$\\	
	\For{$n=0:N-1$}
	{   
		$\delta \x_n := \bar \x_n - \hat\x_n$\\
		Solve QP subproblem Equation (\ref{eq:QuadProgASDDP}) for $\delta \u_n$\\
		$\u_n^*:=\hat\u_n + \delta \u_n$\\
		$\bar \x_{n+1} = \sum_{j}P_{ij}^{(n)}F_n(\bar \x_n,\bu_n^*,w^j)$
	}
	
	\KwOut{$\(\hat \x_n,\hat \bu_n\)_{n=0}^{N} := \(\bar \x_n,\bu_n^*\)_{n=0}^{N}$}	  
\end{algorithm}

\subsubsection{Remarks on Computational Complexity of ASDDP}
In retrospect the value function shown in (\ref{eq:Vn}) is similar to a stochastic variant of DDP presented in \cite{Jacobson1970DDP} in which $V_n$ is explicitly dependent on the stochastic state.  However, here Equation (\ref{eq:Vn}) is not explicitly dependent on the stochastic state due to the fact that ASDDP incorporates the multi-step Markov transition probability $P_{ij}^{(n)}$.  As such, (\ref{eq:Vn}) must only be evaluated for every $w^j \in {W}$, not for every $(w^i,w^j) \in {W}\times{W}$. This significantly reduces the computational complexity of the \textit{backward pass} from $O(|{W}|^2)$ to $O(|{W}|)$ making ASDDP more suitable for real time implementation.

\newpage
\subsection{Average Path Differential Dynamic Programming (APDDP)}\label{section:APDDP}
We now develop \textit{average path differential dynamic programming} (APDDP) to approximately solve Equation (\ref{eq:FHSOCP}).  The problem is re-formulated as
\begin{subequations}\label{eq:FHSOCP_APDDP}
	\begin{align}
	\min_{\bu_0,\bu_1,\dots,\bu_{N-1}} &\sum_{n=0}^{N-1} g_n(\x_n,\x_{n+1},\bu_n,\bar w_n)\Big|\x_0,w_0\\
	\text{subject to}~~	&\bar w_n = \sum_j P_{ij}^{(n)}w^j\\
	& \x_{n+1} = F_n(\x_n,\bu_n,\bar w_n) \\
	&\bar \x_{n+1} = \sum_{j}P_{ij}^{(n)}F_n(\bar \x_n,\bu_n,w^j) \\
	&D_x \bar \x_{n+1} \leq \mathbf c_x \\
	&D_u \bu_n   \leq \mathbf c_u
	\end{align}	
\end{subequations}
Average path differential dynamic programming is identical to the ASDDP method described in Section \ref{section:ASDDP} except the state-control value function is constructed as
\begin{align}\label{eq:Qn_APDDP}
Q_n(\x_n,\bu_n)  &=  g_n(\x_n,\bu_n ,\bar w_n) + V_{n+1}\big( F_n(\x_n,\bu_n,\bar w_n)\big) 
\end{align}
where the average disturbance path is defined as
\begin{align}
\bar w_n = \sum_j P_{ij}^{(n)}w^j
\end{align}

Compared to ASDDP, the primary benefit with APDDP is a significant reduction in computational burden since the summations $\sum_j P_{ij}^{(n)}$ associated with stochastic computations are nearly eliminated during the backward pass.  Through numerical experimentation it was found that APDDP had trouble meeting driver demand when using the same calibrations from ASDDP (i.e. $K_3$ from Equation (\ref{eq:gn}) and $\alpha, \beta$ from Equation (\ref{eq:wset})).  This is likely due to the fact that whereas ASDDP is evaluating all possible values of the disturbance $w_n=w^j, j\in W$ during creation of the state-control value function Equation (\ref{eq:Qn_ASDDP}), APDDP only evaluates the average value $\bar w_n$ during creation of the state-control value function Equation (\ref{eq:Qn_APDDP}).  As a result, APDDP will ignore the impact of disturbance values which deviate from the averaged disturbance value along the horizon.   To remedy this, gains $K_3, \alpha$, and $\beta$ were increased until APDDP was able to satisfy driver demands.  Meeting driver demand is discussed further in a quantitative manner in Section \ref{section:PerformanceMetrics}.  The APDDP algorithm is summarized in Algorithm \ref{Alg:APDDP}.

\subsection{Block Diagram of Stochastic Control Algorithms}
\begin{figure}[h!]
	\centering
	\begin{minipage}{0.8\textwidth}
		\includegraphics [trim = 0mm 90mm 140mm 0mm, clip,width=1.0\textwidth]{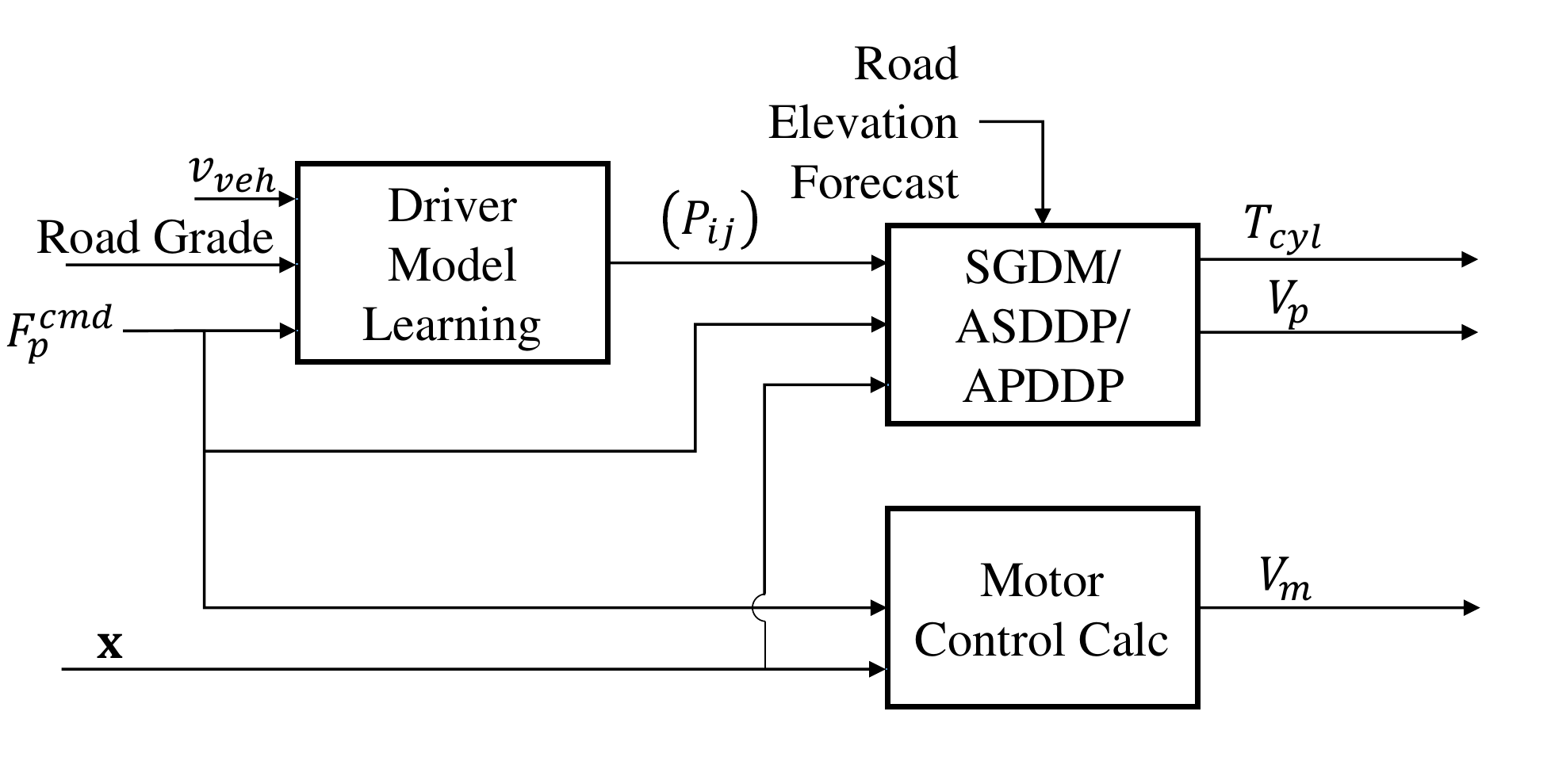}
	\end{minipage}
	\caption{Stochastic algorithm block diagram.}
	\label{fig:StochAlgorithms}
\end{figure}

The implementation of SGDM, ASDDP, and APDDP is shown in Fig. \ref{fig:StochAlgorithms}. 
Each of these algorithms relies on the learned statistical model of driver behavior $\(P_{ij}\)$ to form decisions along the horizon $n=0,1,\dots, N-1$.  The sequence $\(\x_n^*,\u_n^*\)_{n=0}^{N-1}$ is recomputed every $T_s$ seconds.  The motor displacement volume, $V_m$, is updated according to Equation (\ref{eq:MotorDispCalc}).  Using the scaling factors of Equation (\ref{eq:InputScaling}), the inputs $T_{cyl}$ and $V_p$ are formed using the first element from the control sequence  
\begin{align}\label{eq:ControlOutputTs}
\begin{bmatrix}
T_{cyl} \\
V_p
\end{bmatrix} = \begin{bmatrix}
m_1&0 \\
0 & m_2
\end{bmatrix}
\u_0^*
\end{align}
The driver model learning process is described by Equations (\ref{eq:PijEstimator}) and (\ref{eq:wMeasured}),  motor displacement volume calculation is given by Equation (\ref{eq:MotorDispCalc}).

\begin{algorithm}[h!]	
	\setstretch{1.1}
	\caption{APDDP} \label{Alg:APDDP}
	\KwIn{$\x_0,w_0, \(\hat \x_n,\hat \bu_n\)_{n=0}^{N}$}		
	$\hat\x_0:=\x_0, w^i:=w_0$\\
	-----\textit{Backward Pass}-----\\
	$\{Q^{(x)}_n,Q^{(u)}_n,Q^{(xx)}_n,Q^{(uu)}_n,Q^{(ux)}_n\} = 0$ \\
	\For{$n=N-1:0$}{	
		$\bar w_n = \sum_j P_{ij}^{(n)}w^j$\\					
		$\x_{n+1} = F_n(\hat \x_n,\hat \bu_n,\bar w_n)$ \\
		\If{n=N-1}{
			$V_{n+1}^{(x)} = h^{(x)}(\x_{n+1})$, ~
			$V_{n+1}^{(xx)} = h^{(xx)}(\x_{n+1})$ \\
		}
		\Else{
			$V_{n+1}^{(x)} = A + B\big[\x_{n+1}-\hat \x_{n+1}\big]$,~
			$V_{n+1}^{(xx)} = B$
		}	
		$Q^{(x)}_n  = g_n^{(x)}  + V_{n+1}^{(x)}F_n^{(x)}$\\
		$Q^{(u)}_n  = g_n^{(u)}  + V_{n+1}^{(x)}F_n^{(u)}$\\
		$Q^{(xx)}_n  = g_n^{(xx)}  + F_n^{(x)T} V_{n+1}^{(xx)}F_n^{(x)}$\\
		$Q^{(uu)}_n  =g_n^{(uu)}  + F_n^{(u)T} V_{n+1}^{(xx)}F_n^{(u)}$\\
		$Q^{(ux)}_n  = g_n^{(ux)}  + F_n^{(u)T} V_{n+1}^{(xx)}F_n^{(x)}$\\
		$Q^{(xu)}_n = Q^{(ux)T}_n$\\
		
		Modify $Q^{(uu)}_n$ according to Equation (\ref{eq:HessianMod})\\
		$A=Q_n^{(x)}-Q_n^{(u)}\big[Q_n^{(uu)}\big]^{-1}Q_n^{(ux)}$,   $B=Q_n^{(xx)}-Q_n^{(xu)}\big[Q_n^{(uu)}\big]^{-1}Q_n^{(ux)}$ 
	}		
	-----\textit{Forward Pass}----- \\
	\For{$n=0:N-1$}
	{   
		$\delta \x_n := \x_n - \hat\x_n$\\
		Solve QP subproblem Equation (\ref{eq:QuadProgASDDP}) for $\delta \u_n$\\
		$\u_n^*:=\hat\u_n + \delta \u_n$\\
		$\bar \x_{n+1} = \sum_{j}P_{ij}^{(n)}F_n(\bar \x_n,\bu_n^*,w^j)$
	}
	
	\KwOut{$\(\hat \x_n,\hat \bu_n\)_{n=0}^{N} := \( \x_n,\bu_n^*\)_{n=0}^{N}$}	  
\end{algorithm}

\section{Benchmark Strategies}
Two benchmark strategies are provided as a means to evaluate SGDM, ASDDP, and APDDP.  First, a baseline strategy based on instantaneous optimization is representative of that which can be achieved without consideration of upcoming driver demands or road elevation.  Second, a theoretical best strategy is created to demonstrate the best which can be achieved when all cycle information available is provided to the decision making process.  Like SGDM, ASDDP, and APDDP, the baseline strategy is implementable as a real time control algorithm, whereas the theoretical best strategy is not.  

\subsection{Baseline: Instantaneous Optimization}
\begin{figure}[H]
	\centering
	\rotatebox[origin=c]{90}{%
		\begin{minipage}{1.45\textwidth}
			\centering
			\includegraphics [trim = -10mm 20mm 20mm 0mm, clip,width=1\textwidth]{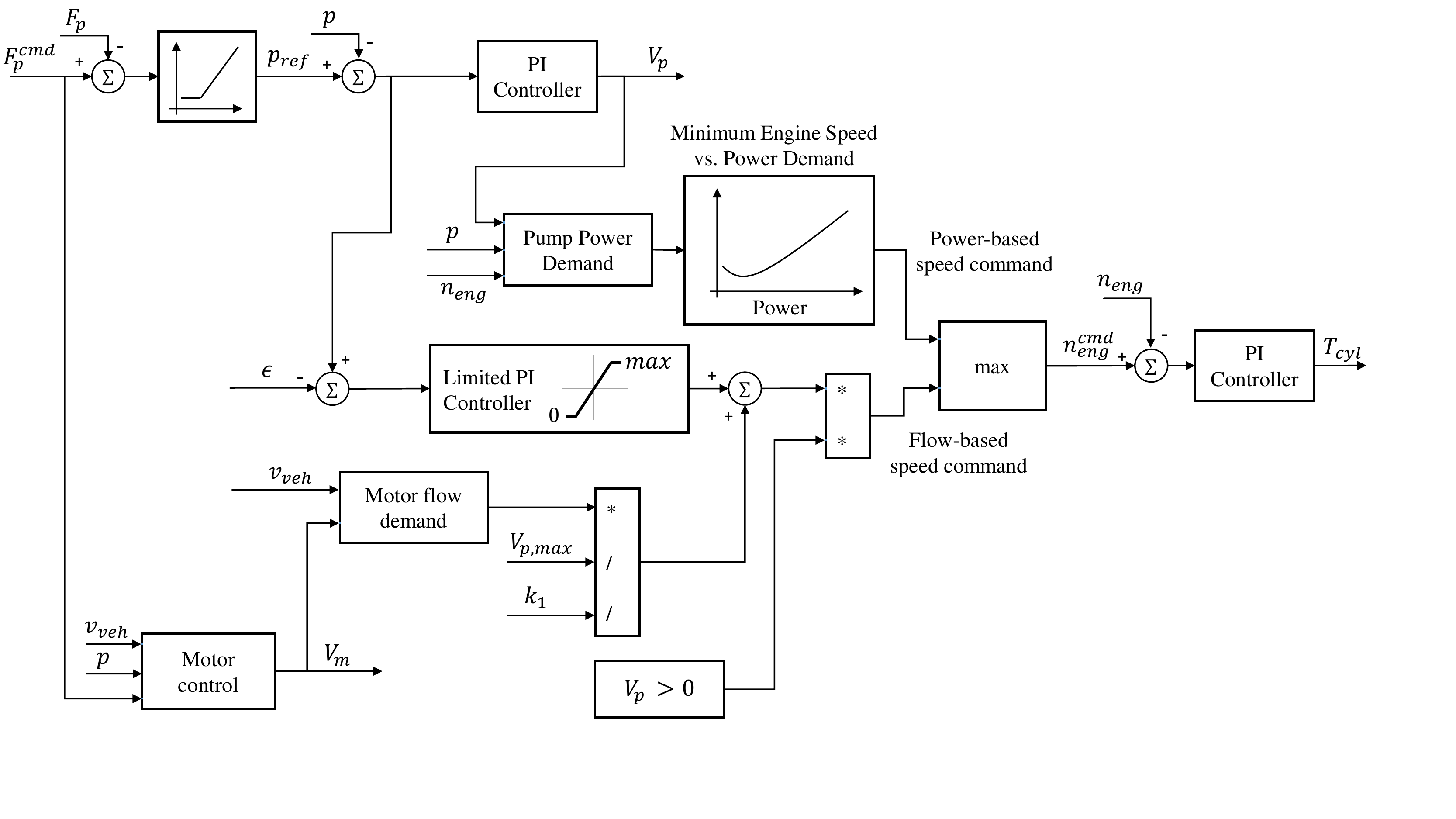}
			\caption{Instantaneous optimization strategy (InstOpt).}
			\label{fig:BaselineStrategy}
		\end{minipage}
	}
\end{figure}

A baseline strategy based on instantaneous optimization (InstOpt) is created, similar to that developed in \cite{SDP4}.  The control inputs are generated to minimize the instantaneous fuel consumption rate considering current operating conditions and neglecting the effect of future driver demands and road elevation.  The strategy is described in Fig. \ref{fig:BaselineStrategy}.  Pump displacement volume is controlled according to a proportional-integral (PI) controller processes to maintain some minimum pressure in the accumulator denoted as $p_{ref}$.  This minimum pressure reference is held fixed at some nominal value and gradually raised if the driver propulsion force demand is not satisfied.  The engine is managed to deliver the minimum speed that can satisfy the power demanded by the pump.  If the accumulator pressure falls to some level $\epsilon$ below $p_{ref}$, engine speed may be commanded to increase according to a limited PI controller process.  A minimum engine speed is set so that the pump can always provide enough flow to satisfy the motor flow demand, unless pump displacement volume is zero in which case this flow-based engine speed command is zero.  The motor displacement is controlled according to Equation (\ref{eq:MotorDispCalc}).  Parameters of the baseline strategy were iteratively calibrated so the strategy performed well on all three drive cycles, with emphasis placed on performance under the UDDS drive cycle.  Once established, these parameters were unchanged from one cycle to the next.  The reference pressure was set to 150 bar, with precharge pressure set to 135 bar (90\% of the reference pressure).  Justification for the 150 bar reference pressure is established with Fig. \ref{fig:FpCmdDistribution} in ~~~~ Section \ref{section:Simulation}.

\subsection{Theoretical Best: Deterministic Differential Dynamic Programming with Driver Forecast}

\begin{figure}[h!]
	\centering
	\begin{minipage}{0.80\textwidth}
		\includegraphics [trim = 0mm 100mm 180mm 0mm, clip,width=1\textwidth]{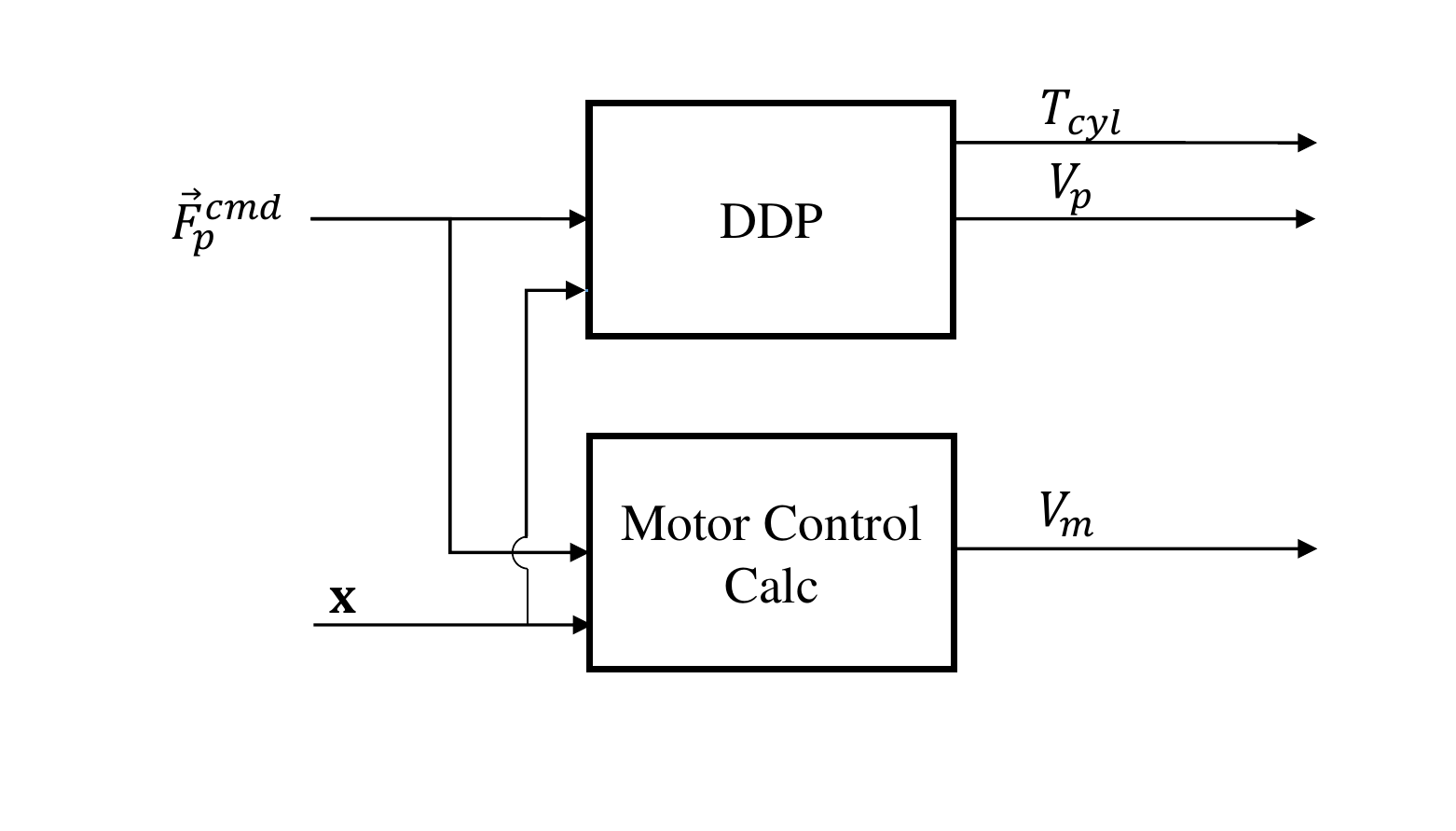}
	\end{minipage} 	
	\caption{Theoretial best strategy: DDP with driver forecast.}
	\label{fig:DDPAlg}
\end{figure}

The classic (deterministic) differential dynamic programming algorithm discussed in Section \ref{section:DDP/iLQR} is used to generate a theoretically best controller to serve as a basis for comparison.  The implementation of \textit{DDP with driver forecast} (DDP for short) is shown in Fig. \ref{fig:DDPAlg}.   
Unlike the stochastic algorithms discussed in Section \ref{section:StochControlFormulations}, DDP has full access to the propulsion force command sequence along the horizon, $\vec F_p^{cmd}$.    Consequently, the DDP algorithm is not actually implementable in practice.  The values for $T_{cyl}$ and $V_p$ are generated every $T_s$ seconds according to Equation (\ref{eq:ControlOutputTs}).  The value for $V_m$ is updated every 0.01 seconds according to Equation (\ref{eq:MotorDispCalc}).



\chapter{SIMULATION}\label{section:Simulation}

\begin{figure}[h!]
	\centering
	\begin{minipage}{1\textwidth}
		\includegraphics [width=1\textwidth]{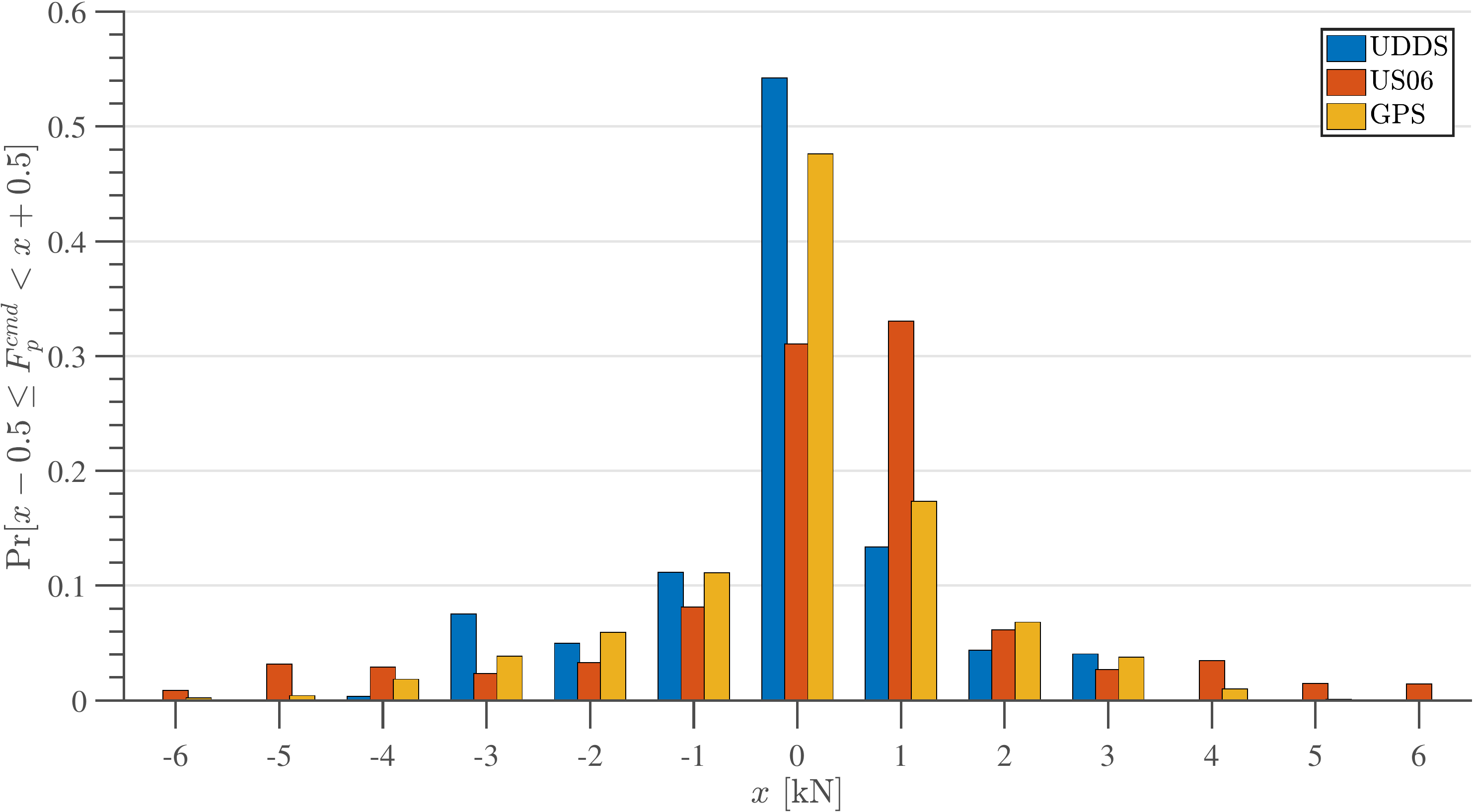}
	\end{minipage}
	\caption{Driver propulsion force command distribution for each drive cycle. }
	\label{fig:FpCmdDistribution}
\end{figure}

Simulation is performed in Matlab Simulink for the series-hybrid configuration shown in Fig. \ref{fig:seriesHHV}.  A mid-size sport utility vehicle is simulated with parameters shown in Table \ref{table:SUVParameters}.  The system is designed so that maximum propulsion force, $F_p^{max}$, can be achieved when differential system pressure is 290 bar when the vehicle is in low gear.  
\begin{table}[h!]
	\caption{Series-Hybrid SUV Parameters.}
	\label{table:SUVParameters}
	\vspace{-2mm}
	\centering 
	\setstretch{1.1}
\begin{tabular}{llll}	
	\textbf{Description} & \textbf{Symbol} & \textbf{Value} & \textbf{Units}  \\
	\hline 
	Vehicle mass & $m_{veh}$ & 2091 & kg \\
	Max eng. power & $P_{eng}^{max}$ & 125 & kW \\
	Max propulsion force & $F_p^{max}$ & 6500 & N \\
	Max vehicle speed & $v_{veh}^{max}$ & 125 & km/h \\
	Dynamic tire radius& $r_{tire}$ & 0.35 & - \\
	Aero drag coefficient& $C_d$ & 1.62 & - \\
	Rolling resistance coefficient& $C_r$ & 0.010 & - \\
	Engine inertia& $I_{eng}$ & 0.5 & kg$\cdot$m$^2$ \\
	Gear ratio 1 & $k_1$ & 1 & - \\
	Gear ratio 2: lo, hi & $k_{2,lo},k_{2,hi}$ & 10, 6.67 & - \\
	Gear ratio 2 lo/hi thresh & $v_{veh,hi}$ & 20 & m/s \\
	Displacement vol. of hyd. pump & $V_p^{max}$ & 63 & cc/rev \\
	Displacement vol. of hyd. motor & $V_m^{max}$ & 50  & cc/rev \\
	Hyd. accumulator precharge vol.& $V_{ha}$ & 50 & L \\
	Hyd. accumulator precharge press.& $p_{ha}$ & 70 & bar \\
	Max differential system press. & $p_{max}$ & 350 & bar \\
	Low-pressure accum press. & $p_{lp}$ & 10 & bar \\
	\hline
\end{tabular} 
\end{table}
The distribution of driver propulsion force command for each of the cycles investigated is shown in Fig. \ref{fig:FpCmdDistribution}.
This distribution indicates the fraction of time the driver spends commanding various levels of propulsion force.  For example, in the UDDS cycle the driver commands a propulsion force between -500 and 500N for approximately 55\% of the cycle.  At the far extreme a propulsion force between 5500 and 6500N is requested during the US06 cycle for approximately 1.4\% of the cycle (8.4 seconds).  Recall that the reference differential system pressure for the baseline strategy InstOpt is $p_{ref}=150$ bar, so that a 3500N propulsion force can be generated in low gear at the reference pressure.  Referring to Fig. \ref{fig:FpCmdDistribution}, a propulsion force of 3500N covers the majority of driving demands for the cycles investigated.  When a propulsion force greater than 3500N is commanded, the baseline strategy will need to increase the differential system pressure as described in Fig. \ref{fig:BaselineStrategy}.

\section{Simulation Setup}
\begin{figure}[h!]
	\centering
	\begin{minipage}{1\textwidth}
		\includegraphics [trim = 0mm 30mm 0mm 30mm, clip,width=1.0\textwidth]{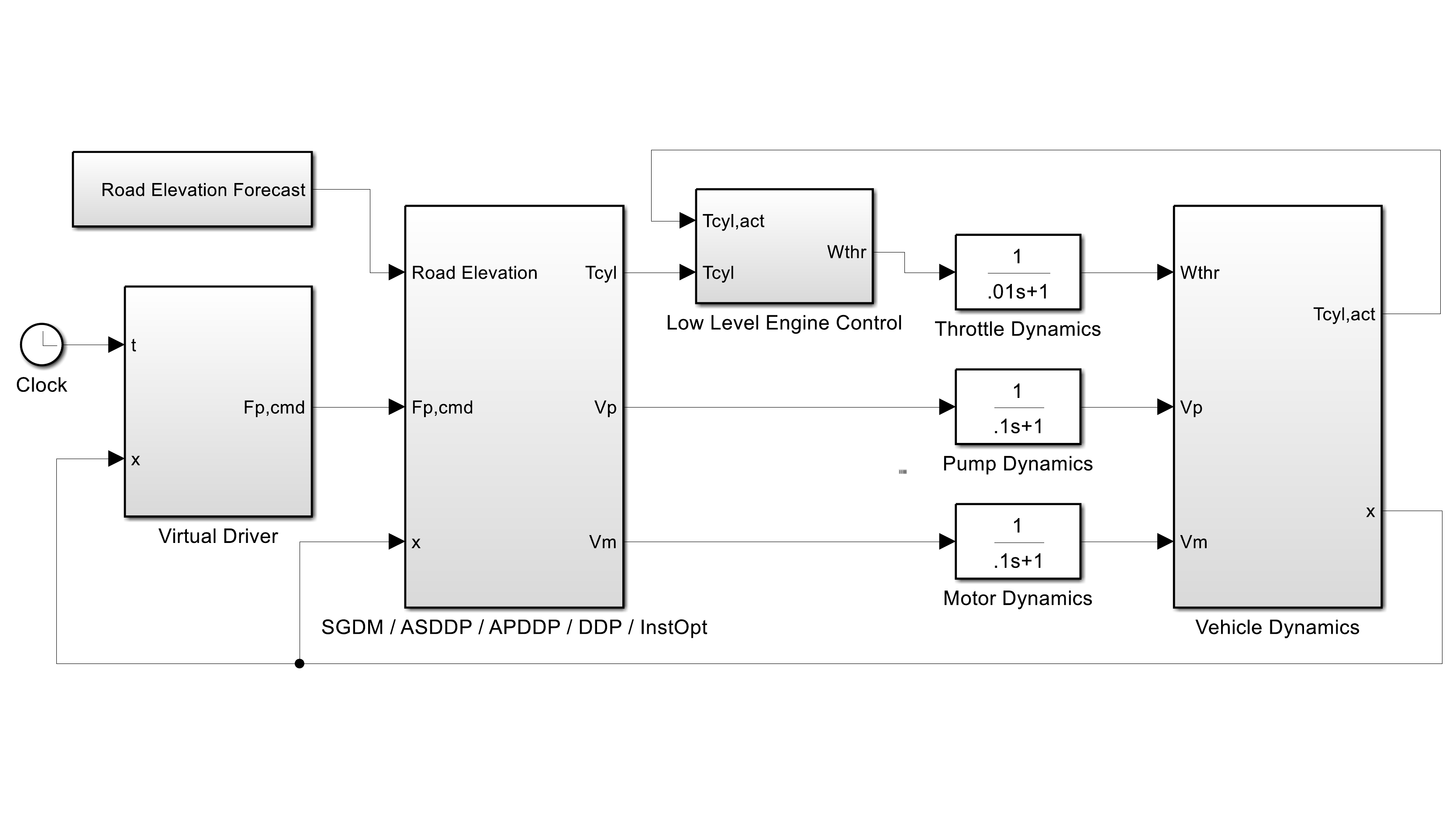}
	\end{minipage}
	\caption{Stochastic algorithm block diagram.}
	\label{fig:SimSetup}
\end{figure}

The simulation configuration is shown in Fig. \ref{fig:SimSetup}.  The vehicle dynamics block contains the engine, vehicle and hydraulics dynamics described in Section \ref{section:SeriesHHVDynamics}.  The algorithm block contains the embedded system model described in Section  
\ref{section:SystemModel} and one of the algorithms described in Chapter \ref{section:PredEnergyManage} (either SGDM, ASDDP, APDDP, DDP, or InstOpt).  The road elevation forecast block described in Section \ref{section:RoadGradeForecasting} provides elevation information along the horizon.  The SGDM, ASDDP, APDDP and DDP algorithms generate control inputs $T_{cyl}$ and $V_p$ every $T_s = 0.1$ seconds and input $V_m$ every 0.01 seconds.  InstOpt generates all three contorl inputs every 0.01 seconds. 

A virtual driver is created which generates propulsion force command $F_p^{cmd}$ along the three drive cycles of Fig. \ref{fig:DriveCycles}.  The virtual driver is a PI controller which tracks the drive cycle's reference vehicle speed $v_{veh}^{ref}$ according to
\begin{align}\label{eq:VirtualDriver}
F_p^{cmd}(t) &=  k_p \left(v_{veh}^{ref}(t)-v_{veh}(t)\right) + \int\limits_{0}^{t}k_i\left(v_{veh}^{ref}(\tau)-v_{veh}(\tau)\right)d\tau
\end{align}
The gains $k_p$ and $k_i$ were tuned so that even a small speed tracking error $v_{veh}^{ref}-v_{veh}$ results in a large propulsion force command.  To ensure excellent speed tracking for all three cycles the penalty $K_3$ from cost rate function Equation (\ref{eq:gn}) is made large so that the tracking of $F_p^{cmd}$ is also excellent, as will be shown.

In the low level engine control block the cylinder torque control input, $T_{cyl}$, is converted into an engine throttle mass flow command, $W_{thr}$, though a simple PI controller
\begin{align}\label{eq:ThrottleController}
W_{thr}(t) &=  k_p \(T_{cyl}(t)-T_{cyl,act}(t)\) + \int\limits_{0}^{t}k_i\(T_{cyl}(\tau)-T_{cyl,act}(\tau)\) d\tau
\end{align}

\section{Cycle Analysis}
In this section some results of the SGDM, ASDDP, DDP and InstOpt algorithms are compared qualitatively.  Reference speed tracking and state / control trajectories are examined.  

\subsection{UDDS Cycle}
\begin{figure}[h] 
	\centering
		\includegraphics [trim = 0mm 0mm 0mm 0mm, clip,width=1.0\textwidth]{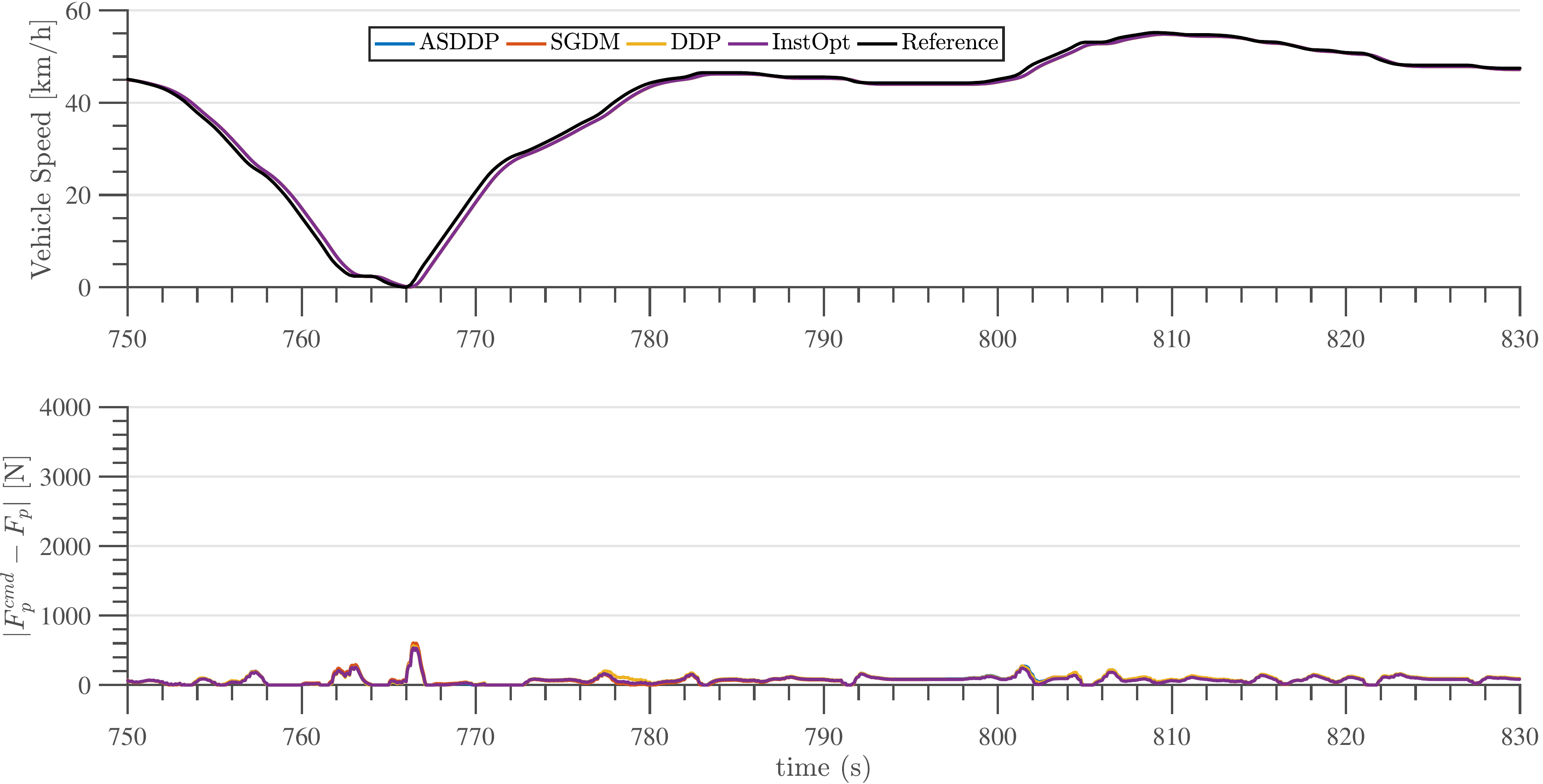}
	\caption{Segment of UDDS Cycle.}
	\label{fig:UDDSResultsVveh}
\end{figure}

A segment of the UDDS drive cycle is shown in Fig. \ref{fig:UDDSResultsVveh}.
This segment corresponds to the driver just finishing a sequence of stop and go driving and beginning a phase of cruising at moderate speed.    The speed and propulsion force tracking are excellent under all four algorithms.

\begin{figure}[H] 
	\centering
		\includegraphics [trim = 0mm 0mm 0mm 0mm, clip,width=1.0\textwidth]{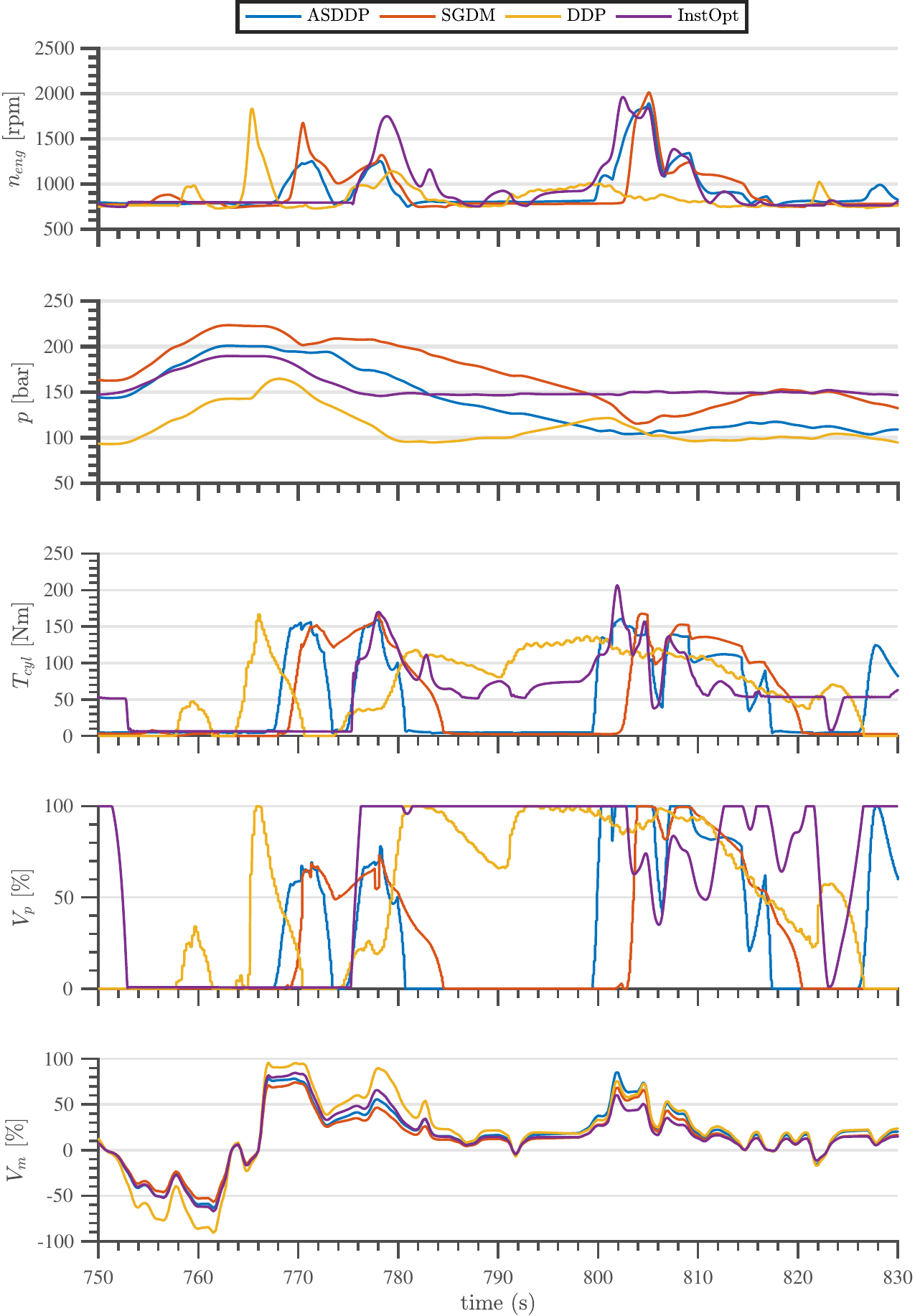}
	\caption{State and control trajectories over segment of UDDS Cycle.}
	\label{fig:UDDSResultsU}
\end{figure}

State and control input trajectories are shown in Fig. \ref{fig:UDDSResultsU}.  The stochastic strategies (SGDM and ASDDP) keep differential system pressure higher during the stop and go driving segment when acceleration demands become large, then lower differential system pressure once the cruising segment begins.  The DDP with driver forecast strategy (DDP), which can foresee upcoming acceleration demands, only raises system pressure briefly to meet the strong acceleration demand near time $t=765$ s.  The baseline strategy based on instantaneous optimization (InstOpt) raises engine speed and differential system pressure in a pattern which is somewhat similar to SGDM and ASDDP.  However, it can be seen that the stochastic strategies have an advantage in that differential system pressure is allowed to drop down as low as 100 bar during the cruising phase where higher pressures are not required (thereby resulting in higher hydraulic displacement volumes and overall improved efficiency).  Comparing the two stochastic strategies, ASDDP tends to adjust $T_{cyl}$ and $V_p$ more rapidly than SGDM, perhaps indicating that ASDDP converges more quickly than SGDM.  

\subsection{US06 Cycle}
\begin{figure}[h] 
	\centering
	\begin{minipage}{.9\textwidth}
		\includegraphics [trim = 0mm 0mm 0mm 0mm, clip,width=1.0\textwidth]{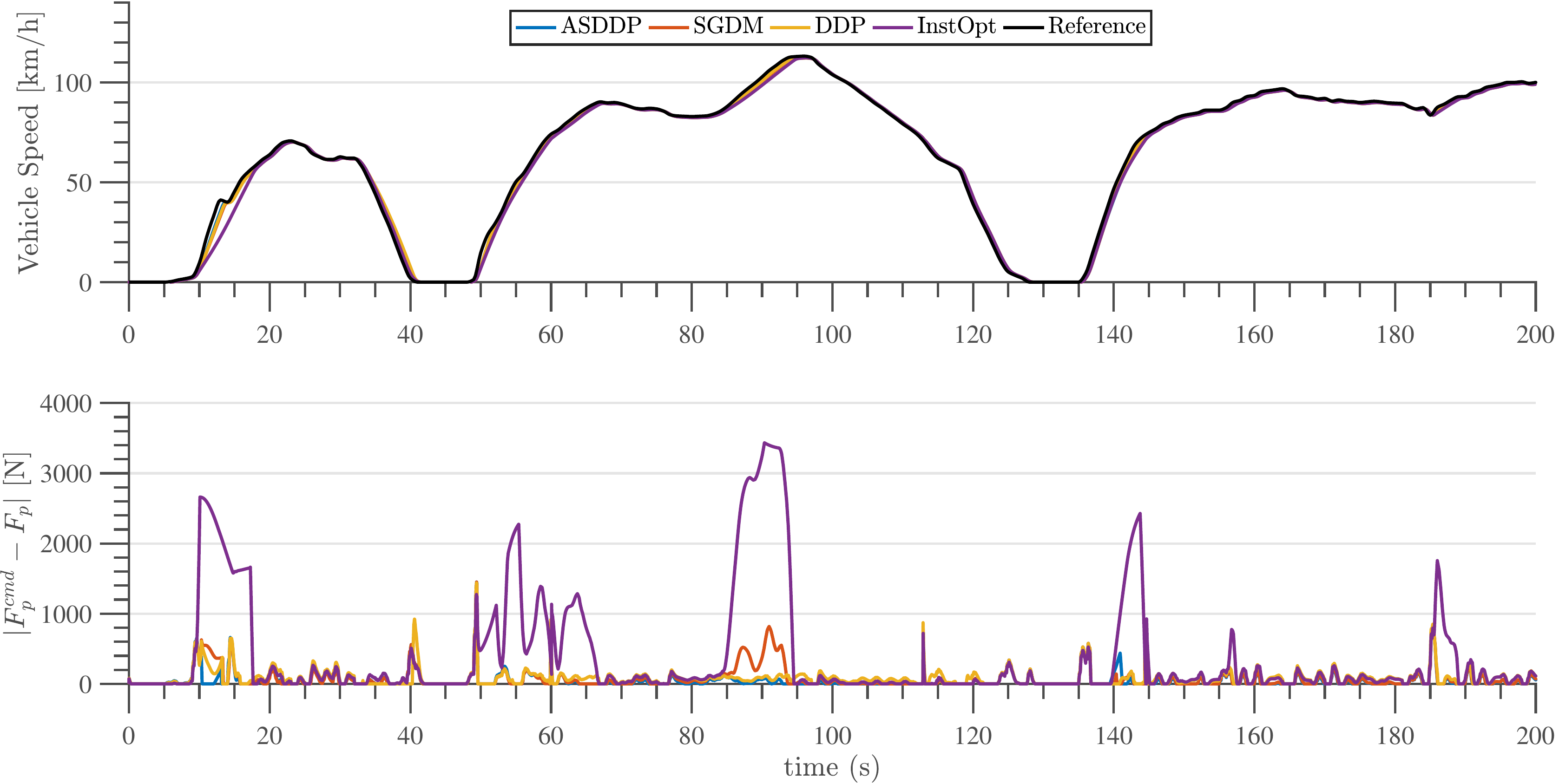}
	\end{minipage} 	
	\caption{Segment of US06 Cycle.}
	\label{fig:US06ResultsVveh}
\end{figure}

A segment of the aggressive US06 drive cycle is shown in Fig. \ref{fig:US06ResultsVveh}.  This segment corresponds to aggressive accelerations near the start of the cycle.  

\begin{figure}[H] 
	\centering
	\begin{minipage}{1\textwidth}
		\includegraphics [trim = 0mm 0mm 0mm 0mm, clip,width=1.0\textwidth]{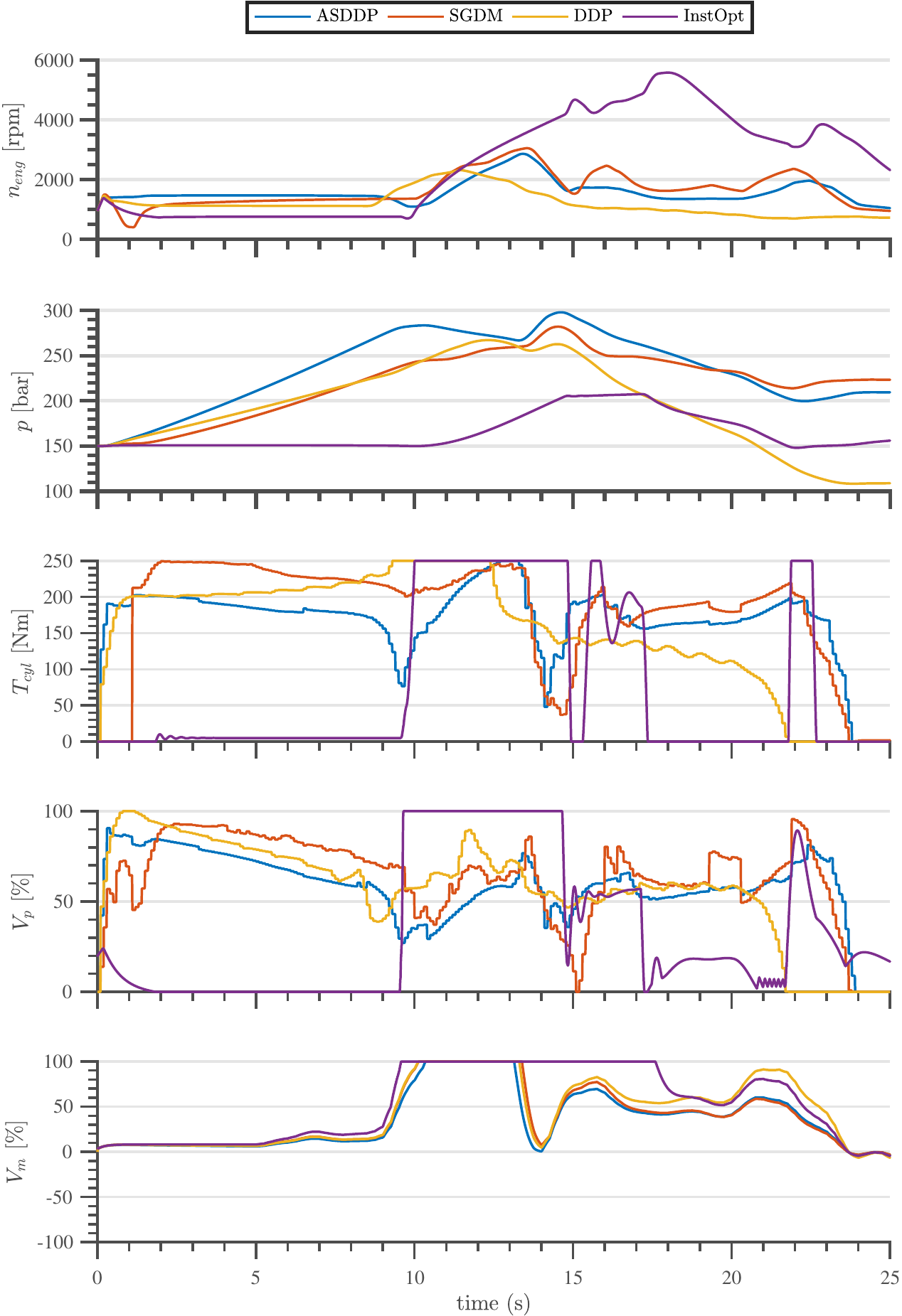}
	\end{minipage} 	
	\caption{State and control trajectories over segment of US06 Cycle.}
	\label{fig:US06ResultsU}
\end{figure}

The speed tracking performance of each algorithm is very good, with the exception of InstOpt.  Large differences between the commanded and actual propulsion force are seen under InstOpt, indicating difficulty meeting the driver demand.  The situation becomes more apparent when the trajectories of engine speed and differential system pressure are examined, shown in Fig. \ref{fig:US06ResultsU}.   It is interesting to note that SGDM, ASDDP, and DDP increase the differential system pressure just before the start of the aggressive acceleration event near time $t= 10$ seconds.  In this way, SGDM, ASDDP and DDP are well positioned to accommodate the driver's aggressive acceleration demand.  The InstOpt strategy, which is provided no information regarding upcoming behavior, maintains differential system pressure at the minimum 150 bar until just before $t = 10$ seconds.  Near $t=10$ seconds, InstOpt rapidly increases $T_{cyl}$ and $V_p$ in an attempt to meet the driver demand.

\subsection{GPS Cycle}
\begin{figure}[h!] 
	\centering
	\begin{minipage}{1\textwidth}
		\includegraphics [trim = 0mm 0mm 0mm 0mm, clip,width=1.0\textwidth]{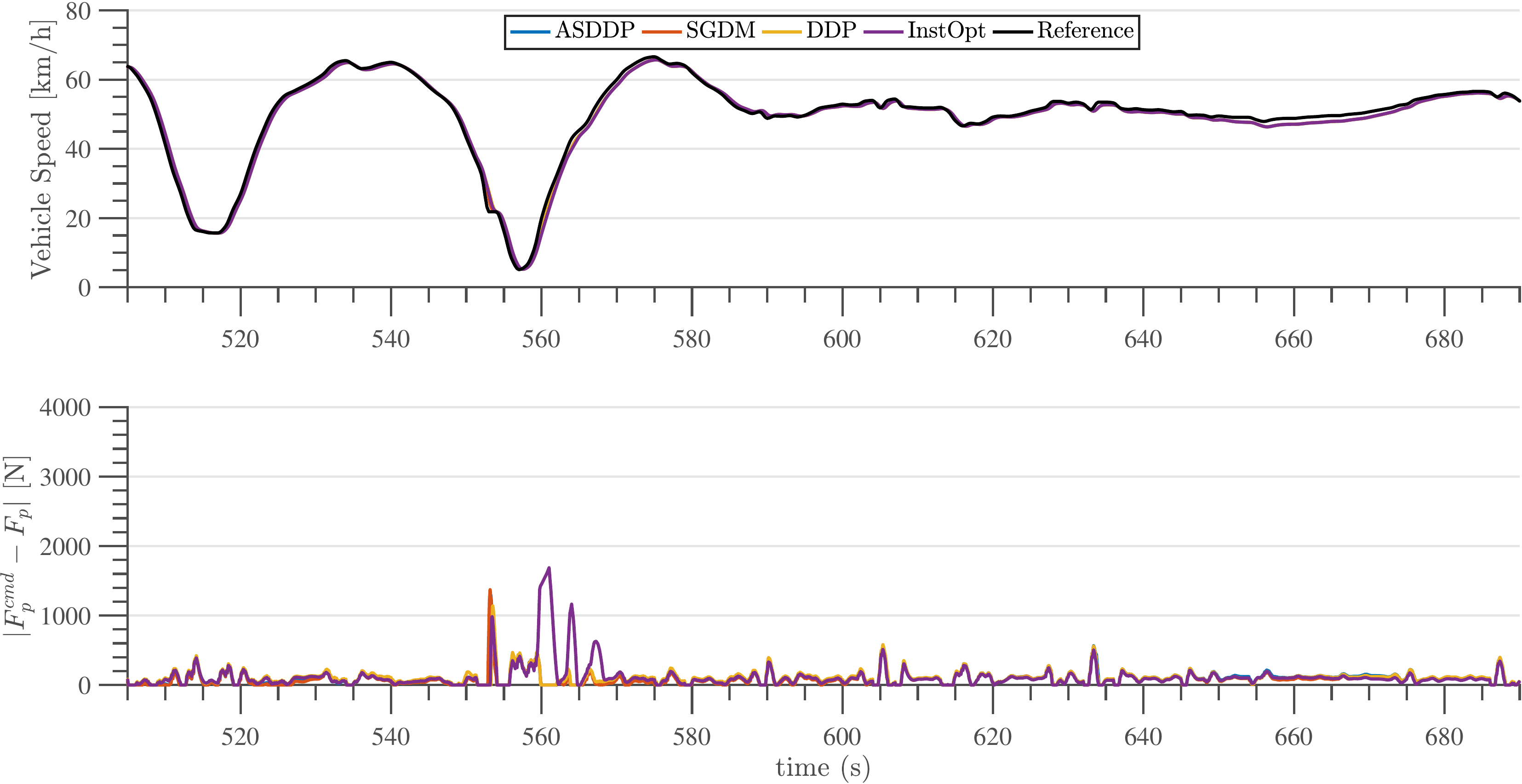}
	\end{minipage} 	
	\caption{Segment of GPS Cycle.}
	\label{fig:GPSResultsVveh}
\end{figure}

\begin{figure}[h!] 
	\centering
	\begin{minipage}{1\textwidth}
		\includegraphics [trim = 0mm 0mm 0mm 0mm, clip,width=1.0\textwidth]{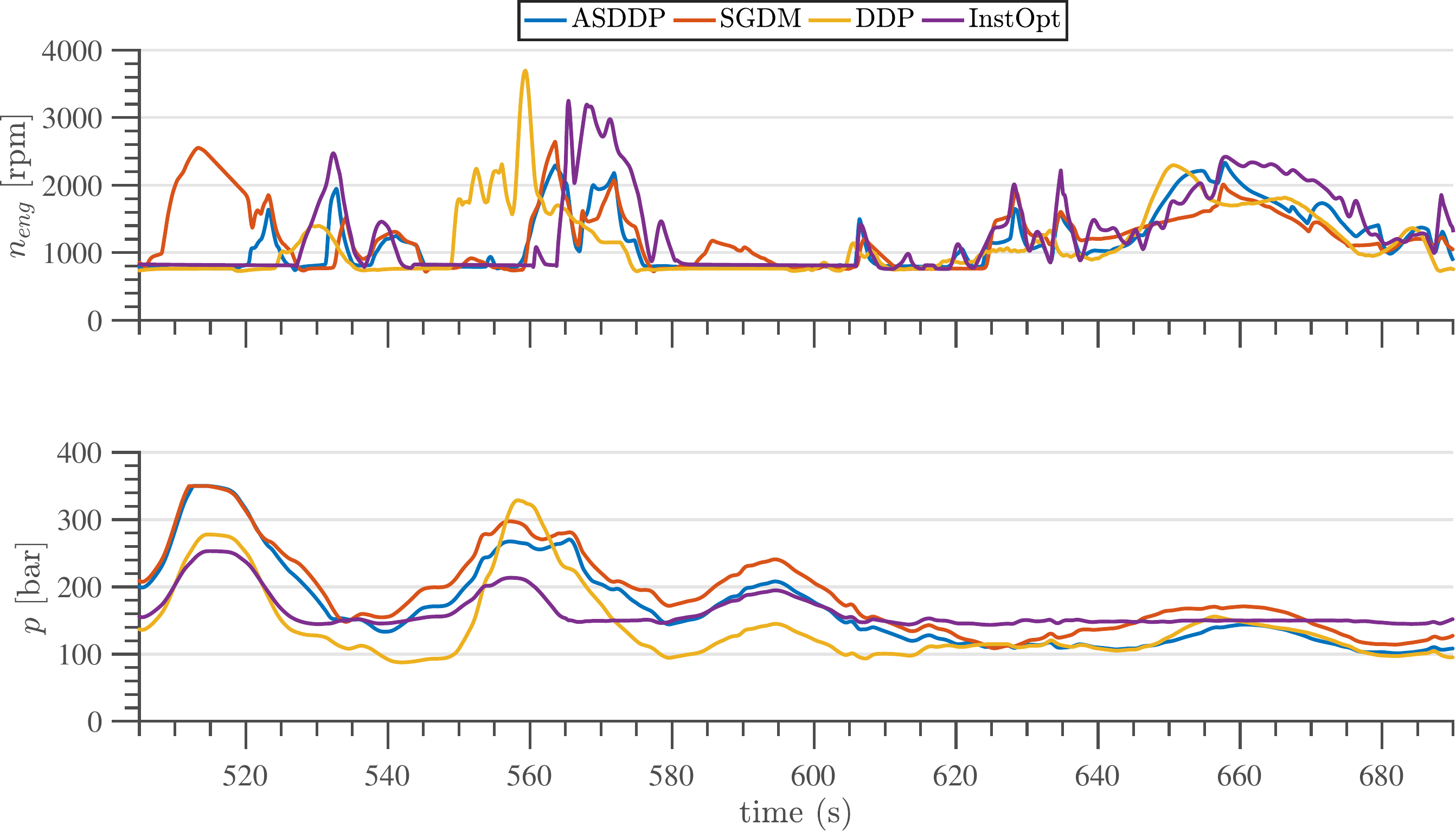}
	\end{minipage} 	
	\caption{Engine speed and differential system pressure over segment of GPS Cycle.}
	\label{fig:GPSResultsX}
\end{figure}

A segment of the GPS drive cycle is shown in Fig. \ref{fig:GPSResultsVveh}.  
This segment corresponds to the driver just finishing a sequence of stop and go driving and beginning a phase of cruising at moderate speed.  
Trajectories of engine speed and differential system pressure are shown in Fig. \ref{fig:GPSResultsX}.  SGDM and ASDDP tend to keep differential system pressure higher during stop and go driving, then lowering differential system pressure during the cruising phase.  Interestingly, ASDDP generates engine speed and differential system pressure trajectories which nearly match DDP during the cruising phase. 

\newpage
\section{Performance Metrics}\label{section:PerformanceMetrics}
To evaluate the performance of each controller quantitatively two metrics are defined.  The first metric is simply the fuel consumed along the entire cycle
\begin{align}
\text{Fuel Consumption} &= \int_{0}^{T} b_f(n_{eng}(t),T_{cyl}(t)) ~dt
\end{align}
where $b_f$ is the fuel consumption rate of the engine described in Fig. \ref{fig:EngMap}.  The second metric indicates how well the driver demand is met along the cycle through a modified speed tracking integral
\begin{align}
\text{Tracking Metric} &= \frac{1}{\text{cycle dist [km]}}\int_{0}^{T} \left|v_{veh}^{ref}(t)-v_{veh}(t)\right|\times \mathds{1}_{V_{m}(t)=V_{m}^{max}}~dt
\end{align}
where the indicator function is given by
\begin{align*}
\mathds{1}_{V_{m}(t)=V_{m}^{max}} &= 
\begin{cases}
1 ~ \text{if}~V_{m}(t)=V_{m}^{max}\\
0~\text{otherwise}
\end{cases}	
\end{align*}
Recall in Section \ref{section:SeriesHHVDynamics} it was shown that the propulsion force is limited by the differential system pressure.  The tracking metric ultimately measures how well a particular controller can anticipate and/or react to the propulsion force commanded by the driver by properly managing the differential system pressure along the drive cycle.  The units of the tracking metric are meters per kilometer, measuring the average distance in meters the vehicle has regressed from the reference cycle per kilometer as a result of insufficient differential system pressure.  The inclusion of the indicator function in the tracking metric definition reduces sensitivity to the virtual driver controller gains  described in Equation (\ref{eq:VirtualDriver}).  A lower tracking metric score indicates better performance.  A score of 0 - 2 m/km indicates that driver demand is (nearly) perfectly met along the entire drive cycle.  A score much greater than 4 m/km (a score of 4 m/km is equivalent to one car length per kilometer) may indicate noticeable discrepancies between commanded and produced propulsion force.

\subsection{Learning Progression}
This section investigates how well SGDM, ASDDP, and APDDP progressively optimize fuel usage and drivability as each cycle is repeated.  Each row of driver model $(P_{ij})$ is initialized to a Gaussian-like distribution, centered around $w^i$.   
On each subsequent run $(P_{ij})$ is adapted to the driver behavior as described in Section \ref{section:LearningDriverBeahvior}.  At the end of each run the elements of $\(P_{ij}\)$ are stored in memory and then used as the initial conditions for the following run.  

\begin{figure}[h!]
	\centering
	\begin{minipage}{1\textwidth}
		\includegraphics [width=1\textwidth]{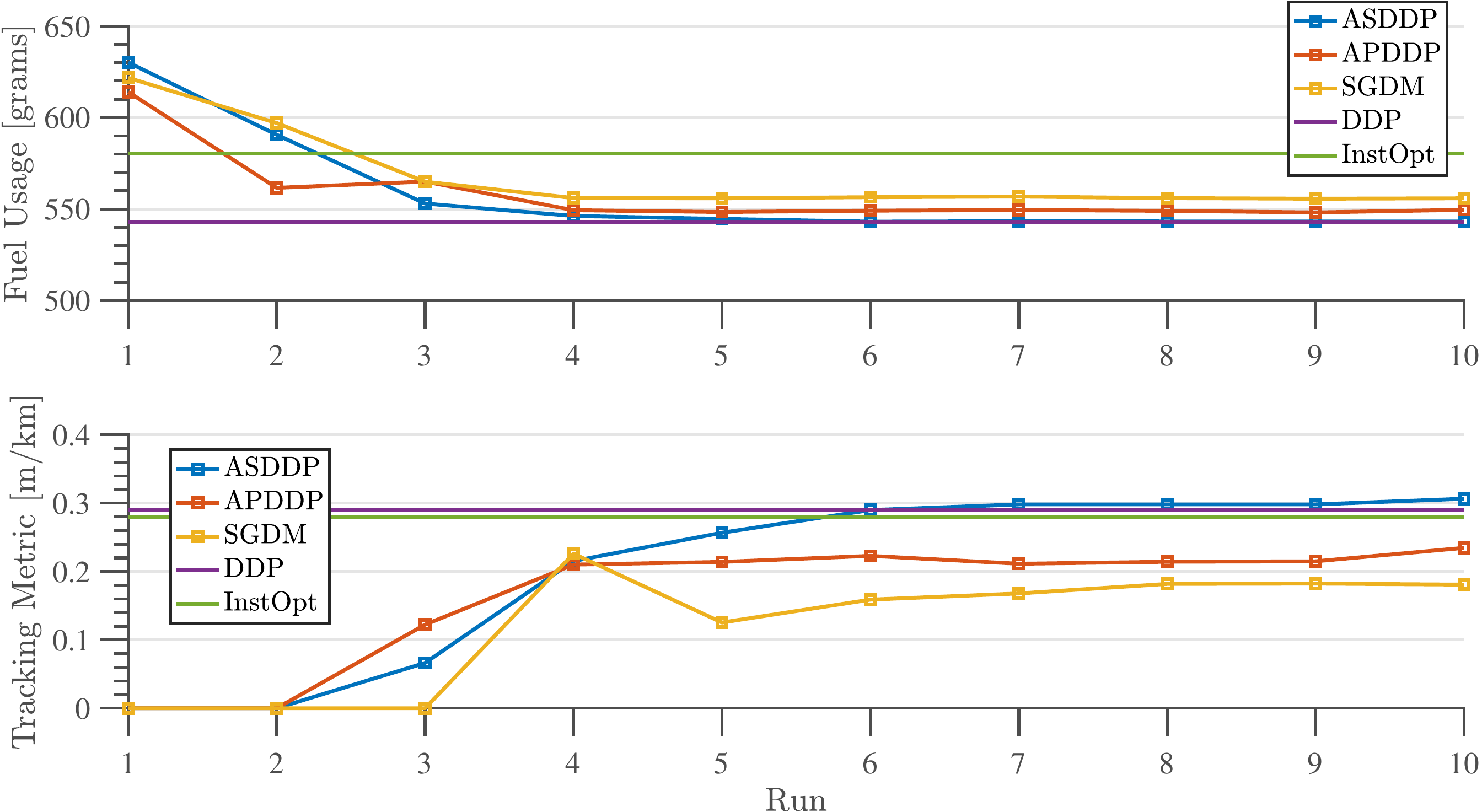}
	\end{minipage}
	\caption{UDDS cycle metrics.}
	\label{fig:progressionUDDS}
\end{figure}

\begin{figure}[H]
	\centering
	\begin{minipage}{1\textwidth} 		
		\includegraphics [width=1\textwidth]{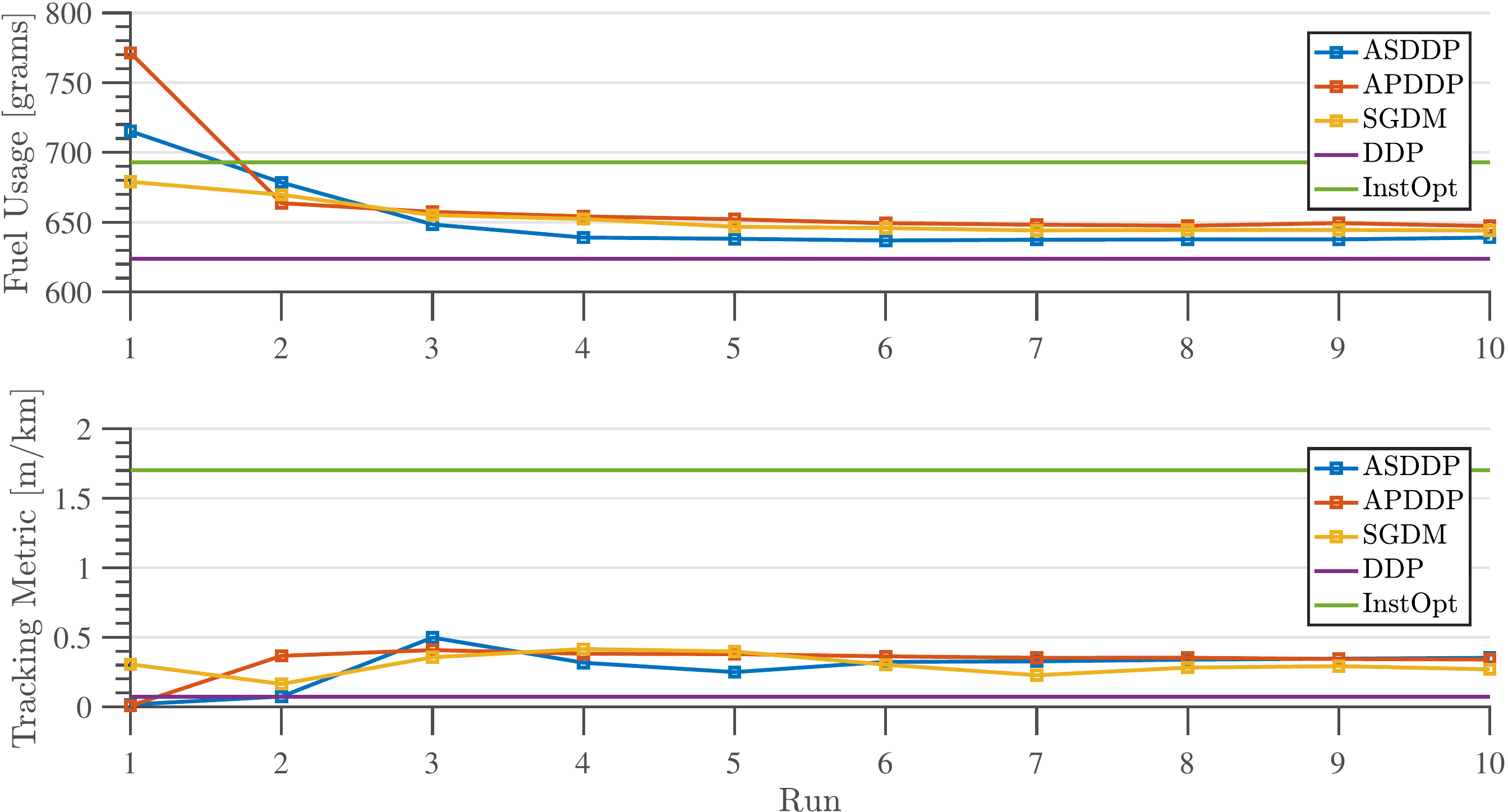}
	\end{minipage}
	\caption{GPS cycle metrics.}
	\label{fig:progressionGPS}
\end{figure}

\begin{figure}[H]
	\centering
	\begin{minipage}{1\textwidth}		
		\includegraphics [width=1\textwidth]{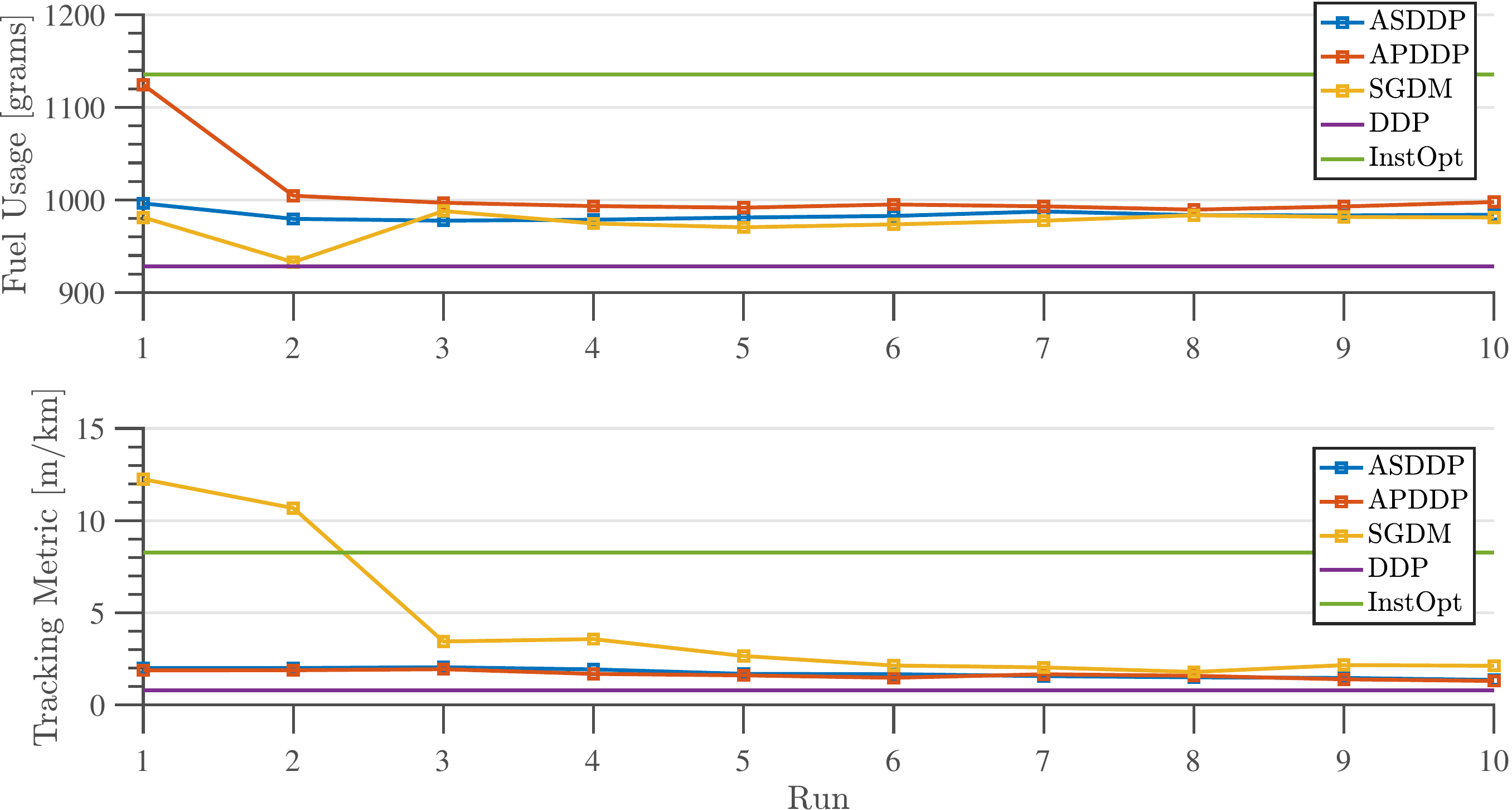}
	\end{minipage}
	\caption{US06 cycle metrics.}
	\label{fig:progressionUS06}
\end{figure}

Learning progression under the UDDS cycle is shown in Fig. \ref{fig:progressionUDDS}.
The results from the DDP and InstOpt benchmark strategies are also plotted, but since these strategies do not adapt to driver behavior their performance metrics are constant across the cycle runs.  As $\(P_{ij}\)$ is adapted to the UDDS drive cycle, fuel usage improves quickly while the tracking metric is increased only slightly (note the scale of the tracking metric).  Interestingly, convergence for both algorithms has nearly been achieved by the end of the fourth run.    Learning progression under the GPS and US06 cycles are shown in Figs. \ref{fig:progressionGPS} and \ref{fig:progressionUS06}.  As with the UDDS cycle, convergence has nearly occurred after the second or third run.  

The final fuel usage and tracking metric results after 10 repeated runs of each cycle are tabulated in Tables \ref{table:SimFuelPercentResults} and \ref{table:SimTrackingResults}.
\begin{table}[H]
	\caption{Fuel usage results, percent relative to DDP.}
	\label{table:SimFuelPercentResults}
	\vspace{-2mm}
	\centering 
	\begin{tabular}{c|c|c|c|c}
		Cycle / Alg & UDDS & US06 & GPS Cycle & GPS Cycle (without rd. gd. forecast)\\	
		\hline DDP       & 100.0\% & 100.0\% & 100.0\% & -\\ 
		SGDM   & 102.4\% & 106.3\% & 102.2\% & 103.4\% \\ 
		ASDDP   & 100.0\% & 105.8\% & 102.3\% & 104.8\%\\ 
		APDDP&	101.2\%&	107.3\%&	103.8\% & 104.2\%\\
		InstOpt & 106.9\% & 122.4\% & 111.1\% & - \\ 		
	\end{tabular} 
\end{table}
\begin{table}[H]
	\caption{Tracking metric results [m/km].}
	\label{table:SimTrackingResults}
	\vspace{-2mm}
	\centering 
	\begin{tabular}{c|c|c|c}
		Cycle / Alg & UDDS & US06 & GPS Cycle\\	
		\hline
		DDP	&0.29&	0.81&	0.07\\
		SGDM&	0.18&	2.02&	0.29\\
		ASDDP&	0.31&	1.36&	0.36\\
		APDDP&	0.23&	1.31&	0.34\\
		InstOpt&	0.28&	8.28&	1.70\\
				
	\end{tabular} 
\end{table}

\subsection{Cross Training}\label{section:CrossTraining}

To better understand the benefit of learning cycle-specific driver behavior, a cross training simulation is performed where each cycle is repeatedly run as in the previous section, but the statistical driver model $(P_{ij})$ is initialized on statistics obtained from other cycles.  The same metrics from the previous section are examined.  In order to simplify the presentation, only the results from the ASDDP and APDDP algorithms are shown.  The results from DDP and InstOpt are also included as reference points.  The progression of the fuel usage and tracking metrics and shown over six runs.  On run zero the driver behavior learning mechanism is frozen so that the effect of running any given cycle on statistics learned from repeatedly running another cycle is determined.  After run zero is complete the driver behavior learning mechanism is allowed to run as normal.  

\begin{figure}[H]
	\centering
	\begin{minipage}{0.85\textwidth}
		\includegraphics [width=1\textwidth]{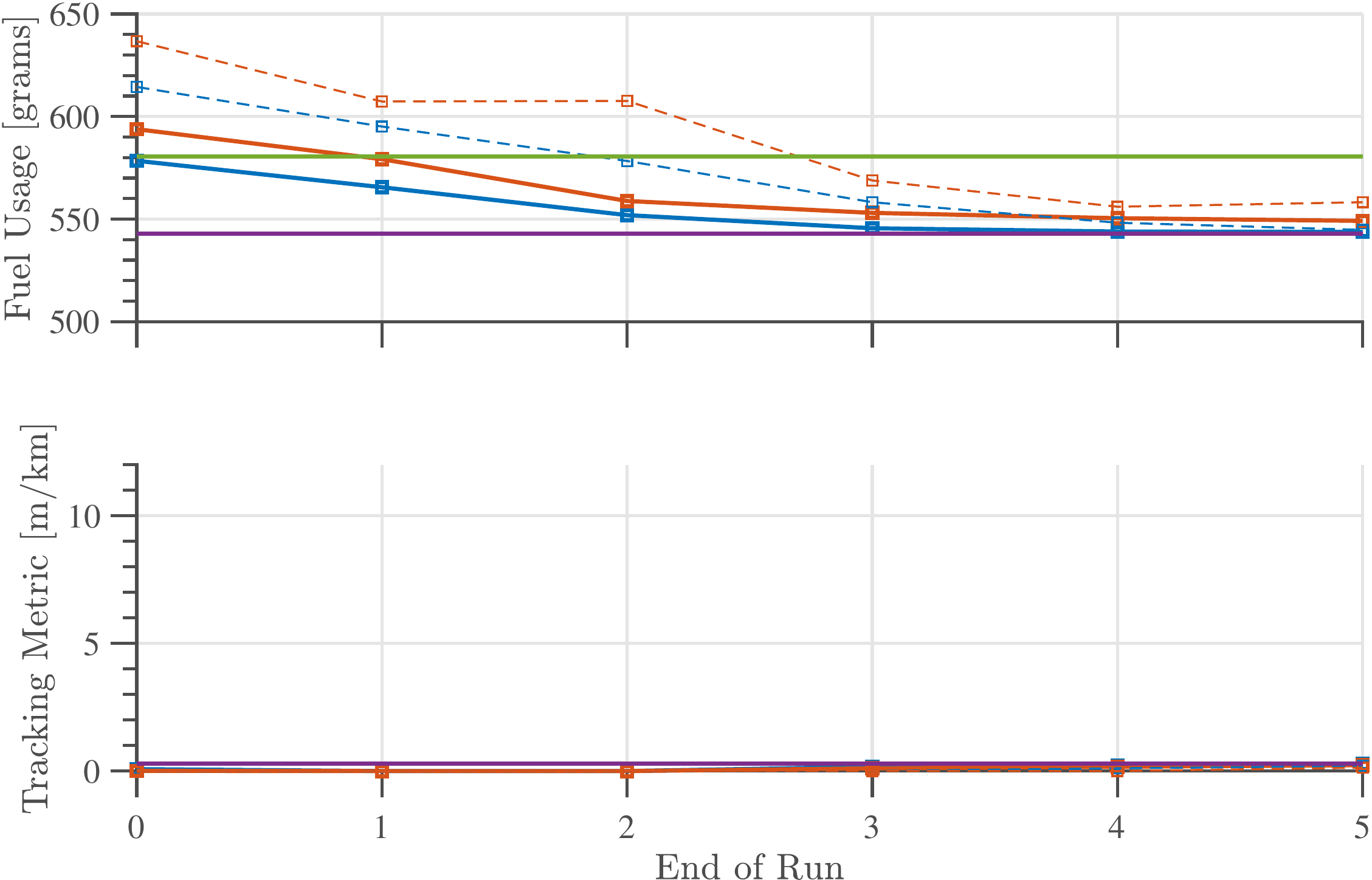}
	\end{minipage}
	\caption{UDDS cycle cross training metrics.  Blue: ASDDP using stats from GPS (solid), US06 (dashed).  Red: APDDP using stats from GPS (solid), US06 (dashed).  Purple: DDP and Green: InstOpt.}
	\label{fig:progressionUDDSCrossTrain}
\end{figure}

The cross trained simulation results for the UDDS cycle are shown in Fig.  \ref{fig:progressionUDDSCrossTrain}.  The blue curves show ASDDP results obtained by initializing $\(P_{ij}\)$ with driver statistics obtained from the GPS and US06 cycles.  Likewise, the red curves show APDDP results obtained in a similar manner.  Interestingly, when $\(P_{ij}\)$ is initialized with US06 statistics (dashed curves) the InstOpt outperforms the ASDDP strategy in terms of fuel usage until during the second run of the UDDS cycle (22-45 minutes) in which driver learning is active.  Similarly, InstOpt outperforms APDDP fuel usage until during the third run of UDDS (45-67 minutes) in which driver learning is active.  This result highlights the importance of adapting to relevant statistics if a stochastic strategy is to be employed.  

\begin{figure}[H]
	\centering
	\begin{minipage}{0.85\textwidth}
		\includegraphics [width=1\textwidth]{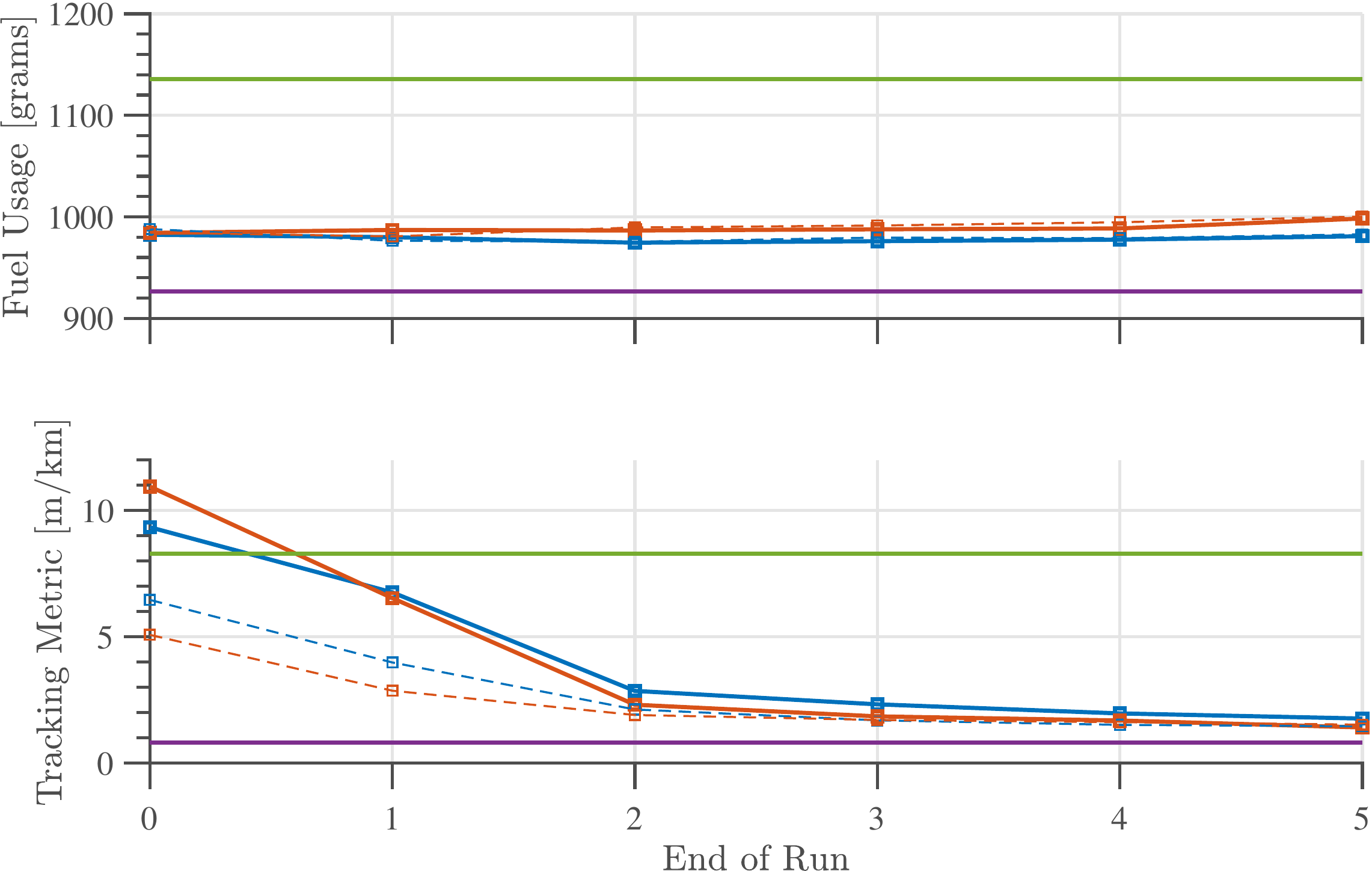}
	\end{minipage} 
	\caption{US06 cycle cross training metrics.  Blue: ASDDP using stats from UDDS (solid), GPS (dashed).  Red: APDDP using stats from UDDS (solid), GPS (dashed).  Purple: DDP and Green: InstOpt.}
	\label{fig:progressionUS06CrossTrain}
\end{figure}

\begin{figure}[H]	
	\centering
	\begin{minipage}{0.85\textwidth} \vspace{1em}
		\includegraphics [width=1\textwidth]{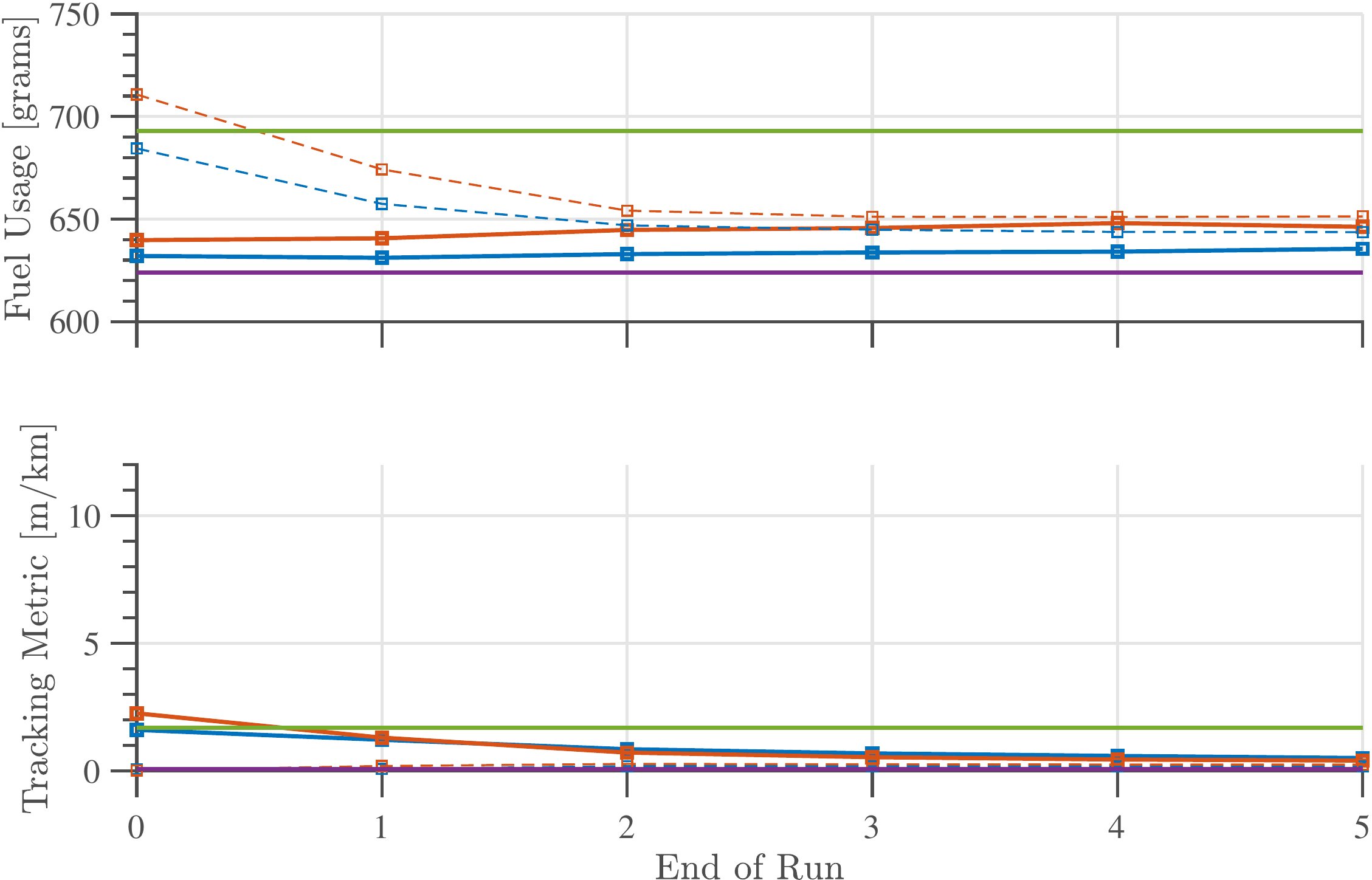}
	\end{minipage}
	\caption{GPS cycle cross training metrics.  Blue: ASDDP using stats from UDDS (solid), US06 (dashed).  Red: APDDP using stats from UDDS (solid), US06 (dashed).  Purple: DDP and Green: InstOpt.}
	\label{fig:progressionGPSCrossTrain}
\end{figure}

The cross trained simulation results for the US06 cycle are shown in Fig.  \ref{fig:progressionUS06CrossTrain}.  Fuel usage results remain relatively constant across the six runs.  However, the tracking metric improves significantly after the first run of the US06 cycle in which driver learning is active (10 minutes).  Cross trained results from the GPS cycle are shown in Fig.  \ref{fig:progressionGPSCrossTrain}.  Regardless of $\(P_{ij}\)$ initialization ASDDP and APDDP outperform InstOpt during the first run in which driver learning is active.

\section{Computation Times}
The average computation times of the three stochastic algorithms are shown in Table \ref{table:CompTimes}.  The values indicate how much faster than real time each algorithm executes.  These values were obtained by running each algorithm in the full simulation setup shown in Fig. \ref{fig:SimSetup} and comparing the simulation run time to elapsed wall-clock time.  The simulations were carried out on a laptop equipped with a 2.6 GHz i7 processor.  ASDDP runs nearly twice as fast as SGDM, and the APDDP runs nealy five times faster than ASDDP.  The massive increase in speed associated with APDDP over the other algorithms can be attributed to the fact that APDDP is not considering the true stochasticity of the problem, resulting in a significantly reduced computational burden.  

\begin{table}[H]
	\caption{Computation times.}
	\label{table:CompTimes}
	\vspace{-2mm}
	\centering 
	\setstretch{1.1}
	\begin{tabular}{c|c}	
		\textbf{Algorithm} & \textbf{Average sim:real time} \\
		\hline 
		SGDM & 3.4:1  \\
		ASDDP & 7:1  \\
		APDDP & 34:1 		
	\end{tabular} 
\end{table}

\chapter{EXPERIMENT}\label{section:Experiment}

An experimental setup  is used to demonstrate the real time potential of the ASDDP algorithm on a processor with limited computational resources.  A secondary objective is to demonstrate a model predictive control approach can successfully control a series hydraulic hybrid using a simplified control-oriented model of the real physics. 

\section{Experimental Hardware}
\begin{figure}[H]
	\centering
	\begin{minipage}{1\textwidth}		
		\includegraphics [trim = 10mm 10mm 10mm 45mm,clip,width=1\textwidth]{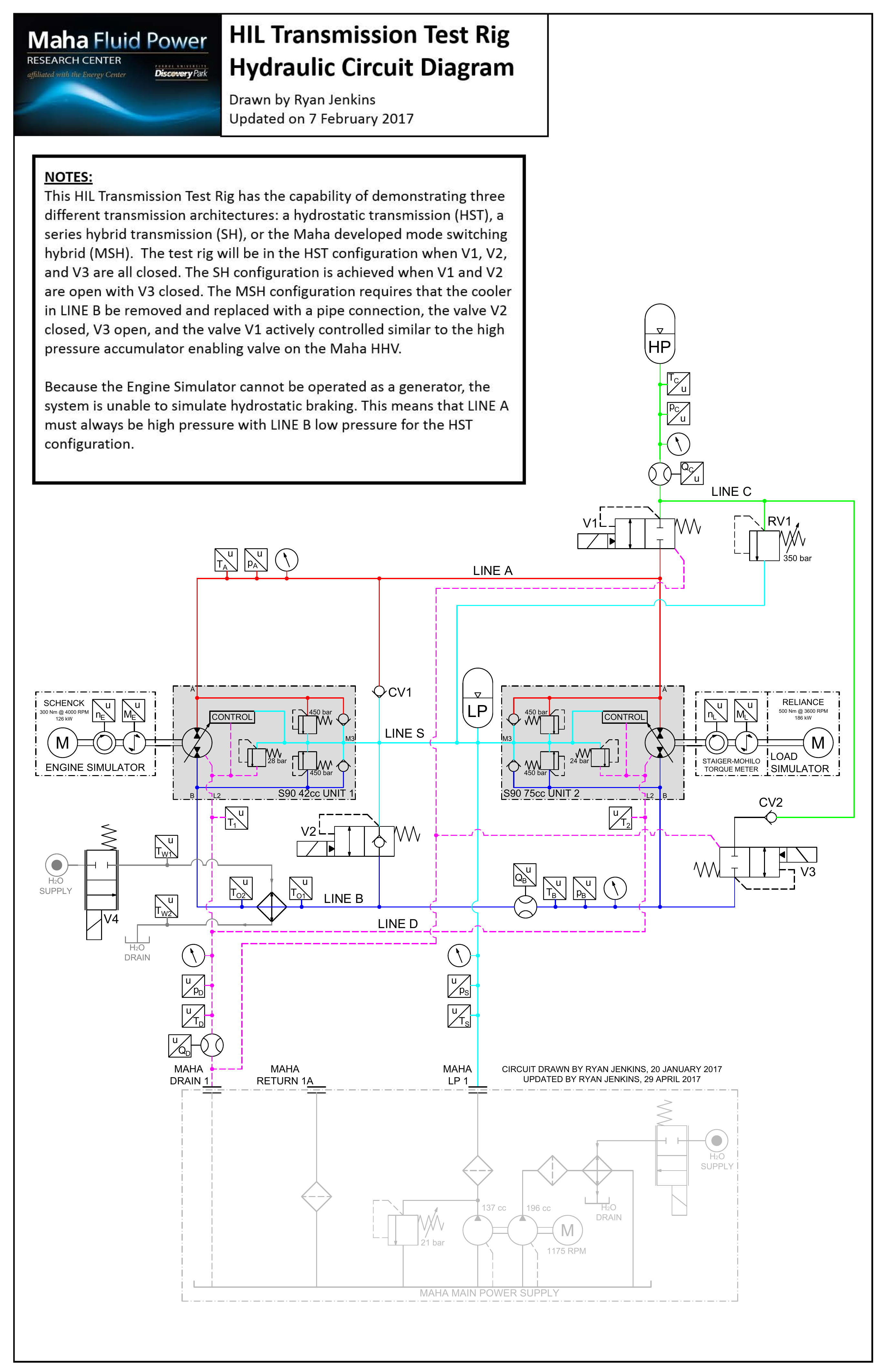}
	\end{minipage}
	\caption{Series hybrid test rig setup at the Maha Fluid Power Research Lab.}
	\label{fig:MahaSeriesHybridTestRig}
\end{figure}

The series hybrid test rig at the Maha Fluid Power Research Center is shown in Fig. \ref{fig:MahaSeriesHybridTestRig}.  An electric motor, referred to as the engine simulator, is directly connected to a hydraulic pump, unit 1.  The engine simulator is a 126 kW Schenck three phase induction motor, capable of providing a 300 Nm torque at 4000 RPM.   Hydraulic unit 1 is a Sauer S90 42 cc/rev variable displacement swash plate type pump.  An electric motor/generator, referred to as the load simulator, is used to simulate vehicle inertia and road load. The load simulator is a 186 kW Reliance motor, capable of producing a 500 Nm torque at 3600 rpm.  A second hydraulic pump/motor is connected directly to the load simulator, referred to as unit 2.  Hydraulic unit 2 is a Sauer S90 75 cc/rev variable displacement swash plate type pump.  The engine and load simulators are coupled to ABB manufactured ACS800 variable frequency drives.  
These drives control the output frequency which facilitates a control over the speed and torque of the two simulators. 
The ABB drives have transient and steady state speed control accuracy better than 0.1 \%.  A hydraulic power supply pressurizes a low pressure line to replace leakage losses, and an accumulator is connected to the high pressure line for energy recovery.  Data acquisition and control was conducted using the cRIO 9074 controller, a product by National Instruments.  The cRIO 9704 has a single core 400 MHz processor and 128 MB of RAM.

\section{Experiment Setup}
The experiment was carried out on the test rig shown in Fig. \ref{fig:MahaSeriesHybridTestRig}.  The load simulator was setup to simulate a lightweight passenger vehicle with parameters listed in Table \ref{table:ExperimentParameters}.
\begin{table}[h!]
	\caption{Series-Hybrid Experiment Parameters.}
	\label{table:ExperimentParameters}
	\vspace{-2mm}
	\centering 
	\setstretch{1.0}
	\begin{tabular}{llll}	
		\textbf{Description} & \textbf{Symbol} & \textbf{Value} & \textbf{Units}  \\
		\hline 
		Vehicle mass & $m_{veh}$ & 1520 & kg \\
		Max propulsion force & $F_p^{max}$ & 4000 & N \\
		Max vehicle speed & $v_{veh}^{max}$ & 60 & km/h \\
		Engine simulator inertia & $I_{eng}$ & 0.38 & kg-m$^2$ \\
		Load simulator inertia & $I_{load}$ & 0.50 & kg-m$^2$ \\
		Virtual axle ratio & $k_{axle}$ & 4:1 & - \\
		Dynamic tire radius& $r_{tire}$ & 0.31 & - \\
		Aero drag coefficient& $C_d$ & 1.62 & - \\
		Rolling resistance coefficient& $C_r$ & 0.010 & - \\
		Displacement vol. of hyd. pump & $V_p^{max}$ & 42 & cc/rev \\
		Displacement vol. of hyd. motor & $V_m^{max}$ & 75  & cc/rev \\
		HP accumulator precharge vol.& $V_{ha}$ & 20 & L \\
		HP accumulator precharge press.& $p_{ha}$ & 80 & bar \\
		LP accumulator precharge vol.& $V_{la}$ & 20 & L \\
		LP accumulator precharge press.& $p_{la}$ & 12 & bar \\
		Max hi pressure & $p_{A,max}$ & 240 & bar \\
		Low-pressure reservoir press. & $p_{lp}$ & 25 & bar \\
		\hline
	\end{tabular} 
\end{table}
The engine simulator is provided a reference speed command generated by the ASDDP algorithm in the following manner.  As described in Section \ref{section:ASDDP}, an optimal state-control sequence $\(\x_n^*,\u_n^*\)_{n=0}^{N-1}$ is generated every $T_s = 0.5$ seconds.  The value $\x_0^*$ is simply the measured state feedback information.  Value $\x_1^*$ is the predicted optimal value of the state at the next horizon time step, where the horizon time is $\Delta t = 1$ second according to Equation (\ref{eq:2ndOrderPropagation}).  The reference engine speed provided to the engine simulator can be computed as the following linearly interpolated value\footnote{At the time of experimentation $n_{eng}^{cmd}$ was implemented with a discrete time first order low pass filter which emulates Equation (\ref{eq:nengcmd_experiments})}
\begin{align}\label{eq:nengcmd_experiments}
	n_{eng}^{cmd} &= n_{eng,0}^* + \(n_{eng,1}^* - n_{eng,0}^*\)\frac{T_s}{\Delta t}
\end{align}
The pump displacement command $V_p$ is generated using Equation (\ref{eq:ControlOutputTs}) and the motor displacement command $V_m$ is generated every 0.01 seconds using Equation (\ref{eq:MotorDispCalc}).

\section{Data-Simulation Comparison}
\begin{figure}[h!]
	\centering
	\begin{minipage}{1\textwidth}
		\includegraphics [trim = 0mm 20mm 0mm 20mm, clip,width=1.0\textwidth]{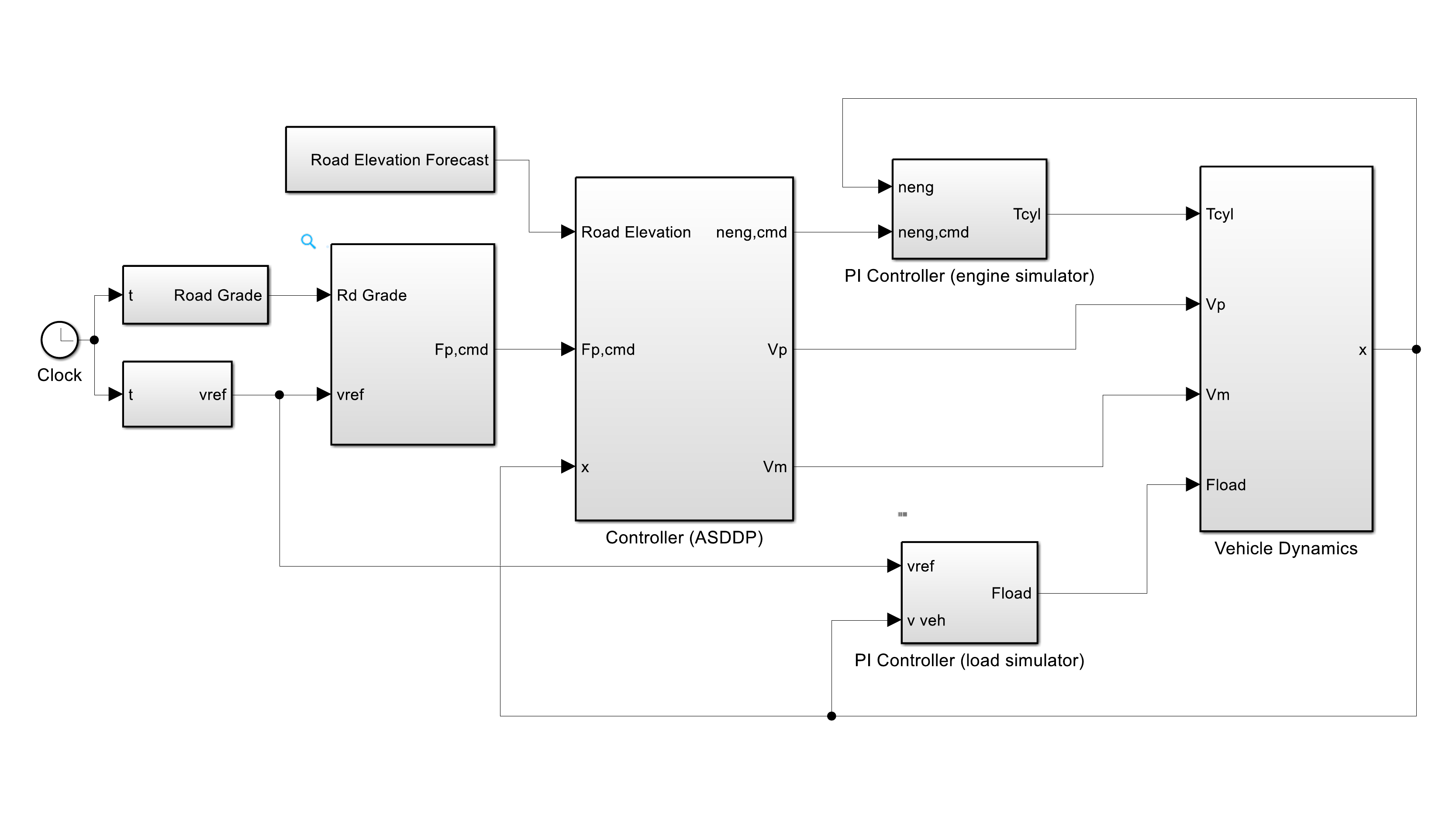}
	\end{minipage}
	\caption{Block diagram of experimental setup.}
	\label{fig:SimSetupExp}
\end{figure}

A simulation is constructed to emulate the test rig setup.  The purpose of this simulation is to validate the modeling equations shown in Chapter 3 and the simulation approach taken in Chapter 6.  The simulation setup is shown in Fig. \ref{fig:SimSetupExp}.
The propulsion force command, $F_p^{cmd}$, is generated completely open loop according to
\begin{align}
F_p^{cmd}  &= m_{veh}a_{veh}^{ref} + \tfrac{1}{2}C_d\rho_{air}({v_{veh}^{ref} })^2+ m_{veh}g\left[C_r cos(\phi)+sin(\phi)\right]
\end{align}
The term $a_{veh}^{ref}$ is a numerical derivative of the vehicle reference speed.  The engine, vehicle, and hydraulic dynamics are the same as given in Section \ref{section:SeriesHHVDynamics}.  The only exceptions are the resistive forces in Equation (\ref{eq:VehicleDynamics}) are replaced with $F_{load}$ created by the load simulator block, and $T_{cyl}$ from Equation (\ref{eq:Tcyl}) is replaced with the value created by the engine simulator block.  The gains of the PI controllers used for the engine and load simulators were tuned to match the performance characteristics of the real electric units.  

\begin{figure}[h!]
	\centering
	\begin{minipage}{1\textwidth}
		\includegraphics [trim = 0mm 0mm 0mm 0mm, clip,width=1.0\textwidth]{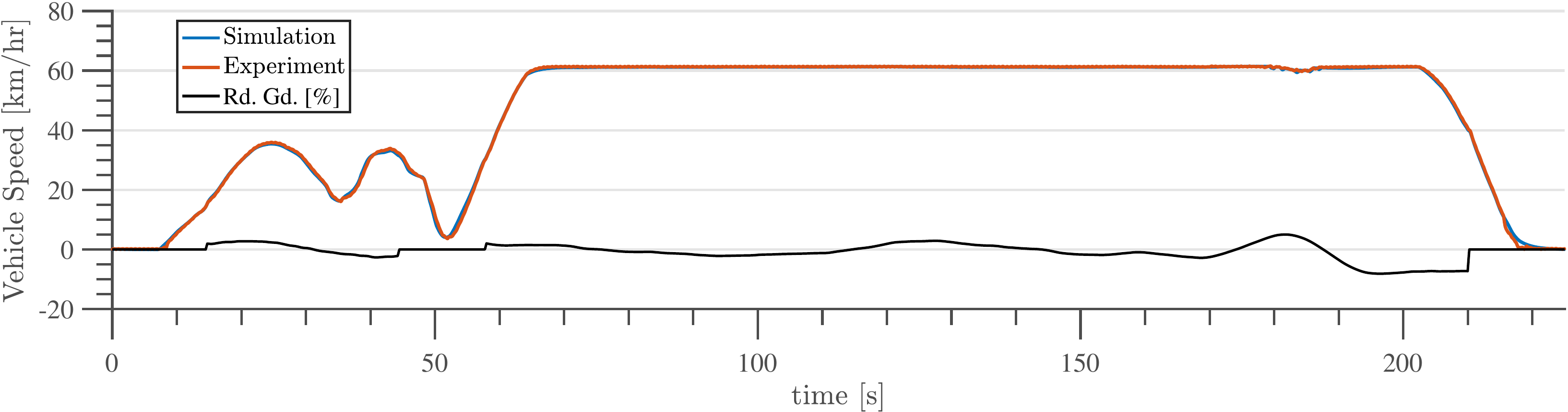}
	\end{minipage}
	\caption{Segment of GPS cycle.}
	\label{fig:SpeedPlotExperiments}
\end{figure}

\begin{figure}[h!]
 	\centering
 	\begin{minipage}{1\textwidth}
 		\includegraphics [trim = 0mm 0mm 0mm 0mm, clip,width=1.0\textwidth]{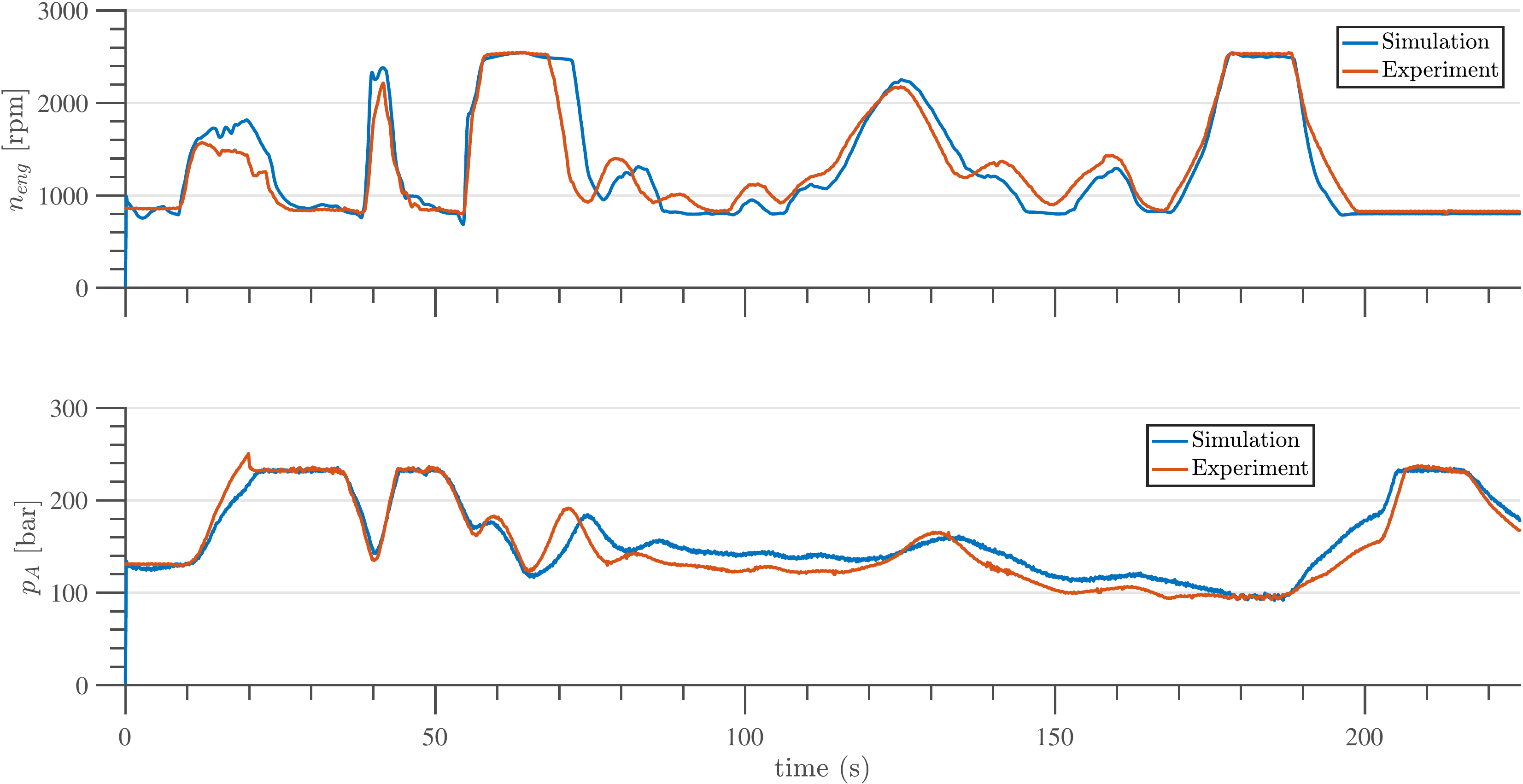}
 	\end{minipage}
 	\caption{Engine speed and high pressure trajectories over segment of GPS cycle.}
 	\label{fig:StatePlotExperiments}
 \end{figure}

The first four minutes of the GPS cycle are carried out in the experiment.  A plot of vehicle speed is shown in Fig. \ref{fig:SpeedPlotExperiments}.
 The vehicle speed profile matches very well between the experiment and simulation.  Engine speed and pressure of the high pressure accumulator are shown in Fig. \ref{fig:StatePlotExperiments}.   

\begin{figure}[ht!]
	\centering
	\begin{minipage}{1\textwidth}
		\includegraphics [trim = 0mm 0mm 0mm 0mm, clip,width=1.0\textwidth]{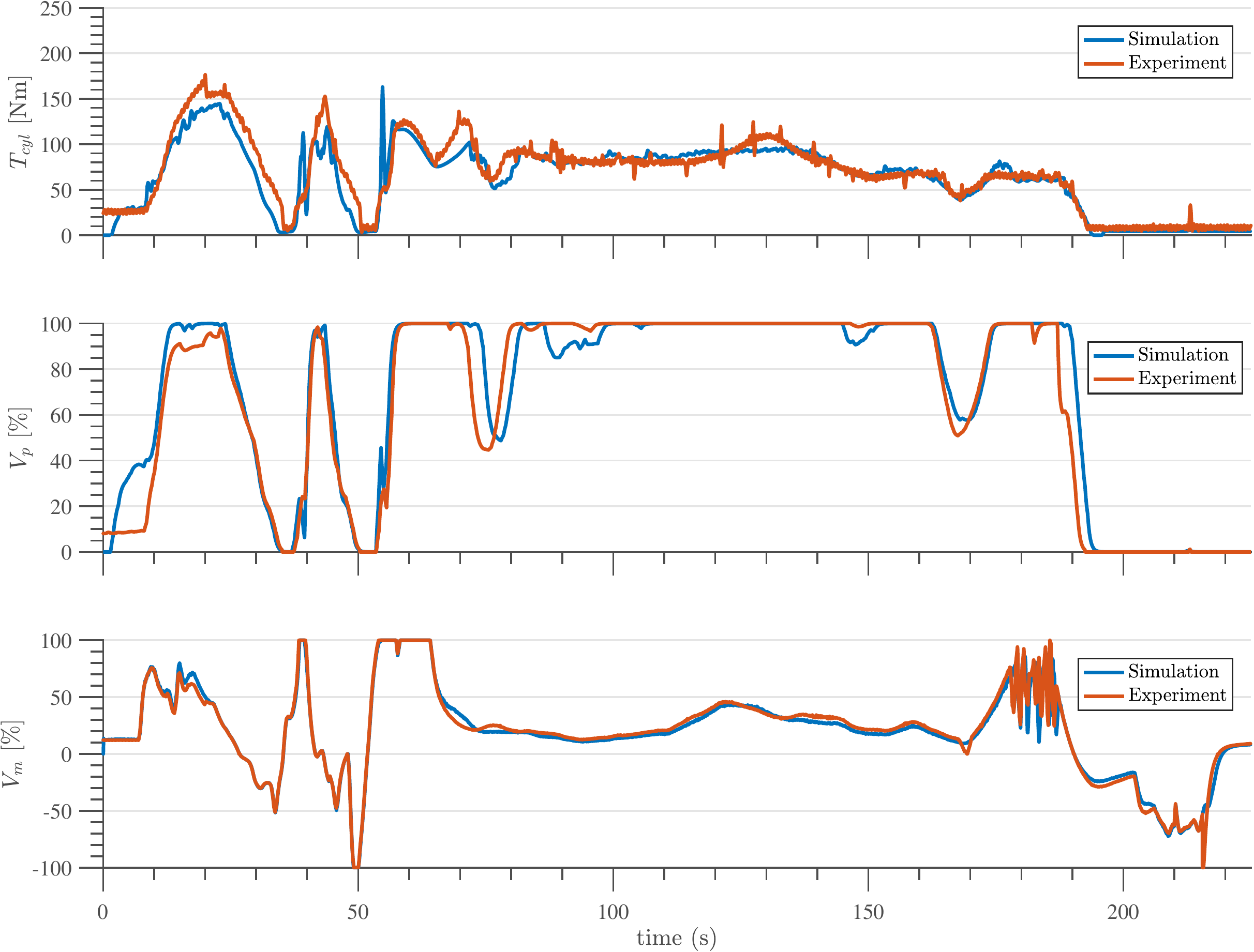}
	\end{minipage}
	\caption{Control input trajectories over segment of GPS cycle.}
	\label{fig:ControlPlotExperiments}
\end{figure}

 \noindent Agreement between the simulation and experimental data is again very good, will some slight deviations seen during periods of vehicle acceleration.   The control inputs are shown in Fig. \ref{fig:ControlPlotExperiments}.  Overall, agreement between simulation and experimental data is very good.     

\begin{figure}[ht!]
	\centering
	\begin{minipage}{1\textwidth}
		\includegraphics [trim = 0mm 0mm 0mm 0mm, clip,width=1.0\textwidth]{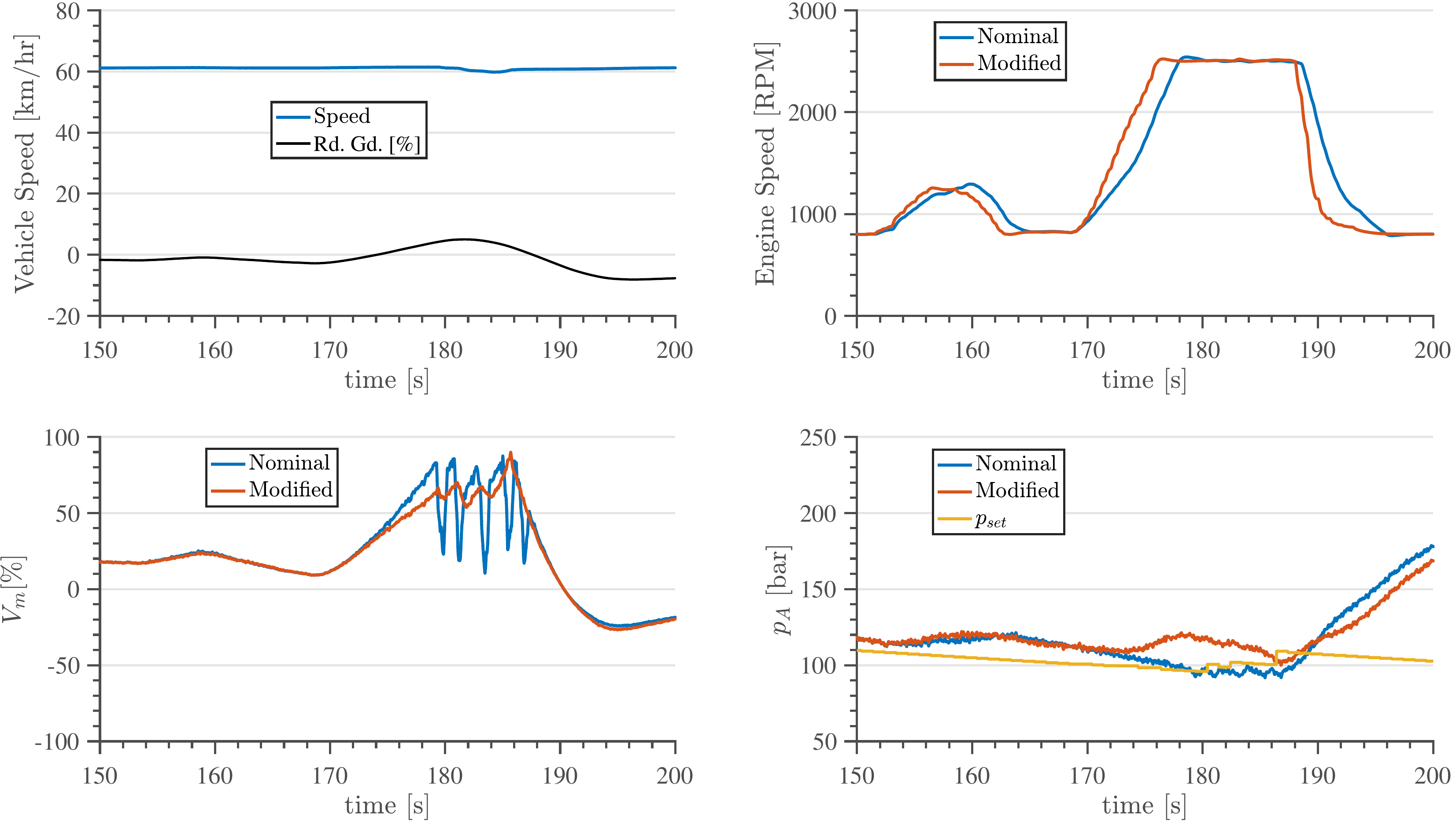}
	\end{minipage}
	\caption{Simulation comparison with $K_1=0.1$ (nominal simulation) and $K_1=0.01$ (modified simulation).}
	\label{fig:ModPlotExperiments}
\end{figure}

Near time 180 seconds a high frequency oscillation is observed in the volumetric displacement of hydraulic unit 2.  It is worthwhile to note this effect is captured nearly perfectly in simulation.  For safety reasons, a small amount of logic was built into the controller which reduces the displacement volume of unit 2 if the high pressure accumulator drops below $p_{set}$ (described by Equation (\ref{eq:pset})).   As shown in Fig. \ref{fig:ModPlotExperiments} the high pressure accumulator drops below $p_{set}$ near time 180 seconds, explaining the rapid adjustments in unit 2 displacement volume.  To investigate this further, the gain $K_1$ from Equation (\ref{eq:gn}), which penalizes changes in engine speed between each horizon timestep, is reduced from a value of 0.1 to 0.01 in simulation.  The comparison between the nominal simulation (with $K_1=0.1$) and the modified simulation (with $K_1=0.01$) is shown in Fig. \ref{fig:ModPlotExperiments}.
Remarkably, the rapid oscillation is eliminated in the modified simulation.  This can be explained considering the differences in engine speed observed in Fig. \ref{fig:ModPlotExperiments}.  In both simulations, ASDDP anticipates the need for a higher engine speed near time 170 seconds in response to the upcoming increase in road grade.  The modified simulation is allowed to increase engine speed at a slightly faster rate, and is therefore able to maintain a pressure in the high pressure accumulator which is above the $p_{set}$ limit.  This phenomenon gives some credence to the predictive abilities of the ASDDP algorithm.  


\chapter{CONCLUSIONS AND FUTURE DIRECTIONS}

Real time optimal control (aka model predictive control aka receding horizon control) is a powerful framework for hybrid vehicle energy management.   It allows us to derive controllers which consider upcoming conditions and past statistics.  By incorporating an adaptive element the controller can be continuously adjusted to maximize performance for the specific operating environment.  

In this work a Markov chain model of driver behavior was employed.  It was shown that the transition probabilities can be adapted in minutes to the drive cycle, even when initialized on values obtained from a cycle with completely incorrect characteristics.  The multi-step transition probabilities were shown to be an effective tool for anticipating driver behavior along a prediction horizon.  Adapting the Markov chain model in real time seems to be critical when employing a stochastic strategy.  As seen in Section \ref{section:CrossTraining}, a poorly tuned statistical model can lead to performance which is worse than a strategy incorporating no statistical information at all.  Three computational methods for real time energy management in a HHV when driver behavior and vehicle route are not known in advance were presented.  When the Markov chain model is correctly adapted to the drive cycle, these methods produce fuel consumption results which are reasonably close to a theoretically best controller which has full access to driver behavior.  Furthermore, each method significantly outperforms a baseline controller which is not provided any statistical driver behavior information.  Road elevation forecasting provides some further gains in fuel reduction, even on a moderately level terrain found in Lafayette, IN.

Of the three computational methods developed in \ref{section:StochControlFormulations}, the ASDDP algorithm seems to provide the most benefit in terms of execution time and fuel consumption results.  Experimental results indicate ASDDP has real time run potential on a resource limited processor.  When executed on a 400 MHz processor with 128 MB of RAM, the ASDDP algorithm successfully controlled a series hybrid test rig.  During the experiment, the controller update timestep was set at $T_s=0.5$ seconds, which is not unreasonable for high level control of a powertrain.  

\section{Future Directions}

\subsection{Adjusting $P_{ij}$ to Driving Indicators}
In this work the Markov chain transition probabilities, $P_{ij}$, are adapted in real time.  However, these values are not altered in response to various indicators such as traffic signals, upcoming traffic congestion, entering / exit a high speed segment of road, etc.  For example, if a red light is being approached the likelihood of a deceleration command in the very near future becomes quite high, regardless of past behavior.  Adjusting matrix $\(P_{ij}\)$ in response to these indicators could provide substantial prediction benefit.  On-board telematics could provide a means to inform the algorithm of upcoming indicators.  

\subsection{MPDDP}
Average path differential dynamic programming (APDDP) developed in Section \ref{section:APDDP} was competitive with ASDDP in terms of fuel consumption but executed in a fraction of the time.  The improved speed of APDDP can be attributed to the fact that each timestep along the horizon APDDP considers only a single disturbance transition, whereas ASDDP considers $|W|$ transitions.  A hybrid algorithm could foresee-ably consider several likely transitions plus several transitions at outer variances of the disturbance path (as seen for example in Fig. \ref{fig:EwnPropagate}) for a total of $1<y<|W|$ transition evaluations.  Such a strategy (possibly \textit{multi-path differential dynamic programming}?) could potentially offer nearly 100\% of the performance benefits of ASDDP at a considerably reduced computational cost.  A mechanism for selecting which transitions to consider at each horizon timestep would be required.   

\subsection{Multi-Stage Markov Chain Modeling}
More can be done in the way of Markov chain modeling.  The Markov chain used in this work was a single-stage model of the form
\begin{align*}
P_{ij} &\triangleq \mathrm{Pr}[w_{n+1} = w^j | w_n = w^i]
\end{align*}
In words, the probability of the next transition is based only on the present disturbance value.  A more sophisticated model could use information about past disturbances to make better predictions about the next transition, such as
\begin{align*}
P_{(i_1,i_2)j} &\triangleq \mathrm{Pr}[w_{n+1} = w^j | w_n = w^{i_1},w_{n-1}=w^{i_2}]
\end{align*}
The hope is that by including more information to the prediction, the prediction becomes more accurate.  The downside is that learning time may increase which could offset prediction benefits (recall the single stage model shown above can be effectively learned in roughly 20-30 minutes).  Additionally, incorporating such a multi-stage model may add computational complexity to the algorithm which needs to be considered.

\bibliography{all}

\appendix

\chapter{DRIVER BEHAVIOR STATISTICS}\label{section:AppDriverBehavior}
\begin{figure}[h!]
	\centering
	\begin{minipage}{1\textwidth}
		\includegraphics [width=1\textwidth]{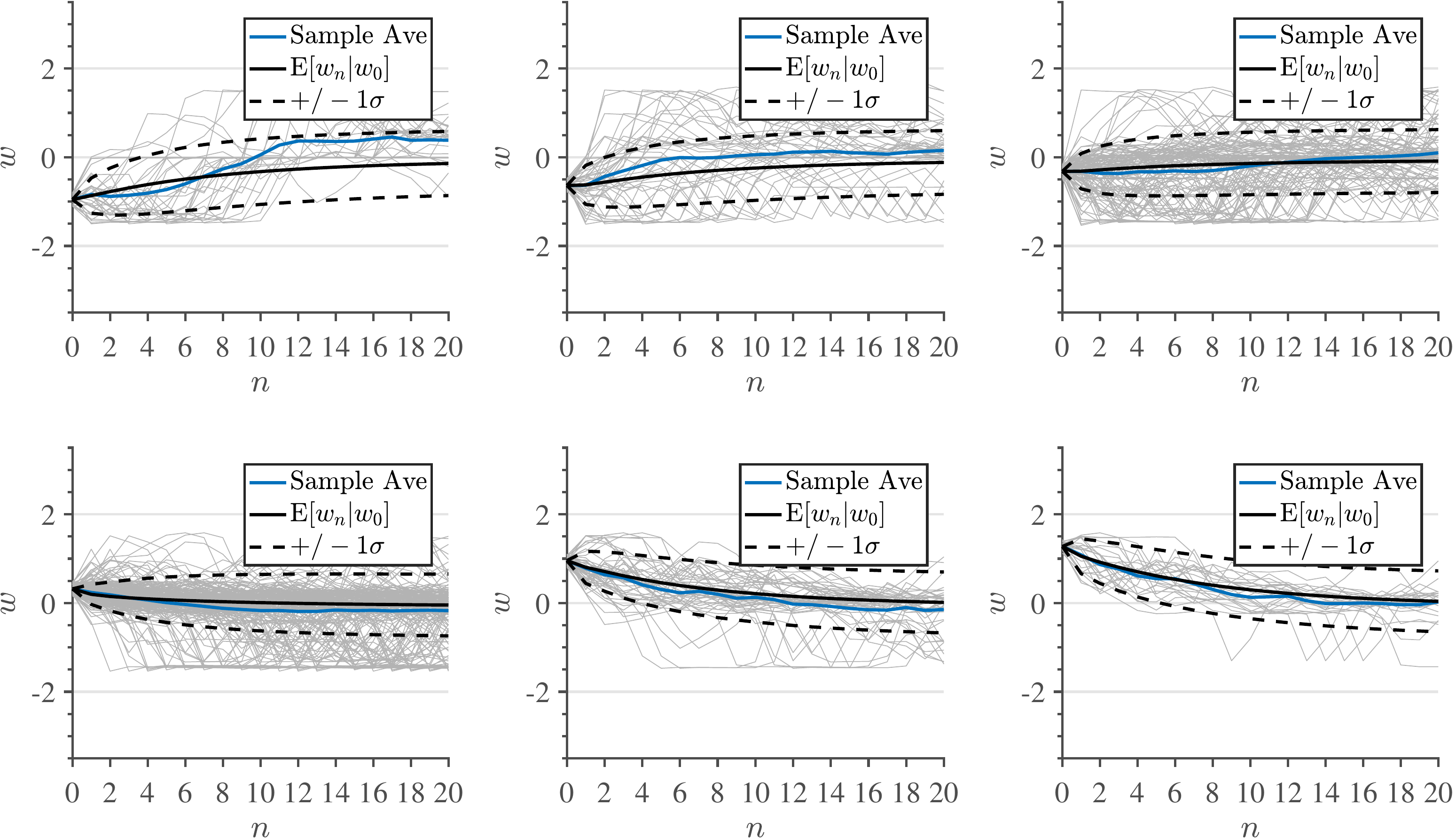}
	\end{minipage} 	
	\caption{Propagation of $\mathbb{E}[w_n|w_0=w^i]$.  Sample paths shown in light grey.  UDDS cycle.}
	\label{fig:UDDS_EwnPropagate}
\end{figure}

\begin{figure}[h!]
	\centering
	\begin{minipage}{1\textwidth}
		\includegraphics [width=1\textwidth]{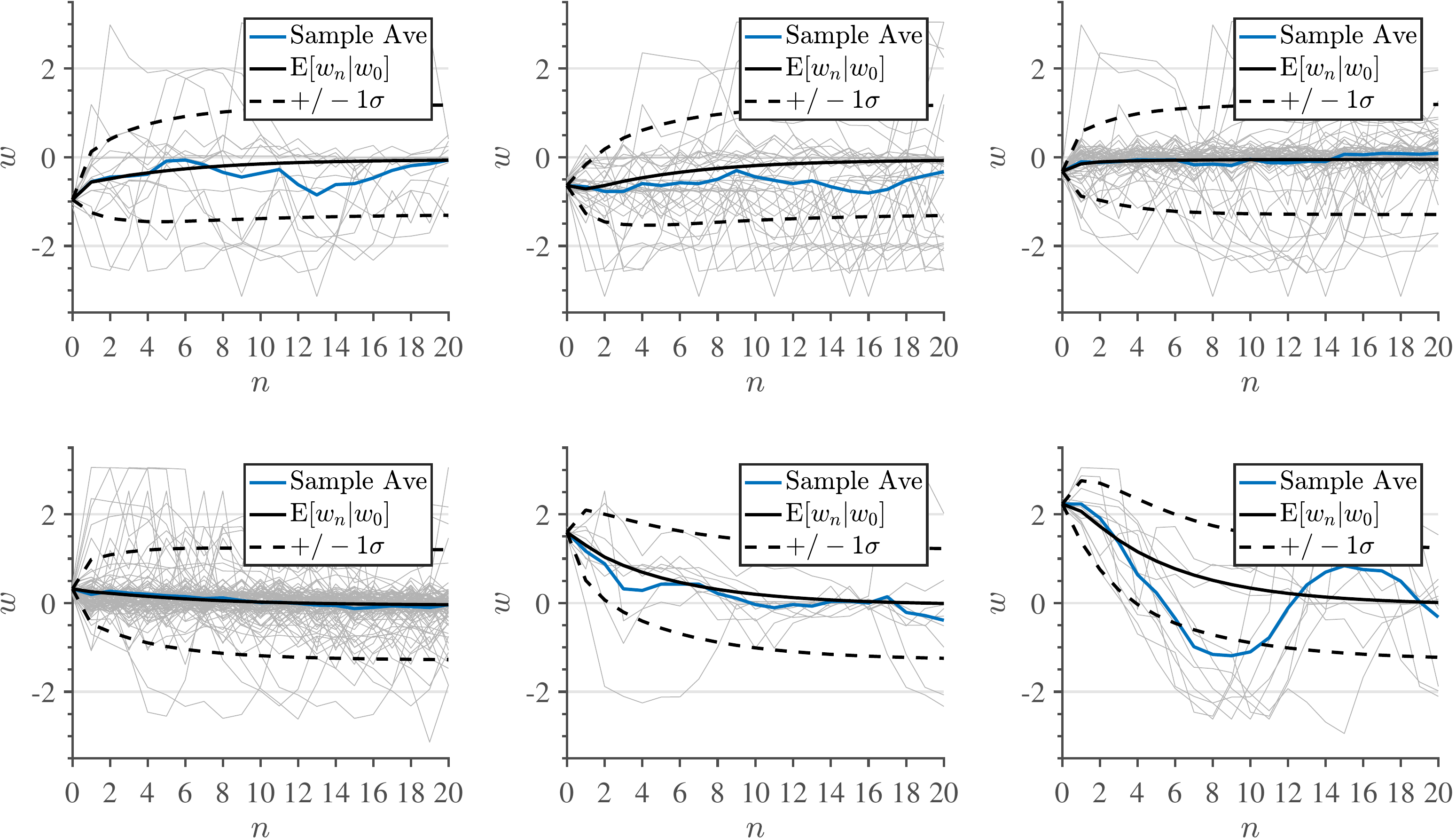}
	\end{minipage} 	
	\caption{Propagation of $\mathbb{E}[w_n|w_0=w^i]$.  Sample paths shown in light grey.  US06 cycle.}
	\label{fig:US06_EwnPropagate}
\end{figure}

\begin{figure}[h!]
	\centering
	\begin{minipage}{1\textwidth}
		\includegraphics [width=1\textwidth]{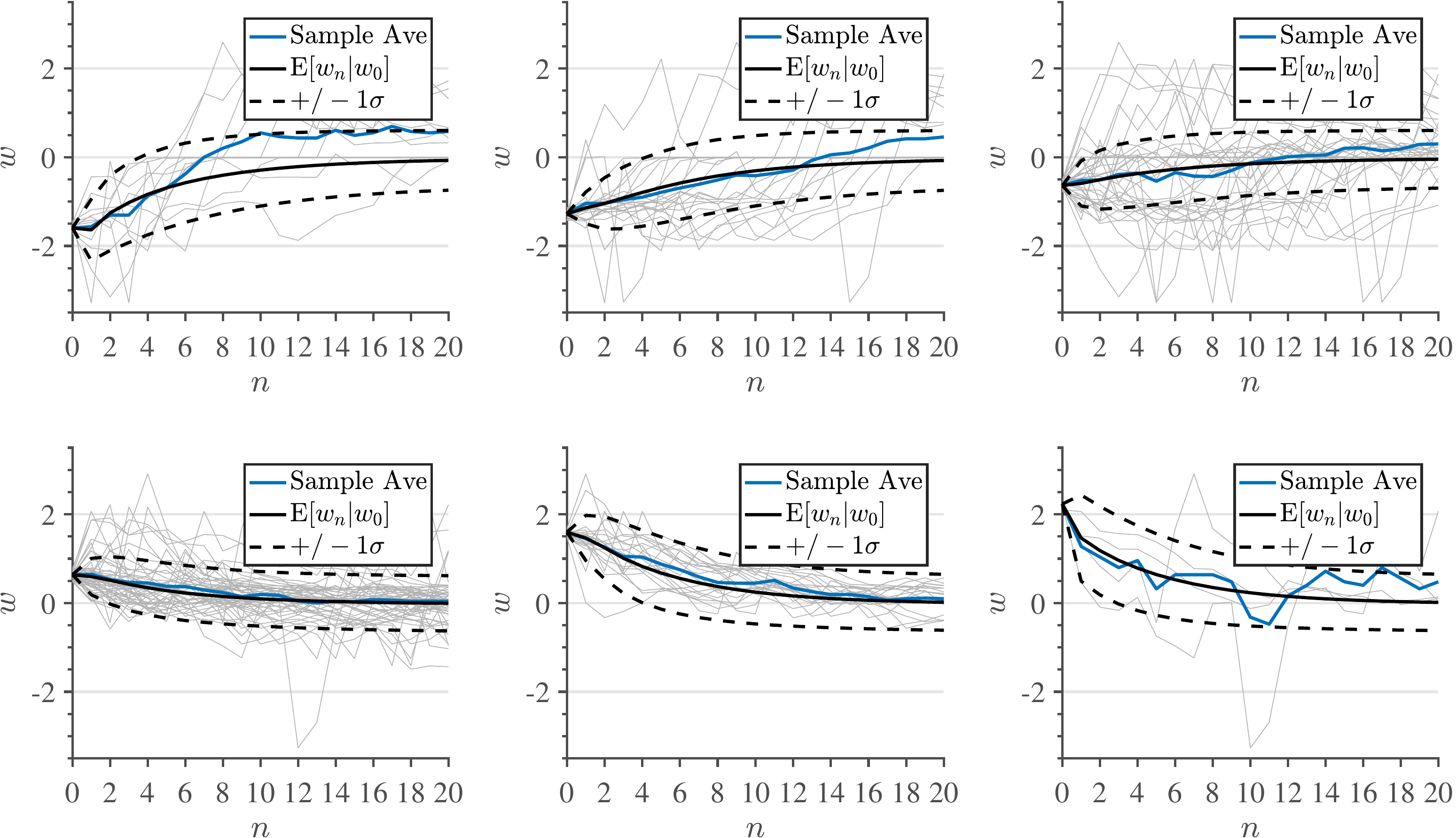}
	\end{minipage} 	
	\caption{Propagation of $\mathbb{E}[w_n|w_0=w^i]$.  Sample paths shown in light grey.  GPS cycle.}
	\label{fig:GPS_EwnPropagate}
\end{figure}

\chapter{VALUE FUNCTION DERIVATION FOR ASDDP}\label{section:VFASDDP}
Define 
\[\bar {\mathbf X}_n(\x,\bu) = \{\x^j | \x^j = F_n(\x,\bu,w^j),~w^j\in W\} \subset \mathbf X\] 
as the set of all states reachable from $\x$ under control input $\bu$ at time $n$.  The finite horizon value function is given by \footnote{Conditional expectation $\mathbb{E}[X]=\sum_y\mathrm{Pr}[Y=y]\mathbb{E}[X|Y=y]$ is used in the second to last equality}

\begin{align}
V_n(\x_n) &=   \min_{\bu_n,...,\bu_{N-1}} \mathbb{E}\Big[ h(\x_N)+\sum_{k=n}^{N-1} g_k(\x_k,\bu_k,w_k) \Big| \x_n, w_0=w^i\Big] \nonumber \\
&=  \min_{\bu_n,...,\bu_{N-1}} \mathbb{E}\left[g_n(\x_n,\bu_n,w_n)+ h(\x_N)+\sum_{k=n+1}^{N-1} g_k(\x_k,\bu_k,w_k) \Big| \x_n, w_0=w^i\right] \nonumber \\
&= \min_{\bu_n} \Bigg\{\mathbb{E}\Big[g_n(\x_n,\bu_n,w_n) \big| \x_n,w_0=w^i \Big]+ \nonumber\\ &~~~~~~~~~~~~\min_{\bu_{n+1},...,\bu_{N-1}}\mathbb{E} \left[ h(\x_N)+\sum_{k=n+1}^{N-1} g_k(\x_k,\bu_k,w_k) \Big| \x_n, w_0=w^i \right]\Bigg\} \nonumber \\
&= \min_{\bu_n} \Bigg\{\mathbb{E}\Big[g_n(\x_n,\bu_n,w_n) \big| \x_n,w_0=w^i \Big]+\nonumber \\
&~~~~~~~~~~~~\sum_{\x^j\in \bar {\mathbf X}_n(\x_n,\bu_n)} \mathrm{Pr}\Big[\x_{n+1}=\x^j \big| \x_n, \bu_n, w_0=w^i\Big] \times \nonumber\\ 
&~~~~~~~~~~~~~~~~~\underbrace{\min_{\bu_{n+1},...,\bu_{N-1}}  \mathbb{E} \left[ h(\x_N)+\sum_{k=n+1}^{N-1} g_k(\x_k,\bu_k,w_k) \Big| \x_{n+1},w_0=w^i \right]}_{V_{n+1}(\x_{n+1})}\Bigg\} \nonumber \\
&= \min_{\bu_n} \sum_j P_{ij}^{(n)} \Big[g_n(\x_n,\bu_n,w^j) + V_{n+1}\big(F_n(\x_n,\bu_n,w^j)\big)\Big]
\end{align}
with boundary condition $V_N(\x) = h(\x)$.  The last equality used the following
\begin{align*}
\mathrm{Pr}\big[\x_{n+1}=\x^j \big| \x_n,\bu_n,w_0=w^i\big] &= \mathrm{Pr}\big[w_n=w^j \big| w_0=w^i\big] \\
&= P_{ij}^{(n)}
\end{align*}
where $\x^j \triangleq F_n(\x_n,\bu_n,w^j) \in \bar {\mathbf X}_n(\x_n,\bu_n)$.  Equation (\ref{eq:Vn}) is equivalent to
\begin{align*}
V_n(\x_n) &=  \min_{\bu_n}\mathbb{E}\Big[ g_n(\x_n,\bu_n ,w_n) + V_{n+1}\big( F_n(\x_n,\bu_n,w_n)\big)\big|\x_n,w_0=w^i\Big]
\end{align*}

\begin{vita}
    Kyle Williams received the degree Bachelor of Science in Mechanical Engineering from Purdue University in May 2005.  He was a summer intern at Caterpillar, Inc. during the summers of 2004 and 2005.  He received his Master's Degree in Mechanical Engineering from Purdue in August 2007.  During the following years he worked for Parker Hannifin and Caterpillar, Inc.  He is currently a control engineer with Caterpillar's Large Power Systems Division.  He received his PhD from Purdue in May 2018.  His interests are predictive control for stochastic systems, with an emphasis on energy and vehicle systems. 
\end{vita}

\makeatletter
\newenvironment{publications}%
{%
	\@@nonchapter{next}{PUBLICATIONS}{y}{0pt}%
}
{\par}
\makeatother

\begin{publications}
	
	\hangindent=3em
	\noindent K. Williams and M. Ivantysynova.  ``Approximate Stochastic Differential Dynamic Programming for Hybrid Vehicle Energy Management", {\sl IEEE Transactions on Control Systems Technology} (submitted)\\
	
	\hangindent=3em
	\noindent K. Williams, R. Kumar and M. Ivantysynova. ``Robust control for a dual stage power split transmission with energy recovery," in {\sl Proceedings of the 6th International Fluid Power Conference}, Dresden, Germany, Vol. 1, pp.127-144, April 2008\\
	
	\hangindent=3em
	\noindent K. Williams and M. Ivantysynova ``Towards an optimal energy management strategy for hybrid hydraulic powertrains based on dual stage power split principle," in {\sl Proceedings of the 5th FPNI PhD Symposium}, Krakow, Poland, pp 27 - 40, July 2008\\
	
	\hangindent=3em
	\noindent R. Kumar, K. Williams and M. Ivantysynova. ``Study of energetic characteristics in power split drives for on-highway trucks and wheel loaders," in {\sl Proceedings of the SAE International Commercial Vehicle Engineering Congress}, Chicago, Illinois, 2007\\
	
	\hangindent=3em
	\noindent K. Williams, ``Energy recovery for hydraulic hybrid power split drives,'' Master's thesis, Purdue University, 2007\\
	
	\hangindent=3em
	\noindent B. Carl, M. Ivantysynova and K. Williams, ``Comparison of Operational Characteristics in Power Split Continuously Variable Transmissions", SAE Technical Paper 2006-01-3468, 2006
	
\end{publications}

\end{document}